\documentclass[11pt,a4paper]{article}
\pdfoutput=1

\usepackage{./jheppub}
\usepackage[T1]{fontenc} 
\usepackage{graphicx}
\usepackage{amsfonts,amsmath,amssymb}
\usepackage{verbatim}
\usepackage{array}
\usepackage{mathrsfs}
\usepackage{mathtools}
\usepackage{yfonts}
\usepackage{dsfont}
\usepackage{bbm}
\usepackage{colonequals}
\usepackage{amscd}
\usepackage{slashed}
\usepackage{extarrows}
\usepackage{float}
\usepackage{afterpage}
\usepackage{subcaption}
\usepackage{wrapfig}
\usepackage{array}

\usepackage{suffix,mathtools,cancel,bbm}
\usepackage{caption}
\usepackage{multirow}
\usepackage{enumerate}

	\usepackage{tikz-cd}
    \usetikzlibrary{decorations.pathmorphing}
\usepackage{scalerel}
\usepackage{stackengine}

\usepackage{colortbl}
\definecolor{mygray}{gray}{0.93}

\usepackage{sectsty}
\allsectionsfont{\boldmath}

\usepackage{comment}

\usepackage{sectsty}
\allsectionsfont{\boldmath}

\usepackage{adjustbox}

\newcommand{\be}{\begin{equation}}
\newcommand{\ee}{\end{equation}}

\newcommand{\ba}{\begin{aligned}}
\newcommand{\ea}{\end{aligned}}

\newcommand{\cA}{\mathcal{A}}
\newcommand{\cB}{\mathcal{B}}
\newcommand{\cC}{\mathcal{C}}

\newcommand{\cF}{\mathcal{F}}
\newcommand{\cG}{\mathcal{G}}
\newcommand{\cH}{\mathcal{H}}
\newcommand{\cI}{\mathcal{I}}

\newcommand{\cK}{\mathcal{K}}

\newcommand{\cN}{\mathcal{N}}

\newcommand{\cP}{\mathcal{P}}

\newcommand{\cU}{\mathcal{U}}
\newcommand{\cV}{\mathcal{V}}
\newcommand{\cW}{\mathcal{W}}
\newcommand{\cX}{\mathcal{X}}

\newcommand{\nn}{\nonumber}

\allowdisplaybreaks

\title{Continuous symmetries and charge measurement of boundary operators in holography}

\author[a]{Ibrahima Bah,}
\author[b]{Federico Bonetti,}
\author[a]{Mufaro Chitoto,}
\author[c]{and Enoch Leung}
\affiliation[a]{William H.~Miller III Department of Physics and Astronomy, Johns Hopkins University,\\ 3400 North Charles Street, Baltimore, MD 21218, U.S.A.}
\affiliation[b]{Departamento de Electromagnetismo y Electr\'onica, Universidad de Murcia,\\ Campus de Espinardo, 30100 Murcia, Spain}
\affiliation[c]{Max Planck Institute for Mathematics in the Sciences,\\ Inselstraße 22, 04103 Leipzig, Germany}

\emailAdd{iboubah@jhu.edu, f.bonetti@um.es, mchitot1@jhu.edu, enoch.leung@mis.mpg.de}

\abstract{We study holographic charge measurement for continuous internal symmetries. Charged boundary operators are characterized by Wilson lines of bulk gauge fields ending on the boundary, while charge measurement is performed using U-shaped defects hanging from the boundary. We derive universal features of this process from a low-energy point of view, and show how the hanging defect picture mimics the thickening regularization of continuous symmetry operators in field theory. Furthermore, we provide explicit top-down realizations in AdS/CFT setups in Type IIB string theory and M-theory, featuring Abelian as well as non-Abelian symmetries. In the case of Type IIB constructions, we analyze the brane dynamics underlying the charge measurement process. Along the way, we also characterize how hanging brane configurations can be regarded as being topological, and demonstrate how tachyon dynamics account for their fusion rules.}


\begin{document}


\maketitle
\flushbottom

\addtocontents{toc}{\protect\setcounter{tocdepth}{2}}


\newpage 

\section{Introduction}\label{sec:introduction}

Symmetries are essential tools that have been very powerful in the study of quantum field theory (QFT) for decades. They provide selection rules and allow for the organization of the spectrum of a theory which leads to a greater understanding of the underlying physics. 
In recent years there have been novel advances in the study of symmetries in QFTs with the advent of generalized symmetries \cite{Gaiotto:2014kfa}. There it was observed that the rich symmetry structure in QFTs can be formulated in terms of topological objects that braid, link and fuse in interesting ways, going beyond the usual group-like symmetry notion, into category theory \cite{Bhardwaj:2017xup,Thorngren:2019iar,Thorngren:2021yso,Bhardwaj:2022yxj,Muller:2025ext,Bah:2025oxi,Kim:2025zdy}.  Along those lines, the AdS/CFT correspondence and holography \cite{Maldacena:1997re,Witten:1998qj} have proven to be important tools in understanding these topological operators from their dual descriptions in string theory. It is known that for an $AdS_{d+1} \times X^{9-d}$ supergravity solution in string theory that is holographically dual to a $d$-dimensional QFT, a {\it global} symmetry in the latter is dual to a {\it gauge} symmetry in the former. Therefore, a natural question is, then, how one can understand the holographic interpretation of such categorical structures. The first step towards establishing this dictionary is to explicitly construct the holographic duals of the topological operators in terms of stringy degrees of freedom residing in the bulk. To this end, much effort, from various perspectives, has gone into realizing the holographic duals of both invertible and non-invertible finite symmetries using D-branes, and also in geometric engineering contexts \cite{Apruzzi:2022rei,GarciaEtxebarria:2022vzq,Heckman:2022muc,Heckman:2022xgu,Etheredge:2023ler,Franco:2024mxa,Gutperle:2024vyp,Knighton:2024noc,Yu:2023nyn,Fernandez-Melgarejo:2024ffg,Dierigl:2023jdp,Bah:2023ymy,Apruzzi:2023uma,Cvetic:2023pgm,Baume:2023kkf,DelZotto:2024tae,Hu:2024zvz,Argurio:2024oym,Zhang:2024oas,Braeger:2024jcj,Bergman:2024its,Caldararu:2025eoj,Bonetti:2024etn,Christensen:2024fiq}. On the other hand, continuous non-Abelian symmetries have been significantly more challenging \cite{Cvetic:2025kdn,Cvetic:2023plv,Garcia-Valdecasas:2023mis,Najjar:2025htp,Bergman:2024aly,Waddleton:2024iiv,Calvo:2025usj,Calvo:2025kjh}.

The corresponding gauge symmetries in the holographic bulk can have two different types of geometric origins in string theory. One of them emerges from the cohomology of the internal space $X^{9-d}$ when considering zero modes of supergravity gauge fields, while the other emerges from the isometries of $X^{9-d}$ when considering zero modes of its metric \cite{Bah:2019rgq,Bah:2019vmq,Bah:2020jas,Bah:2020uev,Bah:2021brs,Mignosa:2026mgf}. Given the disparate natures of these symmetries, it is natural to expect that they may admit qualitatively different string-theoretic realizations, as was argued recently that continuous (Abelian) symmetries associated with cohomology are generated by non-BPS D-branes or fluxbranes \cite{Cvetic:2023plv,Bergman:2024aly,Najjar:2025htp}, whereas those associated with isometries are generated by Kaluza-Klein (KK) monopoles \cite{Cvetic:2025kdn,Calvo:2025usj}. Additionally, it was argued, very recently, in \cite{Mignosa:2026mgf} that the non-BPS D-branes can be regarded as dielectric expansions of non-BPS KK monopoles. In our previous work \cite{Bah:2025vfu}, that was motivated by recent proposals \cite{Calvo:2025kjh,Bergman:2024aly} regarding the use of U-shaped hanging branes to realize continuous symmetries, we used Type IIB string theory in $AdS_5\times S^5$ to argue that the universal way to fully reproduce the symmetry operators imposed by the Gauss's law constraints of the low-energy action involves constructing suitable bound states of BPS KK monopoles and BPS D-branes. One of the main goals of this paper is to further investigate this universal description in more examples. The specific symmetry being generated will be determined by the internal configuration of the D-brane and KK monopole bound state in question. Such a bound state living in 10d string theory then serves as the UV origin of the symmetry generator in the low-energy effective theory on $AdS_{d+1}$. We will provide supplementary details on how to precisely characterize such a hanging brane configuration, and argue that it is indeed a topological operator.

The primary class of examples that we are going to study in this paper is Type IIB supergravity on $AdS_5 \times S^5$ and $AdS_5 \times T^{1,1}$, both of which we generalize to $AdS_5 \times SE_5$. Here $SE_5$ is a 5d Sasaki-Einstein space, which is a circle fibration over a 4d Kähler-Einstein space. Apart from the generically non-trivial cohomology of $SE_5$, the space is also acted on by a $U(1)$ isometry associated with the fiber, so we obtain a rich family of (continuous) Abelian symmetries of both types mentioned earlier. We will explicitly construct brane configurations responsible for generating these symmetries. In particular, for the symmetry arising from the $U(1)$ isometry, we will highlight the importance and subtlety in constructing the operators in a $U(1)$-equivariant manner, distinguishing them from those which are purely cohomological in nature.

In some cases, the Sasaki-Einstein space may also admit non-Abelian isometries.
For instance, $T^{1,1}$ and $S^5$ respectively admit $U(1) \times SU(2) \times SU(2)$ and $SO(6)$ isometry groups (up to global forms). 
We will demonstrate that our prescription generalizes naturally to accommodate such cases. We will also study the dynamics of these hanging branes, which will show that they are indeed topological and stuck to the conformal boundary. Dirichlet boundary conditions on the relevant gauge fields recover the labeling by group elements, such that the operators generate the full non-Abelian group in the dual QFT. Consistency of dualities between string theory and M-theory suggests that one may be able to discuss symmetries associated with isometries of the latter from a similar perspective, where the role of D-branes will be replaced by M5-branes. We will explore this postulation by looking at examples of the form $AdS_4 \times SE_7$ given by consistent truncations of 11d supergravity. 

In all the holographic examples that we explore in this paper, we construct the Wilson line and the symmetry operator that measures its charge. An important observation that we make in Section \ref{sec_Wilson_spheres} is that, for the case of $AdS_{d+1}\times X$ and $S^n\subset X$, we can construct, from branes, a Wilson line charged under the $SO(n+1)$ isometry of the internal $S^n$ sphere factor. This Wilson line has one end on the conformal boundary and extends into the bulk, therefore, the brane from which it is derived needs to also have one leg in the $AdS_{d+1}$ bulk. The action of such a Wilson line is given in \eqref{eqWilson}. To get the symmetry operator that measures the charge of the Wilson line we employ a hanging brane (see Figure \ref{fig_general_HW}). We demonstrate charge measurement of the Wilson line by the symmetry operator via Hanany-Witten transitions of the branes from where these operators originate from respectively. These restrictions then limit what pair of branes we can feasibly use to get the operators. In Section \ref{sec:topdown} we find that to fully reproduce the symmetry structure for Type IIB string theory on $AdS_5\times SE_5$, we need a D3-brane as the Wilson line and a D5-KK bound state as the symmetry operator. Additionally, for M-theory on $AdS_4\times SE_7$, we need an M5-brane as the Wilson line and an M5-KK bound state as the symmetry operator. The details of each construction vary from setup to setup, however, the resulting operators all have a similar form. This is expected from a bottom-up perspective, where the operators are constructed via Gauss's law constraints emanating from the low-energy effective action. We extensively discuss this low-energy point of view in Section \ref{Sec:lowenergy}.

The findings above are a result of studying the topological sector of the worldvolume actions of the branes. Therefore, a natural question one might ask is whether, or not, the dynamical sector of the branes can spoil the results stated above. We find, in Section \ref{Sec:Dynamics}, via an analysis of the dynamical action of the branes, that the symmetry operator is stuck very close to the boundary and is effectively topological. Additionally, the symmetry operators undergo fusion, as expected, when they are pushed to be coincident. This is a result of tachyonic modes that appear when two operators approach each other.
We analyze the role of tachyon condensation in the  process that starts with the two hanging branes apart and, going through partial brane-antibrane annihilation and brane recombination, ends with them sitting on top of each other, see Figures \ref{fusionneww} and \ref{reco2}.

The rest of this paper is organized as follows. In Section \ref{Sec:lowenergy} we 
study aspects of the holographic duals
of global symmetries in field theory by analyzing the low-energy effective action for 
 gauge fields in AdS.
 We discuss bulk Wilson lines, as well as symmetry operators realized, starting from Gauss's law constraints. We make contact with the regularization prescription of \cite{Bah:2024ucp} in field theory, and argue that it is mapped to a universal hanging brane configuration on the AdS side. 
We also show how such a configuration reproduces the expected group-like fusion rules of symmetry operators.
In Section \ref{sec:topdown} 
we illustrate the universal features described in Section \ref{Sec:lowenergy} by means of concrete 
top-down examples of holography. We first discuss the origin of AdS Wilson line operators 
associated to sphere factors in the internal space. We continue with a detailed account of some notable examples of $AdS_5 \times SE_5$ solutions in Type IIB string theory and $AdS_4 \times SE_7$ solutions in M-theory.
Section \ref{Sec:Dynamics} is devoted to a  discussion of the dynamical aspects of the hanging brane configurations that realize symmetry operators. We study brane  fluctuations and   argue for the topological nature of the corresponding operators.
Moreover, we analyze tachyon dynamics to elucidate 
the fusion mechanism for hanging branes.
We conclude in Section \ref{sec_conclusion} with an outlook of possible directions for future research.
The appendices contain further details on some computations performed in the main text, as well as a brief review of equivariant cohomology and its applications to the setups studied in this paper.

\section{Low-energy effective supergravity description}\label{Sec:lowenergy}

This section is devoted to an account of the general features of the holographic setups studied in this work. We focus on the universal aspects of continuous gauge symmetries purely from a low-energy supergravity perspective. In later sections we shall demonstrate how these features originate from concrete top-down constructions in Type IIB string theory and M-theory.

\subsection{Gauge fields in $AdS_{d+1}$}
\label{sec_gauge_field_AdS_general}

Let us consider the low-energy dynamics of light modes in an $AdS_{d+1} \times X^n$ solution in Type IIB string theory or M-theory, where $n=9-d$ or $10-d$ respectively. More precisely, we are interested in analyzing the effective action for the gauge fields in $AdS_{d+1}$. Through the holographic dictionary \cite{Maldacena:1997re}, this is motivated by the fact that we would like to study continuous 0-form global symmetries in field theory.

In terms of the top-down string/M-theory construction, the gauge fields in $AdS_{d+1}$ 
can have different geometric origins. For concreteness, in this work we study the following two possibilities.
\begin{itemize}

    \item Isometries of the internal space. Such gauge fields originate from the off-diagonal components of fluctuations of the higher-dimensional metric along the direction of Killing vector fields on $X^n$. We denote this class of gauge fields as 
    \be \label{eq_gauge_field_isom}
        A^i   \ , \qquad 
        i = 1,\dots, \dim G_{\rm iso} \ , 
    \ee 
    where $G_{\rm iso}$ is the isometry group of $X^n$. We assume $G_{\rm iso}$ to be a compact connected Lie group,\footnote{This condition is sufficient (but not necessary) for the exponential map $\exp: \mathfrak{g} \to G$ to be surjective, which will simplify our subsequent analysis.} and may contain Abelian as well as non-Abelian factors.
    \item Homological cycles in the internal space. If $X^n$ has non-trivial $(p-1)$-cycles in homology, we can get gauge fields in $AdS_{d+1}$ by integrating a higher-dimensional $p$-form field over a $(p-1)$-cycle.\footnote{Cycles of different dimensions, if present, can yield other $q$-form fields in $AdS_{d+1}$. These would correspond to higher-form (and ($-1$)-form) symmetries (see, e.g.~\cite{Bah:2020uev,Bah:2021brs}). In this work, however, we focus on 0-form symmetries only.} For example, if we consider an $AdS_5$ solution in Type IIB string theory, the $C_4$ RR potential  can be integrated over a 3-cycle in the internal 5-manifold to give a gauge field in $AdS_5$. For the purpose of this work, we neglect torsional cycles. We denote this class of gauge fields collectively as
    \be \label{eq_gauge_field_betti}
        A^\alpha   \ , \qquad 
        \alpha = 1,\dots, n_{\rm Betti} \ . 
    \ee 
    We remark that all gauge fields in this class are  Abelian. The label ``Betti'' is related to the fact that non-trivial (non-torsional) cycles in the internal space are counted by Betti numbers.

\end{itemize}
It is convenient for us to introduce a collective notation $A^I$ for gauge fields, with
\be 
A^I = (A^i , A^\alpha) \ . 
\ee
All in all, the gauge fields $A^I$ are associated to the total gauge group $G$ in the low-energy effective gauge theory on $AdS_{d+1}$. The index $I$ runs over the generators of the Lie algebra $\mathfrak g$ of the compact connected Lie group $G$, not necessarily simple or semisimple. 

In some expressions below we also make use of the index-free notation $A$, as a locally defined $\mathfrak g$-valued 1-form. We can write $A$ in terms of the generators $t^I$ as
\be
    A = A^I t_I \ , 
\ee 
where $I=1, \dots, \dim \mathfrak g$. In our conventions, the field strength of the gauge field $A^I$ is given by
\be 
    F^I = dA^I - \tfrac 12 f_{JK}{}^I A^J \wedge A^K \ ,\label{eq_field_strengths}
\ee
where $f_{JK}{}^I$ are the structure constants of $\mathfrak g$.

The low-energy effective action in $AdS_{d+1}$ contains couplings of the form,
\be \label{eq:gauge_field_action_AdS}
    S_{\rm eff} \supset \int_{AdS_{d+1}} \bigg[\!- \frac 12 \, \tau_{IJ} F^I \wedge * F^J + {\rm CS}_{d+1}(A,dA)\bigg] \ . 
\ee 
The quantity $\tau_{IJ} = \tau_{JI}$ is a positive-definite gauge coupling constant.\footnote{More precisely, 
suppose $\mathfrak g$ is of the form $\bigoplus_a \mathfrak g_a \oplus \mathfrak u(1)^{\oplus k}$, where $\mathfrak g_a$ are simple Lie algebras. Within each $\mathfrak g_a$ summand, then, 
$\tau_{IJ}$ is equal to a constant times the Cartan-Killing metric, while within the $u(1)^{\oplus k}$ summand, $\tau_{IJ}$ is a positive-definite symmetric $k \times k$ matrix.}
When $d$ is even, the effective action may also contain a Chern-Simons term ${\rm CS}_{d+1}(A,dA)$. As we shall see, however, we do not need a detailed knowledge of the Chern-Simons terms to carry out our analysis.
 
Some remarks on the action \eqref{eq:gauge_field_action_AdS} are in order. Since our focus is on gauge fields, we have formulated the action on a fixed $AdS_{d+1}$ background. Of course, the full low-energy action contains a dynamical metric in $d+1$ dimensions, but we do not need to keep track of
it in what follows. In a similar way, the low-energy effective action may contain additional fields, such as scalar fields, which we do not include in \eqref{eq:gauge_field_action_AdS}, as they are not necessary for our purposes. Let us stress that keeping only the gauge fields in $AdS_{d+1}$ does not constitute a consistent truncation. This does not invalidate the discussion in this section, which hinges only on the assumption of having a weakly coupled supergravity description in $AdS_{d+1}$.

\subsection{Wilson line operators}\label{Wilson_line_subsection}

The theory \eqref{eq:gauge_field_action_AdS} admits Wilson line operators transforming under the gauge symmetry $G$. As usual, they are constructed from the parallel transport operator of the $G$-connection $A$ (or a holonomy operator if $M^1$ is a closed loop). Explicitly, we consider Wilson lines labeled by a unitary irreducible representation $\mathbf R$ of $G$, which can be written as the trace (in $\mathbf R$) of the holonomy operator,
\be 
    W(\mathbf R; M^1) = {\rm Tr}_{\mathbf R} {\rm P exp} \int_{M^1}  A\ .
\ee 

To set the stage for our subsequent analysis, let us first take a moment to review an alternative construction of the Wilson loop. This approach is based on co-adjoint orbits and their quantization \cite{Balachandran:1977ub,Alekseev:1988vx,Diakonov:1989fc,Stone:1988fu,Alvarez:1989zv}, which presents the Wilson loop operator as a quantum mechanical model supported on $M^1$ and coupled to the bulk gauge field $A$. For reviews, see e.g.~\cite[Part 4, \S 7.7]{deligne1999quantum}, \cite{Beasley:2009mb,Tong:2014yla}. The 1d action reads
\be \label{eq:Wilson_1d_action}
    S_{\rm WL} =   \int_{M^1} \langle q , \chi^{-1} d_A \chi \rangle \qquad \mbox{with} \qquad \chi^{-1} d_A \chi = \chi^{-1} d\chi + \chi^{-1}A \chi \ ,
\ee 
where $\chi$ is a dynamical $G$-valued scalar field on $M^1$, and the 1-form $\chi^{-1} d_A \chi$ takes values in $\mathfrak g$.\footnote{By construction, this 1-form is invariant under a bulk gauge transformation $A \to g(A+d)g^{-1}$ for any $G$-valued 0-form $g$ in $AdS_{d+1}$, provided that the localized scalar field transforms by a left action $\chi \to g\chi$.}

The quantity $q$ in \eqref{eq:Wilson_1d_action} is a constant parameter valued in the dual $\mathfrak g^*$ of the Lie algebra $\mathfrak g$, and the brackets $\langle \cdot , \cdot \rangle$ denote the natural pairing between $\mathfrak g^*$ and $\mathfrak g$. At the quantum level, $q$ is quantized by imposing that the Wilson loop is invariant under gauge transformations of the scalar $\chi$. More explicitly, we demand $q$ to be (dominant) analytically integral, i.e.~$\langle q,H \rangle \in \mathbb{Z}$ for all $H \in \mathfrak{t}$ such that $e^{2\pi H}=1$, where $\mathfrak{t}$ is the Cartan subalgebra associated to the maximal torus $T$ of $G$. It can be shown that the parameter $q$ is indeed in one-to-one correspondence to the choice of the irreducible representation $\mathbf{R}$ for the Wilson line. We refer the reader to Appendix \ref{Wilson_quantization_appendix} for the technical details leading to this conclusion.

The discussion of the above paragraphs is phrased for a generic compact connected Lie group $G$. If $G$ is Abelian, we have considerable simplifications, because the scalar $\chi$ decouples from the bulk gauge field and can be ignored. For example, if $G=U(1)$ we are left with the usual presentation of an Abelian Wilson loop,
\be 
    q \int_{M^1}  A \ .\label{Abelian_Wilson_line}
\ee 
The condition that $q$ be analytically integral in this setting translates simply into $q\in \mathbb Z$. This is the familiar charge quantization, required to ensure invariance under large gauge transformations.\footnote{We normalize Abelian gauge fields in such a way that their field strengths have periods in $2\pi \mathbb Z$.}

A non-Abelian example that will frequently make an appearance in subsequent sections is $G=SO(n+1)$, in which case we can cast the action \eqref{eq:Wilson_1d_action} in the form,
\be \label{SOnWilsonline}
    S_{\rm WL} =  \frac 12 \int_{M^1} q_{ab} (\chi^{-1} d_A \chi)^{ab} \ ,   
\ee
where 
\be
    (\chi^{-1} d_A \chi)^{ab} = (\chi^{-1})^a{}_c \, d \chi^c{}_d \, \delta^{db} + (\chi^{-1})^a{}_c \, A^c{}_d \, \chi^d{}_e \, \delta^{eb} \ .
\ee
Our notation is as follows. Indices $a$, $b = 1,\dots,n+1$ are vector indices of $SO(n+1)$. They are raised and lowered with $\delta$. The matrix $\chi^a{}_b$ is in $SO(n+1)$, i.e.~$\chi^{-1} = \chi^T$. The Cartan-Killing metric induces an isomorphism $\mathfrak g^* \cong \mathfrak g$, and the Lie algebra $\mathfrak g = \mathfrak{so}(n+1)$ can be identified with the vector space of antisymmetric tensors with two indices. Accordingly, we write the parameter $q$ as a constant tensor satisfying $q_{ab} = -q_{ba}$. Recall that $q$ is taken to live in the Cartan subalgebra, so it is of the form,
\be 
    q_{ab} = \small \begin{pmatrix}
    0 & q_1 \\
    -q_1 & 0 \\
    &&0 & q_2 \\
    && -q_2 & 0 \\
    && &&0 & q_3 \\
    && &&-q_3 & 0 \\
    && && && \cdots
    \end{pmatrix} \ . 
    \normalsize
\ee 
If $n$ is even, we have a 0 in the final diagonal entry. The quantities $q_1,\dots,q_k \in \mathbb{Z}$, where $k = \lfloor \frac{n+1}{2} \rfloor$, encode the dominant analytically integral weight of an irreducible representation of $SO(n+1)$.

The case of $G = SU(n)$ can be similarly treated, where $q_{ab} = \text{diag}(q_1,q_2,\dots,q_n)$, with $q_n = -\sum_{i=1}^{n-1} q_i$, takes the form of a traceless diagonal $n \times n$ matrix. For certain low values of $n$, this allows us to construct Wilson lines labeled by spinor representations of $SO(n+1)$, using the exceptional isomorphisms $SO(3) \cong SU(2)/\mathbb{Z}_2$, $SO(4) \cong (SU(2) \times SU(2))/\mathbb{Z}_2$, and $SO(6) \cong SU(4)/\mathbb{Z}_2$. More systematically, for any $n$ one can consider the double cover $Spin(n+1)$ of $SO(n+1)$. It follows that the analytically integral weights $q_1,\dots,q_k$ of $Spin(n+1)$ are either all integral or all half-integral, as opposed to them always being all integral in the $SO(n+1)$ case. The half-integral weights give rise to spinor representations of $SO(n+1)$ in general.

\subsection{Gauss's law constraints}
\label{sec:Gauss_and_sym_op}

Following \cite{Witten:1998wy,Belov:2004ht}, we may study the gauge theory described by \eqref{eq:gauge_field_action_AdS} in the Hamiltonian formalism, with the radial direction of $AdS_{d+1}$ identified with (Euclidean) time. The action does not contain time derivatives of the temporal component of $A^I$. Varying the action with respect to the latter yields the Gauss's law constraint\footnote{We refer here to the classical Gauss's law constraint. A classical treatment is sufficient to our purposes in this work. At the quantum level, the study of Gauss's law constraint can be more involved, see e.g.~\cite{Diaconescu:2003bm,Belov:2004ht,Moore:2004jv,Belov:2006jd,Freed:2006yc,Christensen:2025ktc}.} $\mathbf G_{I} = 0$ where the $d$-form $\mathbf G_I$ is given by
\be \label{eq_Gauss_in_general}
    \mathbf G_{I} = \tau_{IJ}  D* F^J + \mathbf G_{I}^{\rm CS} \ . 
\ee 
In the above expression $*$ is the Hodge star of $AdS_{d+1}$, but $\mathbf G_{I}$ is understood to be evaluated on a slice of $AdS_{d+1}$ at a constant value of the radial coordinate. The symbol $D$ denotes the covariant derivative,
\be 
D*F^I =  d*F^I - f_{JK}{}^I A^J \wedge *F^K \  .
\ee 
The quantity 
$\mathbf G_{I}^{\rm CS}$ 
denotes the contribution to the Gauss's law constraint from the Chern-Simons term in \eqref{eq:gauge_field_action_AdS}, if present, and 
is a polynomial in $F^I$.

Recall that Gauss's law constraints generate bulk gauge transformations. A bulk gauge transformation that is non-trivial near the conformal boundary of $AdS_{d+1}$ implements a global symmetry of the dual field theory. Thus, we can read off the symmetry generators by taking Gauss's law constraints and evaluating them on a slice at constant $AdS_{d+1}$ radius, taking the limit in which we approach the conformal boundary. In the standard AdS/CFT setup, all gauge fields have Dirichlet boundary conditions $A^I=0$ (for theories that are second order in derivatives). Hence, in the limit, the expression for the Gauss's law constraints \eqref{eq_Gauss_in_general} simplifies to 
\be 
\mathbf G_{I} = \tau_{IJ} d*F^J \ .
\ee 
In particular, the contribution $\mathbf G_I^{\rm CS}$ drops out, because it is a polynomial in $F^I$. Crucially, in this limit $\mathbf G_{I}$ is an exact form, since $\tau_{IJ}*F^J$ is globally well-defined.

We are now in a position to identify the symmetry operators for the boundary 0-form global symmetries dual to the gauge fields $A^I$ in $AdS_{d+1}$. To this end, we consider a linear combination $\theta^{I} \mathbf G_{I}$ with constant  $\theta^{I}$ coefficients, and integrate it on a  $d$-chain $\cB^d$ 
inside a radial slice of $AdS_{d+1}$. The boundary of $\cB^d$ is $\partial \cB^d = M^{d-1}$. Using the fact that $\mathbf G_{I}$ is exact at the conformal boundary, and applying Stokes' theorem, we get the quantity
\be 
    U(\theta;M^{d-1}) = \exp \bigg(  i   \theta^{I} \int_{M^{d-1}} \tau_{IJ} *F^{J} \bigg) \ . \label{eq:symmetry_operator}
\ee 
This is the expected form of the symmetry operators from the low-energy analysis.

In \eqref{eq:symmetry_operator}, the quantity $\tau_{IJ}*F^J$ takes values in the dual $\mathfrak g^*$ of the Lie algebra $\mathfrak g$, and is paired with the parameter $\theta^I$ which takes values in $\mathfrak g$.
The operator $U(\theta;M^{d-1})$ implements the $G$ global symmetry transformation associated to the group element $\exp \theta$, where $\exp: \mathfrak {g} \to G$ is the  exponential map. We recall that this map is surjective if $G$ is connected and compact, which is the case in our setup.

Incidentally, we note that it is possible to construct bulk operators associated to the full Gauss's law constraints, including the Chern-Simons terms. This can be achieved via the na\"ive Page charges,
\be 
    \mathbf P_I = \tau_{IJ} *F^J + \mathbf P_I^{\rm CS} \ . 
\ee 
Here $\mathbf P_I^{\rm CS}$ is a locally defined $(d-1)$-form, given as a polynomial in $A^I$, $dA^I$, such that $d\mathbf P_I^{\rm CS} = \mathbf G_I^{\rm CS}$. The quantity $\exp(i \int_{M^{d-1}} \theta^I \mathbf P_I)$ furnishes the na\"ive operator. For it to be gauge-invariant, however, the na\"ive operator must be dressed with a suitable TQFT. The final result is in general a non-invertible operator (see e.g.~\cite{Damia:2022bcd,Choi:2022fgx,Garcia-Valdecasas:2023mis,Fernandez-Melgarejo:2024ffg}).

\subsection{Charge measurement and regularization}
\label{sec:charge_measurement_and_regularization}

In field theory, one can measure the charge of a local operator with respect to the 0-form global symmetry $G$ by linking it with a codimension-1 symmetry operator \eqref{eq:symmetry_operator}. However, as was pointed out by \cite{Bah:2024ucp} (see also \cite{Calvo:2025kjh}), when $G$ is continuous, the heuristic description above fails to hold strictly, in which case divergences arise and need to be regularized. To motivate our universal holographic construction, it will be instructive for us to recall the argument in \cite{Bah:2024ucp}.

Our goal is to analyze  a generic QFT in $d$ dimensions with continuous internal 0-form symmetry $G$, and in particular study correlation functions with an insertion of a symmetry operator for $G$. For ease of exposition, we consider the case $G=U(1)$ and we take the QFT to be that of a complex scalar field~$\phi$. At the classical level, the  $U(1)$ symmetry acts on $\phi$ as $\phi \to e^{i\alpha} \phi$. The Lagrangian contains a canonical kinetic term, as well as a $U(1)$-invariant scalar potential $V=V(\phi^\dagger \phi)$. The Noether current of the $U(1)$ symmetry reads
\be \label{eq_Noether_phi}
    j_\mu  = i (\phi^\dagger \partial_\mu \phi - \phi \partial_\mu \phi^\dagger) \ .
\ee
The na\"ive symmetry operator is constructed by integrating $*j$ along a codimension-1 submanifold of $d$-dimensional spacetime. For definiteness, let us work in a Euclidean setting in flat $d$-dimensional spacetime, with coordinates $(x^1, \dots, x^{d-1},y)$,
and let us insert the symmetry operator on the locus $y=0$,
\be  \label{eq_naive_with_delta}
    \exp \bigg( i \alpha \int_{y=0} d^{d-1} x \, j_y \bigg) =\exp \bigg( i \alpha \int  d^{d-1}x \, dy \, \delta(y) j_y \bigg)  \ . 
\ee
Here $j_y$ is the $\mu=y$ component of the current $j_\mu$. When we insert \eqref{eq_naive_with_delta} in a correlation function, we can undo the insertion by a formal redefinition in the path integral for $\phi$, $\phi^\dagger$,
\be
    \phi'(x,y) = e^{i \alpha \Theta(y)} \phi(x,y) \ , \qquad \phi'{}^\dagger(x,y) = e^{-i \alpha \Theta(y)} \phi(x,y)^\dagger \ ,
\ee
where $\Theta(y)$ is the Heaviside step function. This redefinition, however, runs into difficulties when one attempts to explicitly evaluate correlators using the Ward-Takahashi identity, as it na\"ively generates $\delta(y)^2$ (and higher order) terms which are ill-defined. The presence of these divergent terms calls for the need to regularize the symmetry operator \eqref{eq_naive_with_delta}. 

\begin{figure}
    \centering
\begin{subfigure}{6.5cm}
    \includegraphics[width=0.9\linewidth]{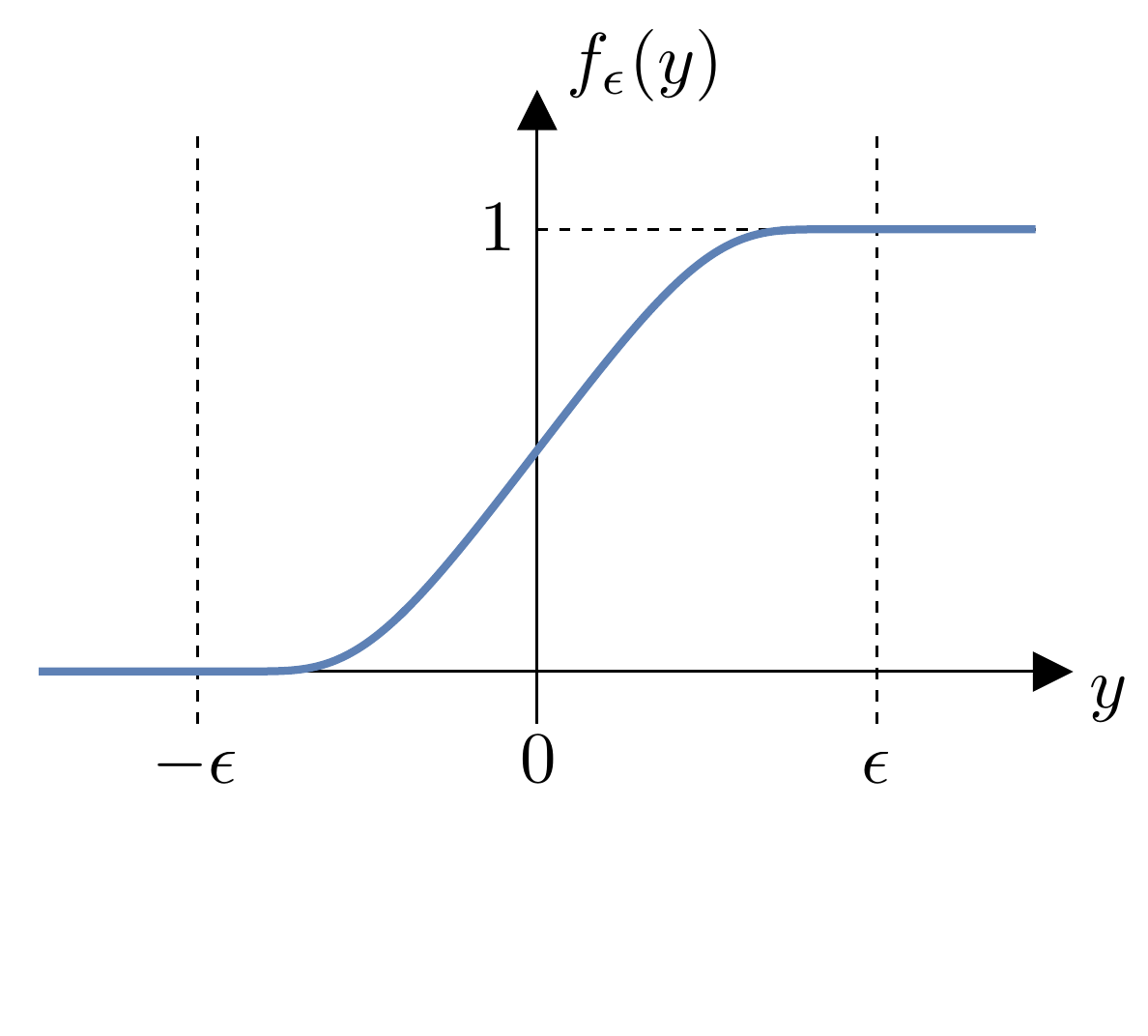}
\end{subfigure}%
\hspace{1cm}
\begin{subfigure}{6cm}
    \centering
    \includegraphics[width=0.9\linewidth]{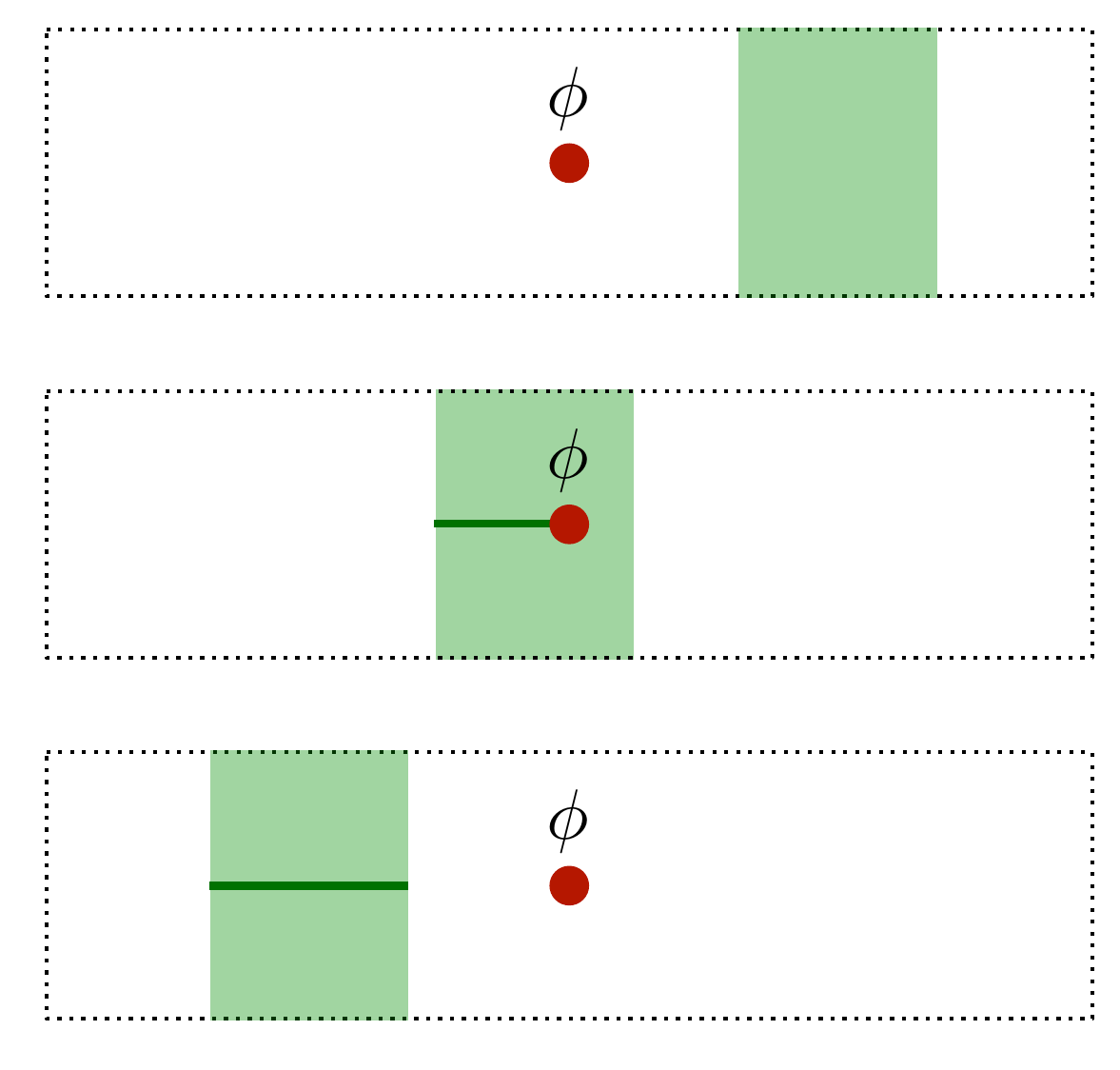}
\end{subfigure}  
    \caption{On the left: plot of the bump function $f_\epsilon(y)$. On the right: a regularized symmetry operator $U_{\rm reg}$ supported on the green strip is gradually moved past a $\phi$ insertion. In the process, a phase is acquired cumulatively by integrating $a_1$ according to equation \eqref{eq_cumulative_phase}. This is depicted pictorially as the green line. Once $U_{\rm reg}$ is moved completely past the $\phi$ insertion, the phase stabilizes.}\label{regularized}
\end{figure}

One natural regularization is to thicken up the symmetry operator as
\be \label{eq_regularized}
    U_{\rm reg} = \exp \int d^{d-1}x \, dy \bigg[i \alpha f_\epsilon'(y) j_y - \Big( \alpha f_\epsilon'(y) \Big)^{\!2} \phi^\dagger \phi\bigg] \ . 
\ee 
The quantity $f_\epsilon(y)$ plays the role of a regulator, and is a 1-parameter family of bump functions of $y$, parametrized by $\epsilon >0$ which characterizes the thickness of the operator. In computing observables or correlators, we work with a small but non-zero $\epsilon$, and take the limit $\epsilon \rightarrow 0$ at the end of the calculation. For any positive $\epsilon$, the function $f_\epsilon(y)$ enjoys the following properties: 
(i) $f_\epsilon (y)$ is defined on $\mathbb R$ and smooth;
(ii) $f_\epsilon(y) =0 $ for all $y \le - \epsilon$;
(iii) $f_\epsilon(y) = 1$ for all $y \ge  \epsilon$;
(iv) $f_\epsilon(y)$ interpolates monotonically between 0 and $1$ for $- \epsilon < y < \epsilon$.\footnote{An explicit example of analytic expression for 
$f_\epsilon(y)$ can be given as follows. Define the function $\varphi : \mathbb R \to \mathbb R$
via $\varphi(t) = e^{-1/t}$ for $t > 0$ and $\varphi(t) = 0$ for $t\le0$, then 
\be 
    f_\epsilon(y) =\frac{\varphi(t)}{\varphi(t) + \varphi(1-t)} \ , \qquad \text{where} \qquad t = \frac{y+\epsilon}{2\epsilon} \ .  \nn 
\ee 
In contrast to the sigmoid function adopted by \cite{Bah:2024ucp}, for our purposes we choose $f_\epsilon(y)$ to have a compact support, such that it has a well-defined width.} See Figure \ref{regularized} on the left.

A key property of the regularized operator \eqref{eq_regularized} is the following. Consider a correlator with an insertion of \eqref{eq_regularized}, accompanied by other insertions of local operators $\phi$ and/or $\phi^\dagger$ at various points, say,
\be \label{eq_my_correlator}
    \langle  U_{\rm reg} \, \phi(x_1,y_1) \dots \phi(x_k,y_k) \, \phi^\dagger (x_{k+1}, y_{k+1}) \dots \phi^\dagger(x_\ell, y_\ell) \rangle \ . 
\ee 
The insertion points $(x_i,y_i)$ may lie at the left of, at the right of, or inside the strip $- \epsilon < y < \epsilon$ where $U_{\rm reg}$ is supported. The operator $U_{\rm reg}$ can be removed from the correlator by means of the following redefinition in the path integral for $\phi$, $\phi^\dagger$,
\be \label{eq_phi_redefinition}
    \phi'(x,y)  =  e^{i \alpha f_\epsilon (y)} \phi(x,y)  \ , \qquad \phi^\dagger{}'(x,y)  =  e^{-i \alpha f_\epsilon (y)} \phi(x,y)^\dagger \ .
\ee
This is related to the fact that $U_{\rm reg}$ is topological even for finite $\epsilon$, as argued in \cite{Bah:2024ucp}. Upon performing the redefinition \eqref{eq_phi_redefinition}, each insertion of $\phi$ and/or $\phi^\dagger$ in the correlator \eqref{eq_my_correlator} acquires a phase, depending on its position $(x_i,y_i)$. If $(x_i,y_i)$ lies to the left of $U_{\rm reg}$ (i.e.~$y_i < - \epsilon$), the phase is trivial. If it lies to the right (i.e.~$y_i > \epsilon$), the phase is $e^{i \alpha}$. If it lies at an intermediate position (i.e.~$- \epsilon < y_i < \epsilon$), the phase is intermediate.

The above construction can be rephrased in terms of a flat connection. More precisely, let us define
\be 
    a_1 = \alpha \, f'_\epsilon(y) dy \ , 
\ee 
which clearly satisfies $da_1=0$. The regularized operator $U_{\rm reg}$ can be expressed in terms of $a_1$.
In particular, the leading term with $j_y$ can be written as
\be \label{eq_regul_leading}
    i \int  a_1 \wedge *j \ .
\ee
The ``seagull'' term with $\phi^\dagger \phi$ in \eqref{eq_regularized} can also be accounted for, by taking the expression for $j$ in \eqref{eq_Noether_phi} and replacing $d\phi$, $d\phi^\dagger$ with $(d+ i a_1)\phi$, $(d- i a_1)\phi^\dagger$. We can regard $a_1$ as a 1-form gauge field on the worldvolume of $U_{\rm reg}$.

It is also useful to note that we can rewrite the path integral redefinition \eqref{eq_phi_redefinition} in terms of $a_1$, namely,
\be \label{eq_cumulative_phase}
    \phi'(x,y) = \phi(x,y) \exp \bigg(i \int^y_{-\epsilon} a_1 \bigg) \ ,
\ee
and similarly for $\phi^\dagger$. The integral inside the exponent is understood to be along the $y$ axis. 
Since $a_1$ is a flat connection, the value of the integral is unaltered if the integration contour is deformed slightly. We also notice that $a_1$ vanishes for $|y| \geq \epsilon$, given the properties of $f_\epsilon(y)$, so the lower limit of integration could be equivalently taken to be $-\infty$.

\subsection{A universal construction of hanging branes}
\label{sec:hanging_branes_intro}

What is the holographic interpretation of the regularization outlined above? Before answering the question, recall that the process of measuring the charge (with respect to a generic group $G$) of local operators in the field theory translates to determining the parameter $q$ (or equivalently, the representation $\mathbf{R}$) of the Wilson line \eqref{eq:Wilson_1d_action}, which ends on the conformal boundary and extends into the $AdS_{d+1}$ bulk. Prior to regularization, the naïve symmetry operator coming from 
Gauss's law constraints is simply a $(d-1)$-dimensional brane localized on the conformal boundary, supported on a codimension-1 manifold $M^{d-1} \subset \partial AdS_{d+1}$. Such an operator acts as a singular insertion along $M^{d-1}$ in the dual field theory, and hence is ill-defined based on the aforementioned argument.

The thickening regularization suggests the existence of a family of $d$-dimensional objects in $AdS_{d+1}$ which reduce to the naïve $(d-1)$-dimensional brane in some ``zero-width'' limit. Most importantly, when we continuously drag the Wilson line across such a finite-width object, we should recover the qualitative behavior as depicted in Figure \ref{regularized}. Inspired by the proposal of \cite{Calvo:2025kjh}, we show that these regularized symmetry operators are realized as U-shaped branes hanging from the conformal boundary of $AdS_{d+1}$. 
In this section we offer a general description of the universal aspects of these hanging brane configurations,
as they can be seen from the low-energy point of view, and demonstrate precisely how they give rise to the desired charge measurement behavior. The specific top-down realization of these features 
is different for different string theory/M-theory solutions.

Let us explain in more detail our universal construction of the hanging brane in $AdS_{d+1}$. We consider a $d$-dimensional brane supported on $\gamma^1 \times M^{d-1}$. Here $M^{d-1} \subset \partial AdS_{d+1}$ as before, while $\gamma^1$ denotes an arc that extends into $AdS_{d+1}$ and connects two points on $\partial AdS_{d+1}$. To make contact with the previous discussion, we can think of these two points as $y = \pm \epsilon$ along the direction orthogonal to $M^{d-1}$. The effective action for this $d$-dimensional object is given by
\be \label{eq_low_energy_U}
    2\pi \int_{\gamma^1 \times M^{d-1}} a_1 \wedge n^{I} \tau_{IJ} *F^J \ ,
\ee
where $\tau_{IJ}$, $*F^I$ are defined as in Section \ref{sec:Gauss_and_sym_op}, and $n^I$ are constant integer parameters. The new ingredient is $a_1$. It is a flat $U(1)$ gauge field (i.e.~$da_1 = 0$) localized on the defect (as opposed to being a bulk field in $AdS_{d+1}$), with a non-trivial parallel transport along $\gamma^1$,
\be \label{eq_a1_integral}
    2\pi\int_{\gamma^1} a_1 = \alpha \ .
\ee
Moreover, to ensure gauge invariance of the resultant operator, $a_1$ satisfies Dirichlet boundary conditions, i.e.~$a_1|_{\partial\gamma^1}=0$, at the endpoints of $\gamma^1$ on the conformal boundary of $AdS_{d+1}$. To be precise, flat connections on the arc $\gamma^1$ are classified by the relative cohomology group $H^1(\gamma^1,\partial\gamma^1;U(1)) = U(1)$, in which relative cocycles vanish at the boundary $\partial\gamma^1$.\footnote{In contrast, the ordinary cohomology group $H^1(\gamma^1;U(1))=0$ is trivial because the arc $\gamma^1$ is contractible.} As a consequence, the parameter $\alpha$ takes values in $U(1)$.

Making use of \eqref{eq_a1_integral}, the hanging brane \eqref{eq_low_energy_U} evaluates to
\be
    \alpha n^I \int_{M^{d-1}} \tau_{IJ} *F^J \ .
\ee
This matches the expected form \eqref{eq:symmetry_operator} from the Gauss's law constraints with the identification 
\be
    \theta^I = \alpha n^I \ . 
\ee
In the top-down derivations from string/M-theory, the $d$-dimensional hanging brane in $AdS_{d+1}$ originates from a higher-dimensional object, wrapping a submanifold in the internal space. The integer parameters $n^I$ encode this wrapping data. (More precise statements can be made as we discuss concrete realizations, see following sections.) Heuristically, we can say that the parallel transport parameter $\alpha$ encodes the ``magnitude'' of the Lie-algebra valued parameter $\theta^I$, while the wrapping data $n^I$ encodes the ``direction'' in which $\theta^I$ points in the Lie algebra. 

\begin{figure}
    \centering
    \includegraphics[width=\linewidth]{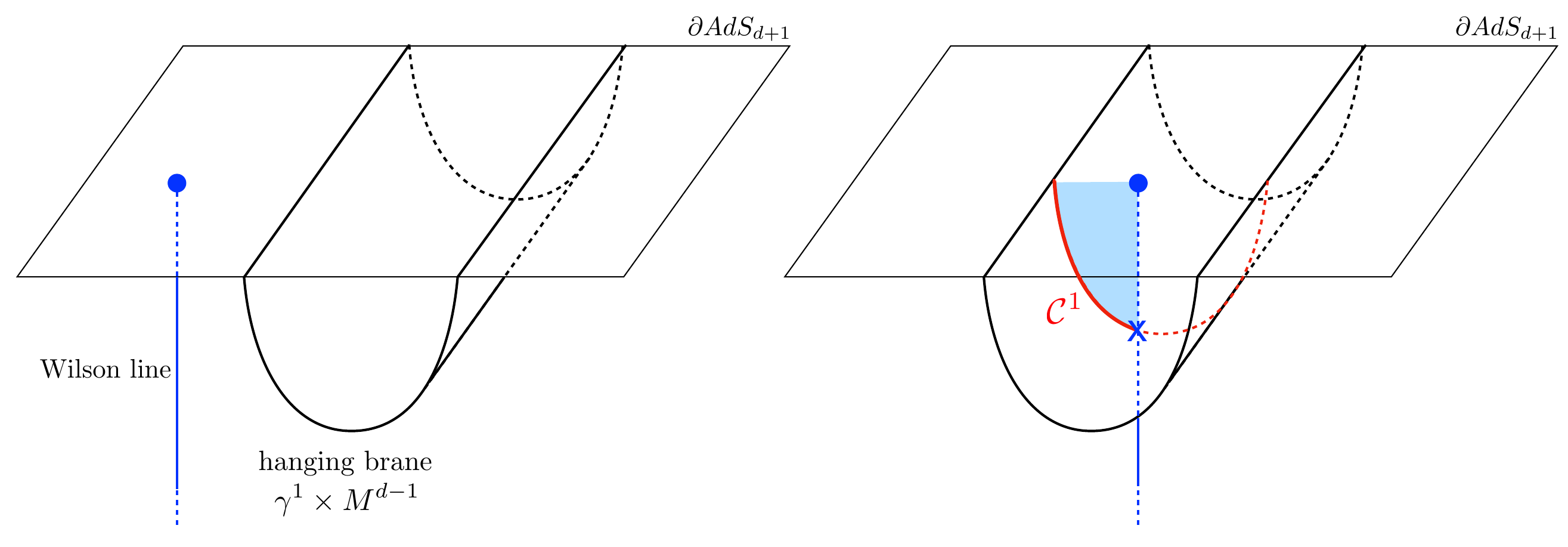}
    \caption{A symmetry operator for a continuous symmetry is  realized as hanging brane with worldvolume $\gamma^1 \times M^{d-1}$, with $M^{d-1}$ a codimension-1 submanifold of $\partial AdS_{d+1}$ and $\gamma^1$ an arc extending into the bulk of $AdS_{d+1}$. If the hanging brane is moved past a Wilson line ending on $\partial AdS_{d+1}$, a Hanany-Witten transition takes place. A new brane is created, depicted as the filled area in light blue. The new brane couples to the gauge field $a_1$ on the hanging brane along the portion $\mathcal C^1$ (depicted in red) of the full arc $\gamma^1$.}
    \label{fig_general_HW}
\end{figure}

As alluded to earlier, the top-down string/M-theory constructions provide a natural mechanism for the interplay between the hanging brane \eqref{eq_low_energy_U} and Wilson lines that end on the conformal boundary of $AdS_{d+1}$, thus reproducing the charge measurement in field theory using the regularized symmetry operator introduced in Section \ref{sec:charge_measurement_and_regularization}. More precisely, we can consider a configuration in which we insert a hanging brane \eqref{eq_low_energy_U} together with a Wilson line extending along the radial direction of $AdS_{d+1}$, and ending on a point on $\partial AdS_{d+1}$
at the left of \eqref{eq_low_energy_U}, see Figure \ref{fig_general_HW}. Starting from this configuration, we imagine moving the hanging brane towards the Wilson line endpoint. As these two objects cross, a Hanany-Witten transition takes place in the top-down construction, see Figure \ref{fig_general_HW}.
The newly created brane is stretched between the hanging U-shaped brane and the Wilson line.

Moreover, the created brane couples to the gauge field $a_1$ on the hanging brane, so it induces a phase factor given by $\int_{\cC^1} a_1$, where the integral is performed over a portion $\cC^1$ of the arc $\gamma^1$ (see Figure \ref{fig_general_HW}). When the hanging brane is moved completely past the Wilson line endpoint, the Hanany-Witten transition is complete. The total phase acquired by means of the $a_1$ coupling 
is $\int_{\gamma^1} a_1=\alpha$. Indeed, this transition corresponds to the process of charge measurement: as it sweeps across the Wilson line, the hanging brane measures the charge of the Wilson line endpoint. We will describe this process in greater detail when we discuss explicit constructions in Type IIB and M-theory in Section \ref{sec:topdown}.

Crucially, through a comparison between \eqref{eq_regul_leading} and \eqref{eq_low_energy_U}, and noting that the global symmetry $G$ in the field theory is realized by the current conservation equation $d \ast j = 0$, whereas on the conformal boundary of $AdS_{d+1}$ it is realized by the Gauss's law constraint $d \ast F = 0$ (applying Dirichlet boundary condition), we observe that the charge measurement processes in the dual pictures are completely analogous. We thus establish the holographic correspondence between the regularized symmetry operator in field theory and the hanging brane in the low-energy effective theory on $AdS_{d+1}$.

Let us also briefly comment on the holographic interpretation of the ``zero-width'' limit of the regulator. As we are going to show, in the top-down string/M-theory constructions, the hanging branes are realized as extended BPS $p$-branes supported on $\gamma^1 \times M^{d-1} \times \Sigma^{p-d+1}$ for some internal submanifold $\Sigma^{p-d+1} \subset X^n$. The U-shaped profile on $\gamma^1$ can be interpreted as a pair of brane and antibrane which recombine in the $AdS_{d+1}$ bulk. One can then understand the ``zero-width'' limit of the hanging brane as the case where these two branes undergo tachyon condensation and decay into a non-BPS $(p-1)$-brane supported on $M^{d-1} \times \Sigma^{p-d+1}$ as constructed by \cite{Bergman:2024aly,Calvo:2025usj}. The dynamical details of the tachyon condensation differ between Type II string theory and M-theory.

\subsection{Hanging brane fusion}\label{top_fusion}

It is instructive to examine how the parallel fusion of the symmetry operator \eqref{eq:symmetry_operator} respects a group-like multiplication rule. For a (compact connected) Lie group $G$ that is generically non-Abelian, the Baker-Campbell-Hausdorff formula asserts that
\begin{align}
    U(\theta;M^{d-1}) \otimes U(\theta';M^{d-1}) & = \exp\bigg((\alpha n + \alpha n')^I X_I + \frac{1}{2} \, \alpha n^I \alpha' (n')^J [X_I,X_J] + \dots\bigg)\nonumber\\
    & = \exp\bigg(\Big(\alpha n + \alpha' n' + \frac{1}{2} \, [\alpha n, \alpha' n'] + \dots\Big)^I X_I\bigg) \ ,\label{eq:fusion_rule}
\end{align}
where we parametrized $\theta^I = \alpha n^I$ as before, and defined the shorthand notation $X_I = i \int_{M^{d-1}} \tau_{IJ} \ast F^J$. We recognize the final expression as the symmetry operator associated with the group element $g'' = g \cdot g' = \exp\theta \cdot \exp\theta'$ under the exponential map. Such an operator is constructed by wrapping a hanging brane on the same worldvolume $\gamma^1 \times M^{d-1}$, yet specified by a different $\mathfrak{g}$-valued parameter $\theta''$.

\begin{figure}[t!]
    \centering
    \includegraphics[width=0.9\linewidth]{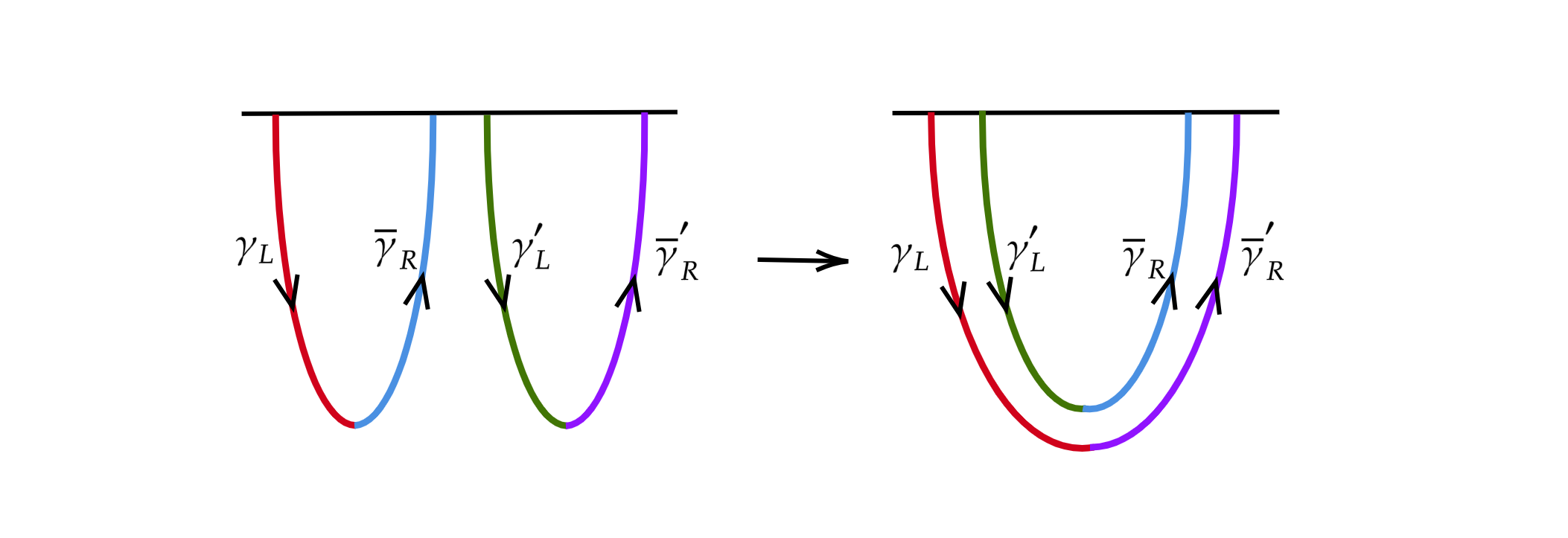}
    \caption{Two hanging branes supported on $\gamma_L \cup \overline{\gamma}_R$ and $\gamma^\prime_L \cup \overline{\gamma}^\prime_R$ respectively split and recombine into two hanging branes on $\gamma_L \cup \overline{\gamma}^\prime_R$ and $\gamma^\prime_L \cup \overline{\gamma}_R$ which are now sitting on top of each other.}
    \label{splitting}
\end{figure}

What is the brane physics underlying the fusion rule \eqref{eq:fusion_rule}? For the sake of illustration, let us first consider the simplest situation where $\theta^I = \alpha n^I$ and $(\theta')^I = \alpha' n^I$, i.e.~they point in the same ``direction'' in the Lie algebra.\footnote{A less restrictive scenario was studied in \cite{Bah:2025vfu}, but the qualitative description therein is essentially the same as that discussed here.} In this case, the fusion rule reduces to
\begin{equation}
    U(\theta;M^{d-1}) \otimes U(\theta';M^{d-1}) = \exp\bigg(i (\alpha + \alpha') n^I \int_{M^{d-1}} \tau_{IJ} \ast F^J\bigg) \, .\label{eq:simplified_fusion_rule}
\end{equation}
As far as the topological couplings of the hanging branes are concerned, we can reproduce the relation above by schematically regarding
\begin{equation}
    U(\theta;M^{d-1}) \sim U_L(\theta_L;M^{d-1}) \circ U_R(\theta_R;M^{d-1})
\end{equation}
as a composite operator with $\theta_L + \theta_R = \theta$, consisting of two branes supported respectively on $\gamma_L \times M^{d-1}$ and $\overline{\gamma}_R \times M^{d-1}$ such that $\gamma_L \cup \overline{\gamma}_R = \gamma$ (hereafter omitting the superscript for ease of notation). See Figure \ref{splitting} for an illustration. For the (flat) gauge field $a_1$ to be well-defined on the arc $\gamma$ supporting the composite hanging brane, we glue the two branes such that their respective gauge fields $a_1|_{\gamma_{L/R}}$ are compatible within the overlap $\gamma_L \cap \overline{\gamma}_R$.\footnote{A precise statement of this gluing condition is provided in Appendix \ref{gluing_condition_appendix}.} Here $\overline{\gamma}_R$ denotes the orientation reversal of $\gamma_R$ to take into account the orientations of $a_1|_{\gamma_{L/R}}$. From a top-down perspective, they correspond to worldvolume gauge fields respectively on a brane and an antibrane. A proper treatment of the string-theoretic origin of the splitting and recombining processes depicted in Figure \ref{splitting} will be presented in Section \ref{dynamic_fusion}.

The resultant configuration is two hanging branes sitting on top of each other, whose center-of-mass mode gives rise to a combined parallel transport,
\begin{equation}
    \alpha_\text{COM} = \int_{\tilde{\gamma}} \tilde{a}_1 + \int_{\tilde{\gamma}'} \tilde{a}_1' = \bigg(\int_{\gamma_L} a_1 - \int_{\gamma_R'} a_1'\bigg) + \bigg(\int_{\gamma_L'} a_1' - \int_{\gamma_R} a_1\bigg) = \alpha + \alpha' \ .\label{eq:com_holonomy}
\end{equation}
Here $\tilde{\gamma} = \gamma_L \cup \overline{\gamma}_R'$ and $\tilde{\gamma}' = \gamma_L' \cup \overline{\gamma}_R$, over which $\tilde{a}_1$ and $\tilde{a}_1'$ are defined. The expression above is equivalent to that from a single hanging brane operator $U(\theta + \theta';M^{d-1})$, thus reproducing the fusion rule \eqref{eq:simplified_fusion_rule}. As a remark, there generally exist relative modes on the coincident hanging branes, which, depending on the details of the underlying brane dynamics, may arise from the enhancement $\mathfrak{u}(1) \oplus \mathfrak{u}(1) \to \mathfrak{u}(2) \cong \mathfrak{u}(1) \oplus \mathfrak{su}(2)$ of the worldvolume gauge algebra. If so, we argue that such non-Abelian effects are negligible as follows. The topological coupling which would contribute to the symmetry operator is first order in the $\mathfrak{su}(2)$-valued gauge field, but its traceless-ness causes the coupling to vanish.

\begin{figure}[t!]
    \centering
    \includegraphics[width=0.9\linewidth]{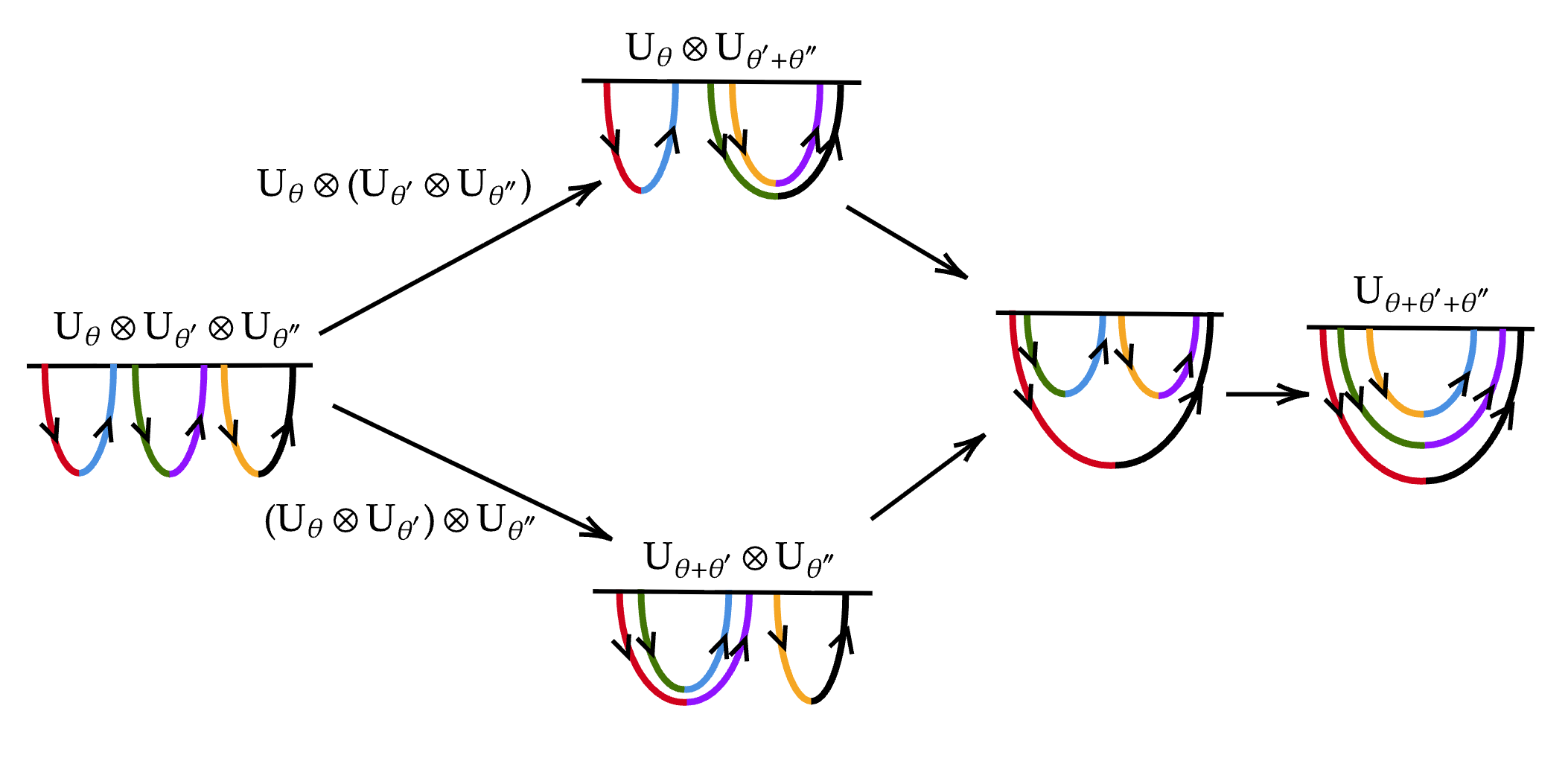}
    \caption{The fusion of the operators $U_{\theta} \equiv U(\theta; M^{d-1})$ is associative, i.e.~$U_\theta\otimes (U_{\theta^\prime}\otimes U_{\theta^{\prime\prime}})= (U_\theta\otimes U_{\theta^\prime})\otimes U_{\theta^{\prime\prime}}=U_{\theta+\theta^\prime+\theta^{\prime\prime}}$. Here we assume that $\theta,\theta',\theta''$ point in the same ``direction'' in the Lie algebra $\mathfrak{g}$. }
    \label{associ}
\end{figure}

It also follows that one can repeat the parallel fusion of hanging branes multiple times, satisfying a group-like multiplication rule. In particular, one can check that the fusion operation is associative, as depicted pictorially in Figure \ref{associ}.

More generally, we can relax the previous assumption that $\theta^I = \alpha n^I$ and $(\theta')^I = \alpha' n^I$. Two ``parallel'' hanging branes, despite sharing the same external support $\gamma^1 \times M^{d-1} \subset AdS_{d+1}$, generically carry labels $n^I$, $(n')^I$ which originate geometrically from internal supports that are not coincident but intersect non-trivially, such that $[n^I,(n')^J] \neq 0$. Hence, the fusion of two such branes manifests as a non-commutative operation in the low-energy effective theory. See \cite{Bah:2025vfu} for a more explicit discussion in the case of the $SO(6)$ symmetry in $AdS_5 \times S^5$.

\section{Top-down constructions in string theory and M-theory}

\label{sec:topdown}

This section starts with a general discussion on the realization of Wilson lines charged under 
gauge fields associated to sphere factors in holographic constructions in string and M-theory. After that, we discuss some examples of 
$AdS_{d+1}$ solutions in Type IIB string theory and M-theory and demonstrate
the string/M-theory realization of the features
described in Section \ref{sec:hanging_branes_intro} from a low-energy perspective. 

\subsection{Wilson lines and sphere factors in holography}
\label{sec_Wilson_spheres}

Let us consider an $AdS_{d+1}$ solution in which the internal space features a sphere factor $S^n$ equipped with its round metric.
In the low-energy $AdS_{d+1}$ effective action we encounter $SO(n+1)$ gauge fields originating from the isometries of $S^n$. In this section we 
focus on the Wilson lines charged under this $SO(n+1)$. For $n\ge 2$, we show that they can be 
realized holographically by branes that wrap a maximal
$S^{n-2} \subset S^n$.
We achieve this by demonstrating how to 
obtain the action \eqref{SOnWilsonline}
for an $SO(n+1)$ Wilson line starting
from the topological couplings on the brane. 
While the specific type of brane
used in the construction depends on the specific solution in string/M-theory,
our derivation relies only on some general ingredients. More precisely, our working assumptions are as follows.
\begin{itemize}
    \item The brane has one external direction in $AdS_{d+1}$.
    \item The brane wraps a maximal $S^{n-2} \subset S^n$.
    \item The $S^n$ factor is supported by a  flux $F_n$, associated to the potential~$C_{n-1}$.
    \item The worldvolume action of the brane includes a topological coupling to $C_{n-1}$ of Wess-Zumino type.
\end{itemize}
These assumptions are met for the $AdS_5 \times S^5$ solution in Type IIB
and for the $AdS_4 \times S^7$ and $AdS_7 \times S^4$ solutions in M-theory.\footnote{For $AdS_4 \times S^7$ the potential $C_{n-1}=C_6$
is the electromagnetic dual to the M-theory 3-form potential~$C_3$.}
We expect our strategy to apply to sphere factors in more general holographic setups.

The sphere $S^n$ can be described in terms of $n+1$ real constrained coordinates $y^a$
satisfying $y^a y_a = 1$.
Here and in what follows $a,b,\dots =1,\dots, n+1$
are vector indices of $SO(n+1)$,
raised and lowered with $\delta_{ab}$, $\delta^{ab}$.
The volume form on $S^n$, normalized to integrate to $1$, can be written as
\be \label{eq_volume_sphere}
V_n = \frac{1}{n! \cV_n} \, \epsilon_{a b_1 \dots b_{n}}
y^a dy^{b_1} \wedge \dots \wedge dy^{b_{n}} \ , \qquad 
\cV_n = \frac{2\pi ^{\frac{n+1}{2}}}{\Gamma(\frac{n+1}{2})} \ .
\ee 
The flux $F_n$ supporting $S^n$ is proportional to $V_n$.
More precisely, due to the presence of $SO(n+1)$ gauge fields (Kaluza-Klein vectors), $F_n$ takes the form
\be \label{eq_Fn_expr}
F_n = N (V_n)^{\rm g}
+ \frac{N}{2(n-1)! \cV_n}
\, \epsilon_{a b_1 \dots b_{n-2} c_1 c_2}
y^a Dy^{b_1} \wedge \dots \wedge Dy^{b_{n-2}} \wedge F^{c_1 c_2}
+ \dots \ ,
\ee 
where $N$ is an integer flux number and $(V_n)^{\rm g}$ is the $n$-form $V_n$ with the replacement $dy^a \to Dy^a$, with
\be 
Dy^a = dy^a + A^{ab} y_b \ . 
\ee 
The 1-forms
$A^{ab} = - A^{ba}$ are the $SO(n+1)$ gauge fields,
and $F^{ab} = dA^{ab} + A^{ac} \wedge A_c{}^b$ is their field strength.
The ellipsis in \eqref{eq_Fn_expr} stand for additional terms that are required for consistency
(for instance, to ensure closure of $F_n$, if $F_n$ satisfies a standard Bianchi identity), but they will not be needed in what follows.\footnote{
The omitted terms, as well as some other 
universal aspects of flux uplift formulae for supersymmetric consistent truncations,
are discussed in 
\cite{Bonetti:2022gsl}.}

As anticipated above, the brane worldvolume is taken to be of the form 
$M^1 \times S^{n-2}$, where
$M^1$ is a line in $AdS_{d+1}$
and $S^{n-2}$ is defined by setting to zero two out of the $n+1$ coordinates $y^a$ of $S^n$. For definiteness, let us consider
\be \label{eq_reference_great_sphere}
S^{n-2} \; : \qquad \{ y^1 = 0 \ , y^2 = 0 \} \ .
\ee 
More precisely, it is also important to consider fluctuations in the position of the brane in the internal $S^n$.
These fluctuations are encoded in a set of real $n+1$ scalar fields $Y^a$, $a=1,\dots, n+1$ living on the brane, governing how its worldvolume is
embedded inside $S^n$.
Just like the coordinates $y^a$, the fields $Y^a$ satisfy
$Y^a Y_a =1$.
Let us introduce coordinates
$(\tau, \sigma^i)$, $i=1,\dots, n-2$ on the worldvolume 
$M^1 \times S^{n-2}$, with $\tau$ parametrizing the $M^1$ factor, and $\sigma^i$ the $S^{n-2}$ factor.
With this notation, we consider the following profile for the scalar fields $Y^a$,
\be \label{eq_brane_scalar_profile}
Y^a(\tau, \sigma) = \chi^a{}_b(\tau) \, Y^b_{(0)}(\sigma) \ . 
\ee 
In the above expression,
the functions $Y^a_{(0)}(\sigma)$ describe a brane that wraps $w$ times around the maximal $S^{n-2} \subset S^n$
defined by the equations \eqref{eq_reference_great_sphere}.\footnote{As a concrete example, the case $S^3 \subset S^5$ can be described by writing
\label{footnote_ecplicit_Y}
\be 
\ba 
Y^1_{(0)}(\sigma) & = 0  \ , &
Y^3_{(0)}(\sigma) & = \sin (\sigma^1) \sin (\sigma^2) \sin (w \sigma^3) \ , &
Y^5_{(0)}(\sigma) & = \sin (\sigma^1) \cos (\sigma^2) \ , \\
Y^2_{(0)}(\sigma) & = 0  \ , &
Y^4_{(0)}(\sigma) & = \sin (\sigma^1) \sin (\sigma^2) \cos (w \sigma^3) \ , &
Y^6_{(0)}(\sigma) & = \cos (\sigma^1) \ , \nn
\ea 
\ee 
where the ranges of the coordinates $\sigma^i$ are
$0 \le \sigma^1 \le \pi$,   
$0 \le \sigma^2 \le \pi$, 
$\sigma^3 \sim \sigma^3 + 2\pi$.
} 
The quantity $\chi^a{}_b$ is a scalar field with values in the group $SO(n+1)$, namely
$\chi^a{}_b \, \chi^c{}_d \, \delta_{ac} = \delta_{bd}$,
or $\chi^T \chi=1$ in matrix notation.
The profile
\eqref{eq_brane_scalar_profile} is motivated by regarding $Y^a_{(0)}(\sigma)$ as describing a vacuum configuration,
while $\chi^a{}_b(\tau)$ captures a collective-coordinate motion of the brane, which rotates its support while preserving the fact that it wraps a maximal $S^{n-2} \subset S^n$. This picture is supported by an analysis of the Dirac-Born-Infeld (DBI) term on the brane, performed in Appendix \ref{Appndx:WilsonDBI}.

Having described the embedding of the brane in $S^{n}$, let us now turn to its topological 
Wess-Zumino-like coupling to the potential $C_{n-1}$
associated to the flux $F_n$.
This coupling is conveniently described using descent.
To this end, we need to write the brane worldvolume 
as the boundary of an auxiliary manifold. More precisely, we assume that the external support $M^1$ of the brane can be written as
$M^1 = \partial N^2$ for  some $N^2$.

In terms of the auxiliary manifold $N^2 \times S^{n-2}$,
 the relevant topological coupling on the brane worldvolume can be written conveniently in terms of $F_n$ as\footnote{In our conventions the periods of $F_n$ are integers.}
\be \label{eq_top_brane_for_Wilson}
S= 2\pi \int_{N^2 \times S^{n-2}} \cP[F_n] \ . 
\ee 
The symbol $\cP$ denotes pullback from the bulk onto the worldvolume of the brane. In particular,
\be 
\cP[y^a] = \chi^a{}_b  Y^b_{(0)} \ , \qquad 
\cP[Dy^a] = d_A\chi^a{}_b Y^b_{(0)} + \chi^a{}_b dY^b_{(0)} \ .
\ee 
In the second expression we have introduced the shorthand notation
\be 
d_A\chi^a{}_b = d\chi^a{}_b + \cP[A^a{}_c] \chi^c{}_b \ ,
\ee 
which contains the pullback of the $AdS_{d+1}$ gauge fields $A^a{}_b$ to the brane.

In order to streamline some intermediate steps of the computation below, it is convenient to introduce the 1-form $\Theta_{ab}$ and the 2-form $\cF_{ab}$ defined as
\be 
\Theta_{ab} = (\chi^{-1} d_A\chi)_{ab} \ , \qquad 
\cF_{ab} = (\chi^{-1} F \chi)_{ab}
\ee 
Here, and in some expressions below, some contracted indices are suppressed and are understood according to   standard matrix multiplication.
We notice that, since $\chi$ is an orthogonal matrix,
$\chi^{-1} = \chi^T$ and
$\Theta_{ab} = - \Theta_{ba}$.

Next, we make use of the expression \eqref{eq_Fn_expr} for $F_n$, performing the pullback with the help of the relations quoted above.
In order to saturate the integral over $S^{n-2}$, we need to select those terms with exactly $(n-2)$ $dY_{(0)}$ factors. This gives us 
\be \label{eq_relevant_for_WL}
\ba 
S = \frac{ 2\pi N}{\cV_n} &\int_{N^2 \times S^{n-2}} 
\bigg[ 
\frac{1}{2(n-1)!} \, \epsilon_{a b_1 b_2 c_1 \dots c_{n-2}} Y_{(0)}^a \cF^{b_1 b_2} \wedge dY_{(0)}^{c_1} \wedge \dots 
\wedge dY_{(0)}^{c_{n-2}}
\\
& - \frac{1}{4(n-2)!}
\, \epsilon_{a_1 \dots a_{n-2} b c_1 c_2} dY_{(0)}^{a_1} \wedge \dots \wedge 
dY_{(0)}^{a_{n-2}} 
\wedge Y_{(0)}^d \Theta_d{}^b \wedge \Theta^{c_1 c_2}
\bigg]  \ . 
\ea 
\ee 
To derive this expression, we have used the fact that 
\be 
\epsilon_{a_1 \dots a_{n+1}}
\chi^{a_1}{}_{b_1} \dots \chi^{a_{n+1}}{}_{b_{n+1}}
= \epsilon_{b_1 \dots b_{n+1}} \ , 
\ee 
which follows from $\det \chi = 1$, as well as a Schouten identity.

The following step is the evaluation of the integrals over $Y_{(0)}$.
To display the result, we introduce the constant tensor $n_{ab}$ satisfying 
\be \label{eq_nab_parameters}
n_{ab} = - n_{ba}  \ , \qquad 
n_{12} = -n_{21} = 1 \ , \qquad 
\text{all other components are zero} \ .
\ee 
We then have
\be \label{eq_max_sphere_integral}
\int_{S^{n-2}} 
Y_{(0)}^{a_1} \wedge dY_{(0)}^{a_2} \wedge\dots \wedge
dY_{(0)}^{a_{n-1}}
= \frac{\cV_{n-2}}{2(n-1)}  \, w \, \epsilon^{a_1 \dots a_{n-1} bc} n_{bc} \ , 
\ee 
where we have included the integer wrapping number $w$.
Using \eqref{eq_max_sphere_integral} in \eqref{eq_relevant_for_WL}, we obtain 
\be 
S = \frac{2\pi N \cV_{n-2} w}{\cV_n (n-1)} \int_{N^2} \frac 12 \, n_{ab} ( 
\cF - \Theta \wedge \Theta
)^{ab} \ .
\ee 
We notice the identity
\be 
\frac{\cV_{n-2}}{\cV_n (n-1)} = \frac{1}{2\pi} \ , 
\ee 
and we recall the expressions of $\cF$ and $\Theta$. The result reads 
\be 
\label{eqWilson}
S = \frac 12 \, Nw \int_{N^2} n_{ab} \Big( \chi^{-1} F \chi - (\chi^{-1} d_A \chi)^2 \Big)^{ab} \ . 
\ee 
The integrand is a total derivative, as follows from the identity
\be 
d(\chi^{-1} d_A \chi) = 
\chi^{-1} F \chi - (\chi^{-1} d_A \chi)^2 \ . 
\ee 
In conclusion, we can recast the integral over $N^2$ as an integral over $M^1 =\partial N^2$. This yields the   action,
\be \label{eq_brane_gives_WL_final}
S = \frac 12 \, N w \int_{M^1} n_{ab} (\chi^{-1} d_A \chi)^{ab}  \ .
\ee 
This 1d action matches the form \eqref{SOnWilsonline} 
of the $SO(n+1)$ Wilson line,
with the identification
\be
q_{ab} =N w n_{ab} \ .
\ee 

We close this section with some comments on the case $n=2$.
The argument laid out formally applies, with the caveat of how to interpret the expression $\cV_0$ that appears in some intermediate steps.
If we apply the general formula for $\cV_n$ given in \eqref{eq_volume_sphere}, we have $\cV_0=2$. This is consistent with the fact that $S^0$ consists of a couple of points.
If instead we wish to consider a brane supported at a single point on $S^2$, 
we can go back to the expression \eqref{eq_relevant_for_WL}, specialized to $n=2$,
\be
\ba 
S = \frac{ 2\pi N}{4\pi} &\int_{N^2 \times {\rm pt}} 
\bigg[ 
\frac{1}{2} \, \epsilon_{a b_1 b_2  } Y_{(0)}^a \cF^{b_1 b_2} 
 - \frac{1}{4 }
\, \epsilon_{ b c_1 c_2}   Y_{(0)}^d \Theta_d{}^b \wedge \Theta^{c_1 c_2}
\bigg]  \ . 
\ea 
\ee 
Instead of computing the  integral using \eqref{eq_max_sphere_integral},
we simply perform the replacement
\be  \label{S0_vs_pt_replacement}
Y_{(0)}^a \to \frac 12 \, w \epsilon^{abc} n_{bc} \ . 
\ee 
The result takes the same form as \eqref{eq_brane_gives_WL_final}, but with an additional overall factor of 1/2.

\subsection{$AdS_5 \times S^5$ in Type IIB}\label{subsec:AdS5S5}

In this subsection we consider Type IIB on $AdS_5 \times S^5$ with $N$ units of RR $G_5$ flux,
which is the holographic dual of 4d $\cN = 4$ $SU(N)$ super Yang-Mills (SYM) theory.
The main results of this subsection were reported in \cite{Bah:2025vfu}. For completeness, we are going to briefly repeat them here. In Appendix \ref{app_AdS5S5} we will also provide supplementary details in the derivation of some claims made therein.

\subsubsection{Low-energy effective action}

Our main focus is on the gauge fields appearing in the $AdS_5$ effective action. The isometries of $S^5$
yield $SO(6)$ gauge fields (Kaluza-Klein vectors).
Since the homology of $S^5$ is non-trivial only in degrees 0 and 5, we do not have any additional $AdS_5$ gauge field originating from expanding a 10d $p$-form potential onto cycles on $S^5$.

Let us recall that Type IIB supergravity admits a consistent truncation on $S^5$ \cite{Cvetic:2000nc,Pilch:2000ue,Cassani:2010uw,Liu:2010sa,Gauntlett:2010vu,Baguet:2015sma}
to 5d maximal  $SO(6)$ gauged supergravity  \cite{Gunaydin:1984qu,Pernici:1985ju,Gunaydin:1985cu}. The truncation does retain the $SO(6)$ gauge fields,
together with the 5d metric and  other bosonic fields (as well as fermions). In what follows, however, we need not make use of the full consistent truncation, and simply restrict our attention to the couplings involving the $SO(6)$ gauge fields.

The relevant terms in the $AdS_5$ effective action are of the general form reported in~\eqref{eq:gauge_field_action_AdS}.
Concretely, they read
\be \label{eq_AdS5S5_eff_action}
S_{\rm eff} = \int_{AdS_5} \bigg[
- \frac {1}{4 } \, \tau F_{ab} \wedge * F^{ab}
- k {\rm CS}_5
\bigg]  \ , 
\ee 
where $a$, $b=1,\dots,6$ are vector indices of $SO(6)$, raised and lowered with $\delta$;
$A^{ab} = - A^{ba}$ are the $SO(6)$ gauge fields;
$F^{ab} = dA^{ab} + A^{ac} \wedge A_c{}^b$ their fields strength; ${\rm CS}_5$ denotes the Chern-Simons coupling, satisfying 
$d{\rm CS}_5 = \tfrac{1}{384 \pi^2} \epsilon_{a_1 \dots a_6} F^{a_1 a_2} \wedge F^{a_3 a_4} \wedge F^{a_5 a_6}$.
The  gauge coupling $\tau$\footnote{This quantity was denoted $g^{-2}$ in \cite{Bah:2025vfu}. Here we use $\tau$ to make contact with the general discussion of Section~\ref{sec_gauge_field_AdS_general}.} and the Chern-Simons level $k$ are determined 
as~\cite{Freedman:1998tz}
\be 
\tau = \frac{N^2}{8\pi^2 L} 
\ ,  \qquad 
k = N^2 -1 \ ,
\ee 
where $L$ denotes the $AdS_5$ radius,
given   as $L^4 = 4\pi g_{\rm s} N (\alpha')^2$.

Let us comment on the 10d origin of the terms
\eqref{eq_AdS5S5_eff_action} in the low-energy effective action.
To this end, let us recall how the $SO(6)$ gauge fields enter the expressions for the 10d metric and RR 5-form flux.
The 10d metric (in Einstein frame) reads
\be \label{eq_AdS5_S5_metric}
ds^2_{10} = ds^2_{AdS_5} + L^2 \delta_{ab} Dy^a Dy^b \ , \qquad 
Dy^a =dy^a + A^{ab} y_b \ .
\ee 
The RR flux $G_5$ is written as
\be \label{eq_AdS5_S5_G5}
G_5 = \cG_5 + *_{10} \cG_5 \, , \qquad
\mathcal \cG_5 = N   
(V_5)^{\rm g} + \tfrac 12 N \tfrac{F^{ab}}{2\pi} \wedge (\omega_{ab})^{\rm g}
\, .
\ee 
Here $*_{10}$ denotes the Hodge star associated to the 10d metric \eqref{eq_AdS5_S5_metric}.
The quantities $y^a$ are constrained coordinates on $S^5$,
satisfying $y^a y_a=1$.
As in Section \ref{sec_Wilson_spheres} we use the notation
$(\dots)^{\rm g}$ as a shorthand for the replacement
$dy^a \to Dy^a$ in a differential form with legs along $S^5$. 
The normalized volume form $V_5$ in the first term of $\cG_5$ is as in \eqref{eq_volume_sphere}.
In the second term in $\cG_5$ we have the 3-forms
$\omega_{ab}$, which satisfy 
\be \label{eq_omega_and_Killing}
\omega_{ab} = 
\tfrac{1}{12 \pi^2} \epsilon_{ab c_1 \dots c_4} y^{c_1} dy^{c_2} \wedge \dots \wedge dy^{c_4} \ , \qquad 
* \omega_{ab} = \tfrac{1}{4\pi^2} dK_{ab} \ , \qquad 
K_{ab} = 2y_{[a} dy_{b]} \ ,
\ee 
where the Hodge star is with respect to the round metric on $S^5$. We notice that the 1-forms $K_{ab}$
are the duals to 
the $SO(6)$ Killing vectors on~$S^5$
(via raising/lowering curved indices with the round $S^5$ metric).

After these preliminaries, we can address the origin of the gauge coupling $\tau$ in \eqref{eq_AdS5S5_eff_action}.
It receives contributions both from the kinetic term for $G_5$ in the Type IIB pseudo-action,
and from the 10d Einstein-Hilbert term.
These contributions are given as 
\cite{Barnes:2005bw,Benvenuti:2006xg},
\be \label{eq:effective_gauge_coupling}
\tau = \tau_{\rm EH} + \tau_{\rm flux}
 \ , \qquad 
\tau_{\rm flux} = 2 \tau_{\rm EH}  
= \frac{N^2}{12\pi^2 L} \ . 
\ee 
As far as the 10d origin of the Chern-Simons term in \eqref{eq_AdS5S5_eff_action} is concerned, it is related to the
R-symmetry anomalies of 4d $\cN = 4$ SYM \cite{Witten:1998qj,Freedman:1998tz}. Building on this fact,
\cite{Bah:2020jas} demonstrates how to reproduce this 
Chern-Simons coupling 
using methods of anomaly inflow onto D3-branes.

\subsubsection{Wilson lines from D3-branes}\label{D3_giant_graviton_subsubsec}

The Type IIB string theory origin of the $AdS_5$ Wilson line operators for the $SO(6)$ gauge fields is given by D3-branes wrapping a maximal $S^3 \subset S^5$. This is in accordance with the general 
discussion of Section \ref{sec_Wilson_spheres},
and with earlier results in the literature 
\cite{Rey:1998ik,McGreevy:2000cw,Grisaru:2000zn,Drukker:2005kx,Benvenuti:2006xg,Gomis:2006sb,Gomis:2006im}.
From the point of view of this paper, the key ingredient is the coupling of the D3-brane to the $C_4$ RR potential for the $G_5$ flux. The expression for the latter is 
given in \eqref{eq_AdS5_S5_G5} above. 
The relevant term is the one with $(V_5)^{\rm g}$,
matching the general form \eqref{eq_Fn_expr}.
As a result, the general derivation of Section
\ref{sec_Wilson_spheres} applies.
The end result is the 1d action \eqref{SOnWilsonline}
for an $SO(n+1)$ Wilson line, specialized to $n=5$,
repeated here for convenience,
\be \label{eq_AdS5S5_WL}
S = \frac 12 \int_{M^1} q_{ab} (\chi^{-1} d_A \chi)^{ab} \ , \qquad 
q_{ab} =  N w n_{ab} \ . 
\ee 
Recall that the parameter $q_{ab} = -q_{ba}$
encodes the representation of the Wilson line.
In the top-down construction, it is given in terms of the 
RR flux $N$,
the wrapping number $w$ of the D3-brane, and $n_{ab} = - n_{ba} \in \{0,\pm 1\}$ which are a set of integer parameters that specify which $S^3 \subset S^5$ is wrapped by the D3-brane.
(For example, 
if  the
$S^3$ is defined by setting $y^1 = 0 =y^2$,
the only independent non-zero entry of $n_{ab}$ is $n_{12}$.)
For simplicity, in what follows we set $w=1$ to have the simple identification $q_{ab} = N n_{ab}$,
but the case of general $w$ can be treated analogously.

\subsubsection{Symmetry operators from hanging 5-branes}

The Type IIB realization of the hanging brane operators discussed in general in Section \ref{sec:hanging_branes_intro} is given 
by a 5-brane wrapping two internal dimensions on $S^5$.
More precisely, the 5-brane consists of a bound states of a D5-brane and a Kaluza-Klein (KK) monopole.
The fact that both types of branes appear mirrors the fact that the total gauge coupling $\tau$ in the $AdS_5$ effective action is the sum of the two contributions $\tau_{\rm flux}$ and $\tau_{\rm EH}$, see \eqref{eq:effective_gauge_coupling}.

The worldvolumes of the D5-brane and the KK monopole are both $\gamma^1 \times M^3 \times \Sigma^2$,
where $\gamma^1$ is an arc in $AdS_5$ and $M^3$ is a codimension-1 subspace of the conformal boundary $\partial AdS_5$, as in Section \ref{sec:hanging_branes_intro}.
The internal support $\Sigma^2$ can be described, 
roughly speaking, as follows.
We use the fact that $S^5$ can be written as a Hopf fibration over a base $\mathbb C \mathbb P^2$,
and we recall that $\mathbb C \mathbb P^2$ has non-trivial homology in degree 2. Then, we think of $\Sigma^2$ as the non-trivial 2-cycle in the 
$\mathbb C \mathbb P^2$ base. A more precise description of $\Sigma^2$ is given in Appendix~\ref{equivariant_cohomology_appendix}.

A derivation of the symmetry operator from the D5-brane and KK monopole was given in \cite{Bah:2025vfu},
making use of  \cite{Eyras:1998hn} for the worldvolume action of a Type IIB KK monopole.
We review it in Appendix \ref{app_AdS5S5}, where we also expand on some useful background material regarding topological couplings on the worldvolume of a KK monopole.
Let us point out here some salient aspects of the construction.

As discussed in general in Section \ref{sec:hanging_branes_intro},
a key feature of the hanging brane picture is the presence of a localized gauge field $a_1$ on the arc 
$\gamma^1$. In the case of $AdS_5 \times S^5$, the top-down realization of the symmetry operator
in terms of a D5-brane and a KK monopole shows the microscopic origin of the gauge field $a_1$. In particular:
\begin{itemize}
    \item for the D5-brane, $a_1$ is identified with the standard Chan-Paton $U(1)$ gauge field on the worldvolume of a D$p$-brane;
    \item for the KK monopole, $a_1$ is identified with a composite gauge field, constructed out of two compact scalars $\varphi$, $\widetilde \varphi$
    living on the KK monopole worldvolume,
    \be 
    a_1 = \tfrac 12 \widetilde \varphi d\varphi - \tfrac 12 \varphi d \widetilde \varphi \ .
    \ee 
\end{itemize}
In both cases, the profile of the field $a_1$
is a flat connection with a non-trivial parallel transport along the arc $\gamma^1$.

The form of the operator realized by the hanging
D5-KK bound state fits with the general expression \eqref{eq:symmetry_operator}. More precisely, it is written as
\be \label{eq_AdS5S5_sym_op}
U = \exp \bigg( \tfrac 12  i \theta^{ab} \int_{M^3} \tau * F_{ab} \bigg) \ , \qquad \text{where} \quad
\theta^{ab} = \alpha m^{ab} \ . 
\ee 
The parameter $\alpha$ is the parallel transport of $a_1$, as in \eqref{eq_a1_integral}.
The constants $m^{ab} = - m^{ba}$ are a set of integer parameters that specify the choice of $\Sigma^2$ inside $S^5$, see Appendix \ref{app_AdS5S5} for more details.

\subsubsection{Charge measurement via Hanany-Witten transition}\label{HW_transition_subsection}

Let us now consider the process in which the hanging brane operator measures the charge of the endpoint of a Wilson line operator that extends along the radial direction of $AdS_5$ and terminates on the conformal boundary $\partial AdS_5$. The general picture from a low-energy point of view was described in Section \ref{sec:hanging_branes_intro}, see also Figure \ref{fig_general_HW}.

In the Type IIB top-down realization, the measuring process happens via a Hanany-Witten transition \cite{Hanany:1996ie}.
The transition involves the D3-brane realizing the Wilson line and the D5-brane realizing the symmetry operator. The KK monopole component of the latter does not participate.

The actual transition involves a U-shaped brane, but its salient features can be understood in terms of simpler transitions, in which the hanging brane is replaced by a pair of straight brane/antibrane. As the first D5-brane is pushed past the D3-brane, an F1-string is created. Upon passing the second D5-brane, it is destroyed. We are left with an F1-string worldvolume stretching between the D5-branes. In terms of the actual U-shaped D5-brane, the F1-string worldvolume is a half disk, bounded by $\gamma^1$ and a segment along the conformal boundary
$\partial AdS_5$.
The simplified Hanany-Witten transition with a ``straightened'' D5-brane is described by the following table,
\begin{center}
\begin{tabular}{|ll|ccccc|ccccc|}
\hline
  &   & \multicolumn{5}{c|}{\rule[-.17cm]{0mm}{0.6cm}$AdS_5$}  & \multicolumn{5}{c|}{
\multirow{2}{*}{$S^5$}
}\\ 
&  & \rule[-.17cm]{0mm}{0.6cm}$z$ & $x^1$ &$x^2$ &$x^3$ &$x^4$ &
\multicolumn{5}{c|}{}
\\\hline 
\rule[-.17cm]{0mm}{0.6cm}Wilson line & D3 & $\mathsf{x}$ &  &  &   &   & 
$\mathsf{x}$ & $\mathsf{x}$ & $\mathsf{x}$ &   &   \\
\rule[-.17cm]{0mm}{0.6cm}Symm.~op. & D5  & $\mathsf{x}$ & $\mathsf{x}$ & $\mathsf{x}$ & $\mathsf{x}$ &  &   &   &  & 
$\mathsf{x}$ & $\mathsf{x}$  \\
\rule[-.17cm]{0mm}{0.6cm} & F1  & $\mathsf{x}$ &   &   & 
  & $\mathsf{x}$ &   &   &   & 
  &    \\ \hline 
\end{tabular}
\end{center}
We use $z$ for the radial coordinate in $AdS_5$.

We recall that a fundamental string ending 
on a D$p$-brane couples with the Chan-Paton gauge field on the D$p$-brane.
For the case at hand, the 
 F1-string created in the 
Hanany-Witten transition
couples to the Chan-Paton
gauge field $a_1$ along the arc $\gamma^1$ in the 
D5-brane worldvolume.
This coupling is captured by the 
following topological term in the exponentiated action,
\begin{equation} \label{eq_F1_contribution}
\exp \bigg[  2\pi i n_{\rm F1} \int_{\gamma^1} a_1 \bigg]
= \exp(  i \alpha n_{\rm F1} )\ , \qquad 
n_{\rm F1} = \Sigma^3 \cdot \Sigma^2 \  ,
\end{equation} 
where $n_{\rm F1}$ is the effective multiplicity of the F1-string, given by the intersection number (inside $S^5$) of the support of the D3-brane ($\Sigma^3$) and that of the D5-brane ($\Sigma^2$).
The phase factor in \eqref{eq_F1_contribution}
encode the charge measurement process, because---as shown in \cite{Bah:2025vfu} and Appendix \ref{app_AdS5S5}---it can be recast as
\be
\exp (\tfrac 12i   \theta^{ab} q_{ab}) \ . 
\ee
Here $q_{ab}$ are the parameters specifying the representation $\mathbf R$ of the Wilson line \eqref{eq_AdS5S5_WL},
while $\theta^{ab}$ are the parameters of the symmetry operator \eqref{eq_AdS5S5_sym_op}. By repeated measurements with different choices of $\theta^{ab}$ parameters
(corresponding to D5-KK-branes with 
different support $\Sigma^2$ and parallel transport parameter $\alpha$), we can reconstruct the parameters $q_{ab}$ and hence the representation $\mathbf R$.

We notice that, in the discussion of the previous paragraphs, we have only made use of the coupling of the F1-string to the Chan-Paton gauge field on the D5-brane. One might wonder if the other couplings in the F1-string effective action can give non-zero contributions. At the level of topological couplings,
the F1-string couples to the NSNS 2-form potential $B_2$.
The latter, however, is zero in the $AdS_5 \times S^5$ solution, and it is not sourced by the 5-brane insertions. As a result, the coupling to $B_2$ can be neglected.\footnote{We expect that the  Nambu-Goto term  on the F1 worldsheet does not contribute to the analysis of charge measurement. It can be accounted for properly by means of a holographic renormalization of the F1 action, but we do not examine this point in detail.} Hence, the string theory analysis above is in agreement with the field theory computation performed in Section \ref{sec:charge_measurement_and_regularization}.

\subsection{$AdS_5 \times T^{1,1}$ in Type IIB}\label{T11_subsec}

Let us consider a similar analysis for Type IIB on $AdS_5 \times T^{1,1}$, again with $N$ units of RR $G_5$ flux, which is holographically dual to a 4d $\mathcal{N}=1$ SCFT with gauge group $SU(N) \times SU(N)$ and with chiral multiplets $A_i, B_i$ ($i=1,2$) transforming respectively in the $(\mathbf{N},\overline{\mathbf{N}})$ and $(\overline{\mathbf{N}},\mathbf{N})$ representations, also known as the Klebanov-Witten theory \cite{Klebanov:1998hh}. Here $T^{1,1} = (SU(2) \times SU(2))/U(1)$ is a homogeneous coset space, where the quotient is performed with the diagonal subgroup of the maximal torus of $SU(2) \times SU(2)$.

To mirror our discussion on $AdS_5 \times S^5$, it is convenient for us to regard the Sasaki-Einstein 5-manifold $T^{1,1}$ as a Hopf fibration $S^1_\psi \hookrightarrow T^{1,1} \to S^2 \times S^2$, and it is topologically equivalent to $S^2 \times S^3$. The continuous symmetry in question is $SO(4) \times U(1)_R \cong (SU(2) \times SU(2))/\mathbb{Z}_2 \times U(1)_R$, where the two $SU(2)$ factors arise from the isometry groups of the 2-spheres in the base of the Hopf fibration (not the homological $S^2$), and the $\mathbb{Z}_2$ quotient is identified with the diagonal subgroup of their centers. Meanwhile, the $U(1)_R$ R-symmetry is associated with the circle fiber, under which $A_i$ and $B_i$ have charge $1/2$.

There is an additional baryonic $U(1)_B$ symmetry coming from the expansion of $C_4 \supset A_B \wedge \omega_3$ onto the harmonic 3-form $\omega_3$ associated with the homological $S^3 \subset T^{1,1}$. The multiplets $A_i$ and $B_i$ respectively have charge $\pm 1$ under this baryonic symmetry. As was studied in \cite{Calvo:2025kjh} (see also \cite{Bergman:2024aly}), the corresponding Wilson lines are radially extending D3-branes wrapping the homological $S^3 \subset T^{1,1}$, and ends on the conformal boundary of $AdS_5$. Similarly to our previous discussion, the symmetry operators can be realized as hanging D5-branes supported on $\gamma^1 \times M^3 \times S^2$, with the last factor being the homological $S^2 \subset T^{1,1}$. For brevity, we do not repeat the analysis of \cite{Calvo:2025kjh} here, but in Section \ref{subsec:AdS5SE5}, we will provide a general prescription to construct operators generating such baryonic symmetries. In the rest of this subsection, we will focus on the Wilson lines and symmetry operators associated with the (continuous) isometry group of $T^{1,1}$.

\subsubsection{Geometry of $T^{1,1}$}\label{T11_geometry_subsubsec}

The line element on $T^{1,1}$ reads
\be \label{eq_T11_line_element}
\ba 
ds^2(T^{1,1}) &= \tfrac 16 (d\theta_1^2 + \sin \theta_1^2 d\phi_1^2)
+ \tfrac 16 (d\theta_2^2 + \sin \theta_2^2 d\phi_2^2)
+ \tfrac 49 D\psi^2 \ , \\
D\psi &= d\psi - \tfrac 12 \cos \theta_1 d\phi_1
 - \tfrac 12 \cos \theta_2 d\phi_2 \ . 
 \ea 
\ee 
In our conventions, $\psi$ has a period of $2\pi$. The radius of $S^1_\psi$ is $\ell = \tfrac 23$. A consistent truncation that retains a $U(1)_\psi$ isometry as well as a vector originating from the (co)homology of $T^{1,1}$ can be found in \cite{Cassani:2010na}.

Let us discuss in more detail some aspects of the base space 
$B^4 = S^2 \times S^2$. The radius of each $S^2$ is $1/\sqrt 6$ to ensure that $B^4$ is Einstein with constant 6. We choose the following basis of harmonic 2-forms on $B^4$,
\be 
\Omega_{2a} = (V_2^{(1)} ,  V_2^{(2)} ) \ , 
\ee 
where $V_2^{(1)}$ is the volume form on the first $S^2$, normalized to integrate to 1, and similarly for $V_2^{(2)}$, such that the intersection pairing on $B^4$ is
\be 
\eta_{ab} = 
\int_{B^4} \Omega_{2a} \wedge \Omega_{2b} = \begin{pmatrix}
    0 & 1 \\
    1 & 0
\end{pmatrix} \ . 
\ee 
The K\"ahler form on $B^4$ can be written as
\be 
J = \frac{2\pi}{3} (V_2^{(1)} +  V_2^{(2)}) \ ,
\ee
associated with which is a 3-form,
\begin{equation}
    \omega_{3\psi} = - \frac{1}{2} \, (V_2^{(1)} + V_2^{(2)}) \wedge \frac{D\psi}{2\pi}  \ ,\label{omega3psi_expression}
\end{equation}
which is dual (through the integration pairing) to a homologically trivial 3-cycle $\Sigma^3_\psi$, constructed by fibering $S^1_\psi$ over the linear combination of 2-spheres dual to $-\frac{1}{2} (V_2^{(1)} +  V_2^{(2)})$.

As mentioned earlier, the total space $T^{1,1}$ is topologically equivalent to $S^2 \times S^3$. The harmonic forms which integrate to 1 over them respectively are
\begin{equation}
    \omega_2 = -\frac{1}{2} \, (V_2^{(1)} - V_2^{(2)}) \ , \qquad \omega_3 = (V_2^{(1)} - V_2^{(2)}) \wedge \frac{D\psi}{2\pi} \ ,\label{T11_harmonic_forms}
\end{equation}
such that $\int_{T^{1,1}} \omega_2 \wedge \omega_3 = 1$.

\subsubsection{Wilson line operators}\label{T11_Wilson_subsubsec}

We can adapt the general discussion in Section \ref{sec_Wilson_spheres} for $n=2$ to study the $SU(2)$ isometries of the 2-spheres at the base $B^4$ of the Hopf fibration of $T^{1,1}$. For clarity, let us label these 2-spheres as $S^2_1$ and $S^2_2$, then we can consider the two 3-spheres,
\begin{equation}
    S^1_\psi \hookrightarrow S^3_1 \to S^2_1 \ , \qquad S^1_\psi \hookrightarrow S^3_2 \to S^2_2 \ .
\end{equation}
The Wilson line operators for the $SU(2)_1$ symmetry are realized by D3-branes with worldvolume $M^1 \times \text{pt}_1 \times S^3_2$, where $\text{pt}_1$ denotes a point on $S^2_1$ (or in other words, a connected component of the codimension-2 sphere $S^0 \subset S^2_1$). For concreteness, we may as well pick $\text{pt}_1$ to be the north pole of $S^2_1$, which plays the role of the vacuum configuration described by $Y^b_{(0)}(\sigma)$ in \eqref{eq_brane_scalar_profile}. The activation of the $\chi^a{}_b(\tau)$ modes on the Wilson line then rotates $\text{pt}_1$ to other points on $S^2_1$ away from the north pole. By the same token, the Wilson line operators for the $SU(2)_2$ symmetry are realized by D3-branes with worldvolume $M^1 \times \text{pt}_2 \times S^3_1$. These two families of Wilson lines are consistent with the known match with dibaryon operators in the Klebanov-Witten quiver gauge theory
\cite{Berenstein:2002ke}. They schematically take the form $\epsilon_{a_1 \dots a_N} \epsilon^{b_1 \dots b_N} (A_{i_1})^{a_1}{}_{b_1} \dots (A_{i_N})^{a_N}{}_{b_N}$ and $\epsilon_{a_1 \dots a_N} \epsilon^{b_1 \dots b_N} (B_{i_1})^{a_1}{}_{b_1} \dots (B_{i_N})^{a_N}{}_{b_N}$, where the indices $a_1$, \dots $b_N$ are (anti)fundamental indices of $SU(N)$, while $i_1$, \dots, $i_N$ are fundamental indices of $SU(2)$.

If one wishes to construct Wilson line operators charged under $SO(3)_i$ rather than $SU(2)_i$ ($i=1,2$), then we argue that the role of $\text{pt}_i$ in the discussion above should be replaced with the codimension-2 sphere $S^0_i \subset S^2_i$, which consists of two disconnected components. Heuristically, this can be seen from \eqref{S0_vs_pt_replacement} where choosing the internal support to be a single point instead of $S^0$ effectively allows the weights $q_{ab}$ to take values in half-integers. Half-integral weights are permitted for $SU(2)$ but not $SO(3)$ representations.\footnote{This observation is analogous to the quantization of the level of Chern-Simons theory, which is schematically $\tfrac{n}{2} \int_{M^3} A \wedge dA$. If the (oriented) manifold $M^3$ admits a spin structure, then $n$ can be odd, otherwise it can only be even. (However, the issue can be cured by invoking a quadratic refinement, see, e.g.~\cite{Hopkins:2002rd,Hsieh:2020jpj,Christensen:2025ktc}.)}

In addition, the Wilson line operators for the $U(1)_\psi$ symmetry are simply realized by D3-branes with worldvolume $M^1 \times \Sigma^3_\psi$, where the 3-manifold $\Sigma^3_\psi$ is defined under \eqref{omega3psi_expression}, i.e.~the $S^1_\psi$-fibration over the 2-manifold that is dual to the Kähler form $J$ on the base $B^4$ (up to a numerical constant).\footnote{The $U(1)_\psi$ gauge field and the $U(1)_R$ gauge field are generally related by $A_\psi = -\frac{2}{3\ell} A_R$ \cite{Bergman:2001qi,Berenstein:2002ke}, where the gauge field $A^R$ has the usual field-theoretic normalization for R-symmetry (because the holomorphic 3-form on the Calabi-Yau cone over the internal 5-manifold $X^5$ has charge $2$). Note that $\ell = \frac{2}{3}$ for $X^5 = T^{1,1}$.} Recall from Section \ref{Wilson_line_subsection} that the worldline scalar $\chi(\tau)$ decouples from $A_\psi$, thus reducing to the standard expression \eqref{Abelian_Wilson_line} for an Abelian Wilson line. We also notice that such a Wilson line is not charged under the baryonic $U(1)_B$ symmetry, and it corresponds to a giant graviton operator in the dual field theory, see \cite{Berenstein:2002ke,Arean:2004mm} for more details.

\subsubsection{Symmetry operators}\label{T11_symmetry_operators_subsubsec}

Similarly to the case of $AdS_5 \times S^5$, the symmetry operators generating the $SU(2)_i$ symmetries ($i=1,2$) are constructed by hanging D5-KK-branes. It is not hard to determine the support of such bound states by demanding them to measure the charge of the corresponding Wilson lines through Hanany-Witten transition as in Section \ref{HW_transition_subsection}. Particularly, the internal part of the worldvolume should intersect non-trivially with that of the Wilson line. It follows that the hanging D5-KK-brane for the, say, $SU(2)_1$ symmetry has a worldvolume $\gamma^1 \times M^3 \times S^2_1$, where $\gamma^1$ is an arc hanging from the conformal boundary, $M^3$ is a codimension-1 submanifold of $\partial AdS_5$, while $S^2_1$ is the first 2-sphere in the base $B^4$. Analogous remarks apply to the hanging D5-KK-branes generating the $SU(2)_2$ symmetry.

In general, one can construct linear combinations of $S^2_1$ and $S^2_2$ over which the D5-KK bound state is supported, thus giving rise to a symmetry operator that simultaneously generates an $SU(2)_1 \times SU(2)_2$ symmetry. However, we observe from \eqref{T11_harmonic_forms} that the (formal) difference, $S^2_1 - S^2_2$, is proportional to the homological $S^2 \subset T^{1,1}$, we may therefore identify such linear combinations as giving rise to a genuine $U(1)_B$ symmetry operator supported on $\gamma^1 \times M^3 \times S^2$, as constructed in \cite{Calvo:2025kjh}. In terms of the dual field theory, this is reflected by the fact that the global structure of (part of) the symmetry group is actually $(SU(2) \times SU(2))/\mathbb{Z}_2 \cong SO(4)$ \cite{Klebanov:1998hh}. The overall $\mathbb{Z}_2$ quotient comes about from the identification of the transformations $(A_i \to -A_i, B_j \to B_j)$ and $(A_i \to A_i, B_j \to -B_j)$, induced by the $U(1)_B$ action $A_i \to e^{i\alpha} A_i, B_j \to e^{-i\alpha} B_j$ with $\alpha=\pi$.

Regarding the $U(1)_\psi$ isometry, one can take the worldvolume of the D5-KK-brane to be $\gamma^1 \times M^3 \times \Sigma^2$ for a 2-cycle $\Sigma^2$ of the base $B^4 = S^2_1 \times S^2_2$, such that the intersection number $\Sigma^3_\psi \cdot \Sigma^2$ is non-zero (note that $\Sigma^2$ needs not be a 2-cycle of $T^{1,1}$ itself). We will revisit this statement in a broader context in Section \ref{SE5_sym_op_subsubsec}.

\subsection{$AdS_5 \times SE_5$ in Type IIB}\label{subsec:AdS5SE5}

The previous analyses for the cases of $AdS_5 \times S^5$ and $AdS_5 \times T^{1,1}$ can be readily generalized to a larger class of Type IIB solutions on $AdS_5 \times X^5$, where $X^5$ denotes the Sasaki-Einstein 5-manifold that is the base of a conical Calabi-Yau threefold. Such setups are realized by a stack of $N$ D3-branes probing the tip of this Calabi-Yau threefold, and are holographically dual to 4d $\mathcal{N}=1$ SCFTs in general. The corresponding Einstein frame metric reads
\begin{equation}
    \begin{gathered}
        ds^2_{10} = ds^2(AdS_5) + L^2 ds^2(X^5) \ ,\\
        ds^2(X^5) = g_{mn} D\xi^m D\xi^n  \ , \qquad D\xi^m = d\xi^m + K_i^m A^i \ .
    \end{gathered}\label{eq_AdS5_X5_metric}
\end{equation}
where the metrics on $AdS_5$ and $X^5$ respectively satisfy $R_{\mu\nu} = -4L^2g_{\mu\nu}$ and $R_{mn} = 4 g_{mn}$, with $L$ being the radius of $AdS_5$. The quantities $K^m_i$ are Killing vectors, with the index $i$ running over the generators of the isometry algebra of $X^5$, satisfying 
\be \label{eq_Lie_algebra_Killing}
[K_i, K_j] = f_{ij}{}^k K_k \ ,
\ee 
where $[-,-]$ denotes the standard Lie bracket of vector fields. For any Sasaki-Einstein 5-manifold $X^5$, a consistent truncation of Type IIB to 5d minimal supergravity exists, in which one retains the $U(1)_\psi$ isometry associated with the Reeb vector \cite{Buchel:2006gb,Gauntlett:2007ma}.

As mentioned in Section \ref{sec_gauge_field_AdS_general}, we will generically consider two kinds of continuous 0-form global symmetries in the field theory that can be attributed either to isometries or homological 3-cycles of $X^5$. The isometry associated with the Reeb vector of the Sasaki-Einstein structure corresponds to the superconformal R-symmetry. Other isometries, if present, are dual to mesonic, flavor, non-R symmetries (provided we choose a basis of additional Killing vectors orthogonal to the Reeb vector \cite{Barnes:2005bw}). In contrast, symmetries arising from homological cycles are dual to baryonic symmetries.

Accordingly, the self-dual 5-form RR flux takes the form,
\begin{equation}
    G_5 = \mathcal{G}_5 + \ast_{10} \mathcal{G}_5 \ , \qquad \mathcal{G}_5 = N (V_5)^{\rm g} + N \tfrac{F^I}{2\pi} \wedge (\omega_{3I})^{\rm g} \ ,\label{eq_mathcalG5_general}
\end{equation}
where $F^I = dA^I - \tfrac{1}{2} f_{JK}{}^I A^J \wedge A^K$, and $I = (i,\alpha)$ is a collective index as in Section \ref{sec_gauge_field_AdS_general}. The superscript ``g'' (which stands for gauging, in the sense of coupling to the fluctuations of the gauge fields in $AdS_5$) is defined such that for any $p$-form $\Lambda_p$ on $X^5$,
\be 
    (\Lambda_p)^{\rm g}  = \frac{1}{p!} \, \Lambda_{m_1 \dots m_p} D\xi^{m_1} \wedge \dots \wedge D\xi^{m_p} \ . 
\ee 
Further details on this operation are reviewed in Appendix \ref{app_gauging_formulae}.

In \eqref{eq_mathcalG5_general}, $V_5$ is the volume form on $X^5$, normalized to integrate to 1. Let us comment on the 3-forms $\omega_{3I}$. For $I=\alpha=1,\dots,n_\text{Betti}$, these are harmonic 3-forms onto which $\mathcal{G}_5$ is expanded. On the other hand, the 3-forms $\omega_{3i}$ for $i=1,\dots,\text{dim}\,G_\text{iso}$ are given by 
\be \label{eq_omega3i}
    \omega_{3i} = - \frac{2\pi}{8 {\rm Vol}(X^5)} *dk_i \ , \qquad k_i = K_i^m g_{mn} d\xi^n \ .
\ee 
Here ${\rm Vol}(X^5)$ is the volume of $X^5$ with its Sasaki-Einstein metric, while $k_i$ is the 1-form obtained from the Killing vector $K^m_i$ by lowering its curved index with the metric on $X^5$, and $*$ denotes the Hodge star operator on $X^5$. An important property obeyed by \eqref{eq_omega3i} is
\be \label{eq_domega3_eq}
    d \omega_{3i} + 2\pi \iota_i V_5 = 0 \ ,
\ee 
where $\iota_i$ denotes the contraction (also known as the interior product) with the Killing vector $K_i$.
We refer the reader to Appendix \ref{app_gauging_formulae} for a derivation of \eqref{eq_domega3_eq}
from \eqref{eq_omega3i}.

\subsubsection{Topology and geometry of $X^5$}

In the case of $X^5 = S^5$, we recorded in \eqref{eq_omega_and_Killing} the expressions for $\omega_{3i}$ associated with the $SO(6)$ Killing vectors on $S^5$. One can similarly determine the expressions for $\omega_{3i}$ associated with the $SO(3) \times SO(3)$ Killing vectors in the case of $X^5 = T^{1,1}$. On more general Sasaki-Einstein 5-manifolds $X^5$, there is always a distinguished $U(1)_\psi$ isometry associated with the Reeb vector. For this reason, and as alluded to in Section \ref{T11_subsec}, let us take a detour to characterize $X^5$ as a Hopf fibration.

Every Sasaki-Einstein space can be written locally as a fibration of the Reeb vector direction over a 4-dimensional K\"ahler-Einstein base. In the case of a regular Sasaki-Einstein space,
a stronger statement holds. Indeed, the space $X^5$ is the total space of a circle bundle over a smooth K\"ahler-Einstein manifold $B^4$ of positive curvature, i.e. 
\be \label{eq_global_bundle}
    S^1_{\psi} \hookrightarrow X^5 \xrightarrow{\pi} B^4 \ . 
\ee 
The metric on $X^5$ is such that $\partial_\psi$ is an isometry and that the radius of the $S^1_\psi$ fiber is constant over the base $B^4$. Importantly, the action of $S^1_\psi$ on $X^5$ is free, meaning that there are no fixed points, which will be a useful property for our subsequent analysis.

Regular Sasaki-Einstein 5-manifolds are classified \cite{friedrich1989einstein}. This is based on the Tian-Yau classification of smooth four-dimensional K\"ahler-Einstein metrics with positive curvature \cite{tian1987kahler,tianyau1987kahler}. The classification includes $S^5$ as a Hopf fibration over $\mathbb C \mathbb P^2$ and $T^{1,1}$ as a circle bundle over $S^2 \times S^2$. The other possible bases
are del Pezzo surfaces $dP_k$, $k=3,\dots,8$ (blowups of $\mathbb C \mathbb P^2$ at $k$ generic points). The corresponding regular Sasaki-Einstein metrics are known to exist but cannot be written down in closed form. See \cite{Gauntlett:2004yd} and the references therein.\footnote{We expect the results of this subsection to apply to quasi-regular Sasaki-Einstein spaces as well. In this case, we still have a fibration as in \eqref{eq_global_bundle}, but the base space $B^4$ is a K\"ahler-Einstein orbifold. We refrain, however, from discussing the quasi-regular case further.}

As before, we will normalize the angular coordinate $\psi$ to have period $2\pi$. We write the metric on $X^5$ as
\be \label{eq_reg_SE_metric}
    ds^2(X^5) = ds^2(B^4) + \ell^2 D\psi^2 \ , \qquad D\psi = d\psi + \cA \ ,
\ee 
where $\ell>0$ is the radius of $S^1_\psi$, and $\cA$ is a connection 1-form on the base $B^4$, whose curvature 2-form satisfies
\be \label{eq_J_vs_cF}
    \ell \cF = 2 J \ ,
\ee 
where $J$ is the K\"ahler form on $B^4$. Moreover, the K\"ahler metric on the base $B^4$ satisfies the Einstein condition $R_{ij} = 6 g_{ij}$, then $J = \tfrac{\pi}{3} c_1(B^4)$ is proportional to the first Chern class of the tangent bundle to $B^4$ \cite{Bergman:2001qi}. Together with \eqref{eq_J_vs_cF}, we obtain
\be 
    \frac{\cF}{2\pi} = \frac{1}{3\ell} \, c_1(B^4) \ ,
\ee 
from which the value of $\ell$ is chosen such that $\cF/2\pi$ has integral periods. For $B^4=\mathbb{CP}^2$, we have $\ell=1$ and $c_1(\mathbb{CP}^2)=3H$ where the hyperplane class $H$ is the generator of $H^2(\mathbb{CP}^2;\mathbb{Z})$. For $B^4=S^2 \times S^2$, we have $\ell=\tfrac{2}{3}$ and $c_1(S^2_1 \times S^2_2)=2(V_2^{(1)}+V_2^{(2)})$ as described in Section \ref{T11_geometry_subsubsec}. For $B^4=dP_k$, we have $\ell=\tfrac{1}{3}$ and $c_1(dP_k)=3\gamma_0 - \sum_{i=1}^k \gamma_i$ where $\gamma_0$ is the hyperplane class of $\mathbb{CP}^2$ and $\gamma_i$ are the exceptional divisors produced by the $k$ blow-ups of 
$\mathbb{CP}^2$ at generic points (see e.g.~\cite{Benvenuti:2006xg}).

\subsubsection{Wilson line operators}\label{SE5_Wilson_line_subsubsec}

With \eqref{eq_J_vs_cF}, we are now in a position to compute $\omega_{3\psi}$ from \eqref{eq_omega3i} in terms of the Kähler form $J$ on $B^4$, i.e.
\begin{equation}
    \omega_{3\psi} = -\frac{\pi\ell}{2\text{Vol}(B^4)} \, \pi^\ast(J) \wedge \frac{D\psi}{2\pi} \ ,
\end{equation}
where $\pi^\ast$ denotes the pullback from $B^4$ to $X^5$ along the circle bundle projection $\pi$ in \eqref{eq_global_bundle}. (In what follows, we sometimes leave $\pi^\ast$ implicit.) We also used the facts that
\be 
    {\rm Vol}(X^5) = 2\pi \ell \, {\rm Vol}(B^4) \ , \qquad J = *_B J \ ,
\ee 
where $*_B$ is the Hodge star operation on $B^4$.

Consequently, Wilson line operators charged under the $U(1)_\psi$ isometry are constructed with D3-branes with worldvolume $M^1 \times \Sigma^3_\psi$, where $\Sigma^3_\psi$ is realized as a circle fibration
\begin{equation}
    S^1_\psi \hookrightarrow \Sigma^3_\psi \to \mathcal{C}^2_\psi \, ,\label{D3psi_internal_support}
\end{equation}
with $\mathcal{C}^2_\psi$ being the 2-cycle of the base $B^4$ (but not necessarily of the total space $X^5$) that is Poincaré-dual to the Kähler form $J$, as we already saw in the case of $X^5=T^{1,1}$ in Section \ref{T11_subsec}, and also in the case of $X^5=S^5$ in Section \ref{subsec:AdS5S5} where $\mathcal{C}^2_\psi$ corresponds to some 2-cycle of the $\mathbb{CP}^2$ base.\footnote{Technically in Section \ref{subsec:AdS5S5} we constructed a larger family of Wilson lines charged under $SO(6)_R \supset U(1)_R$.} This claim can also be verified by explicitly reducing the relevant topological coupling on the D3-brane using \eqref{eq_mathcalG5_general},
\begin{equation}
    \int_{N^2 \times \Sigma^3_\psi} N \, \frac{F^\psi}{2\pi} \wedge \omega_{3\psi} = -\frac{N\pi\ell}{2\text{Vol}(B^4)} \int_{M^1} \frac{A^\psi}{2\pi} \int_{\mathcal{C}^2_\psi} J = -\frac{N\pi\ell}{2\text{Vol}(B^4)} \int_{M^1} \frac{A^\psi}{2\pi} \ ,
\end{equation}
where $\partial N^2 = M^1$. We observe that the integral in the last expression above is indeed the parallel transport of the $U(1)_\psi$ gauge field as anticipated for an Abelian Wilson line.

In the language of \cite{Benvenuti:2006xg}, $\Sigma^3_\psi$ is an example of an invariant submanifold of $X^5$ under $U(1)_\psi$. Two submanifolds $\Sigma^3$, $\Sigma^3{}'$ are considered equivalent if
$\Sigma^3 - \Sigma^3{}'$ is the boundary of an invariant 4-chain. We stress that, depending on $X^5$ and its isometries, we can have an invariant $\Sigma^3$ that is cohomologically trivial.
(This happens when $\Sigma^3$ is the boundary of a 4-chain, but the latter is not invariant under the isometry action.) The option of wrapping a $\Sigma^3$ which is homologically trivial is a deviation
from the more familiar paradigm of branes wrapping homology cycles. The importance of this second option was
highlighted in \cite{Benvenuti:2006xg}, building on \cite{McGreevy:2000cw,Grisaru:2000zn,Herzog:2003dj,Benvenuti:2004wx}. We will comment more on this viewpoint in Section \ref{equivariant_subsubsec}.

The construction of Wilson line operators charged under the subgroup of $G_\text{iso}$ aside from $U(1)_\psi$ follows similar ideas, but the details largely depend on the topology of the Kähler-Einstein base $B^4$. We thus refrain from giving a universal prescription here.

Let us now turn our focus to the symmetries arising from the homological 3-cycles $\mathcal{C}^3$ of $X^5$, which are dual to the harmonic 3-forms $\omega_{3\alpha}$ in \eqref{eq_mathcalG5_general}. In fact, all harmonic 3-forms on $X^5$ can be constructed in the following way \cite{Benvenuti:2006xg}. Suppose $\Lambda_2$ is a harmonic 2-form on the base $B^4$ satisfying $\Lambda_2 \wedge J =0$, then the 3-form $\pi^\ast(\Lambda_2) \wedge D\psi$ is harmonic on $X^5$.\footnote{In this vein, the submanifold $\Sigma^3_\psi$ constructed earlier can be understood as the 3-cycle that is trivialized by the $S^1_\psi$-fibration.} This particularly implies that the Betti numbers are related by
\be \label{eq_Betti_relations}
    b^2(B^4) = b^2(X^5) +  1 \ ,
\ee 
where we used the fact that $b^2=b^3$ on closed 5-manifolds. As a sanity check, we recall that $b^2(\mathbb{CP}^2)=b^2(S^5)+1=1$ and $b^2(S^2 \times S^2)=b^2(T^{1,1})+1=2$. It follows that Wilson line operators charged under such baryonic symmetries are realized by D3-branes with worldvolume $M^1 \times \mathcal{C}^3$, as we saw for the $U(1)_B$ Wilson lines in Section \ref{T11_subsec} where $\mathcal{C}^3$ corresponds to the homological $S^3 \subset T^{1,1}$.

\subsubsection{Connections with equivariant cohomology}\label{equivariant_subsubsec}

More formally, the construction above can be understood in terms of equivariant cohomology. Since we are dealing with regular Sasaki-Einstein manifolds on which the $S^1_\psi$-action is free, there is an isomorphism between the $U(1)$-equivariant cohomology of $X^5$ and the ordinary cohomology of the base $B^4$, i.e.
\begin{equation}
    H^\ast_{U(1)}(X^5;\mathbb{R}) \cong H^\ast(B^4;\mathbb{R}) \, .\label{equivariant_iso}
\end{equation}
Particularly, harmonic 2-forms on $B^4$ can be identified with equivariant 2-cocycles of $X^5$ through the isomorphism above.

In addition, note that we have a Poincaré duality for $B^4$,
\begin{equation}
    H^\ast(B^4;\mathbb{R}) \cong H_{4-\ast}(B^4;\mathbb{R}) \, ,
\end{equation}
so equivariant 2-cocycles of $X^5$ are further mapped to 2-cycles $\mathcal{C}^2$ of $B^4$. One can show that a suitable definition of ``equivariant homology'' of $X^5$ has 3-cycles corresponding precisely to fibrations $S^1_\psi \hookrightarrow \Sigma^3 \to \mathcal{C}^2$ as described previously. As a comparison, the $U(1)_\psi$ Wilson lines can be constructed by D3-branes wrapping $U(1)$-equivariant cycles of $X^5$, whereas the baryonic Wilson lines are D3-branes wrapping ordinary cycles of $X^5$. In Appendix \ref{equivariant_cohomology_appendix}, we review the definition of equivariant (co)homology groups, and compute them directly in several examples of interest to verify \eqref{equivariant_iso}.

\subsubsection{Symmetry operators}\label{SE5_sym_op_subsubsec}

Following similar arguments in Section \ref{T11_symmetry_operators_subsubsec}, the symmetry operators can be realized as hanging D5-KK bound states that are supported on $\gamma^1 \times M^3 \times \Sigma^2$, where $\gamma^1$ is an arc with endpoints on the conformal boundary and $M^3 \subset \partial AdS_5$. Depending on the geometric origin of the symmetry of interest, the choice of $\Sigma^2$ can be slightly different, but in both cases, we always seek for a submanifold which intersects non-trivially with the internal support of the Wilson line. The resultant intersection number then (partially) determines the phase that we will measure through the Hanany-Witten transition as in \eqref{eq_F1_contribution}.

Specifically, for the $U(1)_\psi$ isometry, we can take $\Sigma^2$ to be a 2-cycle of the base $B^4$ such that $\mathcal{C}^2_\psi \cdot \Sigma^2$ yields a non-zero intersection number with respect to $B^4$, where $\mathcal{C}^2_\psi$ is the base of $\Sigma^3_\psi$ as defined in \eqref{D3psi_internal_support}. By pulling back via the projection map $\pi$ in \eqref{eq_global_bundle}, this is equivalent to a pairing $\Sigma^3_\psi \cdot \Sigma^2$ with respect to the total space $X^5$. As discussed in the previous subsections, $\mathcal{C}^2_\psi$ is isomorphic to a $\mathbb{CP}^1 \subset \mathbb{CP}^2$ when $X^5=S^5$, while it is dual to the Kähler form on the base $S^2_1 \times S^2_2$ when $X^5=T^{1,1}$. Hence, in the former case we take $\Sigma^2 \sim \mathbb{CP}^1$ which self-intersects in $\mathbb{CP}^2$, and in the latter we similarly take $\Sigma^2$ to be a linear combination of $S^2_1$ and $S^2_2$.

The case of baryonic symmetries is more straightforward. We simply choose a 2-cycle $\Sigma^2$ of $X^5$ that intersects non-trivially with the 3-cycle $\mathcal{C}^3$. For example, when $X^5=T^{1,1}$, such 2-cycles are all isomorphic to the homological $S^2 \subset T^{1,1}$. It follows that a Wilson line for the $SU(2)_i$ isometry ($i=1,2$) can generally have non-vanishing R-charge and baryonic charge.

\subsection{$AdS_4 \times S^7$ in M-theory}\label{AdS4_S7_subsec}

Via string dualities, one natural question to ask is whether there exist analogous hanging brane configurations which generate continuous symmetries in M-theory holographic backgrounds. In Type II string theory setups, one may argue that the hanging BPS D-branes can equivalently be replaced with non-BPS D-branes localized on the conformal boundary \cite{Bergman:2024aly,Calvo:2025usj}, and likewise for KK monopoles. However, non-BPS states in M-theory (see, e.g.~\cite{Yi:1999hd,Houart:1999bi,Houart:2000vm,Intriligator:2000pk,Loaiza-Brito:2001yer,Gaberdiel:2001ed}) are less understood and, to some extent, conjectural. We therefore argue that hanging BPS brane solutions are much more natural candidates for continuous symmetry operators in M-theory backgrounds.\footnote{Throughout this paper, we use the term ``BPS branes'' to refer to solitonic objects in Type II string theory or M-theory which are tachyon-free and preserve half of the spacetime supersymmetry on flat worldvolumes. For example, in Type IIB string theory these objects include D$p$-branes with $p$ odd, F1-strings, NS5-branes, KK monopoles etc. We do not necessarily take into account the calibration conditions arising from non-flat worldvolumes.} We illustrate below some examples in which such BPS configurations can be realized.

In this subsection we start by considering  the 
 11d supergravity solution  $AdS_4 \times S^7$ with $N$ units of $G_7$ flux through $S^7$. (Here $G_7$ denotes the dual to the 4-form flux $G_4$, as described in greater detail below.)
This solution 
 is holographically dual to the 3d $\mathcal{N}=8$ SCFT on the worldvolume of $N$ coincident M2-branes \cite{Maldacena:1997re}. Our symmetry of interest is the $SO(8)$ isometry of $S^7$. 
Since the homology of $S^7$ is non-trivial only in degrees 0 and 7, we do not have any additional gauge field in $AdS_4$ associated to the expansion of 11d $p$-forms onto cycles in $S^7$.

This setup   can be studied in analogy to our earlier analysis of the Type IIB $AdS_5 \times S^5$ solution in Section \ref{subsec:AdS5S5}.

\subsubsection{Low-energy effective action}

It is known that 11d supergravity admits a consistent truncation on $S^7$ \cite{deWit:1983vq,deWit:1984nz,deWit:1985iy,deWit:1986mz,deWit:1986oxb,Nicolai:2011cy,deWit:2013ija,Hohm:2013pua,Godazgar:2013dma,Hohm:2014qga,Godazgar:2015qia,Varela:2015ywx} to 4d maximal 4d $SO(8)$ gauged supergravity \cite{deWit:1981sst,deWit:1982bul}.
As already mentioned, however,
we need not keep track of the full bosonic content
of maximal supergravity. Rather,
we focus on the $SO(8)$ gauge fields originating from isometries of the $S^7$ internal space. 

Let us start by reviewing the salient features of the 11d metric and flux.
The 11d line element can be written as
\be \label{eq_AdS4S7_metric}
ds^2_{11} = ds^2_{AdS_4} + L_{S^7}^2 \delta_{ab} Dy^a Dy^b  \ ,  \qquad Dy^a = dy^a + A^{ab} y_b\ , 
\ee 
where  $ds^2_{AdS_4}$ 
is the metric on $AdS_4$ of radius $L_{AdS_4}$,
$L_{S^7}$ is the radius of $S^7$,
$a$, $b=1,\dots,8$ are vector indices of $SO(8)$,
and $A^{ab} = -A^{ba}$ are the $SO(8)$ gauge fields.

The $AdS_4 \times S^7$ solution is supported by $N$ units of $G_7$ flux, where $G_7$ is the electromagnetic dual of the 4-form field strength $G_4$. Our conventions for $G_4$, $G_7$ are as follows. The Bianchi identities read\footnote{More precisely, the Bianchi identity for $G_7$ also contains a term 
with $X_8 = \frac{1}{192}(p_1^2 - 4 p_2)$. The presence of this term stems from 
11d topological coupling $C_3 \wedge X_8$ \cite{Duff:1995wd}, which is necessary for the consistency of the 11d low-energy effective action \cite{Witten:1996md}. The presence of $X_8$ in $dG_7$ is also related to the fact that 
 the factor $1/2$ in \eqref{eq_M_Bianchi_both} signals a quadratic refinement. The $X_8$ term, however, does not play an important role in our discussion, and is omitted for simplicity.} 
\be \label{eq_M_Bianchi_both}
dG_4 = 0 \ , \qquad 
dG_7 = \frac 12 \, G_4 \wedge G_4  \ .
\ee 
The duality relation between $G_4$ and $G_7$ is 
\be \label{eq_G4_G7_duality}
(2\pi \ell_{\rm P})^{-3} *_{11} G_4 = G_7 \ , 
\ee 
with $\ell_{\rm P}$ the 11d Planck mass. 
The field strengths $G_4$, $G_7$ can be written in terms of locally defined 3-form and 6-form potentials as
\be 
G_4 = dC_3 \ , \qquad 
G_7 = dC_6 + \frac 12 \, C_3 \wedge dC_3 \ .
\ee 
In the vacuum, $G_7$ is proportional to the volume form on $S^7$. Upon activating the $SO(8)$ Kaluza-Klein vectors, the expression for $G_7$ is (as reviewed e.g.~in \cite{Bonetti:2022gsl})
\be \label{eq_G7_with_vectors}
G_7 = N   
(V_7)^{\rm g} + \tfrac 12 N \tfrac{F^{ab}}{2\pi} \wedge (\omega_{ab})^{\rm g} \ .
\ee 
Here $V_7$ is as in \eqref{eq_volume_sphere} and the 5-forms
$\omega_{ab}$ satisfy 
\be \label{eq_omega_S7}
\omega_{ab} = \tfrac{1}{120 \pi^3} \epsilon_{ab c_1 \dots c_6} y^{c_1} dy^{c_2} \wedge \dots \wedge 
dy^{c_6} \ , \qquad
*\omega_{ab} = \tfrac{1}{2 \pi^3} dK_{ab} \ , \qquad 
K_{ab} = 2y_{[a} dy_{b]} \ .
\ee 
This is the analog of 
\eqref{eq_omega_and_Killing}
in the $AdS_5 \times S^5$ case.\footnote{In general, if we consider $S^n$, the $(n-2)$-forms $\omega_{ab}$ satisfy 
\be 
\omega_{ab} = \tfrac{1}{\cV_{n-2}} \tfrac{1}{(n-2)!} \epsilon_{ab c d_1 \dots d_{n-2}} y^c dy^{d_1} \wedge dy^{d_{n-2}} \ , \qquad 
* \omega_{ab} = \tfrac{1}{2 \cV_{n-2}} dK_{ab} \ , \nn
\ee 
where $\cV_k$ is the volume of the unit $S^k$, see \eqref{eq_Fn_expr}.
}

We are now in a position to discuss
the relevant terms in the low-energy effective action.
They are of the form
\be 
S_{\rm eff} \supset \int - \frac 12 \, \tau F_{ab} \wedge * F^{ab} \ , 
\ee 
where $F^{ab} = dA^{ab} + A^{ac} \wedge A_c{}^b$ is the field strength of the $SO(8)$ gauge fields. 
For completeness, we report the value of the gauge coupling constant $\tau$. It is
given by \cite{Barnes:2005bw}\footnote{In greater detail, $\tau$ is given  as 
\be 
\tau = \frac{L_{S^7}^2}{16\pi G_4 } \ , \qquad 
\frac{1}{G_4} = \frac{\cV_7 L^{7}_{S^7}}{G_{11}} \ , 
\qquad 
\cV_7 = \frac{\pi^4}{3} \ , 
\qquad 
16\pi G_{11} = 2\kappa^2_{11} = \frac{1}{2\pi}(2 \pi 
\ell_{\rm P})^9 \ ,
\nn
\ee 
where 
$L_{S^7}$ is the radius of the $S^7$, $G_4$ is the effective Newton constant in 4d, $G_{11}$ is the Newton constant in 11d, which is often expressed equivalently in terms of $\kappa_{11}$ or the Planck length $\ell_{\rm P}$.
 We also recall that the radii of $S^7$ and $AdS_4$ are given by (see e.g.~\cite{Aharony:1999ti})
\be 
2L_{AdS_4} = L_{S^7} =  \ell_{\rm P} (32\pi^2 N)^{1/6}  \ . \nn 
\ee 
}
\be 
\tau = \frac{N^{3/2}}{3\pi \sqrt 2} \ .
\ee 
As in the case of $AdS_5 \times S^5$, the total gauge coupling $\tau$ originates from two contributions in 11d, one coming from the Einstein-Hilbert term and denoted $\tau_{\rm EH}$, and another associated to the kinetic term for the flux, and denoted $\tau_{\rm flux}$. With this notation, one has \cite{Barnes:2005bw}
\be \label{eq_taus_for_AdS4}
\tau = \tau_{\rm flux} + \tau_{\rm EH} 
\ , \qquad 
\tau_{\rm flux} = 3 \tau_{\rm EH} \ .
\ee

\subsubsection{Wilson line operators}

We now turn to the discussion of Wilson line operators for the $SO(8)$ symmetry.
In the $AdS_4 \times S^7$ setup, these are realized 
by M5-branes with worldvolume $M^1 \times S^5$, where
$S^5 \subset S^7$ is maximal. This fits with the general discussion of Section \ref{sec_Wilson_spheres}, see also \cite{Drukker:2008zx,Giombi:2023vzu,Drukker:2023bip,Zhang:2025yex}.
The derivation of Section \ref{sec_Wilson_spheres} applies because the $S^7$ is supported by $G_7$ flux, and the M5-brane couples to the corresponding potential $C_6$, the electromagnetic dual to the 3-form potential $C_3$ of 11d, see e.g.~\cite{Bandos:1997ui}.

Equivalently, a maximal $S^5 \subset S^7$ can also be described as follows.
Suppose we present $S^7$ as a Hopf fibration of the circle $S^1_{\rm M}$ over a $\mathbb C \mathbb P^3$ base, then $S^5 \subset S^7$ is realized by fibering
$S^1_{\rm M}$ over a $\mathbb C \mathbb P^2$ cycle inside the $\mathbb C \mathbb P^3$ base.

\subsubsection{Symmetry operators}

Let us discuss the salient features of the hanging brane construction in this M-theory setup. 
Our goal is to construct a codimension-1 $SO(8)$ symmetry operator on the 3d boundary, supported on
$M^2$.
We propose that it can be constructed by 
combining two ingredients: an M5-brane and a KK monopole.
This mirrors the fact that the total coupling $\tau$ is the sum of the contributions
$\tau_{\rm flux}$ and $\tau_{\rm EH}$.

We start by discussing the hanging M5-brane.
Its worldvolume is 
 $\gamma^1 \times M^2 \times \Sigma^3$, where $\gamma^1$ is an arc hanging from the conformal boundary of $AdS_4$, and $\Sigma^3$ is some 3d subspace of $S^7$ that we will describe in more detail.\footnote{In analogy to the Type II string theory setups, one may understand the hanging M5-brane configuration as the blowing up of a non-BPS M4-brane \cite{Houart:2000vm,Intriligator:2000pk} localized on the conformal boundary. Tachyonic modes of M2-branes stretched between an M5-brane and an $\overline{\text{M5}}$-brane force them to recombine in the bulk, forming a U-shaped profile. However, we shall not need such an interpretation in our analysis.}

The worldvolume theory of the M5-brane contains  a chiral 2-form gauge field $b_2$ \cite{Witten:1996hc}. 
Although it is not straightforward to directly formulate a Lagrangian description of such a chiral gauge theory, for our purposes we will focus only on the relevant topological couplings.
They can be identified as follows.

In the presence of an M5-brane, the Bianchi identities \eqref{eq_M_Bianchi_both} are modified to \cite{Witten:1995em}
\be 
dG_4 = \delta_5({\rm M5}) \ , \qquad 
dG_7 = \frac 12 \, G_4 \wedge G_4 + X_8 + \mathcal H_3 \wedge \delta_5({\rm  M5}) \ . 
\ee 
Here the closed form $\delta_5({\rm M5})$  is the Poincar\'e dual of the worldvolume of the M5-brane inside 11d spacetime.
The 3-form $\mathcal \cH_3$ is the gauge-invariant combination 
\be 
\cH_3 =db_2 - C_3 \ , 
\ee 
with $b_2$ the chiral 2-form on the M5-brane and $C_3$ pulled back from the bulk.\footnote{For a detailed discussion of this point, we refer the reader to \cite{Monnier:2013rpa}.}
In particular, $\cH_3$ satisfies $d\cH_3 = - G_4$. (This renders the Bianchi identity for $G_7$ consistent with $ddG_7=0$.) 
The presence of the term
$\cH_3 \delta_5(\rm M5)$ in the Bianchi identity for $G_7$
signals a worldvolume coupling of $db_2$ to $C_3$ \cite[eq.~(3.3)]{Townsend:1995af}.
Upon integration by parts, this coupling can be cast~as 
\be \label{eq_our_b2_coupling}
- 2\pi \int_{\cW^6}  b_2 \wedge G_4 \ , 
\ee 
where $\cW_6$ denotes the worldvolume of the M5-brane.

Having reviewed the relevant topological couplings on the M5-brane, we can now describe how a hanging M5-brane with worldvolume $\gamma^1 \times M^2 \times \Sigma^3$ realizes a symmetry operator.
We specialize \eqref{eq_our_b2_coupling} to the case of a hanging M5-brane on $\cW^6 = \gamma^1 \times M^2 \times \Sigma^3$, 
\be \label{eq_6d_action_M5}
- 2\pi \int_{\gamma^1 \times M^2 \times \Sigma^3}   b_2 \wedge G_4 \ . 
\ee 
Next, let us describe $\Sigma^3$. First, we represent $S^7$ as a Hopf fibration of a circle fiber, denoted $S^1_{\rm M}$, over a $\mathbb C \mathbb P^3$ base.
Second, we select
a copy of $\mathbb C \mathbb P^1$, denoted $\mathbb {CP}^1_{\rm M5}$, inside $\mathbb C \mathbb P^2$.
With this notation, 
$\Sigma^3$ is obtained by fibering $S^1_{\rm M}$ over
$\mathbb C \mathbb P^1_{\rm M5}$ (thus,  $\Sigma^3$ is a 3-sphere). The profile for the chiral 2-form $b_2$ is schematically taken to be
\be\label{chiral_2-form_decomposition}
b_2 = a_1 \wedge {\rm vol}(S^1_{\rm M} ) \ , 
\ee 
where ${\rm vol}(S^1_{\rm M} )$ denotes the volume form on 
$S^1_{\rm M}$, normalized to integrate to 1, and 
$a_1$ is a 1-form gauge field, which is flat ($da_1=0$)
and has a non-trivial parallel transport along the arc $\gamma^1$. More precisely, we set 
\be 
\alpha_{\rm M5} = 2\pi\int_{\gamma^1} a_1 \ . 
\ee 
The details of the evaluation of \eqref{eq_6d_action_M5} 
are reported in Appendix \ref{app_AdS4S7}. The result is as follows: the hanging M5-brane
contributes the quantity
\be \label{eq_AdS4_M5_piece}
U_{\rm M5} = \exp \bigg( i
\alpha_{\rm M5} \tau_{\rm flux} \int_{M^2} \tfrac 12 m^{ab} *F_{ab} \bigg) \ . 
\ee 
The constant parameters $m^{ab} = - m^{ba}$ are integers and specify the choice of Hopf fiber $S^1_{\rm M}$ 
in terms of the constrained coordinates $y^a$ on $S^7$, as described in more detail in Appendix~\ref{app_AdS4S7}.

Let us now move on to the contribution of the KK monopole. This is a supersymmetric soliton in 11d supergravity with a 7d worldvolume. In order to engineer the sought-for symmetry defect, the worldvolume of the KK monopole must be of the form $\gamma^1 \times M^2 \times \Sigma^4$,
for a suitable $\Sigma^4 \subset S^7$.
Our proposal for the worldvolume of the KK monopole,
compared with that of the M5-brane, can be summarized in the following table,
\begin{center}
\begin{tabular}{|l|cccc|c|cccccc|}
\hline
     & \multicolumn{4}{c|}{\rule[-.17cm]{0mm}{0.6cm}$AdS_4$}  & \multicolumn{7}{c|}{$S^7$}\\ \cline{6-12}
  & $z$ & $x^1$ &$x^2$ &$x^3$ &\rule[-.17cm]{0mm}{0.6cm}$S^1_{\rm M}$ &
\multicolumn{6}{c|}{$\mathbb C \mathbb P^3$}
\\\hline 
\rule[-.17cm]{0mm}{0.6cm}M5  & $\mathsf{x}$ & $\mathsf{x}$ & $\mathsf{x}$ &   & $\mathsf{x}$ &   &   &  & 
& $\mathsf{x}$ & $\mathsf{x}$  \\
\rule[-.17cm]{0mm}{0.6cm}KK  & $\mathsf{x}$ & $\mathsf{x}$  & $\mathsf{x}$  & 
  & $\mathsf{iso}$ &   &   & $\mathsf{x}$  & $\mathsf{x}$
  &  $\mathsf{x}$ & $\mathsf{x}$\\ \hline 
\end{tabular}
\end{center}
As usual, we use an $\mathsf x$ to mark directions along which an object extends.
Let us recall that one of the four transverse directions to the KK monopole is an isometry of the 11d metric. This preferred transverse direction is indicated with the symbol $\mathsf{iso}$. We choose it to be aligned with the circle fiber $S^1_{\rm M}$ used in defining the worldvolume of the M5-brane.
The table also shows that the internal support $\Sigma^4$ of the KK monopole is a 4-cycle in $\mathbb C \mathbb P^3$. We take it to be a copy of 
$\mathbb {CP}^2$, denoted 
$\mathbb{CP}^2_{\rm KK}$,
inside $\mathbb {CP}^3$.
We also 
assume that $\mathbb{CP}^2_{\rm KK}$ contains
the $\mathbb{CP}^1_{\rm M5}$
associated to the M5-brane. In summary,
\be \label{eq_CP_chain_M}
\mathbb{CP}^1_{\rm M5} \subset 
\mathbb{CP}^2_{\rm KK} \subset 
\mathbb{CP}^3 \  .
\ee 

The worldvolume action of an M-theory KK monopole is given in \cite{Bergshoeff:1998ef}.
It features a localized $U(1)$ gauge field $a_1^{\rm KK}$, analogous to the Chan-Paton gauge field on the worldvolume of a D$p$-brane in Type II. 
The gauge field $a_1^{\rm KK}$ is flat and has a non-trivial parallel transport along the arc $\gamma^1$, encoded in the parameter $\alpha_{\rm KK}$ defined as
\be 
\alpha_{\rm KK} = 2\pi \int_{\gamma^1} a_1^{\rm KK} \ .
\ee 
In Appendix \ref{app_AdS4S7} we study the topological couplings of the hanging KK monopole and show how they yield the contribution 
\be \label{eq_AdS4_KK_piece}
U_{\rm KK} = \exp \bigg( i \ 3 
\alpha_{\rm KK} \tau_{\rm EH} \int_{M^2} \tfrac 12 m^{ab} *F_{ab} \bigg) \ . 
\ee  
The contributions  
\eqref{eq_AdS4_M5_piece} and \eqref{eq_AdS4_KK_piece} have the correct functional form to combine and produce the expected total symmetry operator. 
With the identifications 
\be 
\alpha_{\rm M5} = \alpha \ , \qquad 
\alpha_{\rm KK} = \tfrac 13 \alpha \ , 
\ee 
we arrive at the following expression for the operator,
\be \label{eq_AdS4S7_tot_sym_op}
U  = \exp \bigg( i
\alpha  \tau  \int_{M^2} \tfrac 12 m^{ab} *F_{ab} \bigg) \ . 
\ee 
This is indeed of the form \eqref{eq_AdS5S5_sym_op}
with the identification
$\theta^{ab} = \alpha m^{ab}$.

\subsubsection{Type IIA picture}

We now describe an alternative viewpoint on symmetry operators, based on the reduction from M-theory to Type IIA. More precisely, we can identify the Hopf fiber $S^1_{\rm M}$ with the M-theory circle in the reduction from 11d to 10d (as was anticipated in our notation). In the Type IIA picture, the M5-brane and KK monopole become a D4-brane and a D6-brane, respectively, as described in the following table,
\begin{center}
\begin{tabular}{|l|cccc|cccccc|}
\hline
     & \multicolumn{4}{c|}{\rule[-.17cm]{0mm}{0.6cm}$AdS_4$}  & 
    \multicolumn{6}{c|}{\multirow{2}{*}{$\mathbb C \mathbb P^3$}}
    \\  
  & \rule[-.17cm]{0mm}{0.6cm} $z$ & $x^1$ &$x^2$ &$x^3$   &
\multicolumn{6}{c|}{}
\\\hline 
\rule[-.17cm]{0mm}{0.6cm}D4  & $\mathsf{x}$ & $\mathsf{x}$ & $\mathsf{x}$ &     &   &   &  & 
& $\mathsf{x}$ & $\mathsf{x}$  \\
\rule[-.17cm]{0mm}{0.6cm}D6  & $\mathsf{x}$ & $\mathsf{x}$  & $\mathsf{x}$  & 
    &   &   & $\mathsf{x}$  & $\mathsf{x}$
  &  $\mathsf{x}$ & $\mathsf{x}$\\ \hline 
\end{tabular}
\end{center}
As we have seen around \eqref{eq_CP_chain_M}, in the M-theory picture the supports of the M5-brane and KK monopole are nested inside $\mathbb{CP}^3$. The same structure persists in the Type IIA picture.
The internal support of the D4-brane is $\mathbb{CP}^1_{\rm D4} = \mathbb{CP}^1_{\rm M5}$, the internal support of the D6-brane is $\mathbb{CP}^2_{\rm D6} = \mathbb{CP}^2_{\rm  KK}$, and they satisfy 
\be \label{eq_CP_chain_A}
\mathbb{CP}^1_{\rm D4} \subset 
\mathbb{CP}^2_{\rm D6} \subset 
\mathbb{CP}^3 \  .
\ee 
As we can see, the Type IIA setup features a D4-brane
sitting inside the worldvolume of a D6-brane. This is a configuration that has been studied extensively \cite{Gava:1997jt,Sato:2001su,Larsson:2001wt,Khoze:2003yk},
see also \cite{Callan:1997kz,Howe:1997ue}. An important feature of this configuration is the presence of a worldvolume flux on the D6-brane, with profile equal to the Poincar\'e dual of the D4-brane worldvolume inside the D6-brane.

The relevant topological terms 
on the D4-brane and D6-brane are
(we write them using descent)
\be 
S_{\rm D4} \supset 2\pi \int_{N^6} f_2 \wedge G_4^{\rm IIA} \ , \qquad 
S_{\rm D6} \supset 2\pi \int_{N^8} \tfrac 12 (f_2)^2 \wedge G_4^{\rm IIA}  \ , 
\ee 
with $f_2 =da_1$ the Chan-Paton field strength on each brane and $G_4^{\rm IIA}$ is the RR 4-form flux in Type IIA.
For the D4-brane, 
the  Chan-Paton gauge field has a flat but non-trivial profile over the 
arc~$\gamma^1$. 
This gives us 
a contribution of the form
\be \label{eq_IIA_D4_piece}
2\pi \int_{\gamma^1} a_1 \int_{  M^2 \times \mathbb C\mathbb P^1_{\rm D4}}  G_4^{\rm IIA} \ ,
\ee 
In the case of the D6-brane, the profile of the Chan-Paton field  $a_1$ is the sum of two pieces,
\be 
\text{D6-brane:} \qquad  a_1 = a_1^{(1)} + a_1^{(2)} \ . 
\ee 
The first piece $a_1^{(1)}$ is a flat connection 
with legs along the arc $\gamma^1$ only. It 
furnishes a non-trivial parallel transport parameter $2\pi \int_{\gamma^1} a_1^{(1)}$, as for the D4-brane.
The second piece $a_1^{(2)}$
is a locally defined 1-form 
on $\mathbb{CP}^2_{\rm D6}$ whose field strength 
equals the Poincar\'e dual of
$\mathbb{CP}^1_{\rm D4}$ inside
$\mathbb{CP}^2_{\rm D6}$,
\be 
f_2^{(2)}=da_1^{(2)} = \delta_2(\mathbb C\mathbb P^1_{\rm D4} \subset \mathbb C\mathbb P^2_{\rm D6}) \ .
\ee 
This is the worldvolume flux anticipated in the discussion after \eqref{eq_CP_chain_A}.
Due to the flux contribution $f_2^{(2)}$, 
the D6-brane couples to $G_4^{\rm IIA}$. More precisely, we have a coupling 
\be \label{eq_IIA_D6_piece}
 2\pi \int_{\gamma^1} a_1^{(1)} \int_{M^2 \times \mathbb C \mathbb P^2_{\rm D6}} f_2^{(2)} \wedge G_4^{\rm IIA}
= 2\pi \int_{\gamma^1} a_1^{(1)} \int_{M^2 \times \mathbb C \mathbb P^1_{\rm D4}}  G_4^{\rm IIA} \ . 
\ee 
In summary, 
the coupling \eqref{eq_IIA_D4_piece} from the D4-brane, and the coupling
\eqref{eq_IIA_D6_piece} from the D6-brane have the same functional form. They combine to yield
the total symmetry 
operator~\eqref{eq_AdS4S7_tot_sym_op}.

\subsubsection{Charge measurement via Hanany-Witten transition}

The charge measurement process is realized in M-theory via a Hanany-Witten transition.
This is analogous to the Hanany-Witten transition for $AdS_5 \times S^5$ discussed in Section \ref{subsec:AdS5S5}, and can be summarized 
schematically in the following table,
\begin{center}
\begin{tabular}{|ll|cccc|c|cccccc|}
\hline
  &   & \multicolumn{4}{c|}{\rule[-.17cm]{0mm}{0.6cm}$AdS_4$}  & \multicolumn{7}{c|}{$S^7$}\\ \cline{7-13}
&  & $z$ & $x^1$ &$x^2$ &$x^3$ &\rule[-.17cm]{0mm}{0.6cm}$S^1_{\rm M}$ &
\multicolumn{6}{c|}{$\mathbb C \mathbb P^3$}
\\\hline 
\rule[-.17cm]{0mm}{0.6cm}Wilson line & M5 & $\mathsf{x}$ &  &  &   &  $\mathsf{x}$ & 
$\mathsf{x}$ & $\mathsf{x}$ & $\mathsf{x}$ & $\mathsf{x}$  &  & \\
\rule[-.17cm]{0mm}{0.6cm}Symm.~op. & M5  & $\mathsf{x}$ & $\mathsf{x}$ & $\mathsf{x}$ &   & $\mathsf{x}$ &   &   &  & 
& $\mathsf{x}$ & $\mathsf{x}$  \\
\rule[-.17cm]{0mm}{0.6cm} & M2  & $\mathsf{x}$ &   &   & 
 $\mathsf{x}$ & $\mathsf{x}$ &   &   &   & 
  &   & \\ \hline 
\end{tabular}
\end{center}
When the hanging M5-brane is moved past the M5-brane that realizes  the Wilson line, an M2-brane is created.
Its worldvolume consists of $S^1_{\rm M}$ together with
a half-disk in $AdS_4$, delimited by the arc $\gamma^1$ and by a segment on the conformal boundary.
We have not indicated explicitly the KK monopole component of the symmetry operator, because it does not affect  the Hanany-Witten transition.

The M2-brane created in the transition is an open M2-brane, ending on the hanging M5-brane. The worldvolume theory of an M2-brane ending on an M5-brane contains a coupling to the chiral 2-form $b_2$ on the M5-brane.
This is the analog of the coupling of an F1-string to the Chan-Paton gauge field of the D$p$-brane on which it ends.
Due to the coupling of the M2-brane to $b_2$, 
the net result of the Hanany-Witten transition is the phase factor
\begin{equation}  
\exp \bigg[  2\pi i n_{\rm M2} \int_{\gamma^1 \times S^1_{\rm M}} b_2 \bigg]
= \exp(  i \alpha n_{\rm M2} )\ ,   
\end{equation}
where the effective multiplicity $n_{\rm M2}$ of the
M2-brane is given by the intersection number, inside $\mathbb C \mathbb P^3$, of the supports of the two M5-branes (ignoring their common $S^1_{\rm M}$ directions).

Alternatively, we can consider a Type IIA description of the same Hanany-Witten transition, obtained by reducing along $S^1_{\rm M}$, 
\begin{center}
\begin{tabular}{|ll|cccc|cccccc|}
\hline
  &   & \multicolumn{4}{c|}{\rule[-.17cm]{0mm}{0.6cm}$AdS_4$}  & \multicolumn{6}{c|}{\multirow{2}{*}{$\mathbb C \mathbb P^3$}} \\  
&  & \rule[-.17cm]{0mm}{0.6cm}$z$ & $x^1$ &$x^2$ &$x^3$ &  \multicolumn{6}{c|}{}
\\\hline 
\rule[-.17cm]{0mm}{0.6cm}Wilson line & D4 & $\mathsf{x}$ &  &  &   &   
$\mathsf{x}$ & $\mathsf{x}$ & $\mathsf{x}$ & $\mathsf{x}$  &  & \\
\rule[-.17cm]{0mm}{0.6cm}Symm.~op. & D4  & $\mathsf{x}$ & $\mathsf{x}$ & $\mathsf{x}$ &   &    &   &  & 
& $\mathsf{x}$ & $\mathsf{x}$  \\
\rule[-.17cm]{0mm}{0.6cm} & F1  & $\mathsf{x}$ &   &    & $\mathsf{x}$ &   &   &   & 
  &   & \\ \hline 
\end{tabular}
\end{center}
We have not indicated the D6-brane component of the symmetry operator in the Type IIA picture, because it does not affect the Hanany-Witten transition.

\subsection{$AdS_4 \times SE_7$ in M-theory}\label{subsec:AdS4SE7}

In Section \ref{subsec:AdS5SE5} we saw that the Type IIB solution on $AdS_5 \times S^5$ is a member of the larger family of solutions on $AdS_5 \times SE_5$, with $SE_5$ denoting Sasaki-Einstein 5-manifolds. Other members of the family include $T^{1,1}$, which we studied explicitly in Section \ref{T11_subsec}, and the del Pezzo surfaces $dP_k$, $k=3,\dots,8$. Analogously, the analysis we performed in Section \ref{AdS4_S7_subsec} for $AdS_4 \times S^7$ in M-theory admits a generalization to $AdS_4 \times SE_7$ for generic (regular) Sasaki-Einstein 7-manifolds. The prescription parallels that in Section \ref{subsec:AdS5SE5}, so we will only be brief in the following.

As usual, we parametrize a regular Sasaki-Einstein 7-manifold $X^7$ as a Hopf fibration over a smooth Kähler Einstein manifold $B^6$ of positive curvature, i.e.
\begin{equation}
    S^1_M \hookrightarrow X^7 \xrightarrow{\pi} B^6 \ .
\end{equation}
The fiber is identified as the M-theory circle which, upon being reduced, brings us to the dual Type IIA description of the theory. We have $B^6 = \mathbb{CP}^3$ when $X^7 = S^7$. Similar examples of $B^6$ include $\mathbb{CP}^2 \times \mathbb{CP}^1$ and $\mathbb{CP}^1 \times \mathbb{CP}^1 \times \mathbb{CP}^1$. The corresponding $X^7$, denoted respectively as $M^{3,2}$ and $Q^{1,1,1}$, are homogeneous Sasaki-Einstein manifolds, and are higher-dimensional generalizations of $X^5 = S^5$ with base $B^4 = \mathbb{CP}^2$, and $X^5 = T^{1,1}$ with base $B^4 = \mathbb{CP}^1 \times \mathbb{CP}^1$. In both cases, the M-theory solutions are holographically dual to 3d $\mathcal{N}=2$ theories, see \cite{Duff:1986hr} for a review.

M-theory on $AdS_4 \times X^7$ canonically admits a $U(1)_\psi$ isometry associated with the Reeb vector. The Wilson line operators charged under this symmetry are constructed with M5-branes with worldvolume $M^1 \times \Sigma^5_\psi$, where $\Sigma^5_\psi$ is realized as a circle fibration
\begin{equation}
    S^1_M \hookrightarrow \Sigma^5_\psi \to \mathcal{C}^4_\psi \ ,
\end{equation}
with $\mathcal{C}^4_\psi$ being the 4-cycle of the base $B^6$ that is Poincaré-dual to the Kähler form $J$. As per the discussion in Section \ref{equivariant_subsubsec}, such a cycle on $B^6$ can be regarded as a $U(1)$-equivariant cycle of the total space $X^7$. The symmetry operators of $U(1)_\psi$ are hanging M5-KK bound states, where the M5-brane component is supported on $\gamma^1 \times M^2 \times \Sigma^3$. The internal support $\Sigma^3$ is realized as a circle fibration
\begin{equation}
    S^1_M \hookrightarrow \Sigma^3 \to \Sigma^2 \ ,
\end{equation}
such that $\Sigma^2$ is a 2-cycle which intersects non-trivially with $\mathcal{C}^4_\psi$ in $B^6$. On the other hand, the KK monopole is supported on $\gamma^1 \times M^2 \times \Sigma^4$, with $\Sigma^2 \subset \Sigma^4 \subset B^6$. We can readily check that such configurations generate a $U(1)$ subgroup of the $SO(8)$ symmetry in the $AdS_4 \times S^7$ example that we studied in Section \ref{AdS4_S7_subsec}.

Sasaki-Einstein 7-manifolds generally contain isometries other than $U(1)_\psi$. For example, up to global forms, $M^{3,2}$ has an $SU(3) \times SU(2)$ isometry, while $Q^{1,1,1}$ has an $SU(2) \times SU(2) \times SU(2)$ isometry. 
Each $SU(n)$ factor is associated with the isometry of $\mathbb{CP}^{n-1} \subset B^6$. The relevant brane configurations can be inferred from our previous results for $AdS_5 \times S^5$ and $AdS_5 \times T^{1,1}$ in Type IIB.

As an example, for $X^7 = M^{3,2}$ with base $B^6 = \mathbb{CP}^2 \times \mathbb{CP}^1$, the Wilson line operators charged under $SU(3)$ are M5-branes with worldvolume $M^1 \times \Sigma^5$, where $\Sigma^5$ is a circle fibration
\begin{equation}
    S^1_M \hookrightarrow \Sigma^5 \to \mathbb{CP}^1 \times \mathbb{CP}^1 \ .
\end{equation}
Here the first copy of $\mathbb{CP}^1$ is embedded in $\mathbb{CP}^2 \subset B^6$, described in (projective) coordinates of $\mathbb{C}^3$ by an expression similar to \eqref{eq_brane_scalar_profile},
\be 
    Z^A(\tau, \sigma) = u^A{}_B(\tau) \, Z_{(0)}^B(\sigma) \ , 
\ee 
where $u^A{}_B$ is valued in $SU(3)$, with $A,B = 1,2,3$. This resembles the description of D3-branes as giant gravitons in Section \ref{D3_giant_graviton_subsubsec}, noting that $\mathbb{CP}^2 \cong S^5/S^1_M$. Meanwhile, the Wilson line operators charged under $SU(2)$ are M5-branes on $\gamma^1 \times \Sigma^5$ as well, except that $\Sigma^5$ is now the circle fibration
\begin{equation}
    S^1_M \hookrightarrow \Sigma^5 \to \mathbb{CP}^2 \times \text{pt} \ ,
\end{equation}
where $\text{pt} \subset \mathbb{CP}^1 \subset B^6$. This follows from an argument similar to that in Section \ref{T11_Wilson_subsubsec}. One can also read off the internal support of the dual symmetry operators as hanging M5-KK bound states, by demanding it to intersect non-trivially (in the suitable sense) with that of the Wilson lines. We remark that the example of $X^7 = Q^{1,1,1}$ can be treated in the same fashion, and we refrain from doing so explicitly.

Last but not least, we complete the discussion by commenting on the baryonic symmetries associated with homological 5-cycles $\mathcal{C}^5$ of $X^7$, the presence of which enters $G_7 \supset \tfrac{1}{2} N \tfrac{F^\alpha}{2\pi} \wedge (\omega_{5\alpha})^\text{g}$ as in \eqref{eq_G7_with_vectors} via the harmonic 5-forms $\omega_{5\alpha}$ where $\alpha=1,\dots,b^5(X^7)$. The Wilson line operators charged under such baryonic symmetries are realized by M5-branes with worldvolume $M^1 \times \mathcal{C}^5$, whereas the symmetry operators are constructed by hanging M5-KK bound states similarly to before. In particular, the M5-brane component, which is responsible for the charge measurement via Hanany-Witten transition, is supported on $\gamma^1 \times M^2 \times \Sigma^3$, where $\Sigma^3$ is realized as a circle fibration
\begin{equation}
    S^1_M \hookrightarrow \Sigma^3 \to \Sigma^2 \ ,
\end{equation}
for some $\Sigma^2$ such that it intersects non-trivially with $\mathcal{C}^5$ in $X^7$. Let us emphasize again that, in contrast to the Type IIB examples, the support of the hanging M5-branes in all the aforementioned cases always consists of the M-theory circle $S^1_M$. This is indeed crucial for us to define the parameter $\alpha_\text{M5} = 2\pi \int_{\gamma^1} a_1$ from the chiral 2-form $b_2 \supset a_1 \wedge \text{vol}(S^1_M)$ on the M5-brane as in \eqref{chiral_2-form_decomposition}. It also allows for a consistent interpretation of the M5-branes as D4-branes in the dual Type IIA picture, for both the Wilson lines and symmetry operators.

\section{Dynamical aspects}
\label{Sec:Dynamics}

We devote this section to the analysis of several dynamical aspects of the hanging brane solution in $AdS_{d+1}$ introduced in Section \ref{sec:hanging_branes_intro}, regarding it as a top-down construction in string/M-theory. Importantly, we provide a precise characterization of how the hanging brane becomes topological when localized near the conformal boundary, and analyze the dynamical process underlying the fusion of hanging branes. For concreteness, we focus on a detailed study of the DBI action of the D5-brane in Type IIB string theory, which is applicable to the setups considered in Sections \ref{subsec:AdS5S5} and \ref{subsec:AdS5SE5}. We expect that qualitatively similar conclusions can be drawn for the hanging M5-brane solution in M-theory setups, after taking into account technical subtleties associated with promoting the 1-form Chan-Paton field to the (chiral) 2-form field $b_2$.\footnote{As was argued in \cite{Yi:1999hd}, the chiral 2-form field on the M5-brane combines with the anti-chiral 2-form field on the $\overline{\text{M5}}$-brane to form an ordinary 2-form field, so the M5-$\overline{\text{M5}}$ system does admit a Lagrangian description, as opposed to its individual constituents.} Alternatively, its dual Type IIA description should follow essentially the same analysis that we perform below. We also expect an analysis of the DBI-like action of the KK monopole (which can be found in \cite{Eyras:1998hn}) will lead to qualitatively similar conclusions, but we do not explore the details in this work.

\subsection{Hanging brane profile}\label{dbrane_config}

The hanging brane configuration is motivated by the results of \cite{Bergman:2024aly,Calvo:2025kjh}. From a bottom-up perspective we also see it from the regulator of the insertion of a topological operator as discussed in Section \ref{sec:charge_measurement_and_regularization}. In \cite{Bergman:2024aly} it was proposed that the symmetry operator for the baryonic symmetry from $AdS_5\times T^{1,1}$ can be realized as a (tachyonic) non-BPS D4-brane of Type IIB string theory wrapping the 2-cycle $S^2 \subset T^{1,1}$, and localized at the conformal boundary along a choice of $M^3$.  In \cite{Calvo:2025kjh}, the authors considered a realization of the operator in terms of D5-branes, where they re-interpreted the non-BPS D4-brane as a final state of a D5-$\overline{\text{D}5}$-brane system stretched along the AdS radius.  If the D5-brane and $\overline{\text{D}5}$-brane are slightly separated in the boundary along, say, the $x^3$ direction, then the final state is a hanging D5-brane in a U-shape, where the D5-brane and $\overline{\text{D}5}$-brane recombine in the bulk at some $AdS_5$ radius $z= z_0$. Let us briefly review this configuration.

The action of a D$p$-brane is generally given by 
\begin{equation}
    S_{\text{D$p$}}= T_p \int e^{-\Phi} \sqrt{\det\left[P(G_{ab} + B_{ab} ) + \ell_s^2 f_{ab}\right]} + \mu_p \int C\wedge e^{i B_2 + i\ell_s^2 f_2} \ ,
\end{equation} where
\begin{equation}
    C= \sum_m (-i)^{\frac{p-m+1}{2}} C_m \, .
\end{equation} 
The field strength $f_2=da_1$ is the worldvolume flux for the Chan-Paton gauge field, $P(G_{ab} + B_{ab} )$ is the pullback of the spacetime metric and the $B$-field onto the brane, and $C_m$ are the RR gauge fields. We note that the conventions for the RR gauge fields here match those of \cite{Garousi_2005,Garousi_2008} to make contact with the brane-antibrane systems analyzed there. The hanging D5-brane wraps $\gamma^1 \times M^3 \times \Sigma^2$ with each component extended along the following directions,
\begin{equation}
\label{coordsDp}
    \gamma^1: \{x^3(t),z(t)\} \, , \qquad M^3: \{x^0,x^1,x^2\} \, , \qquad \Sigma^2: \{\theta_1(\vartheta), \phi_1(\varphi), \theta_2(\vartheta), \phi_2(\varphi) \} \ ,
\end{equation}
where $t$ parametrizes the curve $\gamma^1$ and $(\vartheta,\varphi)$ parametrize $\Sigma^2$.
The coordinate $z$ is the radial coordinate on $AdS_5$, while
$x^{0,1,2,3}$ are coordinates along the conformal boundary.
The $T^{1,1}$ coordinates
$\theta_{1,2}$, $\phi_{1,2}$
are defined as in the line element \eqref{eq_T11_line_element}. Our goal is to reduce the D5-brane action with a non-trivial parallel transport for the worldvolume gauge field $a_1$ along $\gamma^1$. It can be shown that this action factorizes into an internal part supported on $\Sigma^2$ and an external part supported on $\gamma^1$, i.e.
\begin{equation}
\label{Eqfactorizes}
    \begin{gathered}
        S = \frac{T_5 L^2 e^{-\Phi} }{3\sqrt{2}} \, \, S_I \cdot S_O \, ,  \\ 
        S_O = V_{M^3}  \int \frac{L^4}{z^4} \sqrt{\left(\frac{dx^3}{dt}\right)^{\!2} + \left(\frac{dz}{dt}\right)^{\!2}} \, dt  \ ,
    \end{gathered}
\end{equation}
where $V_{M^3}$ is the volume factor of $M^3$ (see \eqref{eq:pullback}). We will analyze the external system here and check the internal solution in Appendix \ref{internal-solution}. There, we find that the brane can wind around the internal $S^2\subset T^{1,1}$ an integer number of times which is expected from the homotopy group $\pi_2(S^2)=\mathbb{Z}$, and that the allowed set of internal solutions gives rise to the brane configurations in $T^{1,1}$ discussed in Section \ref{T11_symmetry_operators_subsubsec}.

\begin{figure}[t!]
    \centering
    \includegraphics[width=0.9\linewidth]{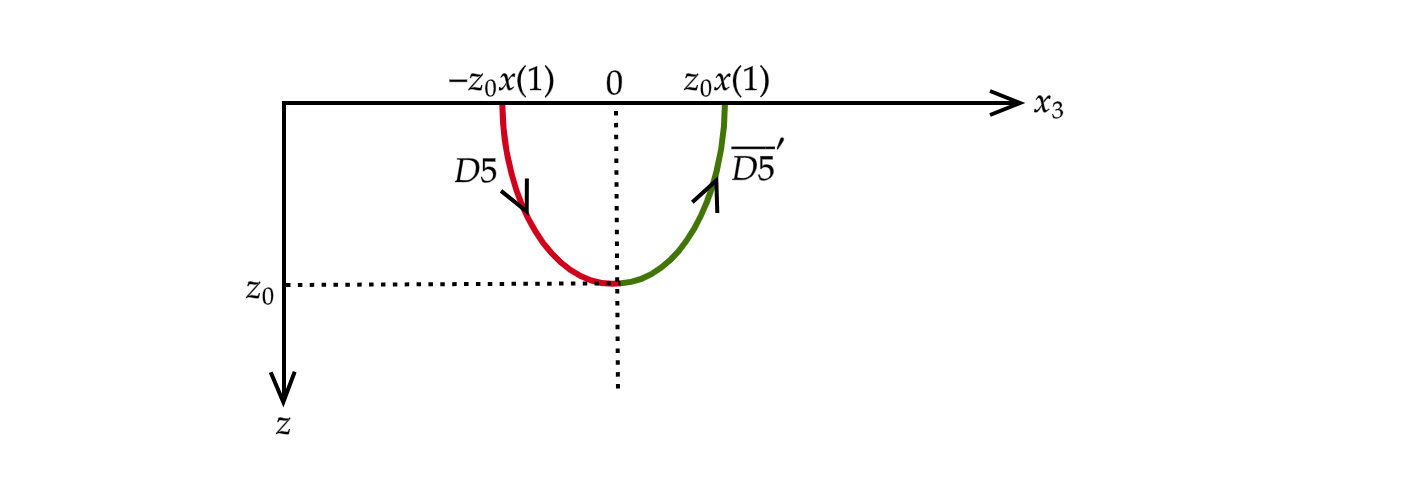}
    \caption{A D5-brane meets a $\overline{\text{D5}}$-brane at $z=z_0$, forming the hanging brane profile. The downward and upward arrows respectively label a brane and an antibrane.}
    \label{profile}
\end{figure}

Let us parametrize $z=z_0 t$ for some arbitrary constant $z_0$, and evaluate the equation of motion for $x^3(t)$ as
\begin{equation}\label{eq:derxh}
   \frac{dx^3}{dt}  =  \pm z_0 \, \frac{t^4}{\sqrt{1-t^8}} \ .
\end{equation}
This imposes that $t \in [0,1]$, where $t=1$ corresponds to $z=z_0$. There are two branches of solutions for $x^3(t)$ that we can write as
\begin{equation}
    x^3_\pm(t) =\pm z_0 \left(x(1) - x(t) \right) \ , \qquad x(t)=\frac{1}{5} \, t^5 \, _2F_1\left(\frac{1}{2},\frac{5}{8};\frac{13}{8};t^8\right) \ ,
\end{equation}
where $_2F_1(-,-;-;-)$ denotes the hypergeometric function. The solutions $x^3_\pm(t)$ start at $\pm z_0 \, x(1)$ when $t=0$ and ends at $0$ when $t=1$, and these two strands connect at $(x^3,z)=(0,z_0)$ to yield a continuous curve for the brane on $\gamma^1$. Figure \ref{profile} illustrates this profile.

We rescale the parameter $t$ to construct a piecewise solution for the hanging D5-brane,
\begin{equation}\label{eq:gammasol}
    x^3_h(t) = \begin{cases} z_0 \, x^3_-(2t) \\ z_0 \, x^3_+(2(1-t)) \end{cases}, \quad z(t) = \begin{cases} 2z_0 t  \\ 2z_0 (1-t)  \end{cases} \quad \mbox{for} \quad \begin{cases}  0\leq t \leq \frac{1}{2} \\ \frac{1}{2}\leq t \leq 1 \end{cases}.  
\end{equation}
Note that the two strands, $x^3_\pm(t)$, can be understood as the profiles of a D5-brane (with $x^3_-(t)$) and an $\overline{\text{D}5}$-brane (with $x^3_+(t)$) which recombine into a single hanging D5-brane at the turning point. The separation between the two branes, or equivalently, the end points of the recombined brane, on the conformal boundary is given by $\Delta x^3 = 2z_0 \, x(1)$.

This interpretation of the two strands as a D5-$\overline{\text{D5}}$ pair can be studied directly by considering the dynamics of two such branes hanging in $AdS_5$.  
The general action for the brane-antibrane tachyonic system is discussed in Appendix \ref{Brane_dynamics_appendix}. Here, when the two branes come close, there is a complex tachyon,  $\tau = \rho e^{i\varphi}$ with $\rho$ its radius and $\varphi$ its phase, that condenses. The attractive force of the two branes vanishes near the boundary and increases with $z$ in the bulk. As a result, the branes recombine at $z=z_0$ and annihilate in the bulk to form the U-shape.  Solving for the hanging brane from the brane-antibrane system, the phase of the tachyon needs not be zero. It fixes the diagonal mode of the Chan-Paton fields of the two branes as
\begin{equation}
\label{diagonal}
   a_1= a_L-a_R = d \varphi \ .
\end{equation}
Importantly, it is a flat connection on the hanging brane whose parallel transport along the arc $\gamma^1$ yields a $U(1)$-valued parameter $\alpha$ as in \eqref{eq_a1_integral}.

\subsection{Fluctuations from metric dependence}\label{subsec:brane_fluctuations}

Now we wish to examine the fluctuations of the hanging brane solution in \eqref{eq:gammasol} induced by the choice of a metric $h_{ab}$ on $M^3$. We parametrize the profile of the brane as
\begin{equation}
    x^3 = x^3_h(t) + s(\sigma^a,z) \ ,
\end{equation}
where $s(\sigma^a,z)$ corresponds to the fluctuations away from the hanging brane background $x^3_h(t)$, and can depend on the worldvolume coordinates $\sigma^a$ and the $AdS_5$ radius $z$. The induced metric on the brane in $AdS_5$ is then
\begin{equation}
    ds^2_{O} =\frac{L^2}{z^2} \left[\left(h_{ab} + \partial_a s \, \partial_b s\right) dx^a dx^b + \left(1 + \left(\partial_z x^3_h + \partial_z s\right)^2 \right) dz^2 \right] \ ,
\end{equation} 
and the Lagrangian for the fluctuations reduces as 
\begin{equation}
    \mathcal{L}_O = \sqrt{h} \sqrt{1+ \left(\partial_z x^3_h\right)^2} \left[1+ \frac{1}{2} \, h^{ab} \partial_a s \, \partial_b s + \frac{2 \partial_z x_h^3 \, \partial_z s + (\partial_z s)^2}{2(1+ \left(\partial_z x^3_h\right)^2)} \right] \ .  
\end{equation} 
We can evaluate the action for the fluctuations by plugging in the solution for $x_h^3$.  Since $\partial_z x_h^3$ is odd across $\gamma^1$ from \eqref{eq:derxh}, the linear $\partial_z s$ term will drop out.  We obtain
\begin{equation}
    S_O = \int \frac{L^4}{z^4}  \left[\frac{z_0^4}{\sqrt{z_0^8-z^8}} \left(1+ \frac{1}{2} \, h^{ab} \partial_a s \, \partial_b s\right) + \frac{\sqrt{z_0^8-z^8}}{2z_0^4} \, (\partial_z s)^2 \right] dz \, dV_{M^3} \ .  
\end{equation} 

For simplicity, let us assume that the fluctuations $s(\sigma^a,z)$ do not depend on $z$, then we can isolate the metric dependence and have the action reduce as
\begin{equation}\label{eq:flaction}
    S_O = \int dV_{M^3} \left[ \Lambda_0 + \frac{1}{2 g^2} \,h^{ab} \partial_a s \, \partial_b s \right] \ ,
\end{equation}
where the constants $\Lambda_0$ and $g^2$ are defined as
\begin{equation}
    \Lambda_0 = \frac{1}{g^2} \coloneqq 2 L^4 \int_{ z_*}^{z_0} \frac{z_0^4 dz}{z^4\sqrt{z_0^8-z^8}} = \frac{2 L^4}{3} \left[\frac{\, _2F_1\left(-\frac{3}{8},\frac{1}{2};\frac{5}{8};(\frac{z_*}{z_0})^8\right)}{z_*^3} -\frac{1}{z_0^3} \frac{\sqrt{\pi } \, \Gamma \left(\frac{5}{8}\right)}{ \Gamma \left(\frac{1}{8}\right)}  \right] \ .
\end{equation}
In evaluating the integral, we imposed a cutoff $z_*<z_0$ which we can take to zero.  We can physically interpret $z_*$ as moving the operator slightly inside the $AdS_5$ space to $z=z_*$. Note that in \eqref{eq:flaction}, $\Lambda_0$ mimics a cosmological constant on the brane, while the second term is the kinetic term of the fluctuations.

\begin{figure}[t!]
    \centering
    \includegraphics[width=0.9\linewidth]{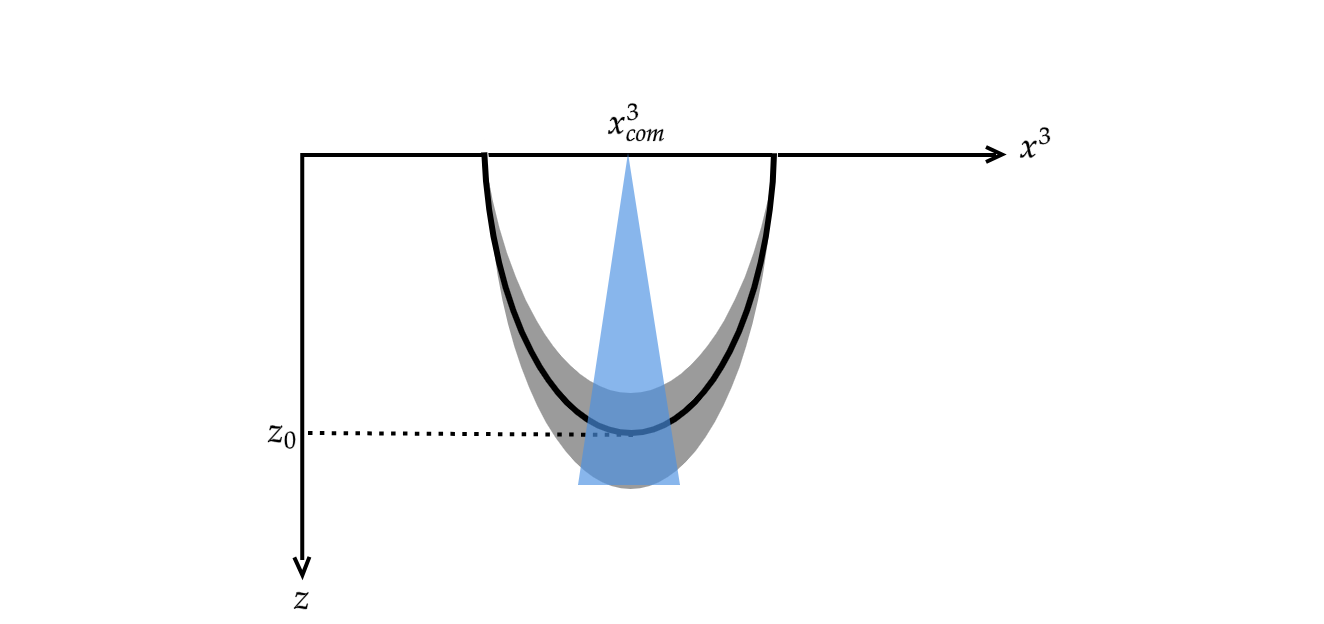}
    \caption{The D5-brane hangs from the conformal boundary and extends into the bulk. The brane is frozen and anchored at the boundary, while the gray shaded region represents the brane fluctuations away from its worldvolume. These fluctuations grow as we go deeper into the bulk. The blue shading represents the brane center-of-mass position, $x^3_{com}$, which is fixed at $z=0$, showing the topological nature of the operator at the conformal boundary. Additionally, the center-of-mass fluctuations grow in the bulk, where it is no longer topological.}
    \label{dynamicsfig}
\end{figure}

The action \eqref{eq:flaction} describes the fluctuations of the center-of-mass mode of the operator. This term is non-topological due to its dependence on the metric of $M^3$. The parameter $g$ is the coupling of the fluctuations. We can consider what happens to the operator when we move it near the boundary, i.e.~$0 \approx z_* \ll z_0$, or when we move it further into the bulk, $z_* \approx z_0$. For these cases we have
\begin{equation}
    \frac{1}{g^2} = \begin{cases} \displaystyle \frac{2 L^4}{3} \bigg(\frac{1}{z_*^3}-\frac{1}{z_0^3} \frac{\sqrt{\pi } \, \Gamma \left(\frac{5}{8}\right)}{\Gamma \left(\frac{1}{8}\right)}\bigg) \, , & 0 \approx z_* \ll z_0 \\[2ex]
    \displaystyle 2 L^4 \bigg(\frac{1}{z_0^{7/2}} \sqrt{\frac{z_0-z_*}{2}}\bigg) \, , & z_* \approx z_0\end{cases} \ .
\end{equation} 
We observe that as the operator is pushed to the boundary where we expect to recover the topological operator, the coupling $g$ approaches $0$ and the center-of-mass mode decouple. Indeed for this case, we observe that the dynamics of the hanging brane does not depend on the metric of $M^3$. See Figure \ref{dynamicsfig}.

On the other hand, when we push the operator into the bulk and near $z=z_0$, the coupling blows up, indicating a strongly coupled regime where the fluctuation description is no longer valid.  This is the regime where the dynamics of the tachyon of the brane and antibrane is important. 

For the rest of the analysis, we will set the fluctuations $s(\sigma^a, z)$ of the center-of-mass mode to zero.

\subsection{Dynamics of brane fusion}
\label{dynamic_fusion}

\begin{figure}[t!]
    \centering
    \includegraphics[width=0.9\linewidth]{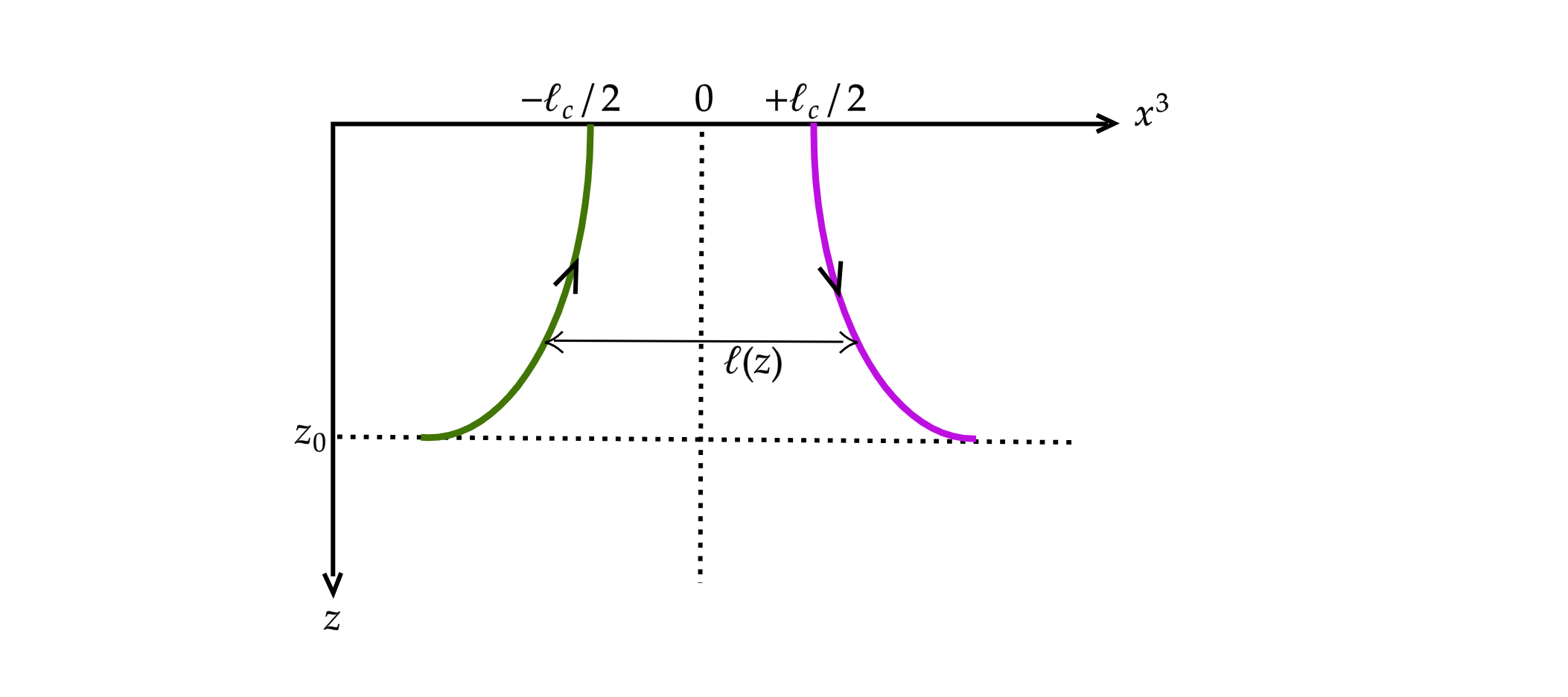}
    \caption{The inner brane-antibrane system.}
    \label{interact3}
\end{figure}

Let us turn our attention to the dynamics of the fusion of hanging branes, whose topological aspects were discussed in Section \ref{top_fusion}. This analysis is carried out in detail in Appendix \ref{Brane_dynamics_appendix}. We study the system where two hanging branes are separated by a distance $2\ell_c$ on the conformal boundary as illustrated in Figure \ref{interact3}. For our purposes, it suffices to focus on the profiles of the two inner antibrane/brane, given respectively by
\begin{equation}
    x^L_3(z) = -\frac{\ell_c}{2} - z_0 x(z) \ , \qquad x_3^R(z) = \frac{\ell_c}{2} + z_0 x(z) \ ,
\end{equation}
where
\begin{equation}
    x(z) = \frac{1}{5} \bigg(\frac{z}{z_0}\bigg)^{\!5} {_2F_1}\bigg(\frac{1}{2},\frac{5}{8};\frac{13}{8};\bigg(\frac{z}{z_0}\bigg)^{\!8}\bigg)
\end{equation}
is obtained by solving the parametric equations \eqref{eq:gammasol}. We are then interested in studying the behavior of the separation
\begin{equation}\label{eq_sepbranes}
    \ell_0(z) \coloneqq x_3^R(z) - x_3^L(z) = \ell_c + 2 z_0 x(z)\ .
\end{equation}
In general, one may also consider fluctuation modes of the branes as in Section \ref{subsec:brane_fluctuations}, but we will hereafter restrict to the hanging brane background solution for simplicity.

We now proceed to show that the dynamics induced by the tachyon on the brane-antibrane system reproduces the brane fusion discussed in Section \ref{top_fusion}. In the following, we are going to study the problem in terms of two different dynamical processes:
\begin{itemize}
    \item as the oppositely oriented inner branes are brought very close to each other, tachyon condensation occurs and produces a D4-brane on the conformal boundary, which can then be re-opened to hanging D5-brane;
    \item we consider the scenario where the two inner branes intersect with each other in the $AdS_5$ bulk (i.e.~the two hanging branes overlap partially but are not coincident), in which case the tachyon induces a brane recombination.
\end{itemize}
At the end of both these processes, the resultant configuration is a stack of two hanging branes with the same orientation. Its center-of-mass mode thus admits a Chan-Paton gauge field which has the desired parallel transport $\alpha_\text{COM} = \alpha + \alpha'$ as anticipated in \eqref{eq:com_holonomy}.

\subsubsection{Touching branes}

The analysis of the dynamics corresponds to keeping track of the tachyonic mode when the two strands are close to each other.  We can evaluate the action and the equation of motion of the tachyon as discussed in Appendix \ref{Brane_dynamics_appendix}.  The complex tachyon is parametrized as $\tau = \rho(x_0,z) e^{i \varphi}$, where the phase couples to the relative gauge field of the two strands, and can be fixed as in \eqref{diagonal}.  With that ansatz and the action expanded to two derivatives, the phase of the tachyon and the gauge fields on the branes decouple from the radius of the tachyon.   

The dynamics of $\rho$ is governed by a Schrödinger problem.  We can then consider a separable solution for the tachyon profile of the following form,
\begin{equation}
    \rho(z,x_0) = z  B(x_0) P(z) \ , \qquad \mbox{with} \qquad \partial_0^2 B + m^2 B =0 \ ,
\end{equation}
where $x_0$ is time and $m^2$ is the mass of the tachyon.  We characterize the time dependence with $B(x_0)$, the eigenfunction of $\frac{\partial^2}{\partial x_0{}^2}$ with eigenvalue given by $-m^2$. 

The radial part $P(z)$ of the wavefunction satisfies a Schr\"{o}dinger-like equation in terms of some $y$-coordinate\footnote{It is explicitly expressed as $y(z)= z \, _2F_1\big(\frac{1}{8},\frac{1}{2};\frac{9}{8};\big(\frac{z}{z_0}\big)^8\big)$.} given by
\begin{equation}\label{Schrodinger_equation}
    -\partial_y^2 P(y) + V_\text{Sch} P(y) = m^2 P(y) \ , 
\end{equation} 
where the Schrödinger potential is
\begin{align}\label{eq:SchPote}
    V_\text{Sch} = \frac{M^2 \ell_0^2(z)}{z^4 F^2(z)}  - \frac{\pi M}{z^2} + \frac{2z_0^8 + z^8}{z^2z_0^8} \ , \qquad F(z) = \frac{dy}{dz} = \frac{z_0^4}{\sqrt{z_0^8-z^8}}\ .
\end{align}
Here, $z_0$ sets the strong coupling scale of the problem, with $y \approx z$ for $z \ll z_0$. The problem is also controlled by a dimensionless parameter
\begin{equation}
    M = \frac{L^2}{\ell_s^2} \gg 1 \ ,\label{M_definition}
\end{equation}
whereas the separation $\ell_0(z)$ is defined in \eqref{eq_sepbranes}.

\begin{figure}[t!]
\label{PotePlots}
     \centering
     \begin{subfigure}{0.45\textwidth}
         \centering
         \includegraphics[width=\linewidth]{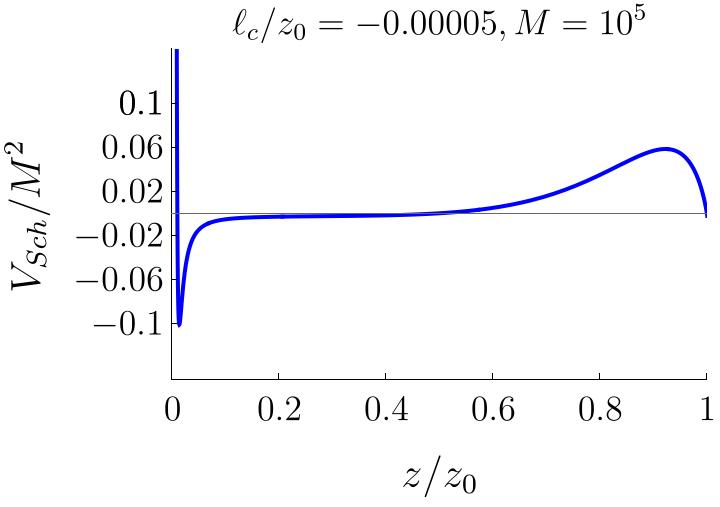}
         \caption{}
         \label{fig:1a}
     \end{subfigure}
     \begin{subfigure}{0.45\textwidth}
         \centering
         \includegraphics[scale=0.5]{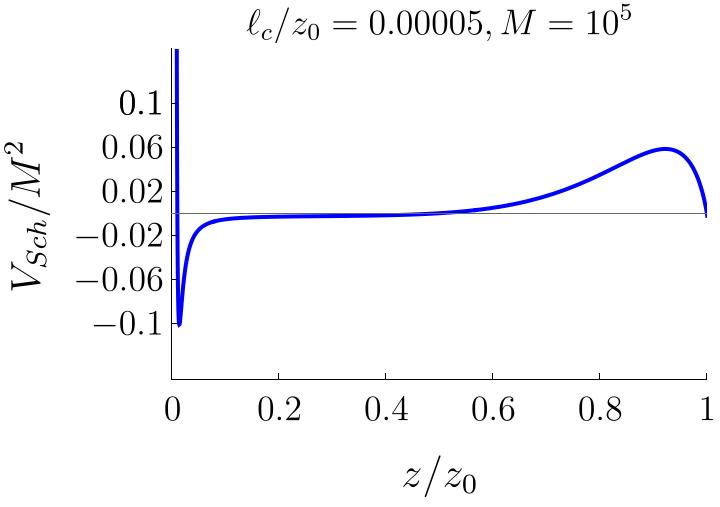}
         \caption{}
         \label{fig:1b}
     \end{subfigure}
         \begin{subfigure}{0.45\textwidth}
         \centering
         \includegraphics[width=\linewidth]{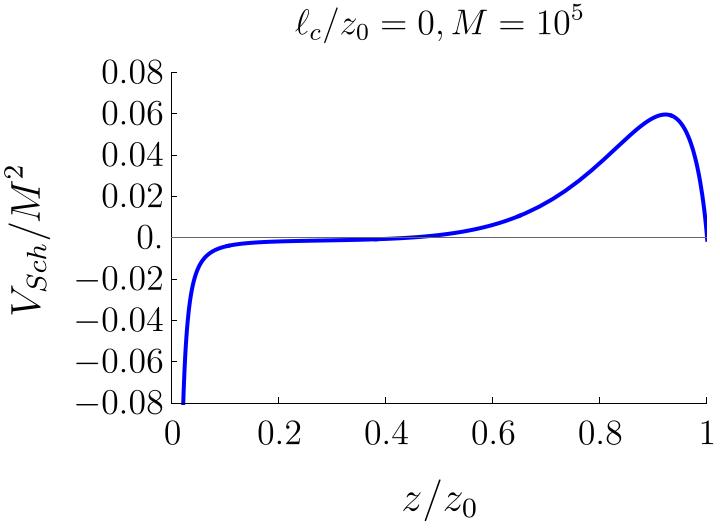}
         \caption{}
         \label{PotePlots}
     \end{subfigure}
     \caption{The Schrödinger potential $V_\text{Sch}$ plotted as a function of $z/z_0$. In (a) the branes are slightly intersecting, in (b) they are very close to touching, and in (c) they just touch but do not intersect. The potential is bounded from below in (a) and (b) but goes to $-\infty$ as $z\rightarrow 0$ in (c).}
     \label{potentialplot}
\end{figure}

We plot the potential in Figure \ref{potentialplot} for very small $\ell_c \ll z_0$ and for $\ell_c =0$.  In the former case, we observe a very negative dip of the potential, but bounded below, in the region near $z\ll z_0$.  This indicates  a discrete spectrum of tachyonic modes. Meanwhile, in the limit where $\ell_c =0$, the dip is unbounded from below, signaling a continuum of tachyonic modes.

\begin{figure}[t!]
    \centering
    \includegraphics[width=0.9\linewidth]{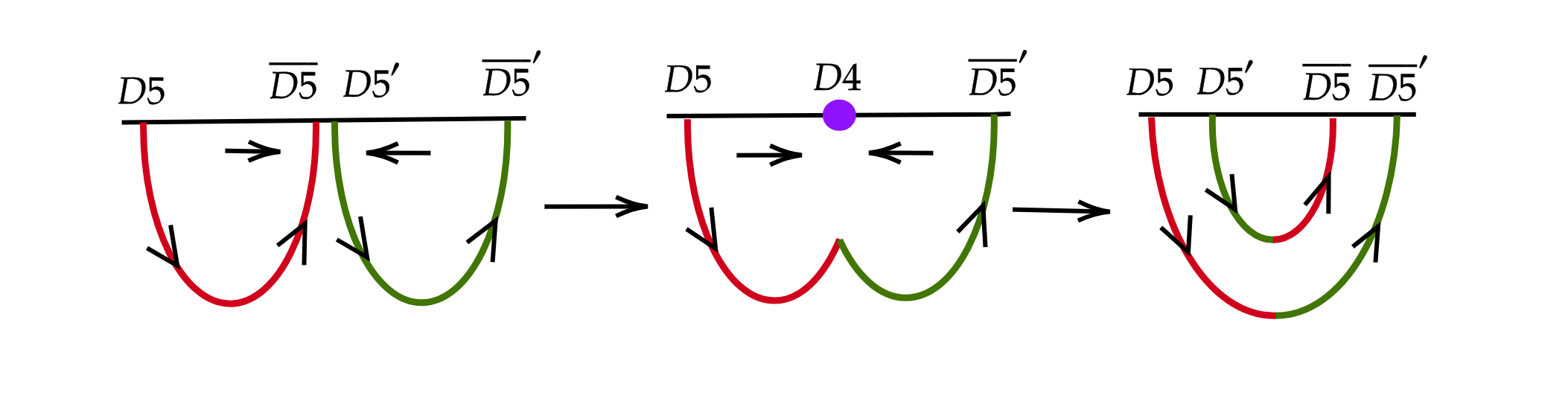}
    \caption{The $\overline{\text{D5}}$-$\text{D5}'$ system condenses into a D4-brane near the boundary when they are brought close together. Continuing to push the hanging branes closer to each other means that we reopen the D4-brane to form the $\text{D5}'$-$\overline{\text{D5}}$ system, giving us a stack of two hanging branes compatible with our desired fusion rules.}
    \label{fusionneww}
\end{figure}

Because tachyonic modes appear when the branes approach each other, there is tachyon condensation that leaves a remnant D4-brane \cite{Sen_1998} on the conformal boundary where the D5-brane and $\overline{\text{D5}}$-brane would meet. This happens as soon as a tachyonic mode exist and before the branes would touch. As we continue to push the boundary topological operators closer to each other, this D4-brane reopens with the orientation of arrows pointing  as illustrated in Figure \ref{fusionneww}.  This orientation is crucial for correctly reproducing the fusion structure in field theory as discussed in Section \ref{top_fusion}.

It is instructive to look at the scale of $\ell_c$ at which the tachyonic modes studied above start to appear. Recall that $\ell_c$ denotes the separation, along the conformal boundary ($z=0$), between the two inner branes. The Schrödinger potential in the limit $z/z_0\ll 1$, is given by
\begin{equation}
\label{eq:approx_Sch}
     V_\text{Sch} \approx  \frac{M^2 \ell_c^2}{z^4}  - \frac{\pi M}{z^2} + \frac{4\ell_cM^2 z}{5z_0^4} \ .
\end{equation}
The turning point of this potential is approximated as the position where the $1/z^2$ term starts dominating the $1/z^4$ term in \eqref{eq:approx_Sch} for $z/z_0\ll 1$, i.e.
\begin{equation}
\label{eqTurningPt}
    \frac{M^2\ell_c^2}{z^4}=\frac{\pi M}{z^2}\Rightarrow z \sim \ell_c M^{1/2} \ .
\end{equation}
We note that the supergravity description inherent in our analysis is valid in the region $\ell_s \ll \ell_c\ll z_0\ll L$, where $L$ denotes the AdS radius. Let us check if the region where the tachyon appears can be consistently described by supergravity. In the calculation of the turning point in the region $z/z_0\ll 1$, given in \eqref{eqTurningPt}, each of the first two terms of the potential \eqref{eq:approx_Sch} dominated the third term. One can show that this is equivalent to the following condition, 
\begin{equation} \label{eq_lc_ineq}
    1 \ll M^{-5/8}z_0 \ell_c^{-1} = \bigg(\frac{z_0}{L}\bigg)^{5/4} \bigg(\frac{\ell_c}{z_0}\bigg)^{1/4} \bigg(\frac{\ell_s}{\ell_c}\bigg)^{5/4} \ ,
\end{equation}
using \eqref{M_definition}. However, we notice that each of the factors on the RHS is much smaller than 1, so the inequality \eqref{eq_lc_ineq} is not compatible with the hierarchy of scales $\ell_s \ll \ell_c\ll z_0\ll L$ required for the validity of the supergravity approximation. Given this apparent problem, one might wonder if we can trust the brane fusion picture we have just established. To answer this question, we turn to a slightly different perspective which sidesteps the issue at hand. We will look at branes that are already intersecting at some  point $z= z_* < z_0$.

\subsubsection{Intersecting branes}

\begin{figure}[t!]
    \centering
    \includegraphics[width=0.9\linewidth]{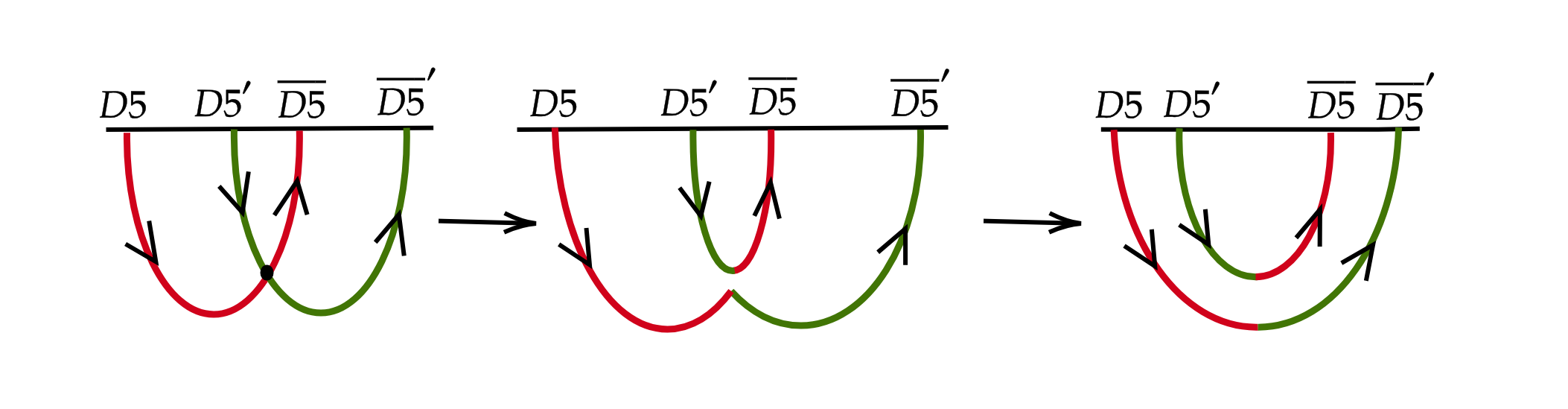}
    \caption{The $\overline{\text{D5}}$-$\text{D5}'$ system annihilates at $z=z_*$, where the branes intersect and recombine into a pair of stacked branes. Like in Figure \ref{fusionneww}, the distance between the boundary anchor points of the $\text{D5}$-$\overline{\text{D5}}$ system and $\text{D5}'$-$\overline{\text{D5}}'$ system are preserved independently.}
    \label{reco2}
\end{figure}

When two hanging branes intersect at $z=z_*$, we can, in this region, express
\begin{equation}
    \ell(z)=\gamma(z_*-z) \ , \qquad \gamma=\frac{2z_*^4}{\sqrt{z_0^8-z_*^8}} \ .
\end{equation}
Let us rewrite \eqref{Schrodinger_equation} as 
\begin{equation}\label{rescaled_Schrodinger_equation}
    \partial_u^2 \widetilde{P}(u) + V_\text{eff} \widetilde{P}(u) = 0 \, , \qquad u = z_* - z \ ,
\end{equation}
where we have defined the rescaled tachyonic mode $\widetilde{P}(z)=P(z)/\sqrt{F(z)}$, with $F(z)= \frac{z_0^4}{\sqrt{z_0^8-z^8}}$. The effective potential $V_\text{eff}$ is given by
\begin{equation}\label{eq:Pote}
    V_\text{eff} =  \left(m^2 + \frac{\pi M }{z^2_*} - \frac{2}{z^2_*} \frac{z_0^8 -7z^8_*}{z_0^8 -z^8_*}\right) \frac{z_0^8}{z_0^8-z^8_*} - \frac{M^2 \gamma^2 u^2 }{z^4_*} \coloneqq \beta - \alpha^2 u^2 \ .
\end{equation}

We obtain a tachyonic mode (i.e.~$m^2 < 0$) in the spectrum when the (rescaled) radial part of the wavefunction is localized and takes the form $\widetilde{P} = \exp(-q u^2 + c)$. Plugging this ansatz into \eqref{rescaled_Schrodinger_equation}, one finds that $\beta = \alpha = 2q$, and
\begin{equation}
    m^2 = -\frac{M}{z_*^2(z_0^8-z_*^8)} \left( \pi z_0^8 - 2 z_*^4 \sqrt{z_0^8-z_*^8} \right) + \frac{2}{z^2_*} \frac{z_0^8 -7z^8_*}{(z_0^8 -z^8_*)^2} < 0 \ .
\end{equation}
The tachyonic instability results in the branes breaking off and recombining at $z=z_*$, where they intersect, via brane-antibrane annihilation \cite{Hashimoto_2003} as shown in Figure \ref{reco2}. Note that in this picture, unlike earlier, we are not probing the region where $\ell_c$ is infinitesimally small and the supergravity description breaks down. More importantly, the analysis here serves as a consistency check on our earlier interpretation that pushing the branes towards each other leads to the opening up of the D4-brane with orientation as in Figure \ref{fusionneww}.

\section{Discussion and outlook}
\label{sec_conclusion}

We have expanded on our proposal in \cite{Bah:2025vfu}, which stated that, in $AdS_5\times S^5$, the $SO(6)$ symmetry operators on the conformal boundary are described by hanging D5-KK bound states in the bulk, to general holographic backgrounds. In particular, we have shown how this framework is realized in $AdS_5\times SE_5$ with $T^{1,1}$ as an example of an $SE_5$, in addition to $S^5$ which we showed in \cite{Bah:2025vfu} and repeated here for completeness. We also described the framework giving rise to symmetry operators for M-theory in $AdS_4\times S^7$, and then had brief discussions for general $AdS_4\times SE_7$ with $M^{3,2}$ and $Q^{1,1,1}$ as examples of $SE_7$ in addition to the aforementioned $S^7$. Starting from the dynamical action of the branes, we demonstrated how the hanging brane profile arises as a solution to the equations of motion. Additionally, we saw that the action of the hanging branes are effectively topological and stuck to the conformal boundary.  We then showed how the dynamics of the branes govern the topological fusion of the symmetry operators. 

As explained in Section \ref{sec:hanging_branes_intro}, the hanging brane construction is universal in generic AdS spaces. The low-energy behavior is uniform across the Type II string theory and M-theory examples we looked at, even though the details differ in each scenario. For instance, consider M-theory on $AdS_7\times S^4$ (an example not covered in this paper in detail). We know from the universal low-energy behavior in Section \ref{sec:hanging_branes_intro} what the operators should look like, but it is still unclear what the detailed high energy setup is. One could try to look at Type IIA string theory, where we know that a Hanany-Witten transition between a D0-brane and a D8-brane, for example, can happen, and then uplift this setup to M-theory on $AdS_7\times S^4$. However, those branes do not have simple lifts to M-theory. There are several ways to possibly attack this question, and we leave this as an open problem for future work. 

It is curious to ask if hanging branes can also be used to realize discrete symmetries. However, in the case of discrete symmetries, the arguments of \cite{Bah:2024ucp} that called for regularization of operators do not apply, therefore, we do not (necessarily) need to use hanging branes to regularize the operators. Nevertheless, it would be interesting to see if hanging branes can still be used there. Another related question is whether we can use hanging branes to realize higher-form symmetries. For example, in the case of 1-form symmetries, the charged objects would then need to be realized by 2d Wilson surfaces and the symmetry operators by $(d-2)$-dimensional objects in the $AdS_{d+1}$ bulk. It will be interesting to see how the hanging branes can be used to reproduce this structure.

We motivated the hanging branes by noting that we need to regularize our symmetry operator by making it have a finite width at the boundary. In the presence of gravity, it was found in \cite{Bah:2024ucp} that this thickening cannot be applied consistently because there is an obstruction in taking the limit where it goes to zero, which means that the operator becomes a dynamical object with a finite width and a finite effective tension. This naturally accommodates our construction where this dynamical object is a hanging brane.\footnote{The fate of symmetry operators in holography is also discussed in \cite{Heckman:2024oot}.}
Another interesting question is how to relate our framework to the SymTFT construction \cite{Ji:2019jhk,Gaiotto:2020iye,Apruzzi:2021nmk,Freed:2022qnc}. Recently, there have been several proposals regarding the SymTFT for continuous symmetries \cite{Brennan:2024fgj,Antinucci:2024zjp,Bonetti:2024cjk,Antinucci:2024bcm, Arbalestrier:2025poq,Bonetti:2025dvm, Jia:2025jmn}. However, none of these feature operators hanging in the bulk akin to our construction, and since we know some sort of regularization is always needed \cite{Bah:2024ucp}, it would be interesting to see if and how one could regularize in the SymTFT picture, and whether there would be a relation to our construction.


\acknowledgments

We would like to thank Oren Bergman, Victor Carmona, Iñaki García Etxebarria, Patrick Jefferson, Francesco Mignosa, Konstantinos Roumpedakis, and Thomas Waddleton for useful discussion and correspondence. IB, MC are supported in part by the Simons Collaboration on Global Categorical Symmetries and also by the NSF grant PHY-2412361. FB is supported by the Program ``Saavedra Fajardo'' 22400/SF/23 from Fundaci\'on S\'eneca de la Regi\'on de Murcia. FB also acknowledges support from Fundaci\'on S\'eneca de la Regi\'on de Murcia FSRM/10.13039/100007801(22581/PI/24), Espa\~na.


\appendix
\addtocontents{toc}{\protect\setcounter{tocdepth}{1}}


\section{Action for  Wilson lines}
\label{Wilson_quantization_appendix}

In this appendix, we review the derivation of the quantization condition of the parameter $q$ in the Wilson loop operator \eqref{eq:Wilson_1d_action}, repeated below for convenience,
\be \label{appeq:Wilson_1d_action}
    S_{\rm WL} = \int_{M^1} \langle q , \chi^{-1} d\chi + \chi^{-1} A \chi \rangle \ .
\ee
The action is invariant under bulk gauge transformations
\begin{equation}
    A \to g A g^{-1} + g dg^{-1}
\end{equation}
for any $G$-valued 0-form $g$ in $AdS_{d+1}$, accompanied by a left action on the scalar field,
\begin{equation}
    \chi \to g \chi \, .
\end{equation}
In addition, it is also invariant under local gauge transformations of $\chi$. Explicitly, under a right action,
\be \label{eq_Wilson_right_action}
    \chi \to \chi h
\ee
for some 0-form $h$ valued in the subgroup $S_q \subset G$ that stabilizes $q$, i.e.
\be \label{eq:stabilizer_subgroup}
    S_q = \{ h \in G | {\rm Ad}^*_g q = q \} \ ,
\ee
the action \eqref{appeq:Wilson_1d_action} transforms as
\be
    S_{\rm WL} \to S_{\rm WL} + \int_{M^1} \langle q , h^{-1}dh \rangle \ .
\ee
Therefore, for $e^{i S_{\rm WL}}$ to be invariant, the parameter $q\in \mathfrak g^*$ must be quantized to ensure 
\be \label{eq_quantization_condition}
    \int _{M^1}\langle q , h^{-1}dh \rangle \in 2\pi  \mathbb Z
\ee
for any $S_q$-valued 0-form $h$ on $M^1$. The notation ${\rm Ad}^*_g$ in \eqref{eq:stabilizer_subgroup} stands for the co-adjoint action of $G$ on $\mathfrak g^*$, defined by the relation
\be 
    \langle {\rm Ad}^*_g q , X \rangle =  \langle q , {\rm Ad}_{g^{-1}} X \rangle \ , 
\ee 
where $X\in \mathfrak g$ and ${\rm Ad}_g X$ is the adjoint action of $G$ on $\mathfrak g$ (i.e.~${\rm Ad}_g X = gXg^{-1}$ if $G$ is a matrix Lie group). The physical phase space of the worldline degrees of freedom is thus $G/S_q$.

It follows that the Wilson loop depends on the constant parameter $q$ only via its co-adjoint orbit. It is also useful to recall that, for $G$ a compact connected Lie group, co-adjoint orbits are in one-to-one correspondence with elements in the fundamental Weyl chamber inside the Cartan subalgebra of $\mathfrak g$
(see e.g.~\cite[Chapter 5, Lemma 3]{kirillov2004lectures}). To see this explicitly, note that given a choice of $q$, one can always conjugate it to a representative (of the co-adjoint orbit) in the fundamental Weyl chamber, i.e.~$q \to {\rm Ad}^*_{g_0}q$ for some suitable constant $g_0 \in G$, by means of a field redefinition,
\begin{equation}
    \chi \to \chi g_0 \ ,
\end{equation}
in the action \eqref{appeq:Wilson_1d_action}.\footnote{Note that the statement is different from that around \eqref{eq_Wilson_right_action}. Here $g_0$ is not necessarily valued in $S_q$, but it is also constant such that $dg_0=0$.}

Let us consider in more detail the quantization condition \eqref{eq_quantization_condition}. As per the previous paragraph, we assume that $q$ takes values in the fundamental Weyl chamber. The stabilizer subgroup $S_q$ always contains the maximal torus $T$ of $G$ as a subgroup, so the condition \eqref{eq_quantization_condition} must hold in particular for $h$ of the form
\be \label{eq_special_h_form}
    h(\tau) = e^{2\pi \tau H} \ , \qquad H\in \mathfrak t \ , \qquad 0 \le \tau \le 1 \ , \qquad e^{2\pi H} = 1 \ .
\ee
In the above expression, $\tau$ is the parameter on $M^1$ and $\mathfrak t$ denotes the Cartan subalgebra associated to $T$. The condition $e^{2\pi H} = 1$ is imposed to ensure that   $h$ is periodic, namely $h(0)=h(1)$. It is easy to evaluate \eqref{eq_quantization_condition} for $h(\tau)$ as in \eqref{eq_special_h_form}, which yields the condition
\be
    \langle q, H \rangle \in \mathbb Z \ . 
\ee
We thus conclude that $q$ satisfies the following property: $\langle q, H \rangle$ is an integer for all $H\in \mathfrak t$ such that $e^{2\pi H} = 1$. An element $q\in \mathfrak t^*$ with this property is called analytically integral.\footnote{An integral weight is a weight which is written as an integral linear combination of fundamental weights. Every analytically integral weight is integral, but the converse does not hold in general. If $G$ is simply connected, then ``analytically integral'' and ``integral'' are equivalent.}

While studying \eqref{eq_quantization_condition} we restricted $h$ to be valued in $T$, but the equation should be satisfied for $h$ taking values in $S_q$. One may wonder if this imposes additional restrictions on $q$. In fact, it does not, as can be seen following \cite[p.~39]{Beasley:2009mb}. The quantization of $q$ stems from the classification of line bundles on the coadjoint orbit $G/S_q$. On the other hand, 
any line bundle on $G/S_q$ pulls back to a line bundle on $G/T$ under the natural projection $\pi: G/T \to G/S_q$ induced by $T \subseteq S_q$, which exhibits $G/T$ as a fiber bundle over $G/S_q$ with fiber $S_q/T$.
Thus, we only need to classify line bundles on $G/T$. The latter are classified by $H^2(G/T;\mathbb Z)$. An argument based on the Leray spectral sequence gives $H^2(G/T;\mathbb Z) \cong {\rm Hom}(T,U(1))$. The latter character group is isomorphic to the lattice of analytically integral weights.

To summarize, the parameter $q$ is a dominant analytically integral weight. (``Dominant'' means lying inside the fundamental Weyl chamber.) Let us now recall the theorem of the highest weight. If $G$ is a compact connected Lie group, then: (i) if $\mathbf R$ is a finite-dimensional irreducible representation of $G$, then $\mathbf R$ has a unique highest weight, which is dominant analytically integral; (ii) if two finite-dimensional irreducible representations have the same highest weight, they are isomorphic; (iii) for each dominant analytically integral weight $q$ there exists a finite-dimensional irreducible representation $\mathbf R$ with highest weight $q$. In conclusion, we see that the data of the irreducible representation $\mathbf R$ labeling the Wilson line is  encoded in a one-to-one way in the parameter $q$ of the action \eqref{appeq:Wilson_1d_action}.

\section{Consistency in brane fusion}\label{gluing_condition_appendix}

The gluing requirement for the worldvolume gauge fields on the brane-antibrane pair described in Section \ref{top_fusion} follows from the Mayer-Vietoris sequence on $\gamma$, which gives us a short exact sequence in relative cohomology,
\begin{multline}
    0 \to H^1(\gamma,\partial\gamma;U(1)) \xrightarrow{\rho} H^1(\gamma_L,\partial\gamma_L;U(1)) \oplus H^1(\overline{\gamma}_R,\partial\overline{\gamma}_R;U(1)) \\
    \xrightarrow{\Delta} H^1(\gamma_L \cap \overline{\gamma}_R,\partial\gamma_L \cap \partial\overline{\gamma}_R;U(1)) \to 0 \ ,\label{Mayer-Vietoris_sequence}
\end{multline}
where $\rho$ and $\Delta$ are respectively restriction and difference maps.\footnote{The sequence is originally a long exact sequence, but in our case we have two zeros that truncate it. Note that the intersection $\gamma_L \cap \overline{\gamma}_R$ is isomorphic to an interval.} The former map reproduces for us the decomposition $\rho(a_1) = a_1|_{\gamma_L} + a_1|_{\overline{\gamma}_R}$. Importantly, $a_1|_{\gamma_L}$ being a relative cocycle implies that it vanishes at both endpoints of the half-arc $\gamma_L$, and similarly for $a_1|_{\overline{\gamma}_R}$. This is simply saying that in order to cut the arc $\gamma$ into $\gamma_L$ and $\overline{\gamma}_R$, one needs to impose Dirichlet boundary conditions on $a_1|_{\gamma_L}$ and $a_1|_{\overline{\gamma}_R}$. Furthermore, $\Delta(\rho(a_1)) = 0$ by exactness, meaning that $a_1|_{\gamma_L}$ agrees with $a_1|_{\overline{\gamma}_R}$ on the overlap $\gamma_L \cap \overline{\gamma}_R$ up to gauge transformations. Analogous remarks apply to the other arc $\gamma' = \gamma_L' \cup \overline{\gamma}_R'$ on which the operator $U(\theta';M^{d-1})$ is supported.

Our goal now is to construct a new pair of hanging branes supported on the arcs $\tilde{\gamma} \cong \gamma_L \cup \overline{\gamma}_R'$ and $\tilde{\gamma}' \cong \gamma_L' \cup \overline{\gamma}_R$, by rearranging the half-arcs at hand. A clean way to analyze this problem is to construct the following diagram using the Mayer-Vietoris sequence \eqref{Mayer-Vietoris_sequence} applied on the disjoint unions $\gamma \sqcup \gamma'$ and $\tilde{\gamma} \sqcup \tilde{\gamma}'$ respectively,
\begin{equation}
    \begin{adjustbox}{max width=\textwidth}
    \begin{tikzcd}
        0 \arrow[r,"="] \arrow[d] & 0 \arrow[d] \\
        H^1(\gamma \sqcup \gamma',\partial\gamma \sqcup \partial\gamma') \arrow[d,"\rho"] \arrow[r,"f^\ast"] & H^1(\tilde{\gamma} \sqcup \tilde{\gamma}',\partial\tilde{\gamma} \sqcup \partial\tilde{\gamma}') \arrow[d,"\tilde{\rho}"] \\
        H^1((\gamma_L \sqcup \gamma_L') \sqcup (\overline{\gamma}_R \sqcup \overline{\gamma}_R'),(\partial\gamma_L \sqcup \partial\gamma_L') \sqcup (\partial\overline{\gamma}_R \sqcup \partial\overline{\gamma}_R')) \arrow[d,"\Delta"] \arrow[r,"\cong"] & H^1((\gamma_L \sqcup \gamma_L') \sqcup (\overline{\gamma}_R' \sqcup \overline{\gamma}_R),(\partial\gamma_L \sqcup \partial\gamma_L') \sqcup (\partial\overline{\gamma}_R' \sqcup \partial\overline{\gamma}_R)) \arrow[d,"\tilde{\Delta}"] \\
        H^1((\gamma_L \sqcup \gamma_L') \cap (\overline{\gamma}_R \sqcup \overline{\gamma}_R'),(\partial\gamma_L \sqcup \partial\gamma_L') \cap (\partial\overline{\gamma}_R \sqcup \partial\overline{\gamma}_R')) \arrow[d] \arrow[r,"f^\ast"] & H^1((\gamma_L \sqcup \gamma_L') \cap (\overline{\gamma}_R' \sqcup \overline{\gamma}_R),(\partial\gamma_L \sqcup \partial\gamma_L') \cap (\partial\overline{\gamma}_R' \sqcup \partial\overline{\gamma}_R)) \arrow[d] \\
        0 \arrow[r,"="] & 0
    \end{tikzcd}
    \end{adjustbox}
\end{equation}
where $f: (\tilde{\gamma} \sqcup \tilde{\gamma}',\partial\tilde{\gamma} \sqcup \partial\tilde{\gamma}') \to (\gamma \sqcup \gamma',\partial\gamma \sqcup \partial\gamma')$ is a continuous map from the pair of new arcs to the pair of old arcs, including their endpoints. The diagram is commutative if we construct $f$ such that it maps the half-arcs of the new curves to the old ones nicely in the following sense,
\begin{equation}
    \begin{gathered}
        f(\gamma_L,\partial\gamma_L) \subseteq (\gamma_L,\partial\gamma_L) \, , \qquad f(\overline{\gamma}_R',\partial\overline{\gamma}_R') \subseteq (\overline{\gamma}_R,\partial\overline{\gamma}_R) \ ,\\
        f(\gamma_L',\partial\gamma_L') \subseteq (\gamma_L',\partial\gamma_L') \, , \qquad f(\overline{\gamma}_R,\partial\overline{\gamma}_R) \subseteq (\overline{\gamma}_R',\partial\overline{\gamma}_R') \ .
    \end{gathered}
\end{equation}
Since all the half-arcs $\gamma_L,\overline{\gamma}_R,\gamma_L',\overline{\gamma}_R'$ are isomorphic to each other, we can simply impose the choice that $f$ is an isomorphism (given by permuting the half-arcs) to make the diagram commutative. In particular, this implies
\begin{equation}
     \tilde{\Delta}(\tilde{\rho}(f^\ast(a_1,a_1'))) = \tilde{\Delta}(f^\ast(\rho(a_1,a_1'))) = f^\ast(\Delta(\rho(a_1,a_1'))) = 0 \ ,
\end{equation}
which means that $a_1|_{\gamma_L}$ agrees with $a_1'|_{\overline{\gamma}_R'}$ on the overlap $\gamma_L \cap \overline{\gamma}_R' \subset \tilde{\gamma}$, and similarly $a_1'|_{\gamma_L'}$ agrees with $a_1|_{\overline{\gamma}_R}$ on the overlap $\gamma_L' \cap \overline{\gamma}_R \subset \tilde{\gamma}'$, i.e.~the gauge fields glue nicely in the new curves $\tilde{\gamma}$ and $\tilde{\gamma}'$, and are indeed restrictions of the new flat connections $(\tilde{a}_1,\tilde{a}_1') = f^\ast(a_1,a_1')$.

\section{Additional details on string/M-theory constructions}

\subsection{Construction of $\mathcal G_5$ in Type IIB}\label{app_gauging_formulae}

Here we report some formulae that are useful in the construction of $\cG_5$ in Section~\ref{subsec:AdS5SE5}.

\subsubsection{The gauging operation}

Let $\Lambda_p$ be a $p$-form on the internal Sasaki-Einstein space $X^5$,
\be \label{eq_gauging_definition}
    \Lambda_p = \frac{1}{p!} \, \Lambda_{m_1 \dots m_p} d\xi^{m_1} \wedge \dots \wedge d\xi^{m_p} \ , 
\ee 
where $\xi^m$ are local coordinates on $X^5$. The gauging operation amounts to coupling $\Lambda_p$ to the external gauge fields $A^i$ associated with isometries of $X^5$, defined such that
\be 
    (\Lambda_p)^{\rm g} = \frac{1}{p!} \, \Lambda_{m_1 \dots m_p} D\xi^{m_1} \wedge D\xi^{m_p} \ , \qquad D\xi^m = d\xi^m + K^m_i A^i \ , 
\ee 
where $K^i_m$ are Killing vectors on $X^5$.

By a direct computation one can verify the identity
\be \label{eq_general_derivative}
    d(\Lambda_p)^{\rm g} + A^i \wedge (\pounds_i \Lambda_p)^{\rm g} = (d\Lambda_p)^{\rm g} + F^i \wedge (\iota_i \Lambda_p)^{\rm g} \ . 
\ee
To check this relation, we make use of the algebra \eqref{eq_Lie_algebra_Killing} of Killing vectors and the expression \eqref{eq_field_strengths} for the field strength $F^i$ of $A^i$. In the construction of $\cG_5$,
the following applications of the general formula \eqref{eq_general_derivative} are useful, 
\be 
\ba 
    d(V_5)^{\rm g} & = F^i \wedge (\iota_i V_5)^{\rm g} \ , \\
    D (\omega_{3i})^{\rm g} := d(\omega_{3i})^{\rm g} + A^j f_{ji}{}^k \omega_{3k} & = (d\omega_{3i})^{\rm g} + F^j \wedge (\iota_j \omega_{3i})^{\rm g} \ , \\ 
    d(\omega_{3\alpha})^{\rm g} & = F^i \wedge (\iota_i \omega_{3\alpha})^{\rm g} \ . 
\ea 
\ee 
We have used the fact that $V_5$ and $\omega_{3\alpha}$ are closed and invariant under the isometries\footnote{Any harmonic 3-form is automatically invariant under all isometry directions. This can seen as follows. From $d\omega_{3\alpha}=0$ we derive $\pounds_i \omega_{3\alpha} = d(\iota_i \omega_{3\alpha})$. Making use of $\nabla_{(m} k_{i|n)} = 0$ and $\nabla^m \omega_{\alpha mnp} = 0$,
we verify $(\pounds_i \omega_{3\alpha})_{mnp} = \nabla^q (k_i \wedge \omega_{3\alpha})_{qmnp}$. We have thus established that the 3-form $\pounds_i \omega_{3\alpha}$ is both exact and co-exact. It follows that $\int_{X^5} (\pounds_i \omega_{3\alpha}) * (\pounds_i \omega_{3\alpha}) = 0$ (no sum over $\alpha$, $i$), which in turn implies $\pounds_i \omega_{3\alpha} = 0$.}, while $\omega_{3i}$ is covariant, i.e.
\be 
    \pounds_i V_5 = 0 = \pounds_i \omega_{3\alpha} \ , \qquad \pounds_i \omega_{3J} = f_{ij}{}^k \omega_{3k} \ . 
\ee 
The last relation can be verified using \eqref{eq_omega3i} and the fact that
\be 
    \pounds_i k_k = [K_i,K_k]^m g_{mn} d\xi^n = f_{ik}{}^k k_k \ .
\ee 

\subsubsection{Properties of the 3-forms $\omega_{3i}$}

Their expressions are presented in \eqref{eq_omega3i}, repeated here for convenience,
\be \label{eq_canonical_omegas2}
    \omega_{3i} = - \frac{2\pi}{ 2t {\rm Vol}(X^5)} * dk_i \ , \qquad k_i = K_i^m g_{mn} d\xi^n \ . 
\ee 
where the Hodge star here is that of the Sasaki-Einstein metric, ${\rm Vol}(X^5)$ is the volume of $X^5$,
and $t$ is the constant in the Einstein condition,
\be \label{eq_Einstein}
    R_{mn} = t g_{mn} \ . 
\ee 
In the main text we fixed $t=4$, but in this appendix we keep $t$ arbitrary. The 3-forms $\omega_{3i}$   satisfy \eqref{eq_domega3_eq} by virtue of the following property of $k_i$ \cite{Benvenuti:2006xg},
\be \label{eq_k_property}
    d*dk_i = 2t *k_i \ , 
\ee 
and of the fact that
\be 
    * k_i = {\rm Vol}(X^5)\, \iota_i V_5  \ .
\ee 
The property \eqref{eq_k_property} can be verified as follows. By unpacking the definitions of $*$ and $d$,
\be \label{eq_Einstein_derivation}
\ba 
    (* d * dk_i)_m & = \frac{1}{4!} \, (d * dk_i)^{n_1 n_2 n_3 n_4} \epsilon_{n_1 n_2 n_3 n_4 m} = \frac{1}{3!} \, \nabla^{n_1}( * dk_i)^{n_2 n_3 n_4} \epsilon_{n_1 n_2 n_3 n_4 m}\\
    & = - \nabla^{n_1}(  dk_i)_{n_1 m} = - 2 \nabla^{n_1} \nabla_{[n_1} k_{i |m]} = - 2 \nabla^{n_1} \nabla_{n_1} k_{i m} \ . 
\ea 
\ee 
In the last step we have used the Killing equation 
\be 
\nabla_m k_{in} + \nabla_n k_{im} = 0 \ . 
\ee 
Now, we consider 
\be 
\nabla_m \nabla^p k_{ip}
- \nabla^p
\nabla_m
k_{ip}
=
g^{pn}[\nabla_m , \nabla_n] k_{ip} = - g^{pn}R^q{}_{pmn} k_{iq}
= -  R^q{}_{m} k_{iq}
= - t k_{im}
\ee 
where we have used the Einstein condition 
\eqref{eq_Einstein}.
If we use once more the Killing equation
we have $\nabla^p k_{ip}=0$ and $\nabla^p \nabla_m k_{ip} = - \nabla^p \nabla_p k_{im}$. Thus, we arrive at 
\be 
\nabla^p \nabla_p k_{im} = - t k_{im} \ .
\ee 
Comparing with
\eqref{eq_Einstein_derivation}, we conclude that $*d*dk_i = 2t k_i$,
from which \eqref{eq_k_property} follows.

\subsection{$AdS_5 \times S^5$ in Type IIB}
\label{app_AdS5S5}

In this subsection we provide additional details on the realization of an $SO(6)$ symmetry operator using a hanging D5-brane and KK monopole. We review some results from \cite{Bah:2025vfu} as well as some aspects of the worldvolume theory of KK monopoles in Type IIB.

\subsubsection{Internal support of the 5-branes}

Both the D5-brane and the KK monopole have a worldvolume of the form $\gamma^1 \times M^3 \times \Sigma^2$. Let us describe the internal support $\Sigma^2 \subset S^5$.
To this end, we write $S^5$ as a Hopf fibration.
Concretely, we introduce the coordinates $(w_1, w_2, \psi)$
on $S^5$ via 
\be \label{eq_Hopf_S5}
\rho e^{i \psi} w_1 = y_1 + i y_2 \ , \qquad 
\rho e^{i \psi} w_2 = y_3 + i y_4 \ , \qquad 
\rho e^{i \psi} = y_5 + i y_6 \ . 
\ee 
Here $w_1$, $w_2$ are unconstrained complex coordinates,
$\psi$ is an angle with period $2\pi$, and the positive quantity $\rho$ is fixed by the constraint $y^a y_a = 1$ to the value 
\be 
\rho^{-2} = 1 + |w_1|^2 + |w_2|^2 \ . 
\ee 
In $(w_1, w_2, \psi)$ coordinates, the round metric on $S^5$
takes the form
\be 
ds^2_{S^5} = ds^2_{\mathbb C \mathbb P^2} + D\psi^2  \ , \qquad 
D\psi = d\psi + \cA \ . 
\ee 
Here $ds^2_{\mathbb C \mathbb P^2}$ is the standard Fubini-Study metric on $\mathbb C \mathbb P^2$
and the connection 1-form $\cA$ 
satisfies 
\be 
d\cA = 2J \ ,
\ee 
with $J$ the K\"ahler form 
  on $\mathbb C \mathbb P^2$. The 1-form $D\psi$ is the dual of the Killing vector $\partial_\psi$. As a result, $D\psi$ can be expanded onto the basis $K_{ab} = 2y_{[a} dy_{b]}$,
\be 
D\psi = \tfrac 12 m^{ab} K_{ab} \ , 
\ee 
where the expansion coefficients are antisymmetric. They are computed to~be 
\be \label{eq_m_from_psi}
m^{ab} = 
\left\{
\begin{array}{ll}
+1 & \text{if $(a,b)=(1,2)$, $(3,4)$, $(5,6)$} \ , \\
-1 & \text{if $(a,b)=(2,1)$, $(4,3)$, $(6,5)$} \ , \\
0 & \text{otherwise}  \ .
\end{array}
\right.  
\ee 
Recalling $*\omega_{ab} = \tfrac{1}{4\pi^2} dK_{ab}$ from \eqref{eq_omega_and_Killing}, we also observe that
\be \label{eq_identity_for_m_omega}
\tfrac 12 m^{ab} \omega_{ab} = \tfrac{1}{4\pi^2} *dD\psi
=  \tfrac{1}{2\pi^2} *J = \tfrac{1}{2\pi^2} J \wedge D\psi
= \tfrac{d\cA}{2\pi} \wedge \tfrac{D\psi}{2\pi} \ .
\ee 
With the notation introduced above, the internal support $\Sigma^2$ is defined as 
\be 
\Sigma^2 = {\rm PD}_{\mathbb C \mathbb P^2} \left[ N \tfrac{d\cA}{2\pi} \right] \ . 
\ee 
It follows that integrals over $\Sigma^2$ can be computed with the formula
\be \label{eq_Sigma2_integrals}
\int_{\Sigma^2} \cX_2 = \int_{S^5} \cX_2 \wedge  N \tfrac{d\cA}{2\pi} \wedge \tfrac{D\psi}{2\pi}
= N \int_{S^5} (\dots) \wedge \tfrac 12 m^{cd} \omega_{cd} \ , 
\ee 
where 
$\cX_2$ is a 2-form and 
in the last step we have made use of \eqref{eq_identity_for_m_omega}.

\subsubsection{D5-brane contribution}

The relevant topological terms in the worldvolume action for the D5-brane are captured by the following
action, written on an auxiliary manifold $N^7$ whose boundary is the brane worldvolume,
\be 
2\pi \int_{N^7} e^{f_2 - B_2} \sum_p G_p \supset 
2\pi \int_{N^7} f_2 \wedge G_5 \ . 
\ee 
Here $f_2 = da_1$ is the field strength of the Chan-Paton
gauge field on the D5-brane,
normalized to have integral fluxes.
Since $G_5$ is closed on shell,\footnote{The $AdS_5 \times S^5$ solution has vanishing $H_3$ and $G_3$ fluxes, and these are not sourced when we turn on the $SO(6)$ gauge fields.}
we can write the relevant topological coupling
as a 6d integral in the form
\be 
2 \pi \int_{\gamma^1 \times M^3 \times \Sigma^2} a_1 \wedge G_5 \ .
\ee 
As anticipated in the main text, the integral over $\gamma^1$ is saturated by a flat profile of $a_1$ with
non-trivial parallel transport
\be 
\alpha_{\rm D5} = 2\pi \int_{\gamma^1} a_1 \ .
\ee 
The relevant terms in $G_5$ are (see \eqref{eq_AdS5_S5_G5})
\be 
G_5 \supset \frac 12 N L^{-1} \frac{*F_{ab}}{2\pi} \wedge * \omega^{ab} \ . 
\ee 
The factor  of the AdS radius $L$ 
comes from the factor $L^2$ in front of the $S^5$ metric
in the 10d line element \eqref{eq_AdS5_S5_metric}.

Combining the above observations, we have
(suppressing wedge products for brevity)
\be 
\ba 
2\pi \int_{\gamma^1 \times M^3 \times \Sigma^2} a_1 G_5
& = \frac{ N \alpha_{\rm D5} }{4\pi L} \int_{M^3} *F^{ab}
\int_{\Sigma^2} *\omega_{ab}
\\
& =\frac{ N^2 \alpha_{\rm D5} }{4\pi L} \int_{M^3} *F^{ab}
\int_{S^5} *\omega_{ab} \wedge \frac 12 m^{cd} \omega_{cd} \ .
\ea 
\ee 
In the last step we have used \eqref{eq_Sigma2_integrals}. 
We proceed with the help of the identity
\be 
\int_{S^5} \omega_{ab} \wedge * \omega^{cd} = \frac{4\pi L \tau_{\rm flux}}{N^2  } \, \delta_{[a}{}^c \delta_{b]}{}^d    \ , \label{eq:omega_orthogonality}
\ee 
which may be verified by direct computation from the expression for $\omega_{ab}$ reported in \eqref{eq_AdS5_S5_G5},
and the value \eqref{eq:effective_gauge_coupling} of $\tau_{\rm flux}$.
In conclusion, the hanging D5-brane gives us the expression
\begin{equation} \label{eq_D5_piece}
    U_\text{D5} = \exp\bigg(\frac 12 i \, \alpha_\text{D5} \, \tau_{\rm flux} \, m^{ab}  \,  \int_{M^3} \ast F_{ab}\bigg) \ . 
\end{equation}

\subsubsection{Intermezzo: topological couplings on the KK monopole}

The topological couplings on the worldvolume of a KK monopole in Type IIB are given in \cite{Eyras:1998hn}.
Here we review the results of this paper, 
using descent to recast them as an 8-form anomaly polynomial.

\paragraph{Preliminaries on the KK monopole worldvolume action.}
The KK monopole of Type IIB is a soliton
with six extended worldvolume directions,
and four transverse directions. 
One of the transverse directions must be an isometry and plays a special role. 
The local line element in ten dimensions can be written as 
\be 
g_{\mu\nu} dx^\mu dx^\nu = h_{mn} dy^m dy^n
+ e^{2\omega} (d\psi + V_m dy^m)^2 \ . 
\ee 
In this section, we use $\mu$, $\nu$ for curved 10d indices, $m$, $n$ for curved 9d indices. 
A $\mu$ index is split as $\mu \to (m, \psi)$.
The isometry direction is $\psi$, associated to the Killing vector
\be 
k^\mu \partial_\mu  = \partial_\psi \ , \qquad \text{hence} \quad 
k^m = 0 \ , \qquad 
k^\psi = 1 \ .
\ee 
The coordinate $\psi$ is an angle with period $2\pi$.
The metric functions $h_{mn}$, $\omega$, $V_m$ are independent of $\psi$.
The norm squared of the Killing vector is
\be 
k \cdot k  = g_{\mu\nu} k^\mu k^\nu = e^{2\omega} \ .
\ee 
The quantity $V_m$ is interpreted as a $U(1)$ gauge field with field strength denoted
\be 
W_2 = dV_1 \ , \quad \text{where} \quad 
V_1 = V_m dy^m \ . 
\ee 
We observe that the 2-form $W_2$ has no legs along the $\psi$ direction,
\be 
\iota_k W_2 = 0 \ . 
\ee 
Here $\iota_k$ denotes the interior product with the vector $k = k^\mu \partial_\mu = \partial_\psi$, which acts on a differential form from the left.\footnote{In contrast, the authors of \cite{Eyras:1998hn} use conventions in which $\iota$ acts from the right.}
We also use the notation $W_{mn}$ for the components of  $W_2$, according to $W_2 = \frac 12 W_{mn} dy^m dy^n$.

To write down the topological couplings on the KK monopole, we need some additional notation.
Let $X^\mu(\xi)$ denote the scalar fields on the worldvolume of the KK monopole which govern its embedding into 10d spacetime. Here 
$\xi^i$, $i=1,\dots,6$, are local coordinates on the KK monopole worldvolume $\cW^6$.
Using $X^\mu$ and the Killing vector $k^\mu$, we construct the  1-forms on $\cW^6$,
\be 
A_1 = -(k \cdot k)^{-1} g_{\mu\nu}k^\mu \partial_i X^\nu d\xi^i \ , \qquad 
DX^\mu = dX^\mu + k^\mu A_1 \ .
\ee
It is useful to observe that
\begin{equation} \label{eq_KKmon_useful}
    \begin{gathered}
        A_1 =  - d X^\psi - V_m d X^m \ , \qquad dA_1 = -\tfrac 12 W_{mn} dX^m \wedge dX^n \ ,\\
        D X^m = d X^m \ , \qquad D X^\psi =  - V_m d X^m \ .
    \end{gathered}
\end{equation}
Among the topological couplings
on the KK monopole, we encounter various terms of the form
\be \label{eq_KKmon_DX_pullback}
\cX_{\mu_1 \dots \mu_p} DX^{\mu_1} \wedge \dots \wedge DX^{\mu_p} \ . 
\ee 
Here $\cX_p$ is a $p$-form in 10d spacetime.
To express \eqref{eq_KKmon_DX_pullback} in a way that is more useful to us, we proceed as follows.
Without loss of generality, the $p$-form $\cX_p$ can be parametrized  as
\be \label{eq_KKmon_param}
\cX_p = \overline \cX_p  + (-1)^{p-1} (\iota_k \cX_p) \wedge (d\psi + V) \ ,
\ee 
where $\overline \cX_p$ is defined by this equation
and satisfies $\iota_k \overline \cX_p=0$.
Combining the parametrization \eqref{eq_KKmon_param} with the relations \eqref{eq_KKmon_useful}, one can verify the identity
\be \label{eq_pullback_comparison}
\cX_{\mu_1 \dots \mu_p} DX^{\mu_1} \wedge \dots \wedge DX^{\mu_p}
= \overline \cX_{\mu_1 \dots \mu_p} dX^{\mu_1} \wedge \dots \wedge dX^{\mu_p} \ . 
\ee 
In other words, performing a modified pullback of
$\cX_p$ with the differentials $DX^\mu$ is equivelent to performing the standard pullback of $\overline \cX_p$ with the standard differentials $dX^\mu$.

Let us recall the localized degrees of freedom  living on the worldvolume of the KK monopole. They are:
\begin{itemize}
    \item Three scalars associated to the transverse directions to the KK monopole. These degrees of freedom are encoded in the scalar fields $X^\mu(\xi)$. Making use of the standard static gauge $X^i(\xi) = \xi^i$, we are naïvely left with four scalar degrees of freedom. One degree of freedom, however, is pure gauge, because it is related to the isometry direction. We refer the reader to \cite{Hull:1990ms,Bergshoeff:1997gy} for further details on this feature of KK monopole worldvolume actions. 
    \item Two compact scalars denoted $\omega_0$, $\widetilde \omega_0$.
    \item One 2-form $\omega_2$ whose field strength $\cK_3$ satisfies a self-duality condition.
\end{itemize}

\paragraph{Topological worldvolume couplings.}
We are now in a position to report the topological couplings on the KK monopole worldvolume.
They are given in equation (3.14) of \cite{Eyras:1998hn}.
In our notation, they read 
\be \label{eq_KK_terms}
\ba 
\frac{1}{2\pi \alpha'} \, \cI_6^{(0)\rm KK} & = d \widetilde \omega_5
+ \frac{1}{2\pi \alpha'} (\iota_k \cN_7)
-(\iota_k \widetilde \cB_6) \wedge d\omega_0
+(\iota_k C_6) \wedge d\widetilde \omega_0
\\
& - \overline C_4 \wedge (\iota_k C_2) \wedge d\omega_0
- \overline C_4 \wedge (\iota_k \cB_2) \wedge d\widetilde \omega_0
\\
& - \frac 12 (\iota_k C_4) \wedge \overline C_2 \wedge d\omega_0
- \frac 12 (\iota_k C_4) \wedge \overline \cB_2 \wedge d\widetilde \omega_0
\\
& - \frac 14 \, \overline C_2 \wedge \overline \cB_2 \wedge (\iota_k C_2) \wedge d\omega_0
+ \frac 14 \, \overline B_2 \wedge \overline C_2 \wedge (\iota_k \cB_2) \wedge d\widetilde \omega_0
\\
& - (2\pi \alpha') \overline C_4 \wedge d\omega_0 \wedge d \widetilde \omega_0
\\
& - \frac 12 (2\pi \alpha') d\omega_2 \wedge \cK_3
+ \frac 12 (2\pi \alpha')^2
d\omega_2 \wedge A_1 \wedge d\omega_0
\wedge d\widetilde \omega_0
\\
& + \frac 12 (2\pi \alpha')^2
\cK_3 \wedge  A_1 \wedge d\omega_0
\wedge d\widetilde \omega_0
\\
& - (2\pi \alpha')^2
d\omega_2 \wedge A_1 \wedge d\omega_0
\wedge d\widetilde \omega_0 \ . 
\ea 
\ee 
Some comments are in order. All terms in the expression above are forms on the worldvolume $\cW^6$.
The quantities $\cN_7$, $\widetilde \cB_5$, $C_6$, $C_4$, $C_2$, $\cB_2$
are bulk fields in ten dimensions, described in greater detail below.
If they appear with a bar, this is understood according to \eqref{eq_KKmon_param}.
All bulk fields (barred and unbarred) are implicitly pulled back to $\cW^6$. The implicit pullback is standard, i.e.~done with $dX^\mu$ as opposed to $DX^\mu$. We have already defined $A_1$, and we have already seen that 
the quantities $\omega_0$, $\widetilde \omega_0$, $\omega_2$, $\cK_3$ are worldvolume fields.

For illustration, we describe in some detail the ``translation'' of one of the terms of (3.14) of \cite{Eyras:1998hn}. The other terms are treated in a similar way.
The term we focus on reads
\be 
30 (DX \dots DX C^{(4)}) (\iota_k C^{(2)}) \partial \omega^{(0)} \ . 
\ee 
This is not written in differential form notation, but rather with implicit, totally antisymmetrized indices. 
We notice the pullback of $C^{(4)}$ performed with $DX^\mu$. 
To translate into our notation, we absorb numerical factors in wedge products.
The factor $(DX \dots DX C^{(4)})$ is treated using the identity \eqref{eq_pullback_comparison}, namely,
we trade the pullback of $C^{(4)}$ with $DX$ with the standard pullback of the barred version
$\overline C^{(4)}$ of $C^{(4)}$.
Moreover, we have to take into account the fact that $(\iota_k C^{(2)})$ switches sign
when we convert from their conventions, in which $\iota$ acts from the right, to our conventions, in which it acts from the left.
Finally, in this work we prefer to write the form degree as a subscript, as opposed to a superscript. 
All in all, the result of the ``translation'' is therefore
\be 
- \overline C_4 \wedge (\iota_k C_2) \wedge d\omega_0 \  . 
\ee 

Next, let us describe the bulk fields entering \eqref{eq_KK_terms}. The 2-form $\cB_2$ is the NS-NS 2-form and $\widetilde \cB_6$ is its electromagnetic dual in ten dimensions.
Similarly, $C_2$ is the R-R 2-form and
$C_6$ its dual. $C_4$ is the R-R 4-form.
The bulk field $\cN_7$ is the electromagnetic dual of the Killing 1-form $k_\mu dx^\mu$, regarded as a 1-form gauge field in ten dimensions.

Let us stress that the authors of
\cite{Eyras:1998hn} use conventions for RR fields that are different from those used in this work. Their conventions can be inferred from the bulk gauge transformations reported in (B.2) of 
\cite{Eyras:1998hn}. 
Those gauge transformations imply that the bulk field strengths are
\be 
\ba 
H_3 & = d\cB_2 \ , \\
F_3 & = dC_2 \ , \\ 
F_5 & = dC_4 + \frac 12 \, \cB_2 \wedge dC_2 - \frac 12 \, C_2 \wedge d\cB_2
 \ , \\
F_7 & = dC_6 - C_4 \wedge d\cB_2 
+ \frac 14 \, \cB_2 \wedge \cB_2 \wedge dC_2 \ , \\
H_7 & = d\widetilde \cB_6 + C_4 \wedge dC_2
+ \frac 14 \, C_2 \wedge C_2 \wedge d\cB_2 \ .  
\ea 
\ee 
As a result, the bulk  Bianchi identities read
\be 
\ba 
dH_3 & = 0 \ , &
dF_3 & = 0 \ , &
dF_5 & = H_3 \wedge F_3 \ , & 
dF_7 & = H_3 \wedge F_5 \ , &
dH_7 & = - F_3 \wedge F_5 \ . 
\ea 
\ee 
For the time being, we use these conventions. At the end of this section, we re-express the final result in our conventions, in which the Bianchi identities for all RR field strengths have a uniform form.

We also need the field strengths of the 
 worldvolume fields $\omega_0$, $\widetilde \omega_0$, $\omega_2$. These are read off from \cite{Eyras:1998hn} via a ``translation'' analogous to the one described above. The result reads
\be 
\ba 
\cK_1 & = d\omega_0 + \frac{1}{2\pi \alpha'} (\iota_k \cB_2) \ , \\ 
\widetilde \cK_1 & = d\widetilde \omega_0 - \frac{1}{2\pi \alpha'} (\iota_k C_2) \ , \\ 
\cK_3 & = d\omega_2 - \frac{1}{2\pi \alpha'} (\iota_k C_4)
+ (2\pi \alpha') A_1 \wedge d\omega_0 \wedge d\widetilde \omega_0
\\
& - \frac 12 
\frac{1}{2\pi \alpha'} \, \overline B_2 
\wedge (\iota C_2)
+ \frac 12 
\frac{1}{2\pi \alpha'} \, \overline C_2 
\wedge (\iota_k \cB_2)
 + d\widetilde \omega_0 \wedge \overline \cB_2
+ d \omega_0 \wedge \overline C_2 \ . 
\ea 
\ee 
The worldvolume Bianchi identities read
\be 
\ba 
d\cK_1 & = - \frac{1}{2\pi \alpha'} (\iota_k H_3) \ , \\
d\widetilde \cK_1 & = + \frac{1}{2\pi \alpha'} (\iota_k F_3) \ , \\
d\cK_3 & = \frac{1}{2 \pi \alpha'} (\iota_k F_5)
+ \overline F_3 \wedge \cK_1
+ \overline H_3 \wedge \widetilde \cK_1
+(2\pi \alpha') W_2 \wedge \widetilde \cK_1 \wedge \cK_1 \ . 
\ea 
\ee 
In verifying these equations we have used the
relation between $dA_1$ and $W_{mn}$, see \eqref{eq_KKmon_useful}.

The final piece of information needed for our purposes is the expression for the field strength
of the bulk field $\iota_k \cN_7$.
We shall denote this field strength as 
$\iota_k \cU_8$.
This object can be reconstructed from the gauge transformations of 
$\iota_k \cN_7$ given in \cite{Eyras:1998hn}. The result reads
\begin{equation}
    \begin{adjustbox}{max width=\textwidth}
        $\begin{aligned}
            - (\iota_k \cU_8) & = d(\iota_k \cN_7) 
            + (\iota_k \cB_2)  d(\iota_k \widetilde \cB_6)
            + (\iota_k C_2) d(\iota_k C_5)
            - \frac 12 (\iota_k C_4) d(\iota_k C_4)
            - (\iota_k \cB_2) (\iota_k C_2) dC_4
            \\
            & + \frac 14 (\iota_k \cB_2) C_2 d(\iota_k C_4)
            - \frac 14 (\iota_k C_2) B_2 d(\iota_k C_4)
            - \frac 34 (\iota_k \cB_2) (\iota_k C_4) dC_2
            + \frac 34 (\iota_k C_2) (\iota_k C_4) d \cB_2
            \\
            & + C_4 (\iota_k \cB_2) d(\iota_k C_2) 
            - C_4 (\iota_k C_2) d(\iota_k \cB_2)
            - \frac 14 \, \cB_2 (\iota_k C_4) d(\iota_k C_2)
            + \frac 14 \, C_2 (\iota_k C_4) d(\iota_k \cB_2)
            \\
            & + \frac 18 (\iota_k \cB_2) (\iota_k C_2) \cB_2 dC_2
            - \frac 18 (\iota_k \cB_2) (\iota_k C_2) C_2 d\cB_2
            + \frac 18 \, \cB_2 C_2 (\iota_k \cB_2) d(\iota_k C_2)
            + \frac 18 \, \cB_2 C_2 (\iota_k C_2) d(\iota_k \cB_2)
            \\
            &  
            + \frac 18 \, C_2 C_2 (\iota_k \cB_2) d(\iota_k \cB_2)
            + \frac 18 \, \cB_2 \cB_2 (\iota_k C_2) d(\iota_k C_2)
            + (\iota_k \cB_2) (\iota_k C_2) (\iota_k C_4) W_2 \ .
        \end{aligned}$
    \end{adjustbox}
\end{equation}
We have suppressed wedge products for brevity. The Bianchi identity for this bulk field strength takes the form
\be 
- d(\iota_k \cU_8)  =(\iota_k F_3)\wedge (\iota_k F_7)  + (\iota_k H_3) \wedge (\iota_k H_7)   - \frac 12 (\iota_k F_5)\wedge (\iota_k F_5) \ . 
\ee

\paragraph{Anomaly polynomial.}
We can now present the anomaly polynomial that encodes the couplings \eqref{eq_KK_terms} via descent.
It is defined as
\be 
\cI_7^{\rm KK} = d \cI_6^{(0)\rm KK} \ . 
\ee 
Making use of the relations recorded above for bulk and worldvolume fields, we are able to cast 
$\cI_7^{\rm KK}$ as a polynomial in fields strengths.
It reads
\be \label{eq_full_KK_poly}
\ba 
\frac{1}{2\pi \alpha'} \, \cI_7^{\rm KK} & = 
- \frac{1}{2 \pi \alpha'} (\iota_k \cU_8)  
+ \frac 12 (\iota_k F_5) \wedge \cK_3
- (\iota_k F_7) \wedge \widetilde \cK_1
+ (\iota_k H_7) \wedge \cK_1
\\
& + \frac 12 (2\pi \alpha')
\overline F_3 \wedge \cK_1 \wedge \cK_3
+ \frac 12 (2\pi \alpha')
\overline H_3 \wedge \widetilde \cK_1 \wedge \cK_3
\\
& + (2\pi \alpha') \overline F_5 \wedge \widetilde \cK_1 \wedge \cK_1
+ \frac 12 (2\pi \alpha') ^2
W_2 \wedge \cK_3 \wedge \widetilde \cK_1 \wedge \cK_1 \ . 
\ea 
\ee 

\paragraph{Translating into our conventions.}
To conclude our analysis of the anomaly polynomial, 
 we perform some further simplifications on
\eqref{eq_full_KK_poly}. First, the factors $2\pi \alpha'$ can be reabsorbed by suitable redefinitions of the worldvolume fields. Second,
we translate the result in our conventions for RR fields. 
In order to distinguish the fields of \cite{Eyras:1998hn}
from our fields,
we put a hat on ours.
Our conventions are
\be 
\widehat H_3 = d\widehat B_2 \ , \qquad 
\widehat G_p = d\widehat C_{p-1} - \widehat H_3 \wedge \widehat C_{p-3} \ . 
\ee 
For the electromagnetic dual of $\widehat B_2$ we use the conventions
\be 
\ba 
\widehat H_7  &= d\widehat B_6
- \frac 12 \, \widehat C_0 \widehat F_7
+ \frac 12 \, \widehat C_2 \widehat F_5
- \frac 12 \, \widehat C_4 \widehat F_3
+ \frac 12 \, \widehat C_6 \widehat F_1 \ ,
\qquad 
d \widehat H_7 =  \widehat F_3 \widehat F_5
 - \widehat F_1 \widehat F_7 \ . 
\ea 
\ee 
The dictionary between the fields of \cite{Eyras:1998hn} and ours is as follows,
\be 
\ba 
\cB_2 & = \widehat B_2 \ , & 
H_3 & = \widehat H_3 \ , \\ 
C_0 & = \widehat C_0 \  , & 
F_1 & = \widehat G_1  \ , \\
C_2 & = \widehat C_2 \ , & 
F_3 & = \widehat G_3 + \widehat C_0 \widehat H_3
\ , \\ 
C_4 & = \widehat C_4 - \tfrac 12 \widehat B_2 \widehat C_2 \ , & 
F_5 & = \widehat G_5 \ , \\
C_6 & = \widehat C_6 - \tfrac 14 \widehat B_2^2 \widehat C_2 \ , &
F_7 & = \widehat G_7 \ , \\
B_6 & =  - \widehat B_6
- \tfrac 12 \widehat C_2 \widehat C_4
- \tfrac 12 \widehat C_0 \widehat C_6
+ \tfrac 14 B_2 \widehat C_2^2 \ , & 
H_7 & = - \widehat H_7 - \widehat C_0 \widehat F_7 \ . 
\ea 
\ee 
We also perform a redefinition of the worldvolume scalars $\omega_0$, $\widetilde \omega_0$. Our scalars are 
denoted $\widehat \omega_0$ and $\widehat{\widetilde \omega}_0$, with field strengths $\widehat \cK_1$ and $\widehat{\widetilde \cK}_1$.
The dictionary between the fields in \cite{Eyras:1998hn} and ours reads
\be 
\widehat \cK_1= \cK_1 \ , \qquad
\widetilde \cK_1 = \widehat{\widetilde \cK}_1 - \widehat C_0 \widehat \cK_1 \ .
\ee

We may now express \eqref{eq_full_KK_poly} in terms of our hatted fields with the dictionary outlined above. In the final result, to avoid clutter, we omit all hats. We find 
\be \label{eq_full_KK_nicer}
\ba 
\cI_7 & = 
-(\iota_k \cU_8)
+ \frac 12 (\iota_k G_5) \wedge \cK_3
- (\iota_k G_7) \wedge \widetilde \cK_1
- (\iota_k H_7) \wedge \cK_1
\\
& + \frac 12  
{\overline G}_3 \wedge \cK_1 \wedge \cK_3
+ \frac 12  
\overline H_3 \wedge \widetilde \cK_1 \wedge \cK_3
\\
& +   \overline G_5 \wedge \widetilde \cK_1 \wedge \cK_1
+ \frac 12  
W_2 \wedge \cK_3 \wedge \widetilde \cK_1 \wedge \cK_1 \ . 
\ea 
\ee 
For completeness, we also record the worldvolume Bianchi identities in our conventions. All fields are understood as hatted, but we do not display the hats,
\be 
\ba 
d\cK_1 & = -  (\iota_k H_3) \ , \\
d\widetilde \cK_1 & = (\iota_k G_3)   
+ G_1 \wedge \cK_1\ , \\
d\cK_3 & =  (\iota_k G_5)  
+ {\overline G}_3 \wedge \cK_1
+  H_3 \wedge \widetilde \cK_1
+  W_2 \wedge \widetilde \cK_1 \wedge \cK_1 \ . 
\ea 
\ee 

\paragraph{Specialization to vanishing $G_1$, $G_3$, $H_3$.}
In a solution of the form $AdS_5 \times SE_5$,
the fluxes $G_3$, $H_3$, $G_1$ are zero. 
Therefore, for our purposes we can
use a simpler expression for the anomaly polynomial on the KK monopole, which reads 
\be 
\ba 
\cI_7 & = 
-(\iota_k \cU_8)
+ \frac 12 (\iota_k G_5) \wedge \cK_3
 +   \overline G_5 \wedge \widetilde \cK_1 \wedge \cK_1
+ \frac 12 \, 
W_2 \wedge \cK_3 \wedge \widetilde \cK_1 \wedge \cK_1 \ . 
\ea 
\ee 
We also report the specialized Bianchi identities for the bulk and worldvolume fields,
\be 
\ba 
dG_5&=0\ , \qquad 
d\cK_1 =0 \ , \qquad 
d \widetilde \cK_1 = 0 \ , \\ 
d\cK_3  &=  (\iota_k G_5)  
+  W_2 \wedge \widetilde \cK_1 \wedge \cK_1 
\ , \qquad 
 d(\iota_k \cU_8)  =   \frac 12 (\iota_k G_5)\wedge (\iota_k G_5) \ .
\ea 
\ee 
We observe that $dG_5 = 0$ implies
\be 
d\overline G_5 + W_2 \wedge (\iota_k G_5) =0 \ , \qquad 
d(\iota_k G_5) = 0 \ , 
\ee 
where we recalled \eqref{eq_KKmon_param} for the definition of $\overline G_5$. We have also used $\pounds_k G_5 = 0$, which is a requirement for a KK monopole (the isometry direction should  be a symmetry direction not only for the metric, but for all fluxes).

\paragraph{Convenient form for topological couplings.}

Our final task is to find a convenient solution to the Bianchi identities (specialized to vanishing $H_3$, $G_3$, $G_1$) and cast $\cI_7^{\rm KK}$ as
\be 
\cI_7^{\rm KK} = d\cI_6^{(0){\rm KK , \ new}} \ .
\ee 
Here $\cI_6^{(0){\rm KK , \ new}}$ is a locally defined 6-form, different from $\cI_6^{(0){\rm KK}}$ in \eqref{eq_KK_terms}. We will construct 
$\cI_6^{(0){\rm KK , \ other}}$ in a way that is
convenient for the study of hanging KK monopoles.

Let us solve the Bianchi identities for $\iota_k G_5$,
$\iota_k \cU_8$,
and $\cK_3$ as 
\be \label{eq_KK_solve_Bianchi}
\iota_k G_5 = dC_3 \ , \quad 
\iota_k \cU_8 = d\cN_6 + \frac 12 \, C_3  \wedge dC_3  \ , \quad 
\cK_3 = d\beta_2 + C_3  + (k\cdot k)^{-1} k \wedge \widetilde \cK_1 \wedge \cK_1 \ ,
\ee 
where 
$C_3$, $\cN_6$ are locally-defined forms in the bulk, 
$\beta_2$ is a locally-defined 2-form on the worldvolume of the KK monopole.
The fields strengths $\iota_k G_5$, $\iota_k \cU_8$,
$\cK_3$ are invariant under the following gauge transformations,
\be \label{eq_Bianchi_sol_gauge}
C_3  \rightarrow C_3 
+ d\Lambda_2 \ , \qquad 
\cN_6  \rightarrow \cN_6  
+ d\Lambda_5 - \frac 12 \, C_3 \wedge d\Lambda_2
\ , \qquad 
\beta_2 \rightarrow \beta_2 + d\lambda_1 - \Lambda_2 \ , 
\ee 
with parameters $\Lambda_2$, $\Lambda_5$ (in the bulk)
and $\lambda_1$ (on the worldvolume of the KK monopole).
Plugging \eqref{eq_KK_solve_Bianchi} into $\cI_7^{\rm KK}$, we get
\be 
\cI_7^{\rm KK} = G_5 \wedge \widetilde \cK_1 \wedge
\cK_1 
+ d \bigg[ 
-\cN_6  
+ \frac 12 \, C_3  \wedge d\beta_2
- \frac 12 (C_3 + d \beta_2) \wedge (k \cdot k)^{-1}k \wedge \widetilde \cK_1 \wedge
\cK_1 
\bigg]  \ .
\ee 
We notice that we have reconstructed the full $G_5$
from its pieces $\overline G_5$ and $\iota_k G_5$.

Next,  we make the following observation. The 7-form $\cI_7^{\rm KK}$, the Bianchi identity for $\cK_3$,
and its solution
depend on $\cK_1$, $\widetilde \cK_1$ only via the combination
$\widetilde K_1 \wedge \cK_1$.
This is a closed 2-form.
We regard it as the field strength of a composite 1-form $U(1)$ gauge field on the
worldvolume of the KK monopole, constructed out of the two compact scalars on the worldvolume.\footnote{We can model gauge-equivalence classes of $U(1)$ $p$-form gauge field by elements
in the differential cohomology group
$\breve H^{p+1}(\cW^6)$,
where $\cW^6$ is the worldvolume of the KK monopole.
In particular, the two compact scalars
on $\cW^6$ are labeled by elements
$\breve k_1$, $\breve{\widetilde k}_1 \in \breve H^1(\cW^6)$.
The composite 1-form gauge field constructed out of these two is modeled by $\breve k_1 \star \breve{\widetilde k}_1 \in \breve H^2(\cW^6)$, where $\star$ denotes the product in differential cohomology.
}
We denote this 1-form gauge field as $v_1$ and we write
\be \label{eq_def_v1}
\widetilde \cK_1 \wedge \cK_1 = dv_1 \ . 
\ee 
Using $v_1$, we can construct
$\cI_6^{(0)\rm {KK, \ new}}$. It takes the form 
\be \label{eq_I6KK}
\cI_6^{(0)\rm {KK, \ new}} = v_1 \wedge G_5 
- \cN_6 
+ \frac 12 \, C_3  \wedge d\beta_2
- \frac 12 (C_3  + d \beta_2) \wedge (k \cdot k)^{-1} k \wedge  dv_1  \ .
\ee 
Under a (small) gauge transformation of the form
\eqref{eq_Bianchi_sol_gauge}, accompanied by $v_1 \rightarrow v_1 + d\lambda_0$, the quantity $\cI_6^{(0)\rm {KK, \ new}}$ varies into a total derivative, as expected from 
descent, 
$\delta \cI_6^{(0)\rm {KK, \ new}} = d\cI_5^{(1){\rm KK , \ new}}$.

\subsubsection{KK monopole contribution}

The relevant term from $\cI_6^{(0)\rm {KK, \ new}}$ in \eqref{eq_I6KK} is described by the six-dimensional action
\be 
2\pi \int_{\gamma^1 \times M^3 \times \Sigma^2} v_1 \wedge G_5 \ , 
\ee 
where we recall that $v_1$ is the composite gauge field defined in \eqref{eq_def_v1}.
The computation is entirely analogous to the D5-brane case. We saturate the $\gamma^1$ integral with $v_1$,
\be 
\alpha_{\rm KK} = 2\pi \int_{\gamma^1} v_1 \ . 
\ee 
We are then left with the evaluation of
\be 
\alpha_{\rm KK} \int_{M^3 \times \Sigma^2} G_5 \ , 
\ee 
which is performed as above. In conclusion, the hanging KK monopole gives  
\begin{equation} \label{eq_KK_piece}
 U_\text{KK} = \exp\bigg( i \, \alpha_\text{KK} \, \tau_{\rm EH} \, m^{ab}    \int_{M^3} \ast F_{ab}\bigg) \, ,
\end{equation}
where we have recalled $\tau_{\rm flux} = 2 \tau_{\rm EH}$, see \eqref{eq:effective_gauge_coupling}.

\subsubsection{Expression for the symmetry operator}

The total expression for the symmetry operator is obtained combining \eqref{eq_D5_piece} and \eqref{eq_KK_piece}, using
\be 
\alpha_{\rm D5} = 2 \alpha_{\rm KK} = \alpha \ .  
\ee 
We get the result
\be \label{eq_S5_final}
U = \exp\bigg(\frac 12 \, i \, \alpha  \, \tau  \, m^{ab}  \,  \int_{M^3} \ast F_{ab}\bigg) \ ,
\ee 
which matches the expectation from the Gauss's law low-energy analysis of Section \ref{sec:Gauss_and_sym_op}.

We observe that our definition of Hopf coordinates on $S^5$ is subordinate to the specific definition of $(w_1, w_2, \psi)$ in terms of $y^a$, see \eqref{eq_Hopf_S5}. Our construction, however, can be repeated for different choices of complex coordinates. In particular, we can generalize \eqref{eq_Hopf_S5} by writing 
\be \label{eq_new_complex}
\ba
\rho e^{i\psi} w_1 &= y_{\sigma(1)} + iy_{\sigma(2)} \ , & 
\rho e^{i\psi} w_2 &= y_{\sigma(3)} + iy_{\sigma(4)} \ , & 
\rho e^{i\psi} &= y_{\sigma(5)} + iy_{\sigma(6)} \ ,
\ea
\ee 
where $\sigma:\{1,\dots,6\} \to \{1,\dots,6\}$ is a permutation.
With this choice, the final result still
takes the form \eqref{eq_S5_final}. The coefficients $m^{ab}$, however, are no longer given as \eqref{eq_m_from_psi}. Rather, their expression is $m^{ab} = m_{(\sigma)}^{ab}$ where
\be \label{eq_m_from_psi_more}
m_{(\sigma)}^{ab} =
\left\{
\begin{array}{ll}
\!\!+1 & \text{\small if $(a,b)=(\sigma(1),\sigma(2))$, $(\sigma(3),\sigma(4))$, $(\sigma(5),\sigma(6))$} \ , \\
\!\!-1 & \text{\small if $(a,b)=(\sigma(2), \sigma(1))$, $(\sigma(4),\sigma(3))$, $(\sigma(6),\sigma(5))$} \ , \\
\!\!0 & \text{\small otherwise}  \ .
\end{array}
\right.  
\ee
We notice that, for the trivial permutation $\sigma = (123456)$,
$m_{(\sigma)}^{ab}$ equals $m^{ab}$ in \eqref{eq_m_from_psi}.

Thanks to the possibility of using different complex coordinates, we can realize symmetry operators $U(\theta)$ of the form \eqref{eq_S5_final} for a wider range of $\theta^{ab}$ parameters. More precisely, we can realize any value 
$\theta^{ab}$ that is of the form $\theta^{ab} = \alpha m_{(\sigma)}^{ab}$ for some $\alpha$ and some permutation~$\sigma$.

Let us now show that, by taking fusion products of operators of the form $U(\theta = \alpha m_{(\sigma)})$, we can realize any $SO(6)$ transformation. To this end, we need to recall the following facts. First, the symmetry operator $U(\theta)$, $\theta \in \mathfrak{so}(6)$, implements the transformation by the group element $\exp \theta \in SO(6)$. Second, the fusion $U(\theta) \, U(\theta')$ gives $U(\theta'')$, where $\exp \theta'' = \exp \theta \exp \theta'$. 
These observations imply that, to prove the desired result, 
we have to verify the following group-theoretical statement: any element $g \in SO(6)$ can be written as a product   
$\exp (\alpha_1 m_{(\sigma_1)}) \dots \exp (\alpha_k m_{(\sigma_k)})$ for suitable $(\alpha_1, \sigma_1)$, \dots, $(\alpha_k , \sigma_k)$.

Let us establish a preliminary result:
for any distinct $a$, $b \in \{1,2,3,4,5,6\}$, 
we can realize a rotation in the $(a,b)$-plane
as a product $\exp (\alpha_1 m_{(\sigma_1)}) \dots \exp (\alpha_k m_{(\sigma_k)})$.
For simplicity, let us consider the $(1,2)$ plane;
the general case is analogous. 
If we use the permutations
$\sigma = (123456)$ and $\sigma = (124365)$, with the same parameter $\alpha$, we realize the $SO(6)$ matrices
\be 
\ba
    \exp(\alpha m_{(123456)})
&= \text{diag}\left(
\left( 
\begin{smallmatrix}
    \cos \alpha & -\sin \alpha \\
    \sin \alpha & \cos \alpha
\end{smallmatrix}
\right)
, 
\left( 
\begin{smallmatrix}
    \cos \alpha & -\sin \alpha \\
    \sin \alpha & \cos \alpha
\end{smallmatrix}
\right)
,
\left( 
\begin{smallmatrix}
    \cos \alpha & -\sin \alpha \\
    \sin \alpha & \cos \alpha
\end{smallmatrix}
\right)
\right) \ , \\
\exp(\alpha m_{(124365)})
&= \text{diag}\left(
\left( 
\begin{smallmatrix}
    \cos \alpha & -\sin \alpha \\
    \sin \alpha & \cos \alpha
\end{smallmatrix}
\right)
,
\left( 
\begin{smallmatrix}
    \cos \alpha & \sin \alpha \\
    -\sin \alpha & \cos \alpha
\end{smallmatrix}
\right)
,
\left( 
\begin{smallmatrix}
    \cos \alpha & \sin \alpha \\
    -\sin \alpha & \cos \alpha
\end{smallmatrix}
\right)
\right)   \ . 
\ea
\ee
Here we have presented an $SO(6)$ matrix as a block-diagonal matrix with three $2\times 2$ blocks.
Taking the product of the $SO(6)$ elements written above, we obtain a rotation in the $(1,2)$-plane of angle $2\alpha$. Since $\alpha$ is arbitrary, we have verified our claim.

Having shown how to realize arbitrary rotations in the $(a,b)$-plane, for any distinct $a$, $b$,
we are finally in a position to argue that any element of $SO(6)$ can be realized. This can be seen using 
the $SO(n)$ analog of the Euler angle parametrization of the most general element of $SO(3)$, see e.g.~\cite{genEuler} for details.

\subsection{$AdS_4 \times S^7$ in M-theory}
\label{app_AdS4S7}

In this section we 
discuss how $SO(8)$ symmetry operators are realized by hanging M5-branes in $AdS_4 \times S^7$.

\subsubsection{Internal support of the M5-brane}

The M5-brane has a worldvolume of the form $\gamma^1 \times M^2 \times \Sigma^3$. Let us describe the internal support $\Sigma^3 \subset S^7$.
This case has close similarities with the $AdS_5 \times S^5$ case discussed above.

We start by writing $S^7$ as a Hopf fibration.
We achieve this by introducing 
 coordinates $(w_1, w_2, w_3,\psi)$
on $S^7$ via 
\be \label{eq_Hopf_S7}
\ba 
\rho e^{i \psi} w_1 &= y_1 + i y_2 \ , & 
\rho e^{i \psi} w_2 &= y_3 + i y_4 \ , & 
\rho e^{i \psi} w_3 &= y_5 + i y_6 \ , & 
\rho e^{i \psi} &= y_7 + i y_8 \ . 
\ea
\ee 
Here $w_1$, $w_2$, $w_3$ are unconstrained complex coordinates,
$\psi$ is an angle with period $2\pi$, and the positive quantity $\rho$ is fixed by the constraint $y^a y_a = 1$ to the value 
\be 
\rho^{-2} = 1 + |w_1|^2 + |w_2|^2 + |w_3|^2\ . 
\ee 
In $(w_1, w_2, w_3,\psi)$ coordinates, the round metric on $S^7$
takes the form
\be \label{eq_S7_is_Hopf}
ds^2_{S^7} = ds^2_{\mathbb C \mathbb P^3} + D\psi^2  \ , \qquad 
D\psi = d\psi + \cA \ . 
\ee 
Here $ds^2_{\mathbb C \mathbb P^3}$ is the standard Fubini-Study metric on $\mathbb C \mathbb P^3$
and the connection 1-form $\cA$ 
satisfies 
\be 
d\cA = 2J \ ,
\ee 
with $J$ the K\"ahler form 
  on $\mathbb C \mathbb P^3$. The 1-form $D\psi$ is the dual of the Killing vector $\partial_\psi$. As a result, $D\psi$ can be expanded onto the basis $K_{ab} = 2y_{[a} dy_{b]}$,
\be 
D\psi = \tfrac 12 m^{ab} K_{ab} \ , 
\ee 
where the expansion coefficients are
\be \label{eq_m_from_psi}
m^{ab} = 
\left\{
\begin{array}{ll}
+1 & \text{if $(a,b)=(1,2)$, $(3,4)$, $(5,6)$, $(7,8)$} \ , \\
-1 & \text{if $(a,b)=(2,1)$, $(4,3)$, $(6,5)$, $(7,8)$} \ , \\
0 & \text{otherwise}  \ .
\end{array}
\right.  
\ee 
Recalling $*\omega_{ab} = \tfrac{1}{2 \pi^3} dK_{ab}$ from \eqref{eq_omega_S7}, we also observe that
\be \label{eq_identity_for_m_omega_S7}
\tfrac 12 m^{ab} \omega_{ab} = \tfrac{1}{2\pi^3} *dD\psi 
= \tfrac{1}{2\pi^3}  
\cdot 2 *J
= \tfrac{1}{2\pi^3}   J^2 \wedge D\psi 
= \left(\tfrac{d\cA}{2\pi} \right)^2 \wedge \tfrac{D\psi}{2\pi} \ . 
\ee 
With the notation introduced above, the internal support $\Sigma^3$ is defined as 
the 3-sphere obtained by fibering the Hopf circle
$S^1_\psi$ over a 2-cycle $\mathbb C \mathbb P^1$ inside the base $\mathbb C \mathbb P^3$, including 
an overall wrapping factor $N$.
In particular, if we integrate a 3-form over $\Sigma^3$, to get a non-zero result we have to assume that the 3-form has a $D\psi$ leg. Keeping this into consideration,
relevant integrals over $\Sigma^3$ can be computed with the formula
\be \label{eq_Sigma3_integrals}
\int_{\Sigma^3} \cX_2 \wedge \tfrac{D\psi}{2\pi} = \int_{S^7} \cX_2 \wedge \tfrac{D\psi}{2\pi} \wedge N   \left(\tfrac{d\cA}{2\pi}\right)^2  
= N  \int_{S^7} \cX_2 \wedge \tfrac 12 m^{cd} \omega_{cd} \ , 
\ee 
where $\cX_2$ is a 2-form and in the last step we have made use of \eqref{eq_identity_for_m_omega_S7}.

\subsubsection{Reduction of topological couplings on the M5-brane}

The relevant topological coupling on the M5-brane can be written as a six-dimensional integral as
\be \label{eq_M5_top_piece}
2\pi \int_{\gamma^1 \times M^2 \times \Sigma^3} -   b_2  \wedge G_4 \ , 
\ee 
see discussion around \eqref{eq_6d_action_M5}.
To saturate the integral over $\gamma^1$ we use one of the legs of the chiral 2-form $b_2$ on the M5-brane.
More precisely, we write
\be 
b_2  = a_1 \wedge \tfrac{D\psi}{2\pi}  \ . 
\ee 
We introduce the notation
\be 
2\pi \int_{\gamma^1} a_1   = \alpha_{\rm M5} \ . 
\ee 
From \eqref{eq_M5_top_piece} we obtain 
\be \label{eq_AdS4_M5_pieceBIS}
-  \alpha_{\rm M5} \int_{  M^2 \times \Sigma^3}  \tfrac{D\psi}{2\pi}  \wedge G_4 \ . 
\ee 
The relevant terms in $G_4$ are
\be \label{eq_integral_to_do_M5}
G_4 = - (2\pi \ell_{\rm P})^{3}*_{11}G_7 \supset
- (2\pi \ell_{\rm P})^{3} *_{11} \left[ 
\tfrac 12 N \tfrac{F^{ab}}{2\pi} \wedge \omega_{ab}
\right] 
= - \tfrac{1}{4\pi} N (2\pi \ell_{\rm P})^{3} L_{S^7}^{-3} *F^{ab} \wedge *\omega_{ab} \ , 
\ee 
where we have used \eqref{eq_G4_G7_duality} and \eqref{eq_G7_with_vectors}. The factor $L_{S^7}^{-3}$ stems from the $L_{S^7}^2$ prefactor of $ds^2_{S^7}$ in the 11d line element \eqref{eq_AdS4S7_metric}.
We have to integrate $* \omega_{ab} \wedge \frac{D\psi}{2\pi}$ over $\Sigma^3$.
This is accomplished using \eqref{eq_Sigma3_integrals},
\be 
\int_{\Sigma^3} *\omega_{ab} \wedge \tfrac{D\psi}{2\pi}
= N \int_{S^7}  * \omega_{ab} \wedge \tfrac 12 m^{cd} \omega_{cd} \  .
\ee 
To proceed, we need the identity\footnote{More generally, if we consider $S^n$, we have
\be 
\int_{S^n} \omega_{ab} \wedge * \omega^{cd} = \frac{4\pi}{(n+1) \cV_{n-2}} \, \delta_a^{[c} \delta_b{}^{d]} \ . \nn
\ee }
\be 
\int_{S^7} \omega_{ab} \wedge * \omega^{cd} = \frac{1}{2\pi^2} \, \delta_{[a}{}^c \delta_{b]}{}^d 
= 
\frac{4}{\pi \sqrt 2 N^{3/2}} \, \tau_{\rm flux}
\, \delta_{[a}{}^c \delta_{b]}{}^d
\ .\label{eq:omega_orthogonalityS7}
\ee 
In the second step, we have used
\be 
\tau = \tau_{\rm flux} + \tau_{\rm EH} = 
\frac{N^{3/2}}{3\pi \sqrt 2}
\ , \qquad 
\tau_{\rm flux} = 3 \tau_{\rm EH} \ , 
\ee 
with the second relation being a special case of equation (3.11) in \cite{Barnes:2005bw}.
The topological term \eqref{eq_AdS4_M5_piece} is then equal to
\be 
\alpha_{\rm M5}
\tau_{\rm flux} \int_{M^2} \tfrac 12 m^{ab} \wedge *F_{ab} \ ,
\ee 
and the M5-brane contribution to the symmetry operators is  
\be 
U_{\rm M5} = \exp \bigg( i
\alpha_{\rm M5} \tau_{\rm flux} \int_{M^2} \tfrac 12 m^{ab} *F_{ab} \bigg) \ .  
\ee

\subsubsection{Topological couplings on a KK monopole in M-theory}

The KK monopole of M-theory is a 7d soliton with four transverse directions. One of the latter must be an isometry of the 11d metric and plays a special role. 
The topological couplings on the worldvolume of an M-theory KK monopole are given in \cite{Bergshoeff:1998ef}.
Here we recast them as an anomaly polynomial using descent. Our discussion in this section is analogous to the discussion of KK monopoles in Type IIB. We shall use a similar notation but give fewer details.

We start by 
recalling the localized degrees of freedom living on the worldvolume of the KK monopole. They are:
\begin{itemize}
    \item Three scalars associated to the transverse directions to the KK monopole.
    Since the KK monopole has four transverse directions, a naïve counting would give four scalars. The scalar associated with the 11d isometry direction, however, is pure gauge \cite{Hull:1990ms,Bergshoeff:1997gy}.
    \item One 1-form gauge field $a_1^{\rm KK}$ with field strength $f_2^{\rm KK}$.
    We notice that the field strength is not given simply as $da_1^{\rm KK}$, see \eqref{eq_f2KK_def} below.
\end{itemize}

To proceed, we parametrize the local 11d line element as
\be 
g_{\mu\nu} dx^\mu dx^\nu = h_{mn} dy^m dy^n
+ e^{2\omega} (d\psi + V_m dy^m)^2 \ . 
\ee 
In this section, we use $\mu$, $\nu$ for curved 11d indices, $m$, $n$ for curved 10d indices. 
A $\mu$ index is split as $\mu \to (m, \psi)$.
The isometry direction is $\psi$, associated to the Killing vector
\be 
k^\mu \partial_\mu  = \partial_\psi  \ .
\ee 
The coordinate $\psi$ is an angle with period $2\pi$.
The metric functions $h_{mn}$, $\omega$, $V_m$ are independent of $\psi$.
The norm squared of the Killing vector is
\be 
k \cdot k  = g_{\mu\nu} k^\mu k^\nu = e^{2\omega} \ .
\ee 
The quantity $V_m$ is interpreted as a $U(1)$ gauge field with field strength denoted
\be 
W_2 = dV_1 \ , \quad \text{where} \quad 
V_1 = V_m dy^m \ . 
\ee 
We observe that the 2-form $W_2$ has no legs along the $\psi$ direction,
\be 
\iota_k W_2 = 0 \ . 
\ee 
Any $p$-form $\cX_p$ in the 11d bulk 
can be decomposed as
\be \label{eq_KKmon_param_BIS}
\cX_p = \overline \cX_p  + (-)^{p-1} (\iota_k \cX_p) \wedge (d\psi + V) \ ,
\ee 
where $\overline \cX_p$ is defined by this equation
and satisfies $\iota_k \overline \cX_p=0$.

We find that the topological worldvolume couplings reported in \cite{Bergshoeff:1998ef}
correspond (in our notation) to the following 8-form anomaly polynomial,
\be \label{eq_KK_Mth_poly}
\cI_8^{\rm KK}  = - (\iota_k \cU_9)  
+ f_2^{\rm KK}\wedge  (\iota_k G_7)  
- \frac 12 (f_2^{\rm KK})^2 \wedge  \overline  G_4 
+ \frac 16 (f_2^{\rm KK})^3 \wedge W_2 \ .
\ee 
Some comments are in order.
The quantities $\cU_9$, $G_7$, $G_4$ are bulk field strengths. We have already defined $G_4$ and $G_7$ in the main text around \eqref{eq_M_Bianchi_both}.
For convenience, we repeat the expressions for $G_4$, $G_7$,
\be 
G_4 = dC_3 \ , \qquad 
G_7 = dC_6 + \frac 12 \, C_3 \wedge dC_3 \ .
\ee 
(The curvature term $X_8$ is omitted because it does not play an important role in our discussion.) 
The quantity $\overline G_4$ is defined from $G_4$ according to formula \eqref{eq_KKmon_param_BIS}.
The 9-form $\mathcal U_9$ is the field strength of an 8-form gauge field $\cN_8$, which is the electromagnetic dual in 11d to the Killing 1-form $k_\mu dx^\mu$, regarded as a 1-form gauge field. The expression for 
$(\iota_k \mathcal U_9)$ can be reconstructed from the gauge transformations of $(\iota_k \cN_8)$ given in 
\cite{Bergshoeff:1998ef}. The result reads
(suppressing wedge products)
\be 
(\iota_k \cU_9)  = - d (\iota_k \cN_8)  
-(\iota_k C_6)  d (\iota_k C_3)  
+ \tfrac 16 (\iota_k C_3)^2 d \overline C_3  
+ \tfrac 16 (\iota_k C_3) \overline C_3  d(\iota_k C_3) 
+ \tfrac 16 (\iota_k C_3)^3 W_2 \ . 
\ee 
The 2-form $f_2^{\rm KK}$ is the field strength of the worldvolume gauge field $a_1^{\rm KK}$
and is defined as
\be \label{eq_f2KK_def}
f_2^{\rm KK} = da_1^{\rm KK}  - (\iota_k C_3) \ . 
\ee 
For convenience, let us collect the Bianchi identities of the bulk and worldvolume field strengths,
\be 
\ba 
d(\iota_k G_7)  &= - \overline G_4 \wedge (\iota_k G_4)  \ , &
d \overline G_7  &= \tfrac 12 \overline G_4 ^2 
- (\iota_k G_7)\wedge    W_2 \ , &
d(\iota_k \cU_9)  &= - (\iota_k G_4) \wedge (\iota_k G_7)  \ , \\ 
d(\iota_k G_4) &= 0 \ , & 
d\overline G_4 &= -(\iota_k G_4)\wedge  W_2  \ , \\
dW_2 &= 0 \ , & 
df_2^{\rm KK} &= - (\iota_k G_4) \ .
\ea 
\ee

\subsubsection{Contribution of the KK monopole to the symmetry operator}
In order to realize the symmetry operator,
we consider a KK monopole with worldvolume 
$\gamma^1 \times M^2 \times \Sigma^4$.
The factors $\gamma^1$, $M^2$ are the same as in the M5-brane worldvolume.
Let us discuss the factor $\Sigma^4$.
To this end, we observe that the 11d
isometry direction for the KK monopole is identified with the $S^1_\psi$ fiber of the Hopf fibration \eqref{eq_S7_is_Hopf} for $S^7$.
The internal support $\Sigma^4$ of the KK monopole is a 4-cycle in the $\mathbb C \mathbb P^3$ base of the Hopf fibration.
More precisely, we take $\Sigma^4$ to be a copy of $\mathbb C \mathbb P^2$, which we denote 
$\mathbb C \mathbb P^2_{\rm KK}$.
Recall that the internal support of the M5-brane
is $\Sigma^3$, which is the fibration of $S^1_\psi$ over a copy of $\mathbb C \mathbb P^1$,
which we now denote as $\mathbb C \mathbb P^1_{\rm M5}$.
We assume that $\mathbb C \mathbb P^2_{\rm KK}$
and $\mathbb C \mathbb P^1_{\rm M5}$ satisfy 
\be 
\mathbb C \mathbb P^1_{\rm M5} \subset 
\mathbb C \mathbb P^2_{\rm KK} \subset \mathbb C \mathbb P ^3 \ .
\ee

The relevant term in the KK monopole 8-form anomaly polynomial  \eqref{eq_KK_Mth_poly}  is 
\be 
\cI_8^{\rm KK}  \supset     
- \frac 12 (f_2^{\rm KK})^2 \wedge  \overline  G_4   \ .
\ee 
The worldvolume field strength $f_2^{\rm KK}$ is expressed as a sum of two contributions,
\be 
f_2^{\rm KK} = f_2^{\rm flux} + \Big[ da_1 -(\iota_k C_3) \Big] \ . 
\ee 
Here, $f_2^{\rm flux}$ is 
the Poincar\'e dual of $\mathbb C \mathbb P^1_{\rm M5}$ inside 
$\mathbb C \mathbb P^2_{\rm KK}$,
\be 
f_2^{\rm flux}  = {\rm PD}_{\mathbb C \mathbb P^2_{\rm KK}} [\mathbb C \mathbb P^1_{\rm M5}] \ . 
\ee 
In particular, $f_2^{\rm flux}$ is closed.
The second contribution $da_1 -(\iota_k C_3)$
contains the bulk $C_3$ and a worldvolume gauge field $a_1$. The latter only has legs along the arc $\gamma^1$, with 
\be 
\alpha_{\rm KK} =  2 \pi \int_{\gamma^1} a_1 \ . 
\ee 

If we analyze the 8-form anomaly polynomial with the form of $f_2^{\rm KK}$ specified above,
and we peel off a derivative to extract the 7d topological couplings, the relevant term is the following,
\be 
2\pi \cI_7^{(0) \rm KK} \supset -2\pi  a_1 \wedge   {\rm PD}_{\mathbb C \mathbb P^2_{\rm KK}} [\mathbb C \mathbb P^1_{\rm M5}] \wedge \overline G_4 \ . 
\ee 
Our task is to integrate this quantity over
$\gamma^1 \times M^2 \times \mathbb C \mathbb P^2_{\rm KK}$.
The factor $a_1$ is saturated by $\gamma^1$, yielding 
\be 
-\alpha_{\rm KK} \int_{M^2 \times \mathbb C \mathbb P^2_{\rm KK}} {\rm PD}_{\mathbb C \mathbb P^2_{\rm KK}} [\mathbb C \mathbb P^1_{\rm M5}] \wedge \overline G_4 \ . 
\ee 
Next, the definition of Poincar\'e dual gives 
\be 
-\alpha_{\rm KK} \int_{M^2 \times \mathbb C \mathbb P^1_{\rm M5}}   \overline G_4 
\ . 
\ee 
A convenient way to perform this integral is to write it as 
\be 
-\alpha_{\rm KK} \int_{M^2 \times \Sigma^3}   \overline G_4  \wedge \tfrac{D\psi}{2\pi}
\ , 
\ee 
where $\Sigma^3$ is obtained fibering $S^1_\psi$ over $\mathbb C \mathbb P^1_{\rm M5}$, and is the same $\Sigma^3$ as in \eqref{eq_Sigma3_integrals}.
Due to the $\tfrac{D\psi}{2\pi}$ factor, we can replace $\overline G_4  = G_4 
+ (\iota_k G_4) \wedge D\psi$ simply with $G_4$.
The resulting integral is 
\be 
-\alpha_{\rm KK} \int_{M^2 \times \Sigma^3}   G_4  \wedge \tfrac{D\psi}{2\pi}
\ . 
\ee 
This is exactly of the same form as the integral
\eqref{eq_integral_to_do_M5}, up to a numerical prefactor. Repeating the same steps as in the derivation above, we arrive at the quantity 
\be 
\alpha_{\rm KK}
\tau_{\rm flux} \int_{M^2} \tfrac 12 m^{ab} \wedge *F_{ab} \ .
\ee 
We trade  $\tau_{\rm flux}$ for $\tau_{\rm EH}$
with the relation $\tau_{\rm flux} = 3 \tau_{\rm EH}$. In conclusion, 
the KK monopole contribution to the symmetry operators is  
\be 
U_{\rm KK} = \exp \bigg( 3 i
 \alpha_{\rm KK} \tau_{\rm EH} \int_{M^2} \tfrac 12 m^{ab} *F_{ab} \bigg) \ .  
\ee

When we combine the M5-brane and KK monopole contributions, the overall prefactor of 
$\int_{M^2} \tfrac 12 m^{ab} *F_{ab}$ in the exponent is 
$i$ times 
\be 
 \alpha_{\rm M5} \tau_{\rm flux} + 3 \alpha_{\rm KK} \tau_{\rm EH}  \ ,
\ee 
where we have recalled \eqref{eq_taus_for_AdS4}.
We are led to the identifications 
\be 
\alpha_{\rm  M5} = \alpha \ , \qquad 
\alpha_{\rm KK} = \tfrac 13 \alpha \ , 
\ee 
in such a way that 
\be
\alpha_{\rm M5} \tau_{\rm flux} + 3 \alpha_{\rm KK} \tau_{\rm EH}
= \alpha ( \tau_{\rm flux} + \tau_{\rm EH} ) = \alpha \tau \ .
\ee
Accordingly,  we get the
 symmetry operator
\be 
U  = \exp \bigg(   i
 \alpha  \tau  \int_{M^2} \tfrac 12 m^{ab} *F_{ab} \bigg) \ ,
\ee 
in agreement with the low-energy analysis.

\section{Brane dynamics}
\label{Brane_dynamics_appendix}

In this appendix, we study the dynamics of the brane-antibrane system in order to show that there are tachyonic modes which appear when the brane and antibrane approach each other. These modes are responsible for the dynamical fusion of the hanging brane operators as discussed in Section \ref{dynamic_fusion}. We proceed by considering the dynamical DBI action $ S_\text{DBI}$ of a brane-antibrane pair which is given by \cite{Garousi_2005},
\begin{align}
\label{BaBAction}
S_\text{DBI} &= - \int d^{p+1} \xi \left( \mathcal{V}^{(1)} \left(\tau, \ell\right) e^{-\Phi(X^{(1)})} \sqrt{- \det \bf{A}^{(1)}} + \mathcal{V}^{(2)} \left(\tau, \ell\right) e^{-\Phi(X^{(2)})} \sqrt{- \det \bf{A}^{(2)}} \right)
  \ .
\end{align}
In this action, $n=(1,2) $ labels the brane or the antibrane whose worldvolume is spanned by the coordinates $\xi^M$. The field $\Phi$ is the bulk dilaton and \( \tau \) is the open string tachyon field which is a complex scalar field charged under the gauge group \( U(1)_1 \times U(1)_2 \) whose gauge fields \( a^{(1)}_M \) lives on the D$p$-brane, while \( a^{(2)}_M \) lives on the $\overline{\text{D}p}$-brane.  For notational convenience, we define the \emph{relative gauge field}, $a_M \equiv a^{(1)}_M - a^{(2)}_M$, which is exactly the Chan-Paton gauge field described in \eqref{diagonal} with the identification $a_L = a^{(1)}$ and $a_R=a^{(2)}$.
The tachyon potential $\mathcal{V(|\tau|,\ell)}$ is given by
\begin{equation}
\mathcal{V}(|\tau|, \ell) = T_p \cdot V(\tau)
\cdot \sqrt{Q(\tau, \ell)} \ , \quad Q(\tau,\ell) =1 + \frac{|\tau|^2}{\ell_s^2} \cdot \ell^i \ell^j\, g_{ij} \ , \quad  V(\tau) = \frac{1}{\cosh(\sqrt{\pi} |\tau |)} \ ,
\end{equation}
where $T_p$ is the D$p$-brane tension,
$\ell^i = X^{(1)i}-X^{(2)i}$ is the separation between the brane and the antibrane in the transverse directions, $g_{ij}$ is the background metric in the transverse space and $\ell = (\ell^i
\ell^jg_{ij})^{1/2}$ is the magnitude of the separation of the branes. In the action \eqref{BaBAction} we have the determinant of $\bf{A}^{(n)}$ given by
\be
\ba
A^{(n)}_{MN} &= P^{(n)}\left[g_{MN}(X^{(n)}) - \frac{|\tau|^2}{\ell_s^2 Q}\, g_{Mi}(X^{(n)})\, \ell^i \ell^j\, g_{jN}(X^{(n)}) \right] \\ & + \ell_s^2 f^{(n)}_{MN} 
+ \frac{1}{Q} \left\{
\frac{\ell_s^2}{2} \left(D_M \tau\, (D_N \tau)^* + D_N \tau\, (D_M \tau)^* \right) \right. 
\\ & + \frac{i}{2} \left( g_{Mi}(X^{(n)}) + \partial_M X^{(n)j} g_{ji}(X^{(n)}) \right) \ell^i \left( \tau (D_N \tau)^* - \tau^* D_N \tau \right) \\ & \left. + \frac{i}{2} \left( \tau (D_M \tau)^* - \tau^* D_M \tau \right) \ell^i \left( g_{iN}(X^{(n)}) - g_{ij}(X^{(n)}) \partial_N X^{(n)j} \right) \right\}.
\ea
\ee
where $P^{(n)}$ signifies the pullback onto the brane, $D_M \tau =  \partial_M \tau - i a_M \tau$ is the covariant derivative acting on the tachyon field and $f_{MN} = \partial_M a_N-\partial_Na_M$ the field strength of the relative gauge field. We will now proceed to study this brane-antibrane action for $p=5$, i.e.~we are looking at a $\text{D5}$-$\overline{\text{D5}}$  pair in an $AdS_5\times T^{1,1}$ background. For definiteness, we work with $T^{1,1}$ as the internal space. However, our results hold for other internal geometries since, between different internal geometries, the AdS bulk action discussed below will only differ by overall constant factors corresponding to internal space volume factors.

\subsection{D5-$\overline{\text{D5}}$ action in $AdS_5 \times T^{1,1}$}
The metric for $AdS_5 \times T^{1,1}$ in Poincaré coordinates with the conformal boundary at $z=0$, where $z$ is the $AdS_5$ radial coordinate, is given by
\begin{gather}
    ds^2_{10} = ds^2_{AdS} + ds^2(T^{1,1}) \ ,\label{AdS5T11metric}\\
    ds^2_{AdS} = \frac{L^2}{z^2} \left( -dx_0^2 + dx_1^2 + dx_2^2 + dx_3^2 + dz^2\right) \ ,\label{AdSmetric}\\
    ds^2(T^{1,1}) = \frac{L^2}{6} \sum_{i=1}^2 (d\theta_i^2 + \sin^2\theta_i \, d\phi_i^2) + \frac{L^2}{9} ( d\psi + \cos\theta_1\, d\phi_1 + \cos\theta_2\, d\phi_2 )^2 \ .\label{T11metric}
\end{gather}
The D5-brane and the $\overline{\text{D}5}$-brane are extended along $(x_0,x_1,x_2,z)$ in $AdS_5$ and along $\Sigma_2$, a sphere inside of $T^{1,1}$. They are coincident in the rest of the internal directions and separated along $x_3$, i.e.~$\ell = x_3^{(1)}-x_3^{(2)}$. 
We work in the supergravity limit where $\ell_s \ll L$ and also take the limit where fluctuations of the brane are small, i.e.~$\partial_a x_3 \ll 1$, for $\xi^a = (x_0,x_2,x_2)$. Finally, with the observation that the tachyon must vanish when $\ell_s \to 0$, we take $\rho$ to be small. Given these assumptions, the action is then expanded up to second order in $\rho$. One immediately finds that the equation of motion for the relative gauge field $a_M$ reads $a_M=\partial_M\varphi$ which fixes the relationship between the tachyon and the gauge field discussed in \eqref{diagonal}. Additionally, we also wish to consider fluctuations of the brane, which are defined as follows,
\begin{equation}
       x_3^{(1,2)}(z,x_0) = x_3^{(1,2)}(z) \pm w(z,x_0) \ , \qquad \ell(x_0,t) = \ell_0(z) + 2w(z,x_0) \ . 
\end{equation}
Here $w(z,x_0)$ is the fluctuation of the relative motion of the two hanging branes, and $\ell_0(z)$ is the separation of the branes as a function of the $AdS_5$ radial coordinate. Given all this, the bulk $AdS_5$ dynamical Lagrangian can now be written down as
\begin{equation}
    \begin{aligned}
        \mathcal{L} &= \frac{2 L^4}{z^4} \left(L_0+L_2 + L_3 \right) \ , \\
        L_0 &= \sqrt{1+ (\partial_z x^{(1,2)}_3(z))^2} \ , \\
        L_2&= \frac{1}{F^3} (\partial_z w)^2 - \frac{1}{F}(\partial_0 w)^2 + \frac{ls^2 z^2}{L^2 F}\left((\partial_z \rho)^2 - F^2 (\partial_0 \rho)^2\right) +\frac{  L^2 \ell_0^2}{ls^2 z^2 F} \rho^2 - \pi \rho^2 F \ , \\
        L_3 &= \frac{12 L^2\ell_0}{ls^2 z^2 F} \,w \,\rho^2 \ .
    \end{aligned}\label{hangingBAB}
\end{equation} 
with $F(z)=\frac{z_0^4}{\sqrt{z_0^8-z^8}}$. The internal $T^{1,1}$ part of the action is studied in Appendix \ref{internal-solution}, which will not affect the analysis here, but will only illustrate how the D5-brane wraps the internal directions. To find the background solution of the Lagrangian above, we turn off all fluctuations, i.e.~$L_2=L_3=0$. The equation of motion is given by 
\begin{equation}
    \frac{d}{dz}\Biggr( \frac{1}{z^4}\frac{\partial_z x_3^{L,R}}{\sqrt{1+(\partial_zx_3^{L/R})^2}}\Biggr) = 0 \ ,
\end{equation}
which can be solved to yield
\begin{equation}
    x_3(z)^{(1,2)}= \pm \frac{\ell_c}{2}\pm \frac{z_0}{5}\bigg(\frac{z}{z_0}\bigg)^{\!5}  {_2F_1}\Bigg(\frac{1}{2},\frac{5}{8};\frac{13}{8};\bigg(\frac{z}{z_0}\bigg)^{\!8}\,\Bigg) \ ,
\end{equation}
where we have defined the constant $z_0$ to be the turning point of the branes. Here we have used the boundary conditions $x_3^{(1)}(z=0) = -\frac{\ell_c}{2}$ and $x_3^{(2)}(z=0) = \frac{\ell_c}{2}$ to specify the positions on the conformal boundary at which the brane and antibrane are anchored as shown in Figure \ref{interact3}. The two strands of hanging branes go into the bulk and are parametrized by
\begin{equation}
    \begin{gathered}
        x^L_3(z)=-\frac{\ell_c}{2}-z_0x(\frac{z}{z_0}) \ , \qquad x_3^R(z) = \frac{\ell_c}{2}+z_0x(\frac{z}{z_0}) \ , \\
        \ell(z) =\ell_c+ 2z_0x(\frac{z}{z_0}) \ , \qquad x(z)=\frac{1}{5}\bigg(\frac{z}{z_0}\bigg)^{\!5}  {_2F_1}\Bigg(\frac{1}{2},\frac{5}{8};\frac{13}{8};\bigg(\frac{z}{z_0}\bigg)^{\!8}\,\Bigg) \ .
    \end{gathered}
\end{equation}
Here we note that  $\ell_0(z)=x_3^{(1)}(z)-x_3^{(2)}(z) = \ell_c+2z_0 x(\tfrac{z}{z_0})$. Let us now reinstate the fluctuations away from the background profile, by turning on $L_1$ and $L_2$ in the Lagrangian, as well as assume a separable solution for the tachyon radius $\rho$ of the following form,
\begin{equation}
    \rho = z \sqrt{F} B(x_0) P(z) \ , \qquad \mbox{with} \qquad \partial_0^2 B(x_0) + m^2 B(x_0) =0 \ ,
\end{equation}
where $B(x_0)$ is an eigenfunction of $\frac{\partial^2}{\partial x_0{}^2}$ with eigenvalue $-m^2$. The equation of motion for $P(z)$ is then given by
\begin{equation}
    \begin{gathered}
        \partial_z^2 P(z) + P(z) V_\text{eff} = \frac{12 M^2 \ell_0(z)}{z^4} \, w P(z) \ , \\
        V_\text{eff} = \left(m^2 + \frac{\pi M }{z^2} - \frac{2}{z^2} \frac{z_0^8 -7z^8}{z_0^8 -z^8}\right) \frac{z_0^8}{z_0^8-z^8} - \frac{M^2 \ell_0^2(z)}{z^4} \ ,
    \end{gathered}\label{Pz_EOM}
\end{equation}
where $M^2 = \frac{L^4}{\ell_s^4}$.  To make progress, we look at the simplified case where we turn off the fluctuation of the relative motion between the two hanging branes, i.e.~$w(z,x_0)=0$, in which case the problem can be turned into the form of a Schrödinger's equation.

\subsection{The Schrödinger problem}

More explicitly, let us make the following redefinition of the tachyon radius $\rho$,
\begin{equation}
    \rho = z  B(x_0) \widetilde{P}(z) \ , \qquad \mbox{with} \qquad \partial_0^2 B + m^2 B =0 \ .
\end{equation}
The equation of motion is now a proper Schrödinger problem,
\begin{equation}
    -\partial_y^2 \widetilde{P}(y) + V_\text{Sch} \widetilde{P}(y) = m^2 \widetilde{P}(y) \ ,
\end{equation}
where we have introduced a new $y$ coordinate and a rescaled wavefunction,
\begin{equation}
    P = \frac{1}{\sqrt{F}} \widetilde{P} \ , \qquad \partial_z = F \partial_y \ , \qquad  y(z)= z \, _2F_1\Bigg(\frac{1}{8},\frac{1}{2};\frac{9}{8};\bigg(\frac{z}{z_0}\bigg)^{\!8} \, \Bigg) \ .  
\end{equation}
The Schrödinger potential is given by
\begin{align}
\label{eq:SchPote}
V_\text{Sch} =\frac{M^2 \ell_0^2(z)}{z^4 F^2}  - \frac{\pi M}{z^2} + \frac{2z_0^8 + z^8}{z^2z_0^8} \ .
 \end{align}
We plot this potential in Figure \ref{potentialplot}, and we see that for $\ell_c = 0$ we have a continuum of tachyonic modes near the boundary $z=0$, and for $\ell_c/z_0\ll 1$ we have a discretum of tachyonic modes near $z=0$. In Section \ref{dynamic_fusion} these tachyonic modes are responsible for the D4-brane that forms near the boundary because of tachyon condensation on the D5-$\overline{\text{D5}}$ system.

\subsection{D5-brane internal solution in $AdS_5\times T^{1,1}$}\label{internal-solution}

We now turn to study the internal $T^{1,1}$ solution. Consider a D5-brane extended along $\gamma^1\times M^3 \times \Sigma^2$ with $\Sigma^2\subset T^{1,1}$. The external support $M^3$ is set to be along the $AdS_5$ directions $(x_0, x_1,x_2)$, while $\gamma^1$ is a curve in the plane spanned by $(x_3, z)$. In the following, we will employ the collective spacetime coordinates $X^{\overline{a}} = (x_0,x_1,x_2)$ and $X^m = (\psi,\theta_1,\phi_1,\theta_2,\phi_2)$. Similarly, denoting the coordinates on the brane as 
\begin{equation}
    \xi^{M}= (x_0,x_1,x_2,t,\vartheta,\varphi) \, ,
\end{equation}
we define $\xi^a=(x_0,x_1,x_2)$ and $\xi^\alpha=(\vartheta,\varphi)$. Because $\vartheta$ and $\varphi$ parametrize an internal sphere, they have the usual ranges $0\leq\vartheta\leq\pi $ and $0\leq\varphi \leq 2\pi$. The brane worldvolume coordinates generally depend on the spacetime coordinates as follows,
\begin{equation}
    \begin{aligned}
        M^3:& \qquad \{x_0,x_1,x_2\} \ , \\
        \gamma^1:& \qquad \{x_3(t),z(t)\} \ , \\
        \Sigma^2:& \qquad \{ \psi(\vartheta,\varphi),\theta_1(\vartheta,\varphi),\phi_1(\vartheta,\varphi),\theta_2(\vartheta,\varphi),\phi_2(\vartheta,\varphi) \} \ ,
    \end{aligned}
\end{equation}
where $t$ parametrizes the hanging brane curve $\gamma^1$.

The worldvolume action of the brane is written as
\begin{equation}
    S = T_5 \int e^{-\Phi} \sqrt{\det\left[P(G)\right]} = T_5 \int e^{-\Phi} \sqrt{\det\left[\frac{\partial X^i}{\partial \xi^M}\frac{\partial X^j}{\partial \xi^N}(G_{ij})\right]} \ ,
\end{equation}
where $P(G)=G_{MN}$ is the pullback of the metric onto the brane worldvolume. This action can be further decomposed as
\begin{equation}
    \begin{adjustbox}{max width=\textwidth}
        $\begin{aligned}
        S &= \label{eq:pullback}
        T_5 \int d^6\xi \;e^{-\Phi} \sqrt{\det\left[\frac{L^2}{z^2}\frac{\partial X^{\overline{a}}}{\partial \xi^a}\frac{\partial X^{\overline{b}}}{\partial \xi^b}\eta_{\overline{a}\overline{b}}\right]}\sqrt{\frac{L^2}{z^2}\left[\left(\frac{\partial x_3}{\partial t}\right)^2 +\left(\frac{\partial z}{\partial t}\right)^2 \right]}\sqrt{\det\left[\frac{\partial X^m}{\partial \xi^\alpha}\frac{\partial X^n}{\partial \xi^\beta}G_{mn}\right]} \\&= \frac{T_5e^{-\Phi}}{3\sqrt{2}} \, V_{M^3}\int dt \, \frac{L^4}{z^4}\sqrt{\left(\frac{\partial x_3}{\partial t}\right)^2 +\left(\frac{\partial z}{\partial t}\right)^2}\int d\vartheta d \varphi\sqrt{\det\left[\frac{\partial X^m}{\partial \xi^\alpha}\frac{\partial X^n}{\partial \xi^\beta}G_{mn}\right]} \\&= \frac{T_5e^{-\Phi}}{3\sqrt{2}} \, V_{M^3} \, S_O\cdot S_I
        \end{aligned}$
    \end{adjustbox}
\end{equation}
with $V_{M^3}=\int dx_0dx_1dx_2\;\sqrt{\det\left[\frac{L^2}{z^2}\frac{\partial X^{\overline{a}}}{\partial \xi^a}\frac{\partial X^{\overline{b}}}{\partial \xi^b}\eta_{\overline{a}\overline{b}}\right]}= \int dx_0dx_1dx_2$. We have also defined the $AdS_5$ bulk action,
\begin{equation}
    S_O= \int dt \, \frac{L^4}{z^4}\sqrt{\left(\frac{\partial x_3}{\partial t}\right)^2 +\left(\frac{\partial z}{\partial t}\right)^2}\ ,
\end{equation}
which is studied in Section \ref{dbrane_config}. In fact, two copies of this action give rise to the background Lagrangian $L_0$ in \eqref{hangingBAB}. Additionally, the internal action is given by
\begin{equation}
\label{internal}
    S_I= \int d\vartheta d \varphi\sqrt{\det\left[\frac{\partial X^m}{\partial \xi^\alpha}\frac{\partial X^n}{\partial \xi^\beta} \, G_{mn}\right]} \ ,
\end{equation}
which is the focus of the rest of this subsection. We note that the metric $G_{mn}$ in the expression above is the $T^{1,1}$ metric, and the Lagrangian density when expanded out reads
\begin{equation}
    \mathcal{L} = (G_{\vartheta\vartheta}\cdot G_{\varphi\varphi}-G_{\vartheta\varphi}^2)^{1/2} =\left(\frac{\partial X^m}{\partial \vartheta}\frac{\partial X^n}{\partial \vartheta}\frac{\partial X^p}{\partial \varphi}\frac{\partial X^q}{\partial \varphi}G_{mn}G_{pq}-\left(\frac{\partial X^m}{\partial \vartheta}\frac{\partial X^n}{\partial \varphi}G_{mn}\right)^{\!2}\right)^{1/2} \ .
\end{equation}

Now we wish to analyze the equations of motion of $X^m(\vartheta,\varphi)$. The first observation we make is that $\mathcal{L}$ does not depend on $\psi$ and $\phi_{i} = (\phi_1,\phi_2)$ so we have three conserved currents,
\begin{equation}
   \partial_{\alpha}j_{\psi}^\alpha = 0 \ , \qquad \partial_{\alpha}j^\alpha_{\phi_1}=0, \qquad \partial_{\alpha}j^\alpha_{\phi_2}=0 \ ,
\end{equation}
where $\xi^\alpha$ runs over $(\vartheta,\varphi)$ and the three currents $j_{\psi}^\alpha , \; j^\alpha_{\phi_i}$ are defined as
\begin{equation}
    \begin{aligned}
        j_{\psi}^\alpha & = \frac{G_{\psi m}}{\mathcal{L}}(G_{\varphi \varphi}\partial_{\vartheta}-G_{\vartheta \varphi}\partial_\varphi\;, \; G_{\vartheta\vartheta}\partial_\varphi-G_{\vartheta\varphi}\partial_\vartheta)X^m \ , \\
        j_{\phi_i}^\alpha & = \frac{G_{\phi_i m}}{\mathcal{L}}(G_{\varphi \varphi}\partial_{\vartheta}-G_{\vartheta \varphi}\partial_\varphi\;, \; G_{\vartheta\vartheta}\partial_\varphi-G_{\vartheta\varphi}\partial_\vartheta)X^m \ .
    \end{aligned}
\end{equation}
The remaining two equations of motion for $\theta_i=(\theta_1,\theta_2)$ are given by 
\begin{multline}
    \partial_\vartheta\left(\frac{G_{\theta_i m}}{\mathcal{L}}(G_{\varphi \varphi}\partial_{\vartheta}-G_{\vartheta \varphi}\partial_\varphi)X^m\right) + \partial_\varphi\left(\frac{G_{\theta_i m}}{\mathcal{L}}(G_{\vartheta\vartheta}\partial_\varphi-G_{\vartheta\varphi}\partial_\vartheta)X^m\right)\\-\frac{\partial_{\theta_i}(G_{mn}G_{pq})}{\mathcal{L}}\left(\partial_\vartheta X^m \partial_\vartheta X^{[n} \partial_\varphi X^{p]} \partial_\varphi X^q\right)=0 \ .
\end{multline}

For simplicity, we orient the brane such that $\theta_i$ and $\phi_i$ lie along $\vartheta$ and $\varphi$ respectively, then our variables become $\{\theta_i(\vartheta),\phi_i(\varphi),\psi\}$. It follows that the $\psi$ terms immediately drop out of the Lagrangian, leaving us with four equations of motion for $\{\theta_i(\vartheta),\phi_i(\varphi)\}$. The $\phi_i$ equations of motion $\partial_\varphi j_{\phi_i}^\varphi = 0$ both give
\begin{equation}
\label{eq:phieom}
    \phi_2^{\prime\prime}(\varphi)\phi_1^\prime(\varphi)-\phi_1^{\prime\prime}(\varphi)\phi_2^\prime(\varphi) = 0 \ .
\end{equation} 
satisfied if both  $n_1,n_2\in \mathbb{Z}$.
The solution to this equation of motion is given by $\phi_1(\varphi)= c_1\cdot\phi_2(\varphi)+c_2$ with $c_1$ and $c_2$ being arbitrary constants. We use the freedom to redefine the coordinate $\varphi$
to write 
\be 
\phi_1(\varphi) = n_1 \varphi \ , \qquad 
\phi_2(\varphi) = n_2 \varphi + c \ ,
\ee 
with $n_1$, $n_2$, $c$ constants.
The coordinates $\varphi$, $\phi_1$, $\phi_2$ all have periods $2\pi$.
As a result, the following quantities must be integral multiples of $2\pi$,
\be 
\phi_1(\varphi + 2\pi) - \phi_1(\varphi)= 2\pi n_1 \ , \qquad 
\phi_2(\varphi + 2\pi) - \phi_2(\varphi)=2\pi n_2 \ .
\ee 
We thus require  $n_1$, $n_2\in \mathbb Z$, which are precisely the winding numbers of the maps $S^1_\varphi \to S^1_{\phi_1}$,
$S^1_\varphi \to S^1_{\phi_2}$, respectively. We note that the arbitrary constants $c$ labels the relative separation of the $\varphi$ to the  $\phi_2$ coordinate system which we have the freedom to set to zero.

Similarly, we redefine $m_1\vartheta = \theta_1(\vartheta)$, but we see that for $m_1\neq 1$, the LHS of this map will have values of $m_1\vartheta$ which either fall outside the range $[0,\pi]$ or do not cover the full range $[0,\pi]$ for $|m_1|<1$. Since these coordinates are supposed to parametrize a 2-sphere, with the range $[0,\pi]$ going from the north pole to the south pole, we are forced to pick $m_1=1$. With this parametrization, the $\theta_2$ equation of motion becomes
\begin{multline}
\label{EOMtheta1}
   2 \left(\theta_2'^2+m_1^2\right) [n_1 \theta_2' \sin (\vartheta  m_1) (2 n_2 \cos \theta_2-n_1 \cos (\vartheta  m_1))+m_1 n_2 \sin \theta_2 (n_2 \cos \theta_2-\\2 n_1 \cos (\vartheta  m_1))]+m_1 \theta_2'' [n_1^2 \cos (2 \vartheta  m_1)-8 n_1 n_2 \cos \theta_2 \cos (\vartheta  m_1)-5 \left(n_1^2+n_2^2\right)\\+n_2^2 \cos (2 \theta_2)]=0 \ .
\end{multline}
Inspired by the forms of $\phi_i$ and $\theta_1$, we employ the ansatz $\theta_2 =m_2\vartheta+b$ where $b$ and $m_2$ are arbitrary constants. The equation of motion reduces to
\be
\ba
\label{EOMtheta}
  2 \left(m_1^2+m_2^2\right) (m_2 n_1 \sin (\vartheta  m_1) (2 n_2 \cos (b+\vartheta  m_2)-n_1 \cos (\vartheta  m_1))\\+m_1 n_2 \sin (b+\vartheta  m_2) (n_2 \cos (b+\vartheta  m_2)-2 n_1 \cos (\vartheta  m_1)))=0 \ .
\ea
\ee
With the following choices of $n_{1,2}$ and $m_{1,2}$, \eqref{EOMtheta} then becomes
\be
\ba
n_1=n_2,\, m_1=\pm m_2 \implies \pm 4m_2^3n_2^3\sin (b) (\cos (b+2 \vartheta  m_2)-2)=0 \ ,\\
n_1=- n_2,\, m_1=\pm m_2 \implies \pm 4m_2^3n_2^3\sin (b) (\cos (b+2 \vartheta  m_2)+2)=0 \ ,
\ea
\ee
which can be solved if we take either $b = 0$ or $b=\pi$. Finally, because we wish to restrict $\theta_2(\vartheta)$ to the range $[0,\pi]$, this then leads us to the family of solutions, 
\begin{equation}
    \begin{gathered}
        \psi(\vartheta, \varphi) = \psi \ ,\\
        \theta_1(\vartheta,\varphi) = \vartheta \ , \qquad \theta_2(\vartheta,\varphi) = \left\{ \begin{array}{l}
         \vartheta \\ 
        \pi-\vartheta
        \end{array} \right.  \ , \\
        \phi_1(\vartheta,\varphi) = n\varphi \ , \quad \phi_2(\vartheta,\varphi) = \pm n\varphi \ .
    \end{gathered}
\end{equation}
where we have used $m_1=\pm m_2=\pm1$, as well as $b=0$ or $\pi$ while keeping the ranges of $\theta_{1,2}$ as $[0,\pi]$. The variables $\psi,n$ are both constants and $n\in \mathbb{Z}$ labels the winding of the $\varphi$ coordinates around the $\phi_i$ coordinates, which shows that the brane can only wrap an $S^2\subset T^{1,1}$ with integer winding numbers. Note that the results follow from the homotopy group $\pi_2(S^2)=\mathbb{Z}$ as expected.

We also observe that \eqref{EOMtheta1} can alternatively be solved by setting $n_2 = 0$
and $\theta_2=\mathrm{const}$. This choice corresponds to a brane that wraps the first $S^2$ factor in the base of $T^{1,1}$, while sitting at a point in the second $S^2$ factor.
By a similar token, one can realize solutions in which the brane sits at a point in the first $S^2$ factor and wraps the second one.

All in all, the solutions studied in this appendix can describe both the homological cycle associated to the baryonic symmetry ($n_1 = 1 = -n_2$), as well as the internal support for realizing the two $SU(2)$ factors in the isometry group of $T^{1,1}$, see Section \ref{T11_symmetry_operators_subsubsec}.

\subsection{D3-brane DBI contribution to the Wilson line in $AdS_5\times S^5$}
\label{Appndx:WilsonDBI}
Let us switch gears to study the DBI action of the D3-brane in $AdS_5\times S^5$. In Section \ref{sec_Wilson_spheres}, the action of the Wilson line was derived from the D3-brane while only considering the topological Wess Zumino part of the action. Therefore, the purpose of this appendix is to show that the DBI action does not spoil that result near the conformal boundary. Recall that the D$3$-brane DBI action is given by 
\begin{equation}
    S_{\text{D}3} = \mu_3\int d^4\xi \;\;\text{Tr}\{e^{-\Phi}[-\det(P(G+B+2\pi \alpha^\prime f_2)]^{1/2}\} \ ,
\end{equation}
where $P$ denotes the pullback to the brane and  $\xi = \{\tau,\sigma^i\}$ are the worldvolume coordinates of the brane. The field $\Phi$ is the dilaton field and we will hereafter turn off both the Kalb-Ramond field $B$ and the Chan Paton field strength $f_2$. The 10d metric is given by 
\begin{equation}
\label{s5metric}
    ds_{10}^2 = ds_{AdS_5}^2+L^2\delta_{ab} DY^aDY^b \ ,
\end{equation}
where $ds_{AdS_5}^2= \frac{L^2}{z^2}\eta_{\mu\nu}dX^\mu dX^\nu$ and $DY^a = dY^a+A_{\mu}{}^{a}{}_bY^bdx^\mu=dY^a+A_{\mu}{}^{a}{}_b Y^bdx^\mu$. Represented as a matrix, we have the metric
\begin{equation}
    G = \begin{pmatrix} \frac{L^2}{z^2}\eta_{\mu\nu}+L^2 \delta_{ab} A_{\mu}{}^{a}{}_e Y^e A_{\nu}{}^{b}{}_d Y^d && L^2\delta_{a b}A_{\mu}{}^{a}{}_{e} Y^e \\  L^2\delta_{a b}A_{\mu}{}^{b}{}_{e} Y^e  && L^2\delta_{ab}
    \end{pmatrix} \ .
\end{equation}

Let us use the brane profile given in \eqref{eq_brane_scalar_profile}, $  Y^a= \chi^{a}{}_{ b}(\tau) \, Y^b_{(0)}(\sigma)$. The scalars $\chi \in SO(6)$ satisfiy $\chi^{-1} = \chi^T$ therefore $ \delta_{cd} = (\chi^{-1}\chi)_{cd} = \chi^{-1}_{cb}\chi^{b}{}_d =  \chi_{bc}\chi^{b}{}_d =\delta_{ab}\chi^a_{\;c}\chi^b_{\; d}$. The D3-brane extends into the $AdS_5$ bulk as $X^\mu=X^\mu{(\tau) = \delta^\mu_z}\tau$, where $\tau = z$ is the $AdS_5$ coordinate on the brane, and $\sigma^i =\sigma^1,\sigma^2,\sigma^3$ are the internal $S^5$ coordinates on the brane. We used diffeomorphism invariance to equate $\tau = z$, and the rest of the coordinates $X^1, X^2, X^3, X^4$ are fixed. We then pull back the metric to the brane. Using $\partial_\tau X^\mu = \delta_\tau^\mu$ and $\eta_{\tau\tau} =\eta_{zz}= 1$, we get three independent terms. The first one is
\begin{equation}
    G_{\tau\tau} = G_{\mu\nu}\partial_\tau X^\mu\partial_\tau X^\nu+2G_{\mu a} \partial_\tau X^\mu\partial_\tau Y^a+G_{ab}\partial_\tau Y^a\partial_\tau Y^b = \frac{L^2}{z^2}+L^2\delta_{ab}\Omega^a_{\;c}\Omega^b_{\; d}Y^c_{(0)}Y^d_{(0)} \ ,
\end{equation}
where we have used the relation $\delta_{b}{}^a = \chi_{b}{}^c\chi^a{}_{c} = (\chi\chi^{-1})_{b}{}^a$ and have defined
\begin{equation} 
    \Omega_{ac} = [\chi^{-1}(\partial_\tau +A)\chi]_{ac}=  ((\chi^{-1})_{ab}(\partial_\tau \chi)^{b}{}_c + \chi^{-1}_{ab}{A_\tau^{\; b}}_d \chi^d_{\;c} = \chi_{ba}\partial_\tau \chi^{b}{}_c + \chi_{ba}{A_\tau^{\; b}}_d \chi^{d}{}_{c} \ .
\end{equation}
The two off-diagonal terms are both equal to
\begin{equation}
    G_{\tau i}= G_{\mu a}\partial_\tau X^{\mu}\partial_{\sigma^i}Y^a +G_{ab}\partial_\tau Y^a\partial_{\sigma^i}Y^b =-L^2\Omega_{cd}Y^c_{(0)} \partial_i Y^d_{(0)} \ ,
\end{equation}
where we use $\partial_i=\partial_{\sigma^i}$. We have also used the antisymmetry of $\Omega_{cd}$ which can be quickly shown using $\partial_\tau \chi^{-1} = -\chi^{-1} (\partial_\tau \chi) \chi^{-1}$ stemming from $0=\partial_\tau(\chi^{-1}\chi) = \partial_\tau \chi^{-1} \chi+\chi^{-1}\partial_\tau \chi$,
\begin{multline}
    \Omega_{cd} = \delta_{ab}\chi^{a}{}_c\partial_{\tau}\chi^{b}{}_d = \delta_{ab}\chi^{a}{}_c\partial_{\tau}(\chi^{-1})_{d}{}^b=-\delta_{ab}\chi^{a}{}_c (\chi^{-1})_{d}{}^f\partial_{\tau}(\chi)_{f}{}^e(\chi^{-1})_{e}{}^b \\ = -\delta_{ab}\chi^{a}{}_c \chi^{f}{}_{d}\partial_{\tau}(\chi)_{f}{}^e \chi^{b}{}_{e} =-\delta_{ce}\chi^{f}{}_d\partial_{\tau}(\chi)_{f}{}^e=-\chi^{f}{}_d\partial_{\tau}(\chi)_{fc}=-\delta_{ab}\chi^{a}{}_d\partial_{\tau}\chi^{b}{}_c = -\Omega_{dc} \ .
\end{multline}
Finally, the third term is given by
\begin{equation}
    G_{ij} = G_{ab}\partial_i Y^a\partial_j Y^b = L^2\delta_{ab} \chi^a_{\; c} \chi^b_{\; d}\partial_i Y^c_{(0)}\partial_jY^d_{(0)}  = L^2\delta_{cd}\partial_i Y^c_{(0)}\partial_jY^d_{(0)} \ .
\end{equation}
The pulled-back metric is written as 
\begin{equation}
   P(G) = \begin{pmatrix}
        \frac{L^2}{z^2}+L^2\delta_{ab}\Omega^{a}{}_{c}\Omega^{b}{}_{d} Y^cY^d && -L^2 \Omega_{cd} Y^c_{(0)}\partial_i Y^d_{(0)}\\ -L^2 \Omega_{cd} Y^c_{(0)}\partial_j Y^d_{(0)} && L^2\delta_{cd}\partial_i Y^c_{(0)}\partial_jY^d_{(0)}
    \end{pmatrix} \ .
\end{equation}
The metric $G$ can be decomposed in powers of $\Omega_{ab}$ as $G = M_0+M$,  where $M_0$ is the term independent of $\Omega_{ab}$, and $M$ comprises the rest of the terms dependent on $\Omega_{ab}$. The resulting $M_0$ and $M$ are given by 
\be
\ba
    M_0 = \begin{pmatrix}
        \frac{L^2}{z^2} && 0\\ 0 && A_{ij}
    \end{pmatrix}\ , && M = \begin{pmatrix}        L^2\delta_{ab}\Omega^{a}{}_{c}\Omega^{b}{}_{d} Y^cY^d && -L^2 \Omega_{cd} Y^c_{(0)}\partial_i Y^d_{(0)}\\ -L^2 \Omega_{cd} Y^c_{(0)}\partial_j Y^d_{(0)} && 0
    \end{pmatrix} \ ,
\ea
\ee
where we have defined $A_{ij}(\sigma) = L^2\delta_{cd}\partial_i Y^c_{(0)}\partial_jY^d_{(0)}$. We also write the inverse of $M_0$ in anticipation of an upcoming calculation,
\begin{equation}
    M_0^{-1}=\begin{pmatrix}
        \frac{z^2}{L^2} && 0 \\0 && (A_{ij})^{-1}
    \end{pmatrix}\ .
\end{equation}

The Lagrangian is proportional to $\sqrt{\det(M_0+M)}$, so we can perform a derivative expansion up to order $\Omega_{ab}^2$,
\begin{equation}
    \begin{adjustbox}{max width=\textwidth}
        $\begin{aligned}
            \sqrt{\det(M_0+M)} & = \sqrt{\det(M_0)}\bigg[1+\frac{1}{2}\,\text{Tr}(M_0^{-1}M)-\frac{1}{4}\,\text{Tr}((M_0^{-1}M)^2)+\frac{1}{8}(\text{Tr}(M_0^{-1}M))^2+\dots\bigg] \\
            & = \frac{L}{z}\sqrt{\det(A_{ij})}\bigg(1+ \frac{z^2}{L^2}\,L^2\Omega_{ab}\Omega_{cd} \delta^{ca}\chi^b_{\; e}\chi^d_{\; f}Y^e_{(0)}Y^f_{(0)}\\& -\frac{1}{2}\frac{z^2}{L^2}\,L^4\Omega_{ab}\Omega_{cd} (A_{ij})^{-1}Y^a_{(0)}Y^c_{(0)}\partial_i Y^b_{(0)}\partial_j Y^d_{(0)}+\dots\bigg) \ .
        \end{aligned}$
    \end{adjustbox}
\end{equation}
We now define and compute the following constants using the explicit expressions of the scalars $Y^a_{(0)}$ in Footnote \ref{footnote_ecplicit_Y},
\be
\ba
    m^{abcd} &= \int_{M^3}d^3\sigma \sqrt{\det(A_{ij})}(A_{ij})^{-1}Y^a_{(0)}Y^c_{(0)}\partial_i Y^b_{(0)}\partial_j Y^d_{(0)}\\
    &= \frac{1}{12} \,\pi^2 L w \bigg[ 
    5\Pi^{a c} \Pi^{b d}
    - 
    \Pi^{a b} \Pi^{c d}
    - \Pi^{c b} \Pi^{a d}
    \bigg] \ ,
    \\
    U^{ab} &= \int_{M^3}d^3\sigma\sqrt{\det(A_{ij})}Y^e_{(0)}Y^f_{(0)} =\frac 12 \,\pi^2 L^3 w \,  \Pi^{ab} \ , \\
    V_{M_3} &= \int_{M^3}d^3\sigma\sqrt{\det(A_{ij})} =2\pi^2 L^3 w \ .
\ea
\ee
In the previous expressions the quantity $\Pi_{ab}$ is defined as
\be 
\Pi_{ab} = \delta_{ab} + n_{ac} n^c{}_b \ ,
\ee 
where $n_{ab} = -n_{ba}$ are the integer parameters that determine the $S^3 \subset S^5$ support of the brane, see discussion around \eqref{eq_nab_parameters}.
We observe that $\Pi_{ab}$ satisfies 
$\Pi_{ab} = \Pi_{ba}$,
$\Pi_{ab} = \Pi_{ac} \Pi^c{}_b$.
Putting all this together, the action becomes
\begin{align}
  S_{\text{D3}}&=\mu_3 \int d^4\xi \; \text{Tr} \{e^{-\Phi}[-\det P(G)]^{1/2}\}  \nonumber\\&= -\mu_3 e^{-\Phi} \int d\tau \; \frac{L}{\tau}\left[V_{M_3}+\frac{\tau^2}{L^2}L^2\left(\delta^{ca}U^{ef}\chi^b_{\; e}(\tau) \chi^d_{\; f}(\tau)-\frac{L^2}{2}m^{abcd}\right)\Omega_{ab}\Omega_{cd}\right] \nonumber\\&= -\mu_3 e^{-\Phi} \int d\tau \; \frac{L}{\tau}\left[V_{M_3}+\tau^2\tilde \beta^{abcd}(\tau)\Omega_{ab}(\tau)\Omega_{cd}(\tau)\right]\nonumber\\&= -\mu_3 e^{-\Phi}\int d\tau \;V_{M_3}\sqrt{g_{\tau\tau}} -\mu_3 e^{-\Phi}\int d\tau \; \tau L\tilde \beta^{abcd}(\tau)\Omega_{ab}(\tau)\Omega_{cd}(\tau) \ ,\label{DBI_final}
\end{align}
with $\tilde\beta^{abcd}(\tau)$ defined as 
\begin{equation}
   \tilde \beta^{abcd}(\tau) =  \delta^{ca}U^{ef}\chi^b_{\; e}(\tau) \chi^d_{\; f}(\tau)-\frac{L^2}{2}\,m^{abcd} \ .
\end{equation}

The first term in \eqref{DBI_final}, which is proportional to the proper length of the Wilson line, can be removed by a counterterm, hence it does not contribute to the group-theoretic properties of the Wilson line. Additionally, the second term in \eqref{DBI_final}, i.e.~the kinetic term $\sim\Omega_{ab}\Omega_{cd}$ of the $\chi^a{}_b(\tau)$ scalars is linear in $\tau$ and can be dropped near the conformal boundary where $\tau\rightarrow 0$. Therefore, only the Wess-Zumino topological term \eqref{eq_brane_gives_WL_final} contributes to the Wilson line action near the boundary, which is our region of interest. 
Finally, we note that this calculation is carried out in $AdS_5\times S^5$ as an explicit example, but our results hold in various other internal geometries because, there, the different $Y^a_{(0)}$ profiles would result in different $V_{M_3}$ and $\tilde \beta^{abcd}$ in \eqref{DBI_final}, but the linear $\tau$ scaling which stems from the $AdS_5$ metric always remains.  Therefore, the DBI action will not contribute to the Wilson line near the conformal boundary in those cases as well. 

\section{A brief review of equivariant cohomology}\label{equivariant_cohomology_appendix}

We describe in this appendix some techniques to compute equivariant cohomology groups, and study a few explicit examples that are relevant to our work.\footnote{For a more detailed exposition, the reader can consult, e.g.~\cite{2007arXiv0709.3615L}. In particular, the relation between the constructions of equivariant cohomology groups using spectral sequences (as we describe here) and using equivariant differential forms is explained clearly therein.} Given a topological space $X$ admitting some $G$-action for a compact Lie group $G$, the {\it $G$-equivariant cohomology} of $X$ with coefficients a commutative ring $R$ is defined as
\begin{equation}
	H_G^\ast(X;R) \coloneqq H^\ast(X_G;R) \, ,
\end{equation}
i.e.~the ordinary cohomology of the homotopy coefficient
\begin{equation}
	X_G \coloneqq EG \times_G X \, ,
\end{equation}
known as the {\it Borel construction} of $X$, where $EG$ is a weakly contractible space admitting a free $G$-action. Such a space is the total space of the universal bundle $G \hookrightarrow EG \to BG$ which classifies principal $G$-bundles, and the base, $BG$, of this bundle is known as the {\it classifying space}. If $G$ also acts freely on $X$, then there is a homotopy equivalence $X_G \simeq X/G$, and so $H^\ast(X_G;R) \cong H^\ast(X/G;R)$ is an isomorphism.

\subsection{Equivariant cohomology through spectral sequence}

Let us present a formal way to compute equivariant cohomology groups through the use of spectral sequences. Note that the homotopy quotient $X_G = EG \times_G X$ is defined as the extension
\begin{equation}
	0 \to X \to EG \times_G X \to EG/G \cong BG \to 0 \, ,
\end{equation}
where $(e,x) \sim (eg,g^{-1}x) \in EG \times X$ are identified under the diagonal $G$-action. We therefore have the {\it Borel fibration}
\begin{equation}
	X \hookrightarrow X_G \to BG
\end{equation}
with fiber $X$ and base $BG$. Our goal is to compute the (ordinary) cohomology of $X_G$.\footnote{Note that the Eilenberg-MacLane space $K(G,1) \simeq BG$ is an infinite-dimensional topological space defined such that its only non-trivial homotopy group is $\pi_1(K(G,1)) = G$. In particular, the corresponding equivariant cohomology groups $H^\ast(X_G;R)$ do not a priori vanish beyond the dimension of the fiber $X$.} This can be achieved using the Leray-Serre spectral sequence for cohomology \cite{Serre:1961abc}, which relates the cohomology groups between the total space, the fiber, and the base in the following manner.

The second page $E_2$ of the spectral sequence consists of entries given by
\begin{equation}
	E_2^{p,q} = H^p(BG;H^q(X;R)) \, ,
\end{equation}
where $p,q \in \mathbb{Z}^{\geq 0}$. Associated with each pair $(p,q)$ is a differential operator acting as a homomorphism $d_2^{p,q} = E_2^{p,q} \to E_2^{p+2,q-1}$, such that the third page $E_3$ has entries $E_3^{p,q} = \text{ker}(d_2^{p,q})/\text{im}(d_2^{p-2,q+1})$.\footnote{With this definition, the entries with negative $p$ or $q$ are implicitly taken to be zero.} In general, the differential on the $r$-th page is a homomorphism
\begin{equation}
	d_r^{p,q} = E_r^{p,q} \to E_r^{p+r,q-r+1} \, ,
\end{equation}
such that
\begin{equation}
	E_{r+1}^{p,q} = \frac{\text{ker}(d_r^{p,q})}{\text{im}(d_r^{p-r,q+r-1})} \, .
\end{equation}
After repeatedly applying the differentials on each page to generate the next page, the entries eventually stabilize to some $E_\infty^{p,q}$. The cohomology $H^\ast(X_G;R)$ of the total space then admits a filtration
\begin{equation}
	E_\infty^{n,0} = F^n_n \subseteq F^n_{n-1} \subseteq \cdots \subseteq F^n_0 = H^n(X_G;R) \, ,
\end{equation}
where the successive subgroups in the sequence above are related by
\begin{equation}
	E_\infty^{p,n-p} = \frac{F^n_p}{F^n_{p+1}} \, .
\end{equation}
To illustrate the technique, we are going to look at a few explicit examples.

\subsubsection{\boldmath The 3-sphere $S^3$}

Consider the 3-sphere as the Hopf fibration $S^1 \hookrightarrow S^3 \to S^2$. There is a $U(1)$-action on $S^3$ induced from the fiber $S^1$. Let us compute the equivariant cohomology $H^\ast_{U(1)}(S^3;\mathbb{Z})$, i.e.~we take $G = U(1)$, $X = S^3$, and $R = \mathbb{Z}$. To do so, recall that
\begin{equation}
	BU(1) \cong \mathbb{CP}^\infty \, ,
\end{equation}
and the cohomology ring of $\mathbb{CP}^\infty$ is $H^\ast(\mathbb{CP}^\infty;\mathbb{Z}) = \mathbb{Z}[c_1]$, where $c_1 \in H^2$ is the first Chern class. In other words, $H^\text{even}(\mathbb{CP}^\infty;\mathbb{Z}) = \mathbb{Z}$ and $H^\text{odd}(\mathbb{CP}^\infty;\mathbb{Z}) = 0$. Moreover, we have $H^0(S^3;\mathbb{Z}) = H^3(S^3;\mathbb{Z}) = \mathbb{Z}$, and zero otherwise.

The $E_2$ page of the relevant spectral sequence is tabulated below.
\begin{center}
	\begingroup
	\setlength{\tabcolsep}{10pt}
	\renewcommand{\arraystretch}{1.25}
	\begin{tabular}{c|ccccccc}
		$3$ & $\mathbb{Z}$ & $0$ & $\mathbb{Z}$ & $0$ & $\mathbb{Z}$ & $0$ & $\cdots$ \\
		$2$ & $0$ & $0$ & $0$ & $0$ & $0$ & $0$ & $\cdots$ \\
		$1$ & $0$ & $0$ & $0$ & $0$ & $0$ & $0$ & $\cdots$ \\
		$0$ & $\mathbb{Z}$ & $0$ & $\mathbb{Z}$ & $0$ & $\mathbb{Z}$ & $0$ & $\cdots$ \\
		\hline
		& $0$ & $1$ & $2$ & $3$ & $4$ & $5$ & $\cdots$ \\
	\end{tabular}
	\endgroup
\end{center}
Since the only non-empty rows are the $0$-th and the third rows, it follows that the only possibly non-trivial differential is $d_4: E_4^{p,q} \to E_4^{p+4,q-3}$, with $E_2^{p,q} = E_3^{p,q} = E_4^{p,q}$. We then have the following $E_5$ page.
\begin{center}
	\begingroup
	\setlength{\tabcolsep}{10pt}
	\renewcommand{\arraystretch}{1.25}
	\begin{tabular}{c|ccccccc}
		$3$ & $\text{ker}(d_4^{0,3})$ & $0$ & $\text{ker}(d_4^{2,3})$ & $0$ & $\text{ker}(d_4^{4,3})$ & $0$ & $\cdots$ \\
		$2$ & $0$ & $0$ & $0$ & $0$ & $0$ & $0$ & $\cdots$ \\
		$1$ & $0$ & $0$ & $0$ & $0$ & $0$ & $0$ & $\cdots$ \\
		$0$ & $\mathbb{Z}$ & $0$ & $\mathbb{Z}$ & $0$ & $\text{coker}(d_4^{0,3})$ & $0$ & $\cdots$ \\
		\hline
		& $0$ & $1$ & $2$ & $3$ & $4$ & $5$ & $\cdots$ \\
	\end{tabular}
	\endgroup
\end{center}
There are no further non-trivial differentials, so $E_5 = E_\infty$. As a result, we conclude that
\begin{equation}
	\begin{gathered}
		H^0_{U(1)}(S^3;\mathbb{Z}) = \mathbb{Z} \, , \qquad H^1_{U(1)}(S^3;\mathbb{Z}) = 0 \, , \qquad H^2_{U(1)}(S^3;\mathbb{Z}) = \mathbb{Z} \, ,\\
		H^{2n+3}_{U(1)}(S^3;\mathbb{Z}) = \text{ker}(d_4^{2n,3}) \, , \qquad H^{2n+4}_{U(1)}(S^3;\mathbb{Z}) = \text{coker}(d_4^{2n,3}) \, .
	\end{gathered}
\end{equation}

One can show that the homomorphism corresponding to the differential $d_4$ is the cup product with the Euler class $e \in H^4(\mathbb{CP}^\infty;\mathbb{Z})$ of the $S^3$-bundle over $\mathbb{CP}^\infty$, known as the {\it Gysin homomorphism}. More explicitly, it acts as
\begin{equation}
	d_4: H^{2n}(\mathbb{CP}^\infty;\mathbb{Z}) \to H^{2n+4}(\mathbb{CP}^\infty;\mathbb{Z}) \, , \qquad x \mapsto e \cup x \, ,
\end{equation}
and further gives rise to a long exact sequence
\begin{equation}
	\resizebox{\textwidth}{!}{$\cdots \to H^{2n+3}(S^3_{U(1)};\mathbb{Z}) \xrightarrow{\pi_\ast} H^{2n}(\mathbb{CP}^\infty;\mathbb{Z}) \xrightarrow{e \, \cup \, (-)} H^{2n+4}(\mathbb{CP}^\infty;\mathbb{Z}) \xrightarrow{\pi^\ast} H^{2n+4}(S^3_{U(1)};\mathbb{Z}) \to \cdots \, ,$}
\end{equation}
where $\pi_\ast$ denotes integration along the fiber $S^3$, and $\pi^\ast$ is the pullback of the fibration $S^3_{U(1)} \xrightarrow{\pi} \mathbb{CP}^\infty$. By exactness, we have
\begin{equation}
	\text{ker}(e \, \cup \, (-)) = \text{im}(\pi_\ast) \, , \qquad \text{im}(e \, \cup \, (-)) = \text{ker}(\pi^\ast) \, ,
\end{equation}
which means that $E_\infty^{2n,3} = \text{ker}(d_4^{2n,3})$ is the subgroup of $H^{2n+3}(S^3_{U(1)};\mathbb{Z})$ surviving under fiber integration, while $E_\infty^{2n+4,0} = \text{coker}(d_4^{2n,3})$ is the quotient of $H^{2n+4}(\mathbb{CP}^\infty;\mathbb{Z}) = \mathbb{Z}$ by the subgroup of elements that pull back trivially to the total space $S^3_{U(1)} = EU(1) \times_{U(1)} S^3$.

The Euler class of the 3-sphere bundle can be constructed explicitly as follows. Since $e$ is a degree-4 class, and the cohomology ring of $\mathbb{CP}^\infty$ is generated by its first Chern class $c_1$ in degree 2, we must have $e = m \, c_1 \cup c_1$ for some integer $m$. Recall that the coordinates on $S^3$ can be parametrized as $(z_1,z_2) \in \mathbb{C}^2$ satisfying the constraint $|z_1|^2 + |z_2|^2 = 1$. The $S^1$-action on $S^3$ is defined by a choice of two integers, $(m_1,m_2)$, such that
\begin{equation}
	\lambda \cdot (z_1,z_2) = (\lambda^{m_1} z_1,\lambda^{m_2} z_2)
\end{equation}
for any $\lambda \in U(1)$, which preserves $|z_1|^2 + |z_2|^2 = 1$. The Euler class is then given by
\begin{equation}
	e = m_1 m_2 \, c_1^2 \, ,
\end{equation}
i.e.~$m = m_1 m_2$. Note that $e$ vanishes if either $m_1$ or $m_2$ is zero (and the $S^1$-action is no longer free), then $\text{ker}(e \, \cup \, (-)) = \mathbb{Z}$ and $\text{im}(e \, \cup \, (-)) = 0$, in which case
\begin{equation}
	H^{2n+3}_{U(1)}(S^3;\mathbb{Z}) = \mathbb{Z} \, ,\qquad H^{2n+4}_{U(1)}(S^3;\mathbb{Z}) = \mathbb{Z} \, .
\end{equation}
On the contrary, if neither of $m_1$ and $m_2$ is zero, then $\text{ker}(e \, \cup \, (-)) = 0$ and $\text{im}(e \, \cup \, (-)) = m\mathbb{Z}$, so we find
\begin{equation}
	H^{2n+3}_{U(1)}(S^3;\mathbb{Z}) = 0 \, ,\qquad H^{2n+4}_{U(1)}(S^3;\mathbb{Z}) = \mathbb{Z}/m\mathbb{Z} = \mathbb{Z}_m \, .
\end{equation}
Equivalently, we can express the cohomology ring as $H^\ast(S^3_{U(1)};\mathbb{Z}) = \mathbb{Z}[c_1]/(mc_1^2)$ where $|c_1|=2$. Note that the $S^1$-action on $S^3$ is free if and only if $m_1 = m_2 = m = 1$, in which case the homotopy quotient $S^3_{U(1)}$ is homotopically equivalent to the quotient $S^3/S^1 \cong S^2$, thus recovering $H^\ast(S^3_{U(1)};\mathbb{Z}) \cong H^\ast(S^3/S^1;\mathbb{Z})$.

\subsubsection{\boldmath The coset space $T^{1,1}$}

Now we replace $S^3$ in the previous computation with the coset space $T^{1,1} = (SU(2) \times SU(2))/U(1)$. For our purposes, it suffices to use the fact that this space can be viewed as a fibration $S^1 \hookrightarrow T^{1,1} \to S^2 \times S^2$, such that it is topologically equivalent to $S^2 \times S^3$, i.e.~$H^0(T^{1,1};\mathbb{Z}) = H^2(T^{1,1};\mathbb{Z}) = H^3(T^{1,1};\mathbb{Z}) = H^5(T^{1,1};\mathbb{Z}) = \mathbb{Z}$, and zero otherwise.

The Borel fibration in question is $T^{1,1} \hookrightarrow T^{1,1}_{U(1)} \to BU(1)$, so the $E_2$ page of the spectral sequence is given by
\begin{center}
	\begingroup
	\setlength{\tabcolsep}{10pt}
	\renewcommand{\arraystretch}{1.25}
	\begin{tabular}{c|ccccccccc}
		$5$ & $\mathbb{Z}$ & $0$ & $\mathbb{Z}$ & $0$ & $\mathbb{Z}$ & $0$ & $\mathbb{Z}$ & $0$ & $\cdots$ \\
		$4$ & $0$ & $0$ & $0$ & $0$ & $0$ & $0$ & $0$ & $0$ & $\cdots$ \\
		$3$ & $\mathbb{Z}$ & $0$ & $\mathbb{Z}$ & $0$ & $\mathbb{Z}$ & $0$ & $\mathbb{Z}$ & $0$ & $\cdots$ \\
		$2$ & $\mathbb{Z}$ & $0$ & $\mathbb{Z}$ & $0$ & $\mathbb{Z}$ & $0$ & $\mathbb{Z}$ & $0$ & $\cdots$ \\
		$1$ & $0$ & $0$ & $0$ & $0$ & $0$ & $0$ & $0$ & $0$ & $\cdots$ \\
		$0$ & $\mathbb{Z}$ & $0$ & $\mathbb{Z}$ & $0$ & $\mathbb{Z}$ & $0$ & $\mathbb{Z}$ & $0$ & $\cdots$ \\
		\hline
		& $0$ & $1$ & $2$ & $3$ & $4$ & $5$ & $6$ & $7$ & $\cdots$ \\
	\end{tabular}
	\endgroup
\end{center}
and the $E_3$ page is
\begin{center}
	\begingroup
	\setlength{\tabcolsep}{10pt}
	\renewcommand{\arraystretch}{1.25}
	\begin{tabular}{c|ccccccccc}
		$5$ & $\mathbb{Z}$ & $0$ & $\mathbb{Z}$ & $0$ & $\mathbb{Z}$ & $0$ & $\mathbb{Z}$ & $0$ & $\cdots$ \\
		$4$ & $0$ & $0$ & $0$ & $0$ & $0$ & $0$ & $0$ & $0$ & $\cdots$ \\
		$3$ & $\text{ker}(d_2^{0,3})$ & $0$ & $\text{ker}(d_2^{2,3})$ & $0$ & $\text{ker}(d_2^{4,3})$ & $0$ & $\text{ker}(d_2^{6,3})$ & $0$ & $\cdots$ \\
		$2$ & $\mathbb{Z}$ & $0$ & $\text{coker}(d_2^{0,3})$ & $0$ & $\text{coker}(d_2^{2,3})$ & $0$ & $\text{coker}(d_2^{4,3})$ & $0$ & $\cdots$ \\
		$1$ & $0$ & $0$ & $0$ & $0$ & $0$ & $0$ & $0$ & $0$ & $\cdots$ \\
		$0$ & $\mathbb{Z}$ & $0$ & $\mathbb{Z}$ & $0$ & $\mathbb{Z}$ & $0$ & $\mathbb{Z}$ & $0$ & $\cdots$ \\
		\hline
		& $0$ & $1$ & $2$ & $3$ & $4$ & $5$ & $6$ & $7$ & $\cdots$ \\
	\end{tabular}
	\endgroup
\end{center}
The $E_4$ page is the same as the $E_3$ page, while the $E_5$ page is
\begin{center}
	\begingroup
	\setlength{\tabcolsep}{10pt}
	\renewcommand{\arraystretch}{1.25}
	\begin{tabular}{c|ccccccccc}
		$5$ & $\text{ker}(d_4^{0,5})$ & $0$ & $\text{ker}(d_4^{2,5})$ & $0$ & $\text{ker}(d_4^{4,5})$ & $0$ & $\text{ker}(d_4^{6,5})$ & $0$ & $\cdots$ \\
		$4$ & $0$ & $0$ & $0$ & $0$ & $0$ & $0$ & $0$ & $0$ & $\cdots$ \\
		$3$ & $\text{ker}(d_4^{0,3})$ & $0$ & $\text{ker}(d_4^{2,3})$ & $0$ & $\text{ker}(d_4^{4,3})$ & $0$ & $\text{ker}(d_4^{6,3})$ & $0$ & $\cdots$ \\
		$2$ & $\mathbb{Z}$ & $0$ & $\text{coker}(d_2^{0,3})$ & $0$ & $\text{coker}(d_4^{0,5})$ & $0$ & $\text{coker}(d_4^{2,5})$ & $0$ & $\cdots$ \\
		$1$ & $0$ & $0$ & $0$ & $0$ & $0$ & $0$ & $0$ & $0$ & $\cdots$ \\
		$0$ & $\mathbb{Z}$ & $0$ & $\mathbb{Z}$ & $0$ & $\text{coker}(d_4^{0,3})$ & $0$ & $\text{coker}(d_4^{2,3})$ & $0$ & $\cdots$ \\
		\hline
		& $0$ & $1$ & $2$ & $3$ & $4$ & $5$ & $6$ & $7$ & $\cdots$ \\
	\end{tabular}
	\endgroup
\end{center}
Similarly, the $E_6$ page is the same as the $E_5$ page, while the $E_7$ page is
\begin{center}
	\begingroup
	\setlength{\tabcolsep}{10pt}
	\renewcommand{\arraystretch}{1.25}
	\begin{tabular}{c|ccccccccc}
		$5$ & $\text{ker}(d_6^{0,5})$ & $0$ & $\text{ker}(d_6^{2,5})$ & $0$ & $\text{ker}(d_6^{4,5})$ & $0$ & $\text{ker}(d_6^{6,5})$ & $0$ & $\cdots$ \\
		$4$ & $0$ & $0$ & $0$ & $0$ & $0$ & $0$ & $0$ & $0$ & $\cdots$ \\
		$3$ & $\text{ker}(d_4^{0,3})$ & $0$ & $\text{ker}(d_4^{2,3})$ & $0$ & $\text{ker}(d_4^{4,3})$ & $0$ & $\text{ker}(d_4^{6,3})$ & $0$ & $\cdots$ \\
		$2$ & $\mathbb{Z}$ & $0$ & $\text{coker}(d_2^{0,3})$ & $0$ & $\text{coker}(d_4^{0,5})$ & $0$ & $\text{coker}(d_4^{2,5})$ & $0$ & $\cdots$ \\
		$1$ & $0$ & $0$ & $0$ & $0$ & $0$ & $0$ & $0$ & $0$ & $\cdots$ \\
		$0$ & $\mathbb{Z}$ & $0$ & $\mathbb{Z}$ & $0$ & $\text{coker}(d_4^{0,3})$ & $0$ & $\text{coker}(d_6^{0,5})$ & $0$ & $\cdots$ \\
		\hline
		& $0$ & $1$ & $2$ & $3$ & $4$ & $5$ & $6$ & $7$ & $\cdots$ \\
	\end{tabular}
	\endgroup
\end{center}
There are no further non-trivial differentials, and so $E_7 = E_\infty$.

Following the previous arguments, we find
\begin{equation}
	\begin{gathered}
		H^0_{U(1)}(T^{1,1};\mathbb{Z}) = \mathbb{Z} \, , \qquad H^1_{U(1)}(T^{1,1};\mathbb{Z}) = 0 \, ,\\
		H^2_{U(1)}(T^{1,1};\mathbb{Z}) = \mathbb{Z} \oplus \mathbb{Z} \, , \qquad H^3_{U(1)}(T^{1,1};\mathbb{Z}) = \text{ker}(d_4^{0,3}) \, ,
	\end{gathered}
\end{equation}
where we made use of the fact that $\text{Ext}(\mathbb{Z},\mathbb{Z})=0$. The equivariant cohomology groups for higher degrees are given by extensions of the form,
\begin{gather}
	0 \to \text{coker}(d_4^{0,3}) \to H^4_{U(1)}(T^{1,1};\mathbb{Z}) \to \text{coker}(d_2^{0,3}) \to 0 \, ,\\
	0 \to \text{ker}(d_4^{2n+2,3}) \to H^{2n+5}_{U(1)}(T^{1,1};\mathbb{Z}) \to \text{ker}(d_6^{2n,5}) \to 0 \, ,\\
	0 \to \text{coker}(d_6^{2n,5}) \to H^{2n+6}_{U(1)}(T^{1,1};\mathbb{Z}) \to \text{coker}(d_4^{2n,5}) \to 0 \, .
\end{gather}

One can modify the standard argument with the Gysin homomorphism in the following manner. The multiplicative structure of the spectral sequence gives us a homomorphism $E_2^{0,3} \times E_2^{2n,0} \to E_2^{2n,3}$, corresponding to the usual cup product $\cup: H^0(\mathbb{CP}^\infty;\mathbb{Z}) \times H^{2n}(\mathbb{CP}^\infty;\mathbb{Z}) \to H^{2n}(\mathbb{CP}^\infty;\mathbb{Z})$, which is an isomorphism since $H^\ast(\mathbb{CP}^\infty;\mathbb{Z})$ is generated only by the first Chern class $c_1$ in degree 2. Therefore, every $x \in E_2^{2n,3} = H^{2n}(\mathbb{CP}^\infty;\mathbb{Z})$ can be expressed as $1 \cup x$ where $1$ is regarded as the unit element in $E_2^{0,3} = H^0(\mathbb{CP}^\infty;\mathbb{Z})$, and the latter $x$ in the cup product is regarded as an element in $E_2^{2n,0} = H^{2n}(\mathbb{CP}^\infty,\mathbb{Z})$.

The differential $d_2$ acts as a graded derivation on $1 \cup x \in E_2^{0,3} \times E_2^{2n,0}$, i.e.
\begin{equation}
	d_2^{2n,3}(1 \cup x) = d_2^{0,3}(1) \cup x + (-1)^{0+3} 1 \cup d_2^{2n,0}(x) = d_2^{0,3}(1) \cup x \, ,
\end{equation}
where we used the fact that $x \in E_2^{2n,0}$ is located at the bottom row and so its differential $d_2^{2n,0}(x)$ automatically vanishes. The relation above can be equivalently interpreted as
\begin{equation}
	d_2^{2n,3}(x) = d_2^{0,3}(1) \cup x
\end{equation}
for $x \in E_2^{2n,3}$. In other words, by defining the Euler class $e \coloneqq d_2^{0,3}(1) \in E_2^{2,2} = H^2(\mathbb{CP}^\infty;\mathbb{Z})$, we can identify the differential $d_2^{2n,3}$ as the cup product with such a class. Similarly to before, if $e = m c_1$ is non-trivial, then
\begin{equation}
	\text{ker}(d_2^{2n,3}) = 0 \, , \qquad \text{coker}(d_2^{2n,3}) = \mathbb{Z}_m \, ,
\end{equation}
and if $e=0$, then
\begin{equation}
	\text{ker}(d_2^{2n,3}) = \mathbb{Z} \, , \qquad \text{coker}(d_2^{2n,3}) = \mathbb{Z} \, .
\end{equation}

Let us first focus on the case where the Euler class $e$ is non-trivial, such that we have the following $E_3=E_4$ page.
\begin{center}
	\begingroup
	\setlength{\tabcolsep}{10pt}
	\renewcommand{\arraystretch}{1.25}
	\begin{tabular}{c|ccccccccc}
		$5$ & $\mathbb{Z}$ & $0$ & $\mathbb{Z}$ & $0$ & $\mathbb{Z}$ & $0$ & $\mathbb{Z}$ & $0$ & $\cdots$ \\
		$4$ & $0$ & $0$ & $0$ & $0$ & $0$ & $0$ & $0$ & $0$ & $\cdots$ \\
		$3$ & $0$ & $0$ & $0$ & $0$ & $0$ & $0$ & $0$ & $0$ & $\cdots$ \\
		$2$ & $\mathbb{Z}$ & $0$ & $\mathbb{Z}_m$ & $0$ & $\mathbb{Z}_m$ & $0$ & $\mathbb{Z}_m$ & $0$ & $\cdots$ \\
		$1$ & $0$ & $0$ & $0$ & $0$ & $0$ & $0$ & $0$ & $0$ & $\cdots$ \\
		$0$ & $\mathbb{Z}$ & $0$ & $\mathbb{Z}$ & $0$ & $\mathbb{Z}$ & $0$ & $\mathbb{Z}$ & $0$ & $\cdots$ \\
		\hline
		& $0$ & $1$ & $2$ & $3$ & $4$ & $5$ & $6$ & $7$ & $\cdots$ \\
	\end{tabular}
	\endgroup
\end{center}
It follows that
\begin{equation}
	\text{ker}(d_4^{2n,3}) = 0 \, , \qquad \text{coker}(d_4^{2n,3}) = \mathbb{Z} \, .
\end{equation}
The $E_5=E_6$ page now reads as
\begin{center}
	\begingroup
	\setlength{\tabcolsep}{10pt}
	\renewcommand{\arraystretch}{1.25}
	\begin{tabular}{c|ccccccccc}
		$5$ & $\text{ker}(d_4^{0,5})$ & $0$ & $\text{ker}(d_4^{2,5})$ & $0$ & $\text{ker}(d_4^{4,5})$ & $0$ & $\text{ker}(d_4^{6,5})$ & $0$ & $\cdots$ \\
		$4$ & $0$ & $0$ & $0$ & $0$ & $0$ & $0$ & $0$ & $0$ & $\cdots$ \\
		$3$ & $0$ & $0$ & $0$ & $0$ & $0$ & $0$ & $0$ & $0$ & $\cdots$ \\
		$2$ & $\mathbb{Z}$ & $0$ & $\mathbb{Z}_m$ & $0$ & $\text{coker}(d_4^{0,5})$ & $0$ & $\text{coker}(d_4^{2,5})$ & $0$ & $\cdots$ \\
		$1$ & $0$ & $0$ & $0$ & $0$ & $0$ & $0$ & $0$ & $0$ & $\cdots$ \\
		$0$ & $\mathbb{Z}$ & $0$ & $\mathbb{Z}$ & $0$ & $\mathbb{Z}$ & $0$ & $\mathbb{Z}$ & $0$ & $\cdots$ \\
		\hline
		& $0$ & $1$ & $2$ & $3$ & $4$ & $5$ & $6$ & $7$ & $\cdots$ \\
	\end{tabular}
	\endgroup
\end{center}
where $d_4^{2n,5}: E_4^{2n,5} \to E_4^{2n+4,2}$ becomes a (group) homomorphism $\mathbb{Z} \to \mathbb{Z}_m$, determined by the choice of an integer $\ell \in \mathbb{Z}_m$ that sends $1 \mapsto \ell$, and generally $a \mapsto \ell a \ \text{mod} \ m$ for any $a \in \mathbb{Z}$. For a given $\ell$, we thus have
\begin{equation}
	\text{ker}(d_4^{2n,5}) = \frac{m}{\gcd(\ell,m)} \, \mathbb{Z} \, , \qquad \text{coker}(d_4^{2n,5}) = \frac{\mathbb{Z}_m}{\mathbb{Z}_{m/\gcd(\ell,m)}} \cong \mathbb{Z}_{\gcd(\ell,m)} \, ,
\end{equation}
with $\gcd(0,m) \coloneqq m$. Finally, we obtain the $E_7=E_\infty$ page below,
\begin{center}
	\begingroup
	\setlength{\tabcolsep}{10pt}
	\renewcommand{\arraystretch}{1.25}
	\begin{tabular}{c|ccccccccc}
		$5$ & $\text{ker}(d_6^{0,5})$ & $0$ & $\text{ker}(d_6^{2,5})$ & $0$ & $\text{ker}(d_6^{4,5})$ & $0$ & $\text{ker}(d_6^{6,5})$ & $0$ & $\cdots$ \\
		$4$ & $0$ & $0$ & $0$ & $0$ & $0$ & $0$ & $0$ & $0$ & $\cdots$ \\
		$3$ & $0$ & $0$ & $0$ & $0$ & $0$ & $0$ & $0$ & $0$ & $\cdots$ \\
		$2$ & $\mathbb{Z}$ & $0$ & $\mathbb{Z}_m$ & $0$ & $\mathbb{Z}_{\gcd(\ell,m)}$ & $0$ & $\mathbb{Z}_{\gcd(\ell,m)}$ & $0$ & $\cdots$ \\
		$1$ & $0$ & $0$ & $0$ & $0$ & $0$ & $0$ & $0$ & $0$ & $\cdots$ \\
		$0$ & $\mathbb{Z}$ & $0$ & $\mathbb{Z}$ & $0$ & $\mathbb{Z}$ & $0$ & $\text{coker}(d_6^{0,5})$ & $0$ & $\cdots$ \\
		\hline
		& $0$ & $1$ & $2$ & $3$ & $4$ & $5$ & $6$ & $7$ & $\cdots$ \\
	\end{tabular}
	\endgroup
\end{center}
where $d_6^{2n,5}: E_6^{2n,5} \to E_6^{2n+6,0}$ becomes a homomorphism $\tilde{m} \mathbb{Z} \to \mathbb{Z}$ with $\tilde{m} = m/\gcd(\ell,m)$, characterized by the choice of an integer $k \in \mathbb{Z}$ that sends $\tilde{m} \mapsto k\tilde{m}$. For generic $k \neq 0$, we thus have
\begin{equation}
	\text{ker}(d_6^{2n,5}) = 0 \, , \qquad \text{coker}(d_6^{2n,5}) = \mathbb{Z}/k\tilde{m}\mathbb{Z} = \mathbb{Z}_{k\tilde{m}} \, ,
\end{equation}
otherwise $\text{ker}(d_6^{2n,5}) = \tilde{m}\mathbb{Z}$ and $\text{coker}(d_6^{2n,5}) = \mathbb{Z}$.

To summarize, when the Euler class $e \coloneqq m c_1 \in H^2(\mathbb{CP}^\infty;\mathbb{Z})$ is non-trivial, the $U(1)$-equivariant cohomology groups of $T^{1,1}$ are given by
\begin{equation}
	\begin{gathered}
		H^0_{U(1)}(T^{1,1};\mathbb{Z}) = \mathbb{Z} \, , \qquad H^1_{U(1)}(T^{1,1};\mathbb{Z}) = 0 \, , \qquad H^2_{U(1)}(T^{1,1};\mathbb{Z}) = \mathbb{Z} \oplus \mathbb{Z} \, ,\\
		H^3_{U(1)}(T^{1,1};\mathbb{Z}) = 0 \, , \qquad 0 \to \mathbb{Z} \to H^4_{U(1)}(T^{1,1};\mathbb{Z}) \to \mathbb{Z}_m \to 0 \, ,\\
		H^{2n+5}_{U(1)}(T^{1,1};\mathbb{Z}) = 0 \, , \qquad 0 \to \mathbb{Z}_{km/\gcd(\ell,m)} \to H^{2n+6}_{U(1)}(T^{1,1};\mathbb{Z}) \to \mathbb{Z}_{\gcd(\ell,m)} \to 0 \, .
	\end{gathered}
\end{equation}
The extension class for $H^4_{U(1)}(T^{1,1};\mathbb{Z})$ corresponds to a choice of 2-cocycle in $H^2(\mathbb{Z}_m,\mathbb{Z}) \cong \mathbb{Z}_m$ in group cohomology. By the same token, the extension class for $H^{2n+6}_{U(1)}(T^{1,1};\mathbb{Z})$ corresponds to a choice of 2-cocycle in $H^2(\mathbb{Z}_{\gcd(\ell,m)},\mathbb{Z}_{km/\gcd(\ell,m)}) \cong \mathbb{Z}_{\gcd(km/\gcd(\ell,m),\gcd(\ell,m))}$.\footnote{When $k=0$, the group $\mathbb{Z}_{km/\gcd(\ell,m)}$ should be replaced with $\mathbb{Z}$.} Note that we recover the equivalence $H^\ast_{U(1)}(T^{1,1};\mathbb{Z}) \cong H^\ast(S^2 \times S^2)$ if and only if we pick $m=1$ and $k=1$. These can be regarded as the conditions for the $S^1$-action on $S^2 \times S^2$, i.e.~the base of the fibration $S^1 \hookrightarrow T^{1,1} \to S^2 \times S^2$, to be free.

For completeness, let us also work out the case where the Euler class $e \in H^2(\mathbb{CP}^\infty;\mathbb{Z})$ is trivial. We have the following $E_3=E_4$ page
\begin{center}
	\begingroup
	\setlength{\tabcolsep}{10pt}
	\renewcommand{\arraystretch}{1.25}
	\begin{tabular}{c|ccccccccc}
		$5$ & $\mathbb{Z}$ & $0$ & $\mathbb{Z}$ & $0$ & $\mathbb{Z}$ & $0$ & $\mathbb{Z}$ & $0$ & $\cdots$ \\
		$4$ & $0$ & $0$ & $0$ & $0$ & $0$ & $0$ & $0$ & $0$ & $\cdots$ \\
		$3$ & $\mathbb{Z}$ & $0$ & $\mathbb{Z}$ & $0$ & $\mathbb{Z}$ & $0$ & $\mathbb{Z}$ & $0$ & $\cdots$ \\
		$2$ & $\mathbb{Z}$ & $0$ & $\mathbb{Z}$ & $0$ & $\mathbb{Z}$ & $0$ & $\mathbb{Z}$ & $0$ & $\cdots$ \\
		$1$ & $0$ & $0$ & $0$ & $0$ & $0$ & $0$ & $0$ & $0$ & $\cdots$ \\
		$0$ & $\mathbb{Z}$ & $0$ & $\mathbb{Z}$ & $0$ & $\mathbb{Z}$ & $0$ & $\mathbb{Z}$ & $0$ & $\cdots$ \\
		\hline
		& $0$ & $1$ & $2$ & $3$ & $4$ & $5$ & $6$ & $7$ & $\cdots$ \\
	\end{tabular}
	\endgroup
\end{center}
and the $E_5=E_6$ page is
\begin{center}
	\begingroup
	\setlength{\tabcolsep}{10pt}
	\renewcommand{\arraystretch}{1.25}
	\begin{tabular}{c|ccccccccc}
		$5$ & $\text{ker}(d_4^{0,5})$ & $0$ & $\text{ker}(d_4^{2,5})$ & $0$ & $\text{ker}(d_4^{4,5})$ & $0$ & $\text{ker}(d_4^{6,5})$ & $0$ & $\cdots$ \\
		$4$ & $0$ & $0$ & $0$ & $0$ & $0$ & $0$ & $0$ & $0$ & $\cdots$ \\
		$3$ & $\text{ker}(d_4^{0,3})$ & $0$ & $\text{ker}(d_4^{2,3})$ & $0$ & $\text{ker}(d_4^{4,3})$ & $0$ & $\text{ker}(d_4^{6,3})$ & $0$ & $\cdots$ \\
		$2$ & $\mathbb{Z}$ & $0$ & $\mathbb{Z}$ & $0$ & $\text{coker}(d_4^{0,5})$ & $0$ & $\text{coker}(d_4^{2,5})$ & $0$ & $\cdots$ \\
		$1$ & $0$ & $0$ & $0$ & $0$ & $0$ & $0$ & $0$ & $0$ & $\cdots$ \\
		$0$ & $\mathbb{Z}$ & $0$ & $\mathbb{Z}$ & $0$ & $\text{coker}(d_4^{0,3})$ & $0$ & $\text{coker}(d_4^{2,3})$ & $0$ & $\cdots$ \\
		\hline
		& $0$ & $1$ & $2$ & $3$ & $4$ & $5$ & $6$ & $7$ & $\cdots$ \\
	\end{tabular}
	\endgroup
\end{center}
where $d_4^{2n,3}: E_4^{2n,3} \to E_4^{2n+4,0}$ is a homomorphism $\mathbb{Z} \to \mathbb{Z}$ characterized by an integer $\ell \in \mathbb{Z}$, while $d_4^{2n,5}: E_4^{2n,5} \to E_4^{2n+4,2}$ is a homomorphism $\mathbb{Z} \to \mathbb{Z}$ characterized by another integer $\ell' \in \mathbb{Z}$. Assuming $\ell,\ell'$ are both non-zero, we have
\begin{equation}
	\text{ker}(d_4^{2n,3}) = 0 \, , \qquad \text{coker}(d_4^{2n,3}) = \mathbb{Z}_\ell \, , \qquad \text{ker}(d_4^{2n,5}) = 0 \, , \qquad \text{coker}(d_4^{2n,5}) = \mathbb{Z}_{\ell'} \, ,
\end{equation}
otherwise $\text{ker}(d_4) = \mathbb{Z}$ and $\text{coker}(d_4) = 0$. The $E_7=E_\infty$ page is then given by
\begin{center}
	\begingroup
	\setlength{\tabcolsep}{10pt}
	\renewcommand{\arraystretch}{1.25}
	\begin{tabular}{c|ccccccccc}
		$5$ & $0$ & $0$ & $0$ & $0$ & $0$ & $0$ & $0$ & $0$ & $\cdots$ \\
		$4$ & $0$ & $0$ & $0$ & $0$ & $0$ & $0$ & $0$ & $0$ & $\cdots$ \\
		$3$ & $0$ & $0$ & $0$ & $0$ & $0$ & $0$ & $0$ & $0$ & $\cdots$ \\
		$2$ & $\mathbb{Z}$ & $0$ & $\mathbb{Z}$ & $0$ & $\mathbb{Z}_{\ell'}$ & $0$ & $\mathbb{Z}_{\ell'}$ & $0$ & $\cdots$ \\
		$1$ & $0$ & $0$ & $0$ & $0$ & $0$ & $0$ & $0$ & $0$ & $\cdots$ \\
		$0$ & $\mathbb{Z}$ & $0$ & $\mathbb{Z}$ & $0$ & $\mathbb{Z}_\ell$ & $0$ & $\mathbb{Z}_\ell$ & $0$ & $\cdots$ \\
		\hline
		& $0$ & $1$ & $2$ & $3$ & $4$ & $5$ & $6$ & $7$ & $\cdots$ \\
	\end{tabular}
	\endgroup
\end{center}
such that the $U(1)$-equivariant cohomology groups of $T^{1,1}$ are
\begin{equation}
	\begin{gathered}
		H^0_{U(1)}(T^{1,1};\mathbb{Z}) = \mathbb{Z} \, , \qquad H^1_{U(1)}(T^{1,1};\mathbb{Z}) = 0 \, , \qquad H^2_{U(1)}(T^{1,1};\mathbb{Z}) = \mathbb{Z} \oplus \mathbb{Z} \, ,\\
		H^3_{U(1)}(T^{1,1};\mathbb{Z}) = 0 \, , \qquad H^4_{U(1)}(T^{1,1};\mathbb{Z}) = \mathbb{Z} \oplus \mathbb{Z}_\ell \, ,\\
		H^{2n+5}_{U(1)}(T^{1,1};\mathbb{Z}) = 0 \, , \qquad 0 \to \mathbb{Z}_\ell \to H^{2n+6}_{U(1)}(T^{1,1};\mathbb{Z}) \to \mathbb{Z}_{\ell'} \to 0 \, .
	\end{gathered}
\end{equation}

One can similarly determine the groups if either $\ell$ or $\ell'$ is zero. In short, we find
\begin{equation}
	\begin{gathered}
		H^3_{U(1)}(T^{1,1};\mathbb{Z}) = \begin{cases} 0 & \ell \neq 0 \, ,\\ \mathbb{Z} & \ell = 0 \, , \end{cases} \qquad H^4_{U(1)}(T^{1,1};\mathbb{Z}) = \begin{cases} \mathbb{Z} \oplus \mathbb{Z}_\ell & \ell \neq 0 \, ,\\ \mathbb{Z} & \ell = 0 \, , \end{cases}\\
		H^{2n+5}_{U(1)}(T^{1,1};\mathbb{Z}) = \begin{cases} 0 & \ell,\ell' \neq 0 \, ,\\ \mathbb{Z} & \ell = 0, \ell' \neq 0 \, ,\\ \frac{\ell}{\gcd(k,\ell)} \, \mathbb{Z} & \ell \neq 0, \ell' = 0 \, ,\\ \mathbb{Z} \oplus \mathbb{Z} & \ell,\ell' = 0 \, , \end{cases}\\
		H^{2n+6}_{U(1)}(T^{1,1};\mathbb{Z}) = \begin{cases} 0 \to \mathbb{Z}_\ell \to H^{2n+6}_{U(1)}(T^{1,1};\mathbb{Z}) \to \mathbb{Z}_{\ell'} \to 0 & \ell,\ell' \neq 0 \, ,\\ \mathbb{Z}_{\ell'} & \ell = 0, \ell' \neq 0 \, ,\\ \mathbb{Z}_{\gcd(k,\ell)} & \ell \neq 0, \ell' = 0 \, ,\\ 0 & \ell,\ell' = 0 \, ,\end{cases}
	\end{gathered}
\end{equation}
where $k \in \mathbb{Z}_\ell$ is an integer specifying the homomorphism $d_6^{2n,6}: \mathbb{Z} \to \mathbb{Z}_\ell$.

\subsubsection{\boldmath The 5-sphere $S^5$}

Another example is to take $X = S^5$, which can be regarded as the fibration $S^1 \hookrightarrow S^5 \to \mathbb{CP}^2$. The $E_2$ page of the relevant spectral sequence is given by
\begin{center}
	\begingroup
	\setlength{\tabcolsep}{10pt}
	\renewcommand{\arraystretch}{1.25}
	\begin{tabular}{c|ccccccccc}
		$5$ & $\mathbb{Z}$ & $0$ & $\mathbb{Z}$ & $0$ & $\mathbb{Z}$ & $0$ & $\mathbb{Z}$ & $0$ & $\cdots$ \\
		$4$ & $0$ & $0$ & $0$ & $0$ & $0$ & $0$ & $0$ & $0$ & $\cdots$ \\
		$3$ & $0$ & $0$ & $0$ & $0$ & $0$ & $0$ & $0$ & $0$ & $\cdots$ \\
		$2$ & $0$ & $0$ & $0$ & $0$ & $0$ & $0$ & $0$ & $0$ & $\cdots$ \\
		$1$ & $0$ & $0$ & $0$ & $0$ & $0$ & $0$ & $0$ & $0$ & $\cdots$ \\
		$0$ & $\mathbb{Z}$ & $0$ & $\mathbb{Z}$ & $0$ & $\mathbb{Z}$ & $0$ & $\mathbb{Z}$ & $0$ & $\cdots$ \\
		\hline
		& $0$ & $1$ & $2$ & $3$ & $4$ & $5$ & $6$ & $7$ & $\cdots$ \\
	\end{tabular}
	\endgroup
\end{center}
and the only non-trivial differential is $d_6: E_6^{p,q} \to E_6^{p+6,q-5}$, which yields the following $E_7$ page.
\begin{center}
	\begingroup
	\setlength{\tabcolsep}{10pt}
	\renewcommand{\arraystretch}{1.25}
	\begin{tabular}{c|ccccccccc}
		$5$ & $\text{ker}(d_6^{0,5})$ & $0$ & $\text{ker}(d_6^{2,5})$ & $0$ & $\text{ker}(d_6^{4,5})$ & $0$ & $\text{ker}(d_6^{6,5})$ & $0$ & $\cdots$ \\
		$4$ & $0$ & $0$ & $0$ & $0$ & $0$ & $0$ & $0$ & $0$ & $\cdots$ \\
		$3$ & $0$ & $0$ & $0$ & $0$ & $0$ & $0$ & $0$ & $0$ & $\cdots$ \\
		$2$ & $0$ & $0$ & $0$ & $0$ & $0$ & $0$ & $0$ & $0$ & $\cdots$ \\
		$1$ & $0$ & $0$ & $0$ & $0$ & $0$ & $0$ & $0$ & $0$ & $\cdots$ \\
		$0$ & $\mathbb{Z}$ & $0$ & $\mathbb{Z}$ & $0$ & $\mathbb{Z}$ & $0$ & $\text{coker}(d_6^{0,5})$ & $0$ & $\cdots$ \\
		\hline
		& $0$ & $1$ & $2$ & $3$ & $4$ & $5$ & $6$ & $7$ & $\cdots$ \\
	\end{tabular}
	\endgroup
\end{center}
Consequently, the $U(1)$-equivariant cohomology groups are
\begin{equation}
	\begin{gathered}
		H^0_{U(1)}(S^5;\mathbb{Z}) = \mathbb{Z} \, , \qquad H^1_{U(1)}(S^5;\mathbb{Z}) = 0 \, , \qquad H^2_{U(1)}(S^5;\mathbb{Z}) = \mathbb{Z} \, ,\\
		H^3_{U(1)}(S^5;\mathbb{Z}) = 0 \, , \qquad H^4_{U(1)}(S^5;\mathbb{Z}) = \mathbb{Z} \, , \\
		H^{2n+5}_{U(1)}(S^5;\mathbb{Z}) = \text{ker}(d_6^{2n,5}) \, , \qquad H^{2n+6}_{U(1)}(S^5;\mathbb{Z}) = \text{coker}(d_6^{2n,5}) \, .
	\end{gathered}
\end{equation}

Similarly to the case of $S^3$, the differential $d_6$ corresponds to the Gysin homomorphism given by the cup product with the Euler class $e \in H^6(\mathbb{CP}^\infty;\mathbb{Z})$. This class is characterized by the $S^1$-action on $S^5 \subset \mathbb{C}^3$, i.e.
\begin{equation}
    \lambda \cdot (z_1,z_2,z_3) = (\lambda^{m_1} z_1,\lambda^{m_2} z_2,\lambda^{m_3} z_3)
\end{equation}
for any $\lambda \in U(1)$, and for some choice of integers $(m_1,m_2,m_3)$, such that
\begin{equation}
    e = m c_1^3 \, ,
\end{equation}
where $m = m_1 m_2 m_3$. Suppose the Euler class is chosen to be non-zero, then the corresponding cohomology ring is given by $H^\ast_{U(1)}(S^5;\mathbb{Z}) = \mathbb{Z}[c_1]/(mc_1^3)$ where $|c_1| = 2$, otherwise we have $H^n_{U(1)}(S^5;\mathbb{Z}) = \mathbb{Z}$ except for $n=1,3$. The $S^1$-action on $S^5$ is free if and only if $m_1 = m_2 = m_3 = m = 1$, in which case we recover $H^\ast_{U(1)}(S^5;\mathbb{Z}) \cong H^\ast(\mathbb{CP}^2;\mathbb{Z})$ by homotopy equivalence, where $\mathbb{CP}^2$ is the base of the fibration $S^1 \hookrightarrow S^5 \to \mathbb{CP}^2$.

\subsubsection{\boldmath The odd-dimensional sphere $S^{2n+1}$}

Our previous analysis can be readily generalized to any odd-dimensional sphere, which can be regarded as the fibration $S^1 \hookrightarrow S^{2n+1} \to \mathbb{CP}^n$. The $S^1$-action on $S^{2n+1} \subset \mathbb{C}^{n+1}$ is
\begin{equation}
    \lambda \cdot (z_1,z_2,\dots,z_{n+1}) = (\lambda^{m_1} z_1,\lambda^{m_2} z_2,\dots,\lambda^{m_{n+1}} z_{n+1})
\end{equation}
for any $\lambda \in U(1)$, and for some choice of integers $(m_1,m_2,\dots,m_{n+1})$, such that the Euler class $e \in H^{2n+2}(\mathbb{CP}^\infty;\mathbb{Z})$ corresponds to
\begin{equation}
    e = m c_1^{n+1} \, ,
\end{equation}
where $m = m_1 m_2 \cdots m_{n+1}$.

If the Euler class is non-zero, then the $U(1)$-equivariant cohomology ring of $S^{2n+1}$ is
\begin{equation}
	H^\ast_{U(1)}(S^{2n+1};\mathbb{Z}) = \mathbb{Z}[c_1]/(mc_1^{n+1}) \, .
\end{equation}
Note that the $S^1$-action is free if and only if $m=1$, thus recovering $H^\ast_{U(1)}(S^{2n+1};\mathbb{Z}) \cong H^\ast(\mathbb{CP}^n;\mathbb{Z})$. On the other hand, if $e=0$, then we find that $H^p_{U(1)}(S^{2n+1};\mathbb{Z}) = \mathbb{Z}$ except for $p=1,3,\dots,2n-1$ where it is vanishing.

\subsubsection{\boldmath The even-dimensional sphere $S^{2n}$}

The case for even-dimensional spheres is qualitatively different from before, since the $S^1$-action on $S^{2n}$ is necessarily not free. In particular, $H^\ast_{U(1)}(S^{2n};\mathbb{Z}) \ncong H^\ast(S^{2n}/S^1;\mathbb{Z})$. Nevertheless, we can recycle some of our previous computations. The non-trivial entries of the $E_2$ page of the spectral sequence are
\begin{equation}
	E_2^{2p,0} = E_2^{2p,2n} = \mathbb{Z}
\end{equation}
for $p \in \mathbb{Z}^{\geq 0}$, so the only possibly non-trivial differential is the Gysin homomorphism,
\begin{equation}
	d_{2n+1}: \mathbb{Z} = E_{2n+1}^{2m,2n} \to E_{2n+1}^{2m+2n+1,0} = 0 \, ,
\end{equation}
corresponding to the cup product with the Euler class $e \in H^{2n+1}(\mathbb{CP}^\infty;\mathbb{Z})$. As the odd cohomology groups of $\mathbb{CP}^\infty$ are trivial, the Euler class automatically vanishes. Consequently, we have
\begin{equation}
	H^{2p < 2n}_{U(1)}(S^{2n};\mathbb{Z}) = \mathbb{Z} \, , \qquad H^{2p+1}_{U(1)}(S^{2n};\mathbb{Z}) = 0 \, , \qquad H^{2p \geq 2n}_{U(1)}(S^{2n};\mathbb{Z}) = \mathbb{Z} \oplus \mathbb{Z} \, .
\end{equation}

\subsection{Equivariant homology and Poincaré duality}

Naïvely, one may define a notion of equivariant homology through an analogous construction,
\begin{equation}
	H_\ast^G(X;R) \stackrel{?}{=} H_\ast(X_G;R) \, .
\end{equation}
However, such a definition is problematic since the homotopy quotient $X_G = EG \times_G X$ is an infinite-dimensional space. We do not attempt to give a proper treatment here, but the basic idea is to break $EG$ up into a filtration of spaces $EG_i$, inducing a finite-dimensional fibration
\begin{equation}
	X \hookrightarrow EG_i \times_G X \to BG_i \, .
\end{equation}
By regarding $X$ as a complex algebraic variety (or generally, a scheme) admitting some $G$-action for a linear algebraic group $G$, along with a linear $G$-representation $V$ satisfying suitable conditions, the equivariant {\it Borel-Moore homology} groups are defined as
\begin{equation}
	H_n^G(X;R) \coloneqq H_{n + 2 \, \text{dim}(V) - 2 \, \text{dim}(G)}(EG_i \times_G X;R)
\end{equation}
for $i \geq \text{dim}_\mathbb{C}(X) - n/2$ (see, e.g.~\cite{1996alg.geom..9018E,1999math......8172G,2018arXiv180903226B}). The latter can then be computed using the Leray-Serre spectral sequence for homology. It turns out that such a definition is independent of the choices of $V$ and $i$.

A class $[Y] \in H_n^G(X;R)$ corresponds to a closed $G$-equivariant subvariety $Y \subseteq X$ with $\text{dim}_\mathbb{C}(Y) = n/2$ (or $\text{dim}_\mathbb{R}(Y) = n$). As with ordinary (co)homology groups, there is a version of {\it Poincaré duality} between equivariant (co)homology groups given by
\begin{equation}
	H^n_G(X;R) \cong H_{2 \, \text{dim}_\mathbb{C}(X) - n}^G(X;R)\label{equivariant_Poincare_duality}
\end{equation}
for any $X$ that is smooth and closed \cite{1996alg.geom..9018E,10.1307/mmj/1030132709}. This is a special case of the cap product
\begin{equation}
	\cap: H_p^G(X;R) \times H^q_G(X;R) \to H_{p-q}^G(X;R) \, , \qquad (\alpha,\beta) \mapsto \alpha \cap \beta \, ,
\end{equation}
which can be interpreted as an integration of equivariant cocycles against equivariant cycles. In particular, integrating an $n$-cocycle $\alpha$ against the fundamental class $[X] \in H_{2 \, \text{dim}_\mathbb{C}(X)}^G(X;R)$ yields a cycle $[Y] = \alpha \cap [X]$ with $\text{dim}_\mathbb{C}(Y) = \text{dim}_\mathbb{C}(X) - n/2$.

Interestingly, the Poincaré duality described above is related to the ordinary version if the $G$-action on $X$ is free. Recall that in this case, there is a homotopy equivalence $X_G \simeq X/G$, then
\begin{equation}
	H^n_G(X;R) \cong H^n(X/G;R) \cong H_{\text{dim}_\mathbb{R}(X/G) - n}(X/G;R) \, ,
\end{equation}
where the second isomorphism is the statement of Poincaré duality between ordinary (co)homology groups, using the fact that the orbit space $X/G$ is smooth when $G$ acts freely. This further establishes an isomorphism
\begin{equation}
	H_n^G(X;R) \cong H_{n + \text{dim}_\mathbb{R}(X/G) - 2 \, \text{dim}_\mathbb{C}(X)}(X/G;R)\label{equivariant_homology_ordinary_homology_isomorphism}
\end{equation}
between the equivariant homology of $X$ and the ordinary homology of $X/G$.

\subsubsection{Examples}

Rather than computing the equivariant Borel-Moore homology groups directly using a spectral sequence, we can apply Poincaré duality on the results we obtained earlier for the equivariant cohomology groups. For simplicity, let us hereafter restrict to the cases where the $S^1$-action acts freely on the spaces of interest.

The odd-dimensional sphere $S^1 \hookrightarrow S^{2n+1} \to \mathbb{CP}^n$ can be regarded as a closed subspace embedded in $\mathbb{C}^{n+1}$, defined by the set of points $(z_1,\dots,z_{n+1}) \in \mathbb{C}^{n+1}$ satisfying $\sum_{i=1}^{n+1} |z_i|^2 = 1$. In this sense, we take $\text{dim}_\mathbb{C}(S^{2n+1}) = n+1$.\footnote{It is perhaps more proper to regard $S^{2n+1}$ as a real algebraic variety defined by $\sum_{i=1}^{2n+2} x_i^2 = 1$ with $(x_1,\dots,x_{2n+2}) \in \mathbb{R}^{2n+2}$, and then compute the analogous equivariant Borel-Moore homology groups \cite{Anderson_Fulton_2023}. For our purposes, it suffices to use the complex version as introduced above.} The equivariant Poincaré duality \eqref{equivariant_Poincare_duality} tells us that the non-vanishing $U(1)$-equivariant homology groups of $S^{2m+1}$ are
\begin{equation}
	H_{2(n+1-p)}^{U(1)}(S^{2n+1};\mathbb{Z}) \cong H^{2p}_{U(1)}(S^{2n+1};\mathbb{Z}) = \mathbb{Z}
\end{equation}
for $0 \leq p \leq n$. For example, when $n=2$, we have
\begin{equation}
	H_2^{U(1)}(S^5;\mathbb{Z}) \cong H_4^{U(1)}(S^5;\mathbb{Z}) \cong H_6^{U(1)}(S^5;\mathbb{Z}) = \mathbb{Z} \, .\label{S5_equivariant_homology}
\end{equation}
The complex dimensions of these generators as equivariant subvarieties of $\mathbb{C}^3$ are respectively 1, 2, and 3. In other words, the real dimensions of such equivariant cycles as submanifolds $Y \subset S^5$ are respectively 1, 3, and 5. We may identify them as the fibrations,
\begin{equation}
	S^1 \hookrightarrow Y^1 \to \ast \, , \qquad S^1 \hookrightarrow Y_3 \to x \, , \qquad S^1 \hookrightarrow Y^5 \to x^2 \, ,
\end{equation}
where $x \in H_2(\mathbb{CP}^2;\mathbb{Z})$ denotes the generator of $\mathbb{CP}^2$ satisfying $x^3=0$, and $S^1$ can be understood as the fiber of the fibration $S^1 \hookrightarrow S^5 \to \mathbb{CP}^2$. This presentation manifests the isomorphism \eqref{equivariant_homology_ordinary_homology_isomorphism}, i.e.
\begin{equation}
	H_n^{U(1)}(S^5;\mathbb{Z}) \cong H_{n-2}(\mathbb{CP}^2;\mathbb{Z}) \, ,
\end{equation}
where $\text{dim}_\mathbb{R}(\mathbb{CP}^2)=4$ and $\text{dim}_\mathbb{C}(S^5)=3$. It is obvious that the equivariant homology groups \eqref{S5_equivariant_homology} correspond to the following ordinary homology groups of the orbit space $S^5/S^1 \cong \mathbb{CP}^2$,
\begin{equation}
	H_0(\mathbb{CP}^2;\mathbb{Z}) \cong H_2(\mathbb{CP}^2;\mathbb{Z}) \cong H_4(\mathbb{CP}^2;\mathbb{Z}) = \mathbb{Z} \, ,
\end{equation}
whose generators are precisely the bases of the fibrations, $(\ast,x,x^2) \in H_\ast(\mathbb{CP}^2;\mathbb{Z})$, as we described above.

The coset space $S^1 \hookrightarrow T^{1,1} \to S^2 \times S^2$ can be studied in a similar fashion. Through \eqref{equivariant_Poincare_duality}, we find that
\begin{equation}
	\begin{gathered}
		H_2^{U(1)}(T^{1,1};\mathbb{Z}) \cong H^4_{U(1)}(T^{1,1};\mathbb{Z}) = \mathbb{Z} \, ,\\
		H_4^{U(1)}(T^{1,1};\mathbb{Z}) \cong H^2_{U(1)}(T^{1,1};\mathbb{Z}) = \mathbb{Z} \oplus \mathbb{Z} \, ,\\
		H_6^{U(1)}(T^{1,1};\mathbb{Z}) \cong H^0_{U(1)}(T^{1,1};\mathbb{Z}) = \mathbb{Z} \, ,
	\end{gathered}
\end{equation}
whose generators respectively have complex dimensions 1, 2, and 3, corresponding to submanifolds $Y \subset T^{1,1}$ with real dimensions 1, 3, and 5. Accordingly, we may identify them as the fibrations,
\begin{equation}
	S^1 \hookrightarrow Y^1 \to \ast \, , \qquad S^1 \hookrightarrow Y^3_1 \to x_1 \, , \qquad S^1 \hookrightarrow Y^3_2 \to x_2 \, , \qquad S^1 \hookrightarrow Y^5 \to x_1 x_2 \, ,
\end{equation}
where $x_1,x_2 \in H_2(S^2 \times S^2) = \mathbb{Z} \oplus \mathbb{Z}$ are the fundamental classes of the two copies of $S^2$. Just like before, these equivariant homology groups map to the following ordinary homology groups of the orbit space $T^{1,1}/S^1 \cong S^2 \times S^2$,
\begin{equation}
	H_0(S^2 \times S^2;\mathbb{Z}) = \mathbb{Z} \, , \qquad H_2(S^2 \times S^2;\mathbb{Z}) = \mathbb{Z} \oplus \mathbb{Z} \, , \qquad  H_4(S^2 \times S^2;\mathbb{Z}) = \mathbb{Z} \, ,
\end{equation}
whose generators are precisely the bases of the aforementioned fibrations.

Let us also illustrate how the cap product, or equivalently, equivariant integration works. Focusing on the example of $T^{1,1}$, suppose we denote the generator of the first factor of $H^2_{U(1)}(T^{1,1};\mathbb{Z}) = \mathbb{Z} \oplus \mathbb{Z}$ as $\alpha$. One can then integrate it against, say, its Poincaré dual $[Y^3_1] \in H_4^{U(1)}(T^{1,1};\mathbb{Z})$, by computing the cap product
\begin{equation}
	\int_{Y^3_1} \alpha \coloneqq \alpha \cap [Y^3_1] = [Y^1] \in H_2^{U(1)}(T^{1,1};\mathbb{Z}) = \mathbb{Z} \, ,
\end{equation}
where $\text{dim}_\mathbb{C}([Y^1]) = 1$. Note that the counterpart of $[Y^1]$ in ordinary homology is nothing but the base of the fibration $S^1 \hookrightarrow Y^1 \to \ast$, i.e.~the point $\ast \in H_0(S^2 \times S^2;\mathbb{Z})$. This is consistent with the philosophy that ``integrating an $n$-cocycle against an $n$-cycle yields a number,'' as one usually obtains in ordinary (co)homology.

\subsection{Relation to equivariant K-theory}

Formally, D-brane charges in Type II string theories are classified by topological K-theory \cite{Minasian:1997mm,Witten:1998cd,Maldacena:2001xj}.\footnote{If there exists a cohomologically non-trivial NSNS 3-form flux $H_3$, then such charges are instead classified by twisted K-theory.} Let us briefly comment on the relation between the cohomology and the K-theory of the spaces studied in this work, particularly in the equivariant setting.

We start by recalling that in the non-equivariant case, there is an isomorphism between the rational K-theory and the rational cohomology of any finite CW complex $X$,
\begin{equation}
    K^\ast(X) \otimes \mathbb{Q} \cong H^\ast(X;\mathbb{Q}) \, ,
\end{equation}
given by the Chern character map \cite{AtiyahHirzebruch1961}. Specifically, it is a natural ring homomorphism which sends
\begin{equation}
    \text{ch}: \begin{cases} K^0(X) \to H^\text{even}(X;\mathbb{Q}) = \bigoplus_{n \geq 0} H^{2n}(X;\mathbb{Q}) \, ,\\[1ex] K^1(X) \to H^\text{odd}(X;\mathbb{Q}) = \bigoplus_{n \geq 0} H^{2n+1}(X;\mathbb{Q}) \, . \end{cases}
\end{equation}
This isomorphism, however, no longer holds at the level of integer coefficients.

To compute $K^\ast(X)$, one uses the Atiyah-Hirzebruch spectral sequence \cite{AtiyahHirzebruch1961}, whose second page $E_2$ consists of entries
\begin{equation}
    E_2^{p,q} = H^p(X;K^q(\text{pt})) \Rightarrow K^{p+q}(X) \, .
\end{equation}
For complex K-theory, we have
\begin{equation}
    K^q(\text{pt}) = \begin{cases} \mathbb{Z} & q \ \text{even,} \\ 0 & q \ \text{odd,} \end{cases}
\end{equation}
so all the odd rows of $E_2$ identically vanish. As a result, the first non-trivial differential is $d_3: E_3^{p,q} \to E_3^{p+3,q-2}$ on the third page $E_3$, given explicitly by
\begin{equation}
    d_3 = \beta \circ \text{Sq}^2 \circ (\text{mod} \ 2): H^p(X;\mathbb{Z}) \to H^{p+3}(X;\mathbb{Z}) \, ,
\end{equation}
where $\beta: H^i(X;\mathbb{Z}_2) \to H^{i+1}(X;\mathbb{Z})$ is the Bockstein homomorphism associated with the short exact sequence $0 \to \mathbb{Z} \to \mathbb{Z} \to \mathbb{Z}_2 \to 0$, and $\text{Sq}^2: H^i(X;\mathbb{Z}_2) \to H^{i+2}(X;\mathbb{Z}_2)$ is the Steenrod square operation. All other non-trivial differentials are necessarily those on odd pages $E_{2n+1}$, and they are constructed with similar stable cohomology operations.

It follows that K-theory and cohomology match for any $X$ of dimension $n \leq 2$, since all differentials can only act trivially. Another case where they agree is when $X$ is torsion-free, such that $H^\ast(X;\mathbb{Z})$ is a free Abelian group. Here all the stable cohomology operations involved in the differentials act trivially on torsion-free elements. We also do not have to worry about possibly non-trivial group extensions (as we saw previously when working with the Leray-Serre spectral sequence), since the extensions must split when only free groups are present. Nevertheless, it is important to note that the match between $K^\ast(X)$ and $H^\ast(X;\mathbb{Z})$ in the cases above holds only at the level of Abelian groups, but not rings as with rational coefficients.

Let us move on to the equivariant setting. As discussed earlier, equivariant cohomology $H^\ast_G(X)$ is defined to be the ordinary cohomology of the Borel construction $X_G = EG \times_G X$. Equivariant K-theory $K_G(X)$, on the other hand, is a classification of $G$-equivariant vector bundles over $X$, and it is generally not isomorphic to the ordinary K-theory of $X_G$, also known as the Borel-equivariant K-theory $K_G^\text{Bor}(X) \coloneqq K(X_G)$.

In general, the Atiyah-Segal completion theorem \cite{10.4310/jdg/1214428815} asserts that there is an isomorphism upon completion,
\begin{equation}
    K_G^\ast(X)_{\hat{I}} \cong K^\ast(X_G) \, ,
\end{equation}
where $I$ denotes the augmentation ideal of the representation ring $R(G)$ of $G$. However, throughout this work we are primarily interested in regular Sasaki-Einstein spaces $X$ on which $G = U(1)$ acts freely, such that $X_G \cong X/G = B$ reduces to the Kähler-Einstein base. In this case, we indeed have an isomorphism between the equivariant K-theory of $X$ and the ordinary K-theory of the base,
\begin{equation}
    K_G^\ast(X) \cong K^\ast(B) \, ,
\end{equation}
similarly to the relation between equivariant cohomology and ordinary cohomology under free $G$-actions.

The question of comparing between $K_{U(1)}^\ast(X)$ and $H_{U(1)}^\ast(X)$ thus boils down to a comparison between $K^\ast(B)$ and $H^\ast(B)$. Just like before, at the level of rational coefficients, there is an isomorphism of rings,
\begin{equation}
    K_{U(1)}^\ast(X) \otimes \mathbb{Q} \cong H_{U(1)}^\ast(X;\mathbb{Q}) \, ,
\end{equation}
which fails to hold if we switch to integer coefficients. In spite of that, we can still compare $K^\ast(B)$ and $H^\ast(B;\mathbb{Z})$ as Abelian groups.

For a regular Sasaki-Einstein manifold $X$ of (real) dimension $n$, the base $B$ is a Kähler-Einstein manifold of dimension $n-1$, satisfying the Einstein condition,
\begin{equation}
    R = \lambda g \, ,
\end{equation}
where $R$ denotes the Ricci tensor and $g$ denotes the metric. Moreover, the Einstein constant $\lambda$ is strictly positive, making $B$ a Fano manifold.\footnote{For quasi-regular Sasaki-Einstein manifolds (e.g.~$X = Y^{p,q}$ or $L^{p,q,r}$), the base $B$ is a Fano orbifold.}

At $n=3$, the only Fano curve is $B \cong \mathbb{CP}^1$, which has no torsion, so we can immediately conclude that $K_{U(1)}^\ast(S^3) \cong H_{U(1)}^\ast(S^3;\mathbb{Z})$. At $n=5$, Fano surfaces, also known as del Pezzo surfaces, are completely classified. Those which admit Kähler-Einstein metrics are isomorphic either to $\mathbb{CP}^2$, $\mathbb{CP}^1 \times \mathbb{CP}^1$, or $\mathbb{CP}^2$ blown up at $3 \leq k \leq 8$ generic points. All such manifolds are torsion-free, and hence $K_{U(1)}^\ast(X) \cong H_{U(1)}^\ast(X;\mathbb{Z})$. We therefore see that as far as D-branes in Type IIB string theory on $AdS_5 \times SE_5$ are concerned, working with equivariant cohomology or equivariant K-theory are the same at the level of Abelian groups.

At $n=7$, Fano threefolds are also completely classified \cite{V_A_Iskovskih_1977,Mori_Mukai_1981,Mori_Mukai_2003}, but $B$ may admit torsion in degrees $3$ and $4$. There is then a possibly non-trivial differential $d_3: H^3(B;\mathbb{Z}) \to H^6(B;\mathbb{Z})$.\footnote{Note that the higher differentials vanish automatically due to degree reasons.} Even if it acts trivially, the resulting group extensions from the spectral sequence do not necessarily split, so we do not have an isomorphism between $K_{U(1)}^\ast(X)$ and $H_{U(1)}^\ast(X;\mathbb{Z})$ in general. We note, though, that the explicit examples of Sasaki-Einstein 7-manifolds (i.e.~$S^7$, $M^{3,2}$, and $Q^{1,1,1}$) considered in Section \ref{subsec:AdS4SE7} all have torsion-free bases, thereby the isomorphism does hold. In any case, these computations apply to M-theory on $AdS_7 \times SE_7$, where the relevant probe objects are M2/5-branes rather than D-branes. The charges of the former are typically thought of to be classified by ordinary cohmology (or some version of elliptic cohomology \cite{Kriz:2004xj,Sati:2010ss}) but not K-theory, so it is somewhat less illuminating to formulate our construction in K-theoretic terms.

\newpage


\bibliographystyle{JHEP}
\bibliography{refs}

\providecommand{\href}[2]{#2}\begingroup\raggedright\begin{thebibliography}{100}

\bibitem{Gaiotto:2014kfa}
D.~Gaiotto, A.~Kapustin, N.~Seiberg and B.~Willett, \emph{{Generalized Global Symmetries}}, \href{https://doi.org/10.1007/JHEP02(2015)172}{\emph{JHEP} {\bfseries 02} (2015) 172} [\href{https://arxiv.org/abs/1412.5148}{{\ttfamily 1412.5148}}].

\bibitem{Bhardwaj:2017xup}
L.~Bhardwaj and Y.~Tachikawa, \emph{{On finite symmetries and their gauging in two dimensions}}, \href{https://doi.org/10.1007/JHEP03(2018)189}{\emph{JHEP} {\bfseries 03} (2018) 189} [\href{https://arxiv.org/abs/1704.02330}{{\ttfamily 1704.02330}}].

\bibitem{Thorngren:2019iar}
R.~Thorngren and Y.~Wang, \emph{{Fusion category symmetry. Part I. Anomaly in-flow and gapped phases}}, \href{https://doi.org/10.1007/JHEP04(2024)132}{\emph{JHEP} {\bfseries 04} (2024) 132} [\href{https://arxiv.org/abs/1912.02817}{{\ttfamily 1912.02817}}].

\bibitem{Thorngren:2021yso}
R.~Thorngren and Y.~Wang, \emph{{Fusion category symmetry. Part II. Categoriosities at c = 1 and beyond}}, \href{https://doi.org/10.1007/JHEP07(2024)051}{\emph{JHEP} {\bfseries 07} (2024) 051} [\href{https://arxiv.org/abs/2106.12577}{{\ttfamily 2106.12577}}].

\bibitem{Bhardwaj:2022yxj}
L.~Bhardwaj, L.E.~Bottini, S.~Schafer-Nameki and A.~Tiwari, \emph{{Non-invertible higher-categorical symmetries}}, \href{https://doi.org/10.21468/SciPostPhys.14.1.007}{\emph{SciPost Phys.} {\bfseries 14} (2023) 007} [\href{https://arxiv.org/abs/2204.06564}{{\ttfamily 2204.06564}}].

\bibitem{Muller:2025ext}
L.~M{\"u}ller, \emph{{On the Higher Categorical Structure of Topological Defects in Quantum Field Theories}},  \href{https://arxiv.org/abs/2505.04761}{{\ttfamily 2505.04761}}.

\bibitem{Bah:2025oxi}
I.~Bah, E.~Leung and T.~Waddleton, \emph{{On the Physics of Higher Condensation Defects}},  \href{https://arxiv.org/abs/2506.04346}{{\ttfamily 2506.04346}}.

\bibitem{Kim:2025zdy}
S.~Kim, O.~Sela and Z.~Sun, \emph{{Higher structure of non-invertible symmetries from Lagrangian descriptions}}, \href{https://doi.org/10.1007/JHEP05(2026)120}{\emph{JHEP} {\bfseries 05} (2026) 120} [\href{https://arxiv.org/abs/2509.20540}{{\ttfamily 2509.20540}}].

\bibitem{Maldacena:1997re}
J.M.~Maldacena, \emph{{The Large $N$ limit of superconformal field theories and supergravity}}, \href{https://doi.org/10.4310/ATMP.1998.v2.n2.a1}{\emph{Adv. Theor. Math. Phys.} {\bfseries 2} (1998) 231} [\href{https://arxiv.org/abs/hep-th/9711200}{{\ttfamily hep-th/9711200}}].

\bibitem{Witten:1998qj}
E.~Witten, \emph{{Anti de Sitter space and holography}}, \href{https://doi.org/10.4310/ATMP.1998.v2.n2.a2}{\emph{Adv. Theor. Math. Phys.} {\bfseries 2} (1998) 253} [\href{https://arxiv.org/abs/hep-th/9802150}{{\ttfamily hep-th/9802150}}].

\bibitem{Apruzzi:2022rei}
F.~Apruzzi, I.~Bah, F.~Bonetti and S.~Schafer-Nameki, \emph{{Noninvertible Symmetries from Holography and Branes}}, \href{https://doi.org/10.1103/PhysRevLett.130.121601}{\emph{Phys. Rev. Lett.} {\bfseries 130} (2023) 121601} [\href{https://arxiv.org/abs/2208.07373}{{\ttfamily 2208.07373}}].

\bibitem{GarciaEtxebarria:2022vzq}
I.~Garc{\'\i}a~Etxebarria, \emph{{Branes and Non-Invertible Symmetries}}, \href{https://doi.org/10.1002/prop.202200154}{\emph{Fortsch. Phys.} {\bfseries 70} (2022) 2200154} [\href{https://arxiv.org/abs/2208.07508}{{\ttfamily 2208.07508}}].

\bibitem{Heckman:2022muc}
J.J.~Heckman, M.~H{\"u}bner, E.~Torres and H.Y.~Zhang, \emph{{The Branes Behind Generalized Symmetry Operators}}, \href{https://doi.org/10.1002/prop.202200180}{\emph{Fortsch. Phys.} {\bfseries 71} (2023) 2200180} [\href{https://arxiv.org/abs/2209.03343}{{\ttfamily 2209.03343}}].

\bibitem{Heckman:2022xgu}
J.J.~Heckman, M.~Hubner, E.~Torres, X.~Yu and H.Y.~Zhang, \emph{{Top down approach to topological duality defects}}, \href{https://doi.org/10.1103/PhysRevD.108.046015}{\emph{Phys. Rev. D} {\bfseries 108} (2023) 046015} [\href{https://arxiv.org/abs/2212.09743}{{\ttfamily 2212.09743}}].

\bibitem{Etheredge:2023ler}
M.~Etheredge, I.~Garcia~Etxebarria, B.~Heidenreich and S.~Rauch, \emph{{Branes and symmetries for $ \mathcal{N} $ = 3 S-folds}}, \href{https://doi.org/10.1007/JHEP09(2023)005}{\emph{JHEP} {\bfseries 09} (2023) 005} [\href{https://arxiv.org/abs/2302.14068}{{\ttfamily 2302.14068}}].

\bibitem{Franco:2024mxa}
S.~Franco and X.~Yu, \emph{{Generalized symmetries in 2D from string theory: SymTFTs, intrinsic relativeness, and anomalies of non-invertible symmetries}}, \href{https://doi.org/10.1007/JHEP11(2024)004}{\emph{JHEP} {\bfseries 11} (2024) 004} [\href{https://arxiv.org/abs/2404.19761}{{\ttfamily 2404.19761}}].

\bibitem{Gutperle:2024vyp}
M.~Gutperle, Y.-Y.~Li, D.~Rathore and K.~Roumpedakis, \emph{{Non-invertible symmetries in S$_{N}$ orbifold CFTs and holography}}, \href{https://doi.org/10.1007/JHEP09(2024)110}{\emph{JHEP} {\bfseries 09} (2024) 110} [\href{https://arxiv.org/abs/2405.15693}{{\ttfamily 2405.15693}}].

\bibitem{Knighton:2024noc}
B.~Knighton, V.~Sriprachyakul and J.~Vo{\v{s}}mera, \emph{{Topological defects and tensionless holography}}, \href{https://doi.org/10.1007/JHEP07(2025)083}{\emph{JHEP} {\bfseries 07} (2025) 083} [\href{https://arxiv.org/abs/2406.03467}{{\ttfamily 2406.03467}}].

\bibitem{Yu:2023nyn}
X.~Yu, \emph{{Noninvertible symmetries in 2D from type IIB string theory}}, \href{https://doi.org/10.1103/PhysRevD.110.065008}{\emph{Phys. Rev. D} {\bfseries 110} (2024) 065008} [\href{https://arxiv.org/abs/2310.15339}{{\ttfamily 2310.15339}}].

\bibitem{Fernandez-Melgarejo:2024ffg}
J.J.~Fernandez-Melgarejo, G.~Giorgi, D.~Marques and J.A.~Rosabal, \emph{{Noninvertible symmetries in type IIB supergravity}}, \href{https://doi.org/10.1103/PhysRevD.111.066024}{\emph{Phys. Rev. D} {\bfseries 111} (2025) 066024} [\href{https://arxiv.org/abs/2407.09402}{{\ttfamily 2407.09402}}].

\bibitem{Dierigl:2023jdp}
M.~Dierigl, J.J.~Heckman, M.~Montero and E.~Torres, \emph{{R7-branes as charge conjugation operators}}, \href{https://doi.org/10.1103/PhysRevD.109.046004}{\emph{Phys. Rev. D} {\bfseries 109} (2024) 046004} [\href{https://arxiv.org/abs/2305.05689}{{\ttfamily 2305.05689}}].

\bibitem{Bah:2023ymy}
I.~Bah, E.~Leung and T.~Waddleton, \emph{{Non-invertible symmetries, brane dynamics, and tachyon condensation}}, \href{https://doi.org/10.1007/JHEP01(2024)117}{\emph{JHEP} {\bfseries 01} (2024) 117} [\href{https://arxiv.org/abs/2306.15783}{{\ttfamily 2306.15783}}].

\bibitem{Apruzzi:2023uma}
F.~Apruzzi, F.~Bonetti, D.S.W.~Gould and S.~Schafer-Nameki, \emph{{Aspects of categorical symmetries from branes: SymTFTs and generalized charges}}, \href{https://doi.org/10.21468/SciPostPhys.17.1.025}{\emph{SciPost Phys.} {\bfseries 17} (2024) 025} [\href{https://arxiv.org/abs/2306.16405}{{\ttfamily 2306.16405}}].

\bibitem{Cvetic:2023pgm}
M.~Cveti{\v{c}}, J.J.~Heckman, M.~H{\"u}bner and E.~Torres, \emph{{Generalized symmetries, gravity, and the swampland}}, \href{https://doi.org/10.1103/PhysRevD.109.026012}{\emph{Phys. Rev. D} {\bfseries 109} (2024) 026012} [\href{https://arxiv.org/abs/2307.13027}{{\ttfamily 2307.13027}}].

\bibitem{Baume:2023kkf}
F.~Baume, J.J.~Heckman, M.~H{\"u}bner, E.~Torres, A.P.~Turner and X.~Yu, \emph{{SymTrees and Multi-Sector QFTs}}, \href{https://doi.org/10.1103/PhysRevD.109.106013}{\emph{Phys. Rev. D} {\bfseries 109} (2024) 106013} [\href{https://arxiv.org/abs/2310.12980}{{\ttfamily 2310.12980}}].

\bibitem{DelZotto:2024tae}
M.~Del~Zotto, S.N.~Meynet and R.~Moscrop, \emph{{Remarks on geometric engineering, symmetry TFTs and anomalies}}, \href{https://doi.org/10.1007/JHEP07(2024)220}{\emph{JHEP} {\bfseries 07} (2024) 220} [\href{https://arxiv.org/abs/2402.18646}{{\ttfamily 2402.18646}}].

\bibitem{Hu:2024zvz}
S.~Hu, \emph{{Nontrivial bundles and defect operators in n-form gauge theories}}, \href{https://doi.org/10.1007/JHEP12(2024)171}{\emph{JHEP} {\bfseries 12} (2024) 171} [\href{https://arxiv.org/abs/2404.03406}{{\ttfamily 2404.03406}}].

\bibitem{Argurio:2024oym}
R.~Argurio, F.~Benini, M.~Bertolini, G.~Galati and P.~Niro, \emph{{On the symmetry TFT of Yang-Mills-Chern-Simons theory}}, \href{https://doi.org/10.1007/JHEP07(2024)130}{\emph{JHEP} {\bfseries 07} (2024) 130} [\href{https://arxiv.org/abs/2404.06601}{{\ttfamily 2404.06601}}].

\bibitem{Zhang:2024oas}
H.Y.~Zhang, \emph{{K-theoretic Global Symmetry in String-constructed QFT and T-duality}}, \href{https://doi.org/10.1007/JHEP05(2026)118}{\emph{JHEP} {\bfseries 05} (2026) 118} [\href{https://arxiv.org/abs/2404.16097}{{\ttfamily 2404.16097}}].

\bibitem{Braeger:2024jcj}
N.~Braeger, V.~Chakrabhavi, J.J.~Heckman and M.~H{\"u}bner, \emph{{Generalized symmetries of nonsupersymmetric orbifolds}}, \href{https://doi.org/10.1103/PhysRevD.111.066015}{\emph{Phys. Rev. D} {\bfseries 111} (2025) 066015} [\href{https://arxiv.org/abs/2404.17639}{{\ttfamily 2404.17639}}].

\bibitem{Bergman:2024its}
O.~Bergman and F.~Mignosa, \emph{{String theory and the SymTFT of 3d orthosymplectic Chern-Simons theory}}, \href{https://doi.org/10.1007/JHEP04(2025)047}{\emph{JHEP} {\bfseries 04} (2025) 047} [\href{https://arxiv.org/abs/2412.00184}{{\ttfamily 2412.00184}}].

\bibitem{Caldararu:2025eoj}
A.~Caldararu, T.~Pantev, E.~Sharpe, B.~Sung and X.~Yu, \emph{{Noninvertible symmetries in the B model TFT}}, \href{https://doi.org/10.1016/j.geomphys.2025.105653}{\emph{J. Geom. Phys.} {\bfseries 218} (2025) 105653} [\href{https://arxiv.org/abs/2504.02023}{{\ttfamily 2504.02023}}].

\bibitem{Bonetti:2024etn}
F.~Bonetti, M.~Del~Zotto and R.~Minasian, \emph{{SymTFTs and non-invertible symmetries of 6d (2,0) SCFTs of type D from M-theory}}, \href{https://doi.org/10.1007/JHEP02(2025)156}{\emph{JHEP} {\bfseries 02} (2025) 156} [\href{https://arxiv.org/abs/2412.07842}{{\ttfamily 2412.07842}}].

\bibitem{Christensen:2024fiq}
F.B.~Christensen, \emph{{Holography and discrete theta angles for disconnected gauge groups}},  \href{https://arxiv.org/abs/2412.19887}{{\ttfamily 2412.19887}}.

\bibitem{Cvetic:2025kdn}
M.~Cveti{\v{c}}, J.J.~Heckman, M.~H{\"u}bner and C.~Murdia, \emph{{Metric Isometries, Holography, and Continuous Symmetry Operators}}, \href{https://doi.org/10.1103/t68l-n5gg}{\emph{Phys. Rev. D} {\bfseries 112} (2025) 106020} [\href{https://arxiv.org/abs/2501.17911}{{\ttfamily 2501.17911}}].

\bibitem{Cvetic:2023plv}
M.~Cveti{\v{c}}, J.J.~Heckman, M.~H{\"u}bner and E.~Torres, \emph{{Fluxbranes, generalized symmetries, and Verlinde{\textquoteright}s metastable monopole}}, \href{https://doi.org/10.1103/PhysRevD.109.046007}{\emph{Phys. Rev. D} {\bfseries 109} (2024) 046007} [\href{https://arxiv.org/abs/2305.09665}{{\ttfamily 2305.09665}}].

\bibitem{Garcia-Valdecasas:2023mis}
E.~Garc{\'\i}a-Valdecasas, \emph{{Non-invertible symmetries in supergravity}}, \href{https://doi.org/10.1007/JHEP04(2023)102}{\emph{JHEP} {\bfseries 04} (2023) 102} [\href{https://arxiv.org/abs/2301.00777}{{\ttfamily 2301.00777}}].

\bibitem{Najjar:2025htp}
M.~Najjar, \emph{{Modified instanton sum and 4-group structure in 4d $\mathcal{N}=1$$SU(M)$ SYM from holography}},  \href{https://arxiv.org/abs/2503.17108}{{\ttfamily 2503.17108}}.

\bibitem{Bergman:2024aly}
O.~Bergman, E.~Garcia-Valdecasas, F.~Mignosa and D.~Rodriguez-Gomez, \emph{{Non-BPS branes and continuous symmetries}}, \href{https://doi.org/10.1007/JHEP02(2025)066}{\emph{JHEP} {\bfseries 02} (2025) 066} [\href{https://arxiv.org/abs/2407.00773}{{\ttfamily 2407.00773}}].

\bibitem{Waddleton:2024iiv}
T.~Waddleton, \emph{{$U(1)$$R$-Symmetry Topological Operators from Branes in Holography}}, \href{https://doi.org/10.1007/JHEP11(2025)104}{\emph{JHEP} {\bfseries 11} (2025) 104} [\href{https://arxiv.org/abs/2408.14542}{{\ttfamily 2408.14542}}].

\bibitem{Calvo:2025usj}
H.~Calvo, F.~Mignosa and D.~Rodriguez-Gomez, \emph{{R-symmetries, anomalies and non-invertible defects from non-BPS branes}}, \href{https://doi.org/10.1007/JHEP10(2025)107}{\emph{JHEP} {\bfseries 10} (2025) 107} [\href{https://arxiv.org/abs/2506.13859}{{\ttfamily 2506.13859}}].

\bibitem{Calvo:2025kjh}
H.~Calvo, F.~Mignosa and D.~Rodriguez-Gomez, \emph{{Continuous symmetry defects and brane/anti-brane systems}}, \href{https://doi.org/10.1007/JHEP06(2025)196}{\emph{JHEP} {\bfseries 06} (2025) 196} [\href{https://arxiv.org/abs/2503.04892}{{\ttfamily 2503.04892}}].

\bibitem{Bah:2019rgq}
I.~Bah, F.~Bonetti, R.~Minasian and E.~Nardoni, \emph{{Anomalies of QFTs from M-theory and Holography}}, \href{https://doi.org/10.1007/JHEP01(2020)125}{\emph{JHEP} {\bfseries 01} (2020) 125} [\href{https://arxiv.org/abs/1910.04166}{{\ttfamily 1910.04166}}].

\bibitem{Bah:2019vmq}
I.~Bah and F.~Bonetti, \emph{{Anomaly Inflow, Accidental Symmetry, and Spontaneous Symmetry Breaking}}, \href{https://doi.org/10.1007/JHEP01(2020)117}{\emph{JHEP} {\bfseries 01} (2020) 117} [\href{https://arxiv.org/abs/1910.07549}{{\ttfamily 1910.07549}}].

\bibitem{Bah:2020jas}
I.~Bah, F.~Bonetti, R.~Minasian and P.~Weck, \emph{{Anomaly Inflow Methods for SCFT Constructions in Type IIB}}, \href{https://doi.org/10.1007/JHEP02(2021)116}{\emph{JHEP} {\bfseries 02} (2021) 116} [\href{https://arxiv.org/abs/2002.10466}{{\ttfamily 2002.10466}}].

\bibitem{Bah:2020uev}
I.~Bah, F.~Bonetti and R.~Minasian, \emph{{Discrete and higher-form symmetries in SCFTs from wrapped M5-branes}}, \href{https://doi.org/10.1007/JHEP03(2021)196}{\emph{JHEP} {\bfseries 03} (2021) 196} [\href{https://arxiv.org/abs/2007.15003}{{\ttfamily 2007.15003}}].

\bibitem{Bah:2021brs}
I.~Bah, F.~Bonetti, E.~Leung and P.~Weck, \emph{{M5-branes probing flux backgrounds}}, \href{https://doi.org/10.1007/JHEP10(2022)122}{\emph{JHEP} {\bfseries 10} (2022) 122} [\href{https://arxiv.org/abs/2111.01790}{{\ttfamily 2111.01790}}].

\bibitem{Mignosa:2026mgf}
F.~Mignosa and D.~Rodriguez-Gomez, \emph{{Non-Abelian R-symmetry and dielectric branes}},  \href{https://arxiv.org/abs/2601.23217}{{\ttfamily 2601.23217}}.

\bibitem{Bah:2025vfu}
I.~Bah, F.~Bonetti, M.~Chitoto and E.~Leung, \emph{{Non-Abelian Symmetry Operators from Hanging Branes in $AdS_5 \times S^5$}},  \href{https://arxiv.org/abs/2510.19812}{{\ttfamily 2510.19812}}.

\bibitem{Bah:2024ucp}
I.~Bah, P.~Jefferson, K.~Roumpedakis and T.~Waddleton, \emph{{Symmetry Operators and Gravity}}, \href{https://doi.org/10.21468/SciPostPhys.19.4.116}{\emph{SciPost Phys.} {\bfseries 19} (2025) 116} [\href{https://arxiv.org/abs/2411.08858}{{\ttfamily 2411.08858}}].

\bibitem{Balachandran:1977ub}
A.P.~Balachandran, S.~Borchardt and A.~Stern, \emph{{Lagrangian and Hamiltonian Descriptions of Yang-Mills Particles}}, \href{https://doi.org/10.1103/PhysRevD.17.3247}{\emph{Phys. Rev. D} {\bfseries 17} (1978) 3247}.

\bibitem{Alekseev:1988vx}
A.~Alekseev, L.D.~Faddeev and S.L.~Shatashvili, \emph{{Quantization of symplectic orbits of compact Lie groups by means of the functional integral}}, \href{https://doi.org/10.1016/0393-0440(88)90031-9}{\emph{J. Geom. Phys.} {\bfseries 5} (1988) 391}.

\bibitem{Diakonov:1989fc}
D.~Diakonov and V.Y.~Petrov, \emph{{A Formula for the Wilson Loop}}, \href{https://doi.org/10.1016/0370-2693(89)91062-9}{\emph{Phys. Lett. B} {\bfseries 224} (1989) 131}.

\bibitem{Stone:1988fu}
M.~Stone, \emph{{Supersymmetry and the Quantum Mechanics of Spin}}, \href{https://doi.org/10.1016/0550-3213(89)90408-2}{\emph{Nucl. Phys. B} {\bfseries 314} (1989) 557}.

\bibitem{Alvarez:1989zv}
O.~Alvarez, I.M.~Singer and P.~Windey, \emph{{Quantum Mechanics and the Geometry of the Weyl Character Formula}}, \href{https://doi.org/10.1016/0550-3213(90)90278-L}{\emph{Nucl. Phys. B} {\bfseries 337} (1990) 467}.

\bibitem{deligne1999quantum}
P.~Deligne, P.~Etingof, D.S.~Freed, L.C.~Jeffrey, D.~Kazhdan, J.W.~Morgan et~al., \emph{Quantum Fields and Strings: A Course for Mathematicians: Volume 2}, vol.~2, American Mathematical Society, Providence (1999).

\bibitem{Beasley:2009mb}
C.~Beasley, \emph{{Localization for Wilson Loops in Chern-Simons Theory}}, \href{https://doi.org/10.4310/ATMP.2013.v17.n1.a1}{\emph{Adv. Theor. Math. Phys.} {\bfseries 17} (2013) 1} [\href{https://arxiv.org/abs/0911.2687}{{\ttfamily 0911.2687}}].

\bibitem{Tong:2014yla}
D.~Tong and K.~Wong, \emph{{Monopoles and Wilson Lines}}, \href{https://doi.org/10.1007/JHEP06(2014)048}{\emph{JHEP} {\bfseries 06} (2014) 048} [\href{https://arxiv.org/abs/1401.6167}{{\ttfamily 1401.6167}}].

\bibitem{Witten:1998wy}
E.~Witten, \emph{{AdS/CFT correspondence and topological field theory.}}, \href{https://doi.org/10.1088/1126-6708/1998/12/012}{\emph{JHEP} {\bfseries 12} (1998) 012} [\href{https://arxiv.org/abs/hep-th/9812012}{{\ttfamily hep-th/9812012}}].

\bibitem{Belov:2004ht}
D.~Belov and G.W.~Moore, \emph{{Conformal blocks for AdS(5) singletons}},  \href{https://arxiv.org/abs/hep-th/0412167}{{\ttfamily hep-th/0412167}}.

\bibitem{Diaconescu:2003bm}
E.~Diaconescu, G.W.~Moore and D.S.~Freed, \emph{{The M theory three form and E(8) gauge theory}},  \href{https://arxiv.org/abs/hep-th/0312069}{{\ttfamily hep-th/0312069}}.

\bibitem{Moore:2004jv}
G.W.~Moore, \emph{{Anomalies, Gauss laws, and Page charges in M-theory}}, \href{https://doi.org/10.1016/j.crhy.2004.12.005}{\emph{Comptes Rendus Physique} {\bfseries 6} (2005) 251} [\href{https://arxiv.org/abs/hep-th/0409158}{{\ttfamily hep-th/0409158}}].

\bibitem{Belov:2006jd}
D.~Belov and G.W.~Moore, \emph{{Holographic Action for the Self-Dual Field}},  \href{https://arxiv.org/abs/hep-th/0605038}{{\ttfamily hep-th/0605038}}.

\bibitem{Freed:2006yc}
D.S.~Freed, G.W.~Moore and G.~Segal, \emph{{Heisenberg Groups and Noncommutative Fluxes}}, \href{https://doi.org/10.1016/j.aop.2006.07.014}{\emph{Annals Phys.} {\bfseries 322} (2007) 236} [\href{https://arxiv.org/abs/hep-th/0605200}{{\ttfamily hep-th/0605200}}].

\bibitem{Christensen:2025ktc}
F.B.~Christensen, I.~Garc{\'\i}a~Etxebarria and E.~Leung, \emph{{Anomaly-induced vanishing of brane partition functions}}, \href{https://doi.org/10.1007/JHEP02(2026)177}{\emph{JHEP} {\bfseries 02} (2026) 177} [\href{https://arxiv.org/abs/2510.19935}{{\ttfamily 2510.19935}}].

\bibitem{Damia:2022bcd}
J.A.~Damia, R.~Argurio and E.~Garcia-Valdecasas, \emph{{Non-invertible defects in 5d, boundaries and holography}}, \href{https://doi.org/10.21468/SciPostPhys.14.4.067}{\emph{SciPost Phys.} {\bfseries 14} (2023) 067} [\href{https://arxiv.org/abs/2207.02831}{{\ttfamily 2207.02831}}].

\bibitem{Choi:2022fgx}
Y.~Choi, H.T.~Lam and S.-H.~Shao, \emph{{Non-invertible Gauss law and axions}}, \href{https://doi.org/10.1007/JHEP09(2023)067}{\emph{JHEP} {\bfseries 09} (2023) 067} [\href{https://arxiv.org/abs/2212.04499}{{\ttfamily 2212.04499}}].

\bibitem{Bonetti:2022gsl}
F.~Bonetti, R.~Minasian, V.V.~Camell and P.~Weck, \emph{{Consistent truncations from the geometry of sphere bundles}}, \href{https://doi.org/10.1007/JHEP05(2023)156}{\emph{JHEP} {\bfseries 05} (2023) 156} [\href{https://arxiv.org/abs/2212.08068}{{\ttfamily 2212.08068}}].

\bibitem{Cvetic:2000nc}
M.~Cvetic, H.~Lu, C.N.~Pope, A.~Sadrzadeh and T.A.~Tran, \emph{{Consistent SO(6) reduction of type IIB supergravity on S**5}}, \href{https://doi.org/10.1016/S0550-3213(00)00372-2}{\emph{Nucl. Phys. B} {\bfseries 586} (2000) 275} [\href{https://arxiv.org/abs/hep-th/0003103}{{\ttfamily hep-th/0003103}}].

\bibitem{Pilch:2000ue}
K.~Pilch and N.P.~Warner, \emph{{N=2 supersymmetric RG flows and the IIB dilaton}}, \href{https://doi.org/10.1016/S0550-3213(00)00656-8}{\emph{Nucl. Phys. B} {\bfseries 594} (2001) 209} [\href{https://arxiv.org/abs/hep-th/0004063}{{\ttfamily hep-th/0004063}}].

\bibitem{Cassani:2010uw}
D.~Cassani, G.~Dall'Agata and A.F.~Faedo, \emph{{Type IIB supergravity on squashed Sasaki-Einstein manifolds}}, \href{https://doi.org/10.1007/JHEP05(2010)094}{\emph{JHEP} {\bfseries 05} (2010) 094} [\href{https://arxiv.org/abs/1003.4283}{{\ttfamily 1003.4283}}].

\bibitem{Liu:2010sa}
J.T.~Liu, P.~Szepietowski and Z.~Zhao, \emph{{Consistent massive truncations of IIB supergravity on Sasaki-Einstein manifolds}}, \href{https://doi.org/10.1103/PhysRevD.81.124028}{\emph{Phys. Rev. D} {\bfseries 81} (2010) 124028} [\href{https://arxiv.org/abs/1003.5374}{{\ttfamily 1003.5374}}].

\bibitem{Gauntlett:2010vu}
J.P.~Gauntlett and O.~Varela, \emph{{Universal Kaluza-Klein reductions of type IIB to N=4 supergravity in five dimensions}}, \href{https://doi.org/10.1007/JHEP06(2010)081}{\emph{JHEP} {\bfseries 06} (2010) 081} [\href{https://arxiv.org/abs/1003.5642}{{\ttfamily 1003.5642}}].

\bibitem{Baguet:2015sma}
A.~Baguet, O.~Hohm and H.~Samtleben, \emph{{Consistent Type IIB Reductions to Maximal 5D Supergravity}}, \href{https://doi.org/10.1103/PhysRevD.92.065004}{\emph{Phys. Rev. D} {\bfseries 92} (2015) 065004} [\href{https://arxiv.org/abs/1506.01385}{{\ttfamily 1506.01385}}].

\bibitem{Gunaydin:1984qu}
M.~Gunaydin, L.J.~Romans and N.P.~Warner, \emph{{Gauged N=8 Supergravity in Five-Dimensions}}, \href{https://doi.org/10.1016/0370-2693(85)90361-2}{\emph{Phys. Lett. B} {\bfseries 154} (1985) 268}.

\bibitem{Pernici:1985ju}
M.~Pernici, K.~Pilch and P.~van Nieuwenhuizen, \emph{{Gauged N=8 D=5 Supergravity}}, \href{https://doi.org/10.1016/0550-3213(85)90645-5}{\emph{Nucl. Phys. B} {\bfseries 259} (1985) 460}.

\bibitem{Gunaydin:1985cu}
M.~Gunaydin, L.J.~Romans and N.P.~Warner, \emph{{Compact and Noncompact Gauged Supergravity Theories in Five-Dimensions}}, \href{https://doi.org/10.1016/0550-3213(86)90237-3}{\emph{Nucl. Phys. B} {\bfseries 272} (1986) 598}.

\bibitem{Freedman:1998tz}
D.Z.~Freedman, S.D.~Mathur, A.~Matusis and L.~Rastelli, \emph{{Correlation functions in the CFT(d) / AdS(d+1) correspondence}}, \href{https://doi.org/10.1016/S0550-3213(99)00053-X}{\emph{Nucl. Phys. B} {\bfseries 546} (1999) 96} [\href{https://arxiv.org/abs/hep-th/9804058}{{\ttfamily hep-th/9804058}}].

\bibitem{Barnes:2005bw}
E.~Barnes, E.~Gorbatov, K.A.~Intriligator and J.~Wright, \emph{{Current correlators and AdS/CFT geometry}}, \href{https://doi.org/10.1016/j.nuclphysb.2005.10.013}{\emph{Nucl. Phys. B} {\bfseries 732} (2006) 89} [\href{https://arxiv.org/abs/hep-th/0507146}{{\ttfamily hep-th/0507146}}].

\bibitem{Benvenuti:2006xg}
S.~Benvenuti, L.A.~Pando~Zayas and Y.~Tachikawa, \emph{{Triangle anomalies from Einstein manifolds}}, \href{https://doi.org/10.4310/ATMP.2006.v10.n3.a4}{\emph{Adv. Theor. Math. Phys.} {\bfseries 10} (2006) 395} [\href{https://arxiv.org/abs/hep-th/0601054}{{\ttfamily hep-th/0601054}}].

\bibitem{Rey:1998ik}
S.-J.~Rey and J.-T.~Yee, \emph{{Macroscopic strings as heavy quarks in large N gauge theory and anti-de Sitter supergravity}}, \href{https://doi.org/10.1007/s100520100799}{\emph{Eur. Phys. J. C} {\bfseries 22} (2001) 379} [\href{https://arxiv.org/abs/hep-th/9803001}{{\ttfamily hep-th/9803001}}].

\bibitem{McGreevy:2000cw}
J.~McGreevy, L.~Susskind and N.~Toumbas, \emph{{Invasion of the giant gravitons from Anti-de Sitter space}}, \href{https://doi.org/10.1088/1126-6708/2000/06/008}{\emph{JHEP} {\bfseries 06} (2000) 008} [\href{https://arxiv.org/abs/hep-th/0003075}{{\ttfamily hep-th/0003075}}].

\bibitem{Grisaru:2000zn}
M.T.~Grisaru, R.C.~Myers and O.~Tafjord, \emph{{SUSY and goliath}}, \href{https://doi.org/10.1088/1126-6708/2000/08/040}{\emph{JHEP} {\bfseries 08} (2000) 040} [\href{https://arxiv.org/abs/hep-th/0008015}{{\ttfamily hep-th/0008015}}].

\bibitem{Drukker:2005kx}
N.~Drukker and B.~Fiol, \emph{{All-genus calculation of Wilson loops using D-branes}}, \href{https://doi.org/10.1088/1126-6708/2005/02/010}{\emph{JHEP} {\bfseries 02} (2005) 010} [\href{https://arxiv.org/abs/hep-th/0501109}{{\ttfamily hep-th/0501109}}].

\bibitem{Gomis:2006sb}
J.~Gomis and F.~Passerini, \emph{{Holographic Wilson Loops}}, \href{https://doi.org/10.1088/1126-6708/2006/08/074}{\emph{JHEP} {\bfseries 08} (2006) 074} [\href{https://arxiv.org/abs/hep-th/0604007}{{\ttfamily hep-th/0604007}}].

\bibitem{Gomis:2006im}
J.~Gomis and F.~Passerini, \emph{{Wilson Loops as D3-Branes}}, \href{https://doi.org/10.1088/1126-6708/2007/01/097}{\emph{JHEP} {\bfseries 01} (2007) 097} [\href{https://arxiv.org/abs/hep-th/0612022}{{\ttfamily hep-th/0612022}}].

\bibitem{Eyras:1998hn}
E.~Eyras, B.~Janssen and Y.~Lozano, \emph{{Five-branes, K K monopoles and T duality}}, \href{https://doi.org/10.1016/S0550-3213(98)00575-6}{\emph{Nucl. Phys. B} {\bfseries 531} (1998) 275} [\href{https://arxiv.org/abs/hep-th/9806169}{{\ttfamily hep-th/9806169}}].

\bibitem{Hanany:1996ie}
A.~Hanany and E.~Witten, \emph{{Type IIB superstrings, BPS monopoles, and three-dimensional gauge dynamics}}, \href{https://doi.org/10.1016/S0550-3213(97)00157-0}{\emph{Nucl. Phys. B} {\bfseries 492} (1997) 152} [\href{https://arxiv.org/abs/hep-th/9611230}{{\ttfamily hep-th/9611230}}].

\bibitem{Klebanov:1998hh}
I.R.~Klebanov and E.~Witten, \emph{{Superconformal field theory on three-branes at a Calabi-Yau singularity}}, \href{https://doi.org/10.1016/S0550-3213(98)00654-3}{\emph{Nucl. Phys. B} {\bfseries 536} (1998) 199} [\href{https://arxiv.org/abs/hep-th/9807080}{{\ttfamily hep-th/9807080}}].

\bibitem{Cassani:2010na}
D.~Cassani and A.F.~Faedo, \emph{{A Supersymmetric consistent truncation for conifold solutions}}, \href{https://doi.org/10.1016/j.nuclphysb.2010.10.010}{\emph{Nucl. Phys. B} {\bfseries 843} (2011) 455} [\href{https://arxiv.org/abs/1008.0883}{{\ttfamily 1008.0883}}].

\bibitem{Berenstein:2002ke}
D.~Berenstein, C.P.~Herzog and I.R.~Klebanov, \emph{{Baryon spectra and AdS /CFT correspondence}}, \href{https://doi.org/10.1088/1126-6708/2002/06/047}{\emph{JHEP} {\bfseries 06} (2002) 047} [\href{https://arxiv.org/abs/hep-th/0202150}{{\ttfamily hep-th/0202150}}].

\bibitem{Hopkins:2002rd}
M.J.~Hopkins and I.M.~Singer, \emph{{Quadratic functions in geometry, topology, and M theory}}, {\emph{J. Diff. Geom.} {\bfseries 70} (2005) 329} [\href{https://arxiv.org/abs/math/0211216}{{\ttfamily math/0211216}}].

\bibitem{Hsieh:2020jpj}
C.-T.~Hsieh, Y.~Tachikawa and K.~Yonekura, \emph{{Anomaly Inflow and p-Form Gauge Theories}}, \href{https://doi.org/10.1007/s00220-022-04333-w}{\emph{Commun. Math. Phys.} {\bfseries 391} (2022) 495} [\href{https://arxiv.org/abs/2003.11550}{{\ttfamily 2003.11550}}].

\bibitem{Bergman:2001qi}
A.~Bergman and C.P.~Herzog, \emph{{The Volume of some nonspherical horizons and the AdS / CFT correspondence}}, \href{https://doi.org/10.1088/1126-6708/2002/01/030}{\emph{JHEP} {\bfseries 01} (2002) 030} [\href{https://arxiv.org/abs/hep-th/0108020}{{\ttfamily hep-th/0108020}}].

\bibitem{Arean:2004mm}
D.~Arean, D.E.~Crooks and A.V.~Ramallo, \emph{{Supersymmetric probes on the conifold}}, \href{https://doi.org/10.1088/1126-6708/2004/11/035}{\emph{JHEP} {\bfseries 11} (2004) 035} [\href{https://arxiv.org/abs/hep-th/0408210}{{\ttfamily hep-th/0408210}}].

\bibitem{Buchel:2006gb}
A.~Buchel and J.T.~Liu, \emph{{Gauged supergravity from type IIB string theory on Y**p,q manifolds}}, \href{https://doi.org/10.1016/j.nuclphysb.2007.03.001}{\emph{Nucl. Phys. B} {\bfseries 771} (2007) 93} [\href{https://arxiv.org/abs/hep-th/0608002}{{\ttfamily hep-th/0608002}}].

\bibitem{Gauntlett:2007ma}
J.P.~Gauntlett and O.~Varela, \emph{{Consistent Kaluza-Klein reductions for general supersymmetric AdS solutions}}, \href{https://doi.org/10.1103/PhysRevD.76.126007}{\emph{Phys. Rev. D} {\bfseries 76} (2007) 126007} [\href{https://arxiv.org/abs/0707.2315}{{\ttfamily 0707.2315}}].

\bibitem{friedrich1989einstein}
T.~Friedrich and I.~Kath, \emph{Einstein manifolds of dimension five with small first eigenvalue of the dirac operator}, {\emph{Journal of differential geometry} {\bfseries 29} (1989) 263}.

\bibitem{tian1987kahler}
G.~Tian, \emph{On k{\"a}hler-einstein metrics on certain k{\"a}hler manifolds with c 1 (m)> 0}, {\emph{Inventiones mathematicae} {\bfseries 89} (1987) 225}.

\bibitem{tianyau1987kahler}
G.~Tian and S.-T.~Yau, \emph{K{\"a}hler-einstein metrics on complex surfaces with c 1> 0}, {\emph{Communications in mathematical physics} {\bfseries 112} (1987) 175}.

\bibitem{Gauntlett:2004yd}
J.P.~Gauntlett, D.~Martelli, J.~Sparks and D.~Waldram, \emph{{Sasaki-Einstein metrics on S**2 x S**3}}, \href{https://doi.org/10.4310/ATMP.2004.v8.n4.a3}{\emph{Adv. Theor. Math. Phys.} {\bfseries 8} (2004) 711} [\href{https://arxiv.org/abs/hep-th/0403002}{{\ttfamily hep-th/0403002}}].

\bibitem{Herzog:2003dj}
C.P.~Herzog and J.~Walcher, \emph{{Dibaryons from exceptional collections}}, \href{https://doi.org/10.1088/1126-6708/2003/09/060}{\emph{JHEP} {\bfseries 09} (2003) 060} [\href{https://arxiv.org/abs/hep-th/0306298}{{\ttfamily hep-th/0306298}}].

\bibitem{Benvenuti:2004wx}
S.~Benvenuti, A.~Hanany and P.~Kazakopoulos, \emph{{The Toric phases of the Y**p,q quivers}}, \href{https://doi.org/10.1088/1126-6708/2005/07/021}{\emph{JHEP} {\bfseries 07} (2005) 021} [\href{https://arxiv.org/abs/hep-th/0412279}{{\ttfamily hep-th/0412279}}].

\bibitem{Yi:1999hd}
P.~Yi, \emph{{Membranes from five-branes and fundamental strings from Dp branes}}, \href{https://doi.org/10.1016/S0550-3213(99)00191-1}{\emph{Nucl. Phys. B} {\bfseries 550} (1999) 214} [\href{https://arxiv.org/abs/hep-th/9901159}{{\ttfamily hep-th/9901159}}].

\bibitem{Houart:1999bi}
L.~Houart and Y.~Lozano, \emph{{Type II branes from brane - anti-brane in M theory}}, \href{https://doi.org/10.1016/S0550-3213(00)00065-1}{\emph{Nucl. Phys. B} {\bfseries 575} (2000) 195} [\href{https://arxiv.org/abs/hep-th/9910266}{{\ttfamily hep-th/9910266}}].

\bibitem{Houart:2000vm}
L.~Houart and Y.~Lozano, \emph{{Brane descent relations in M theory}}, \href{https://doi.org/10.1016/S0370-2693(00)00317-8}{\emph{Phys. Lett. B} {\bfseries 479} (2000) 299} [\href{https://arxiv.org/abs/hep-th/0001170}{{\ttfamily hep-th/0001170}}].

\bibitem{Intriligator:2000pk}
K.A.~Intriligator, M.~Kleban and J.~Kumar, \emph{{Comments on unstable branes}}, \href{https://doi.org/10.1088/1126-6708/2001/02/023}{\emph{JHEP} {\bfseries 02} (2001) 023} [\href{https://arxiv.org/abs/hep-th/0101010}{{\ttfamily hep-th/0101010}}].

\bibitem{Loaiza-Brito:2001yer}
O.~Loaiza-Brito and A.M.~Uranga, \emph{{The Fate of the type I nonBPS D7-brane}}, \href{https://doi.org/10.1016/S0550-3213(01)00505-3}{\emph{Nucl. Phys. B} {\bfseries 619} (2001) 211} [\href{https://arxiv.org/abs/hep-th/0104173}{{\ttfamily hep-th/0104173}}].

\bibitem{Gaberdiel:2001ed}
M.R.~Gaberdiel and S.~Schafer-Nameki, \emph{{NonBPS D branes and M theory}}, \href{https://doi.org/10.1088/1126-6708/2001/09/028}{\emph{JHEP} {\bfseries 09} (2001) 028} [\href{https://arxiv.org/abs/hep-th/0108202}{{\ttfamily hep-th/0108202}}].

\bibitem{deWit:1983vq}
B.~de~Wit and H.~Nicolai, \emph{{On the Relation Between $d=4$ and $d=11$ Supergravity}}, \href{https://doi.org/10.1016/0550-3213(84)90387-0}{\emph{Nucl. Phys. B} {\bfseries 243} (1984) 91}.

\bibitem{deWit:1984nz}
B.~de~Wit, H.~Nicolai and N.P.~Warner, \emph{{The Embedding of Gauged $N=8$ Supergravity Into $d=11$ Supergravity}}, \href{https://doi.org/10.1016/0550-3213(85)90128-2}{\emph{Nucl. Phys. B} {\bfseries 255} (1985) 29}.

\bibitem{deWit:1985iy}
B.~de~Wit and H.~Nicolai, \emph{{Hidden Symmetry in $d=11$ Supergravity}}, \href{https://doi.org/10.1016/0370-2693(85)91030-5}{\emph{Phys. Lett. B} {\bfseries 155} (1985) 47}.

\bibitem{deWit:1986mz}
B.~de~Wit and H.~Nicolai, \emph{{$d=11$ Supergravity With Local SU(8) Invariance}}, \href{https://doi.org/10.1016/0550-3213(86)90290-7}{\emph{Nucl. Phys. B} {\bfseries 274} (1986) 363}.

\bibitem{deWit:1986oxb}
B.~de~Wit and H.~Nicolai, \emph{{The Consistency of the S**7 Truncation in D=11 Supergravity}}, \href{https://doi.org/10.1016/0550-3213(87)90253-7}{\emph{Nucl. Phys. B} {\bfseries 281} (1987) 211}.

\bibitem{Nicolai:2011cy}
H.~Nicolai and K.~Pilch, \emph{{Consistent Truncation of d = 11 Supergravity on AdS$_4 \times S^7$}}, \href{https://doi.org/10.1007/JHEP03(2012)099}{\emph{JHEP} {\bfseries 03} (2012) 099} [\href{https://arxiv.org/abs/1112.6131}{{\ttfamily 1112.6131}}].

\bibitem{deWit:2013ija}
B.~de~Wit and H.~Nicolai, \emph{{Deformations of gauged SO(8) supergravity and supergravity in eleven dimensions}}, \href{https://doi.org/10.1007/JHEP05(2013)077}{\emph{JHEP} {\bfseries 05} (2013) 077} [\href{https://arxiv.org/abs/1302.6219}{{\ttfamily 1302.6219}}].

\bibitem{Hohm:2013pua}
O.~Hohm and H.~Samtleben, \emph{{Exceptional Form of D=11 Supergravity}}, \href{https://doi.org/10.1103/PhysRevLett.111.231601}{\emph{Phys. Rev. Lett.} {\bfseries 111} (2013) 231601} [\href{https://arxiv.org/abs/1308.1673}{{\ttfamily 1308.1673}}].

\bibitem{Godazgar:2013dma}
H.~Godazgar, M.~Godazgar and H.~Nicolai, \emph{{Generalised geometry from the ground up}}, \href{https://doi.org/10.1007/JHEP02(2014)075}{\emph{JHEP} {\bfseries 02} (2014) 075} [\href{https://arxiv.org/abs/1307.8295}{{\ttfamily 1307.8295}}].

\bibitem{Hohm:2014qga}
O.~Hohm and H.~Samtleben, \emph{{Consistent Kaluza-Klein Truncations via Exceptional Field Theory}}, \href{https://doi.org/10.1007/JHEP01(2015)131}{\emph{JHEP} {\bfseries 01} (2015) 131} [\href{https://arxiv.org/abs/1410.8145}{{\ttfamily 1410.8145}}].

\bibitem{Godazgar:2015qia}
H.~Godazgar, M.~Godazgar, O.~Kr{\"u}ger and H.~Nicolai, \emph{{Consistent 4-form fluxes for maximal supergravity}}, \href{https://doi.org/10.1007/JHEP10(2015)169}{\emph{JHEP} {\bfseries 10} (2015) 169} [\href{https://arxiv.org/abs/1507.07684}{{\ttfamily 1507.07684}}].

\bibitem{Varela:2015ywx}
O.~Varela, \emph{{Complete $D=11$ embedding of SO(8) supergravity}}, \href{https://doi.org/10.1103/PhysRevD.97.045010}{\emph{Phys. Rev. D} {\bfseries 97} (2018) 045010} [\href{https://arxiv.org/abs/1512.04943}{{\ttfamily 1512.04943}}].

\bibitem{deWit:1981sst}
B.~de~Wit and H.~Nicolai, \emph{{N=8 Supergravity with Local SO(8) x SU(8) Invariance}}, \href{https://doi.org/10.1016/0370-2693(82)91194-7}{\emph{Phys. Lett. B} {\bfseries 108} (1982) 285}.

\bibitem{deWit:1982bul}
B.~de~Wit and H.~Nicolai, \emph{{N=8 Supergravity}}, \href{https://doi.org/10.1016/0550-3213(82)90120-1}{\emph{Nucl. Phys. B} {\bfseries 208} (1982) 323}.

\bibitem{Duff:1995wd}
M.J.~Duff, J.T.~Liu and R.~Minasian, \emph{{Eleven-dimensional origin of string/string duality: a one-loop test*}}, \href{https://doi.org/10.1201/9781482268737-17}{\emph{Nucl. Phys. B} {\bfseries 452} (1995) 261} [\href{https://arxiv.org/abs/hep-th/9506126}{{\ttfamily hep-th/9506126}}].

\bibitem{Witten:1996md}
E.~Witten, \emph{{On flux quantization in $M$-theory and the effective action}}, \href{https://doi.org/10.1016/S0393-0440(96)00042-3}{\emph{J. Geom. Phys.} {\bfseries 22} (1997) 1} [\href{https://arxiv.org/abs/hep-th/9609122}{{\ttfamily hep-th/9609122}}].

\bibitem{Aharony:1999ti}
O.~Aharony, S.S.~Gubser, J.M.~Maldacena, H.~Ooguri and Y.~Oz, \emph{{Large N field theories, string theory and gravity}}, \href{https://doi.org/10.1016/S0370-1573(99)00083-6}{\emph{Phys. Rept.} {\bfseries 323} (2000) 183} [\href{https://arxiv.org/abs/hep-th/9905111}{{\ttfamily hep-th/9905111}}].

\bibitem{Drukker:2008zx}
N.~Drukker, J.~Plefka and D.~Young, \emph{{Wilson loops in 3-dimensional N=6 supersymmetric Chern-Simons Theory and their string theory duals}}, \href{https://doi.org/10.1088/1126-6708/2008/11/019}{\emph{JHEP} {\bfseries 11} (2008) 019} [\href{https://arxiv.org/abs/0809.2787}{{\ttfamily 0809.2787}}].

\bibitem{Giombi:2023vzu}
S.~Giombi and A.A.~Tseytlin, \emph{{Wilson Loops at Large N and the Quantum M2-Brane}}, \href{https://doi.org/10.1103/PhysRevLett.130.201601}{\emph{Phys. Rev. Lett.} {\bfseries 130} (2023) 201601} [\href{https://arxiv.org/abs/2303.15207}{{\ttfamily 2303.15207}}].

\bibitem{Drukker:2023bip}
N.~Drukker and O.~Shahpo, \emph{{Vortex loop operators and quantum M2-branes}}, \href{https://doi.org/10.21468/SciPostPhys.17.1.016}{\emph{SciPost Phys.} {\bfseries 17} (2024) 016} [\href{https://arxiv.org/abs/2312.17091}{{\ttfamily 2312.17091}}].

\bibitem{Zhang:2025yex}
X.-Y.~Zhang, Y.~Jiang and J.-B.~Wu, \emph{{Holographic Wilson loop one-point functions in ABJM theory}}, \href{https://doi.org/10.1007/JHEP11(2025)008}{\emph{JHEP} {\bfseries 11} (2025) 008} [\href{https://arxiv.org/abs/2508.00281}{{\ttfamily 2508.00281}}].

\bibitem{Bandos:1997ui}
I.A.~Bandos, K.~Lechner, A.~Nurmagambetov, P.~Pasti, D.P.~Sorokin and M.~Tonin, \emph{{Covariant action for the superfive-brane of M theory}}, \href{https://doi.org/10.1103/PhysRevLett.78.4332}{\emph{Phys. Rev. Lett.} {\bfseries 78} (1997) 4332} [\href{https://arxiv.org/abs/hep-th/9701149}{{\ttfamily hep-th/9701149}}].

\bibitem{Witten:1996hc}
E.~Witten, \emph{{Five-brane effective action in $M$-theory.}}, \href{https://doi.org/10.1016/S0393-0440(97)80160-X}{\emph{J. Geom. Phys.} {\bfseries 22} (1997) 103} [\href{https://arxiv.org/abs/hep-th/9610234}{{\ttfamily hep-th/9610234}}].

\bibitem{Witten:1995em}
E.~Witten, \emph{{Five-branes and M-theory on an orbifold}}, \href{https://doi.org/10.1201/9781482268737-19}{\emph{Nucl. Phys. B} {\bfseries 463} (1996) 383} [\href{https://arxiv.org/abs/hep-th/9512219}{{\ttfamily hep-th/9512219}}].

\bibitem{Monnier:2013rpa}
S.~Monnier, \emph{{Global gravitational anomaly cancellation for five-branes}}, \href{https://doi.org/10.4310/ATMP.2015.v19.n3.a5}{\emph{Adv. Theor. Math. Phys.} {\bfseries 19} (2015) 701} [\href{https://arxiv.org/abs/1310.2250}{{\ttfamily 1310.2250}}].

\bibitem{Townsend:1995af}
P.K.~Townsend, \emph{{D-branes from M-branes}}, \href{https://doi.org/10.1201/9781482268737-14}{\emph{Phys. Lett. B} {\bfseries 373} (1996) 68} [\href{https://arxiv.org/abs/hep-th/9512062}{{\ttfamily hep-th/9512062}}].

\bibitem{Bergshoeff:1998ef}
E.~Bergshoeff, E.~Eyras and Y.~Lozano, \emph{{The Massive Kaluza-Klein monopole}}, \href{https://doi.org/10.1016/S0370-2693(98)00501-2}{\emph{Phys. Lett. B} {\bfseries 430} (1998) 77} [\href{https://arxiv.org/abs/hep-th/9802199}{{\ttfamily hep-th/9802199}}].

\bibitem{Gava:1997jt}
E.~Gava, K.S.~Narain and M.H.~Sarmadi, \emph{{On the bound states of p-branes and (p+2)-branes}}, \href{https://doi.org/10.1016/S0550-3213(97)00508-7}{\emph{Nucl. Phys. B} {\bfseries 504} (1997) 214} [\href{https://arxiv.org/abs/hep-th/9704006}{{\ttfamily hep-th/9704006}}].

\bibitem{Sato:2001su}
M.~Sato, \emph{{BPS bound states of D6-branes and lower dimensional D-branes}}, \href{https://doi.org/10.1142/S0217751X01005304}{\emph{Int. J. Mod. Phys. A} {\bfseries 16} (2001) 4069} [\href{https://arxiv.org/abs/hep-th/0101226}{{\ttfamily hep-th/0101226}}].

\bibitem{Larsson:2001wt}
H.~Larsson, \emph{{A Note on half supersymmetric bound states in M-theory and type IIA}}, \href{https://doi.org/10.1088/0264-9381/19/10/311}{\emph{Class. Quant. Grav.} {\bfseries 19} (2002) 2689} [\href{https://arxiv.org/abs/hep-th/0105083}{{\ttfamily hep-th/0105083}}].

\bibitem{Khoze:2003yk}
V.V.~Khoze, \emph{{From branes to branes}}, \href{https://doi.org/10.1016/j.nuclphysb.2004.01.030}{\emph{Nucl. Phys. B} {\bfseries 682} (2004) 217} [\href{https://arxiv.org/abs/hep-th/0311065}{{\ttfamily hep-th/0311065}}].

\bibitem{Callan:1997kz}
C.G.~Callan and J.M.~Maldacena, \emph{{Brane death and dynamics from the Born-Infeld action}}, \href{https://doi.org/10.1016/S0550-3213(97)00700-1}{\emph{Nucl. Phys. B} {\bfseries 513} (1998) 198} [\href{https://arxiv.org/abs/hep-th/9708147}{{\ttfamily hep-th/9708147}}].

\bibitem{Howe:1997ue}
P.S.~Howe, N.D.~Lambert and P.C.~West, \emph{{The Selfdual string soliton}}, \href{https://doi.org/10.1016/S0550-3213(97)00750-5}{\emph{Nucl. Phys. B} {\bfseries 515} (1998) 203} [\href{https://arxiv.org/abs/hep-th/9709014}{{\ttfamily hep-th/9709014}}].

\bibitem{Duff:1986hr}
M.J.~Duff, B.E.W.~Nilsson and C.N.~Pope, \emph{{Kaluza-Klein Supergravity}}, \href{https://doi.org/10.1016/0370-1573(86)90163-8}{\emph{Phys. Rept.} {\bfseries 130} (1986) 1}.

\bibitem{Garousi_2005}
M.R.~Garousi, \emph{D-brane anti-d-brane effective action and brane interaction in open string channel}, \href{https://doi.org/10.1088/1126-6708/2005/01/029}{\emph{Journal of High Energy Physics} {\bfseries 2005} (2005) 029–029}.

\bibitem{Garousi_2008}
M.R.~Garousi and E.~Hatefi, \emph{On wess–zumino terms of brane–antibrane systems}, \href{https://doi.org/10.1016/j.nuclphysb.2008.01.024}{\emph{Nuclear Physics B} {\bfseries 800} (2008) 502–516}.

\bibitem{Sen_1998}
A.~Sen, \emph{Tachyon condensation on the brane antibrane system}, \href{https://doi.org/10.1088/1126-6708/1998/08/012}{\emph{Journal of High Energy Physics} {\bfseries 1998} (1998) 012–012}.

\bibitem{Hashimoto_2003}
K.~Hashimoto and S.~Nagaoka, \emph{Recombination of intersecting d-branes by local tachyon condensation}, \href{https://doi.org/10.1088/1126-6708/2003/06/034}{\emph{Journal of High Energy Physics} {\bfseries 2003} (2003) 034–034}.

\bibitem{Heckman:2024oot}
J.J.~Heckman, M.~H{\"u}bner and C.~Murdia, \emph{{On the holographic dual of a topological symmetry operator}}, \href{https://doi.org/10.1103/PhysRevD.110.046007}{\emph{Phys. Rev. D} {\bfseries 110} (2024) 046007} [\href{https://arxiv.org/abs/2401.09538}{{\ttfamily 2401.09538}}].

\bibitem{Ji:2019jhk}
W.~Ji and X.-G.~Wen, \emph{{Categorical symmetry and noninvertible anomaly in symmetry-breaking and topological phase transitions}}, \href{https://doi.org/10.1103/PhysRevResearch.2.033417}{\emph{Phys. Rev. Res.} {\bfseries 2} (2020) 033417} [\href{https://arxiv.org/abs/1912.13492}{{\ttfamily 1912.13492}}].

\bibitem{Gaiotto:2020iye}
D.~Gaiotto and J.~Kulp, \emph{{Orbifold groupoids}}, \href{https://doi.org/10.1007/JHEP02(2021)132}{\emph{JHEP} {\bfseries 02} (2021) 132} [\href{https://arxiv.org/abs/2008.05960}{{\ttfamily 2008.05960}}].

\bibitem{Apruzzi:2021nmk}
F.~Apruzzi, F.~Bonetti, I.~Garc{\'\i}a~Etxebarria, S.S.~Hosseini and S.~Schafer-Nameki, \emph{{Symmetry TFTs from String Theory}}, \href{https://doi.org/10.1007/s00220-023-04737-2}{\emph{Commun. Math. Phys.} {\bfseries 402} (2023) 895} [\href{https://arxiv.org/abs/2112.02092}{{\ttfamily 2112.02092}}].

\bibitem{Freed:2022qnc}
D.S.~Freed, G.W.~Moore and C.~Teleman, \emph{{Topological symmetry in quantum field theory}},  \href{https://arxiv.org/abs/2209.07471}{{\ttfamily 2209.07471}}.

\bibitem{Brennan:2024fgj}
T.D.~Brennan and Z.~Sun, \emph{{A SymTFT for continuous symmetries}}, \href{https://doi.org/10.1007/JHEP12(2024)100}{\emph{JHEP} {\bfseries 12} (2024) 100} [\href{https://arxiv.org/abs/2401.06128}{{\ttfamily 2401.06128}}].

\bibitem{Antinucci:2024zjp}
A.~Antinucci and F.~Benini, \emph{{Anomalies and gauging of U(1) symmetries}}, \href{https://doi.org/10.1103/PhysRevB.111.024110}{\emph{Phys. Rev. B} {\bfseries 111} (2025) 024110} [\href{https://arxiv.org/abs/2401.10165}{{\ttfamily 2401.10165}}].

\bibitem{Bonetti:2024cjk}
F.~Bonetti, M.~Del~Zotto and R.~Minasian, \emph{{SymTFTs for continuous non-Abelian symmetries}}, \href{https://doi.org/10.1016/j.physletb.2025.140010}{\emph{Phys. Lett. B} {\bfseries 871} (2025) 140010} [\href{https://arxiv.org/abs/2402.12347}{{\ttfamily 2402.12347}}].

\bibitem{Antinucci:2024bcm}
A.~Antinucci, F.~Benini and G.~Rizi, \emph{{Holographic Duals of Symmetry Broken Phases}}, \href{https://doi.org/10.1002/prop.202400172}{\emph{Fortsch. Phys.} {\bfseries 72} (2024) 2400172} [\href{https://arxiv.org/abs/2408.01418}{{\ttfamily 2408.01418}}].

\bibitem{Arbalestrier:2025poq}
A.~Arbalestrier, R.~Argurio and L.~Tizzano, \emph{{U(1) gauging, continuous TQFTs, and higher symmetry structures}}, \href{https://doi.org/10.21468/SciPostPhys.19.2.032}{\emph{SciPost Phys.} {\bfseries 19} (2025) 032} [\href{https://arxiv.org/abs/2502.12997}{{\ttfamily 2502.12997}}].

\bibitem{Bonetti:2025dvm}
F.~Bonetti, M.~Del~Zotto and R.~Minasian, \emph{{SymTFT for Continuous Symmetries: Non-linear Realizations and Spontaneous Breaking}}, \href{https://doi.org/10.1007/JHEP05(2026)048}{\emph{JHEP} {\bfseries 05} (2026) 048} [\href{https://arxiv.org/abs/2509.10343}{{\ttfamily 2509.10343}}].

\bibitem{Jia:2025jmn}
Q.~Jia, R.~Luo, J.~Tian, Y.-N.~Wang and Y.~Zhang, \emph{{Symmetry Topological Field Theory for Flavor Symmetry}},  \href{https://arxiv.org/abs/2503.04546}{{\ttfamily 2503.04546}}.

\bibitem{kirillov2004lectures}
A.A.~Kirillov, \emph{Lectures on the Orbit Method}, vol.~64, American Mathematical Soc., Providence (2004).

\bibitem{Hull:1990ms}
C.M.~Hull and B.J.~Spence, \emph{{The Geometry of the gauged sigma model with Wess-Zumino term}}, \href{https://doi.org/10.1016/0550-3213(91)90342-U}{\emph{Nucl. Phys. B} {\bfseries 353} (1991) 379}.

\bibitem{Bergshoeff:1997gy}
E.~Bergshoeff, B.~Janssen and T.~Ortin, \emph{{Kaluza-Klein monopoles and gauged sigma models}}, \href{https://doi.org/10.1016/S0370-2693(97)00946-5}{\emph{Phys. Lett. B} {\bfseries 410} (1997) 131} [\href{https://arxiv.org/abs/hep-th/9706117}{{\ttfamily hep-th/9706117}}].

\bibitem{genEuler}
D.K.~Hoffman, R.C.~Raffenetti and K.~Ruedenberg, \emph{Generalization of euler angles to n-dimensional orthogonal matrices}, \href{https://doi.org/10.1063/1.1666011}{\emph{J. Math. Phys. (N.Y.)} {\bfseries 13} (1972) 528}.

\bibitem{2007arXiv0709.3615L}
M.~{Libine}, \emph{{Lecture Notes on Equivariant Cohomology}}, \href{https://doi.org/10.48550/arXiv.0709.3615}{\emph{arXiv e-prints} (2007) arXiv:0709.3615} [\href{https://arxiv.org/abs/0709.3615}{{\ttfamily 0709.3615}}].

\bibitem{Serre:1961abc}
J.-P.~Serre, \emph{Homologie singulière des espaces fibrès}, {\emph{Annals of Mathematics} {\bfseries 54} (1951) 425}.

\bibitem{1996alg.geom..9018E}
D.~{Edidin} and W.~{Graham}, \emph{{Equivariant intersection theory}}, \href{https://doi.org/10.48550/arXiv.alg-geom/9609018}{\emph{arXiv e-prints} (1996) alg} [\href{https://arxiv.org/abs/alg-geom/9609018}{{\ttfamily alg-geom/9609018}}].

\bibitem{1999math......8172G}
W.~{Graham}, \emph{{Positivity in equivariant Schubert calculus}}, \href{https://doi.org/10.48550/arXiv.math/9908172}{\emph{arXiv Mathematics e-prints} (1999) math/9908172} [\href{https://arxiv.org/abs/math/9908172}{{\ttfamily math/9908172}}].

\bibitem{2018arXiv180903226B}
A.~{Bingham}, M.~{Bilen Can} and Y.~{Ozan}, \emph{{A Filtration on Equivariant Borel-Moore Homology}}, \href{https://doi.org/10.48550/arXiv.1809.03226}{\emph{arXiv e-prints} (2018) arXiv:1809.03226} [\href{https://arxiv.org/abs/1809.03226}{{\ttfamily 1809.03226}}].

\bibitem{10.1307/mmj/1030132709}
M.~Brion, \emph{{Poincaré duality and equivariant (co)homology.}}, \href{https://doi.org/10.1307/mmj/1030132709}{\emph{Michigan Mathematical Journal} {\bfseries 48} (2000) 77 }.

\bibitem{Anderson_Fulton_2023}
D.~Anderson and W.~Fulton, \emph{Equivariant homology},  in \emph{Equivariant Cohomology in Algebraic Geometry}, Cambridge Studies in Advanced Mathematics, p.~311–325, Cambridge University Press (2023).

\bibitem{Minasian:1997mm}
R.~Minasian and G.W.~Moore, \emph{{K theory and Ramond-Ramond charge}}, \href{https://doi.org/10.1088/1126-6708/1997/11/002}{\emph{JHEP} {\bfseries 11} (1997) 002} [\href{https://arxiv.org/abs/hep-th/9710230}{{\ttfamily hep-th/9710230}}].

\bibitem{Witten:1998cd}
E.~Witten, \emph{{D-branes and K-theory}}, \href{https://doi.org/10.1088/1126-6708/1998/12/019}{\emph{JHEP} {\bfseries 12} (1998) 019} [\href{https://arxiv.org/abs/hep-th/9810188}{{\ttfamily hep-th/9810188}}].

\bibitem{Maldacena:2001xj}
J.M.~Maldacena, G.W.~Moore and N.~Seiberg, \emph{{D-brane instantons and K theory charges}}, \href{https://doi.org/10.1088/1126-6708/2001/11/062}{\emph{JHEP} {\bfseries 11} (2001) 062} [\href{https://arxiv.org/abs/hep-th/0108100}{{\ttfamily hep-th/0108100}}].

\bibitem{AtiyahHirzebruch1961}
M.F.~Atiyah and F.~Hirzebruch, \emph{Vector bundles and homogeneous spaces},  in \emph{Differential Geometry}, vol.~III of \emph{Proceedings of Symposia in Pure Mathematics}, pp.~7--38, American Mathematical Society (1961).

\bibitem{10.4310/jdg/1214428815}
M.F.~Atiyah and G.B.~Segal, \emph{{Equivariant $K$-theory and completion}}, \href{https://doi.org/10.4310/jdg/1214428815}{\emph{Journal of Differential Geometry} {\bfseries 3} (1969) 1 }.

\bibitem{V_A_Iskovskih_1977}
V.A.~Iskovskih, \emph{Fano 3-folds. i}, \href{https://doi.org/10.1070/IM1977v011n03ABEH001733}{\emph{Mathematics of the USSR-Izvestiya} {\bfseries 11} (1977) 485}.

\bibitem{Mori_Mukai_1981}
S.~Mori and S.~Mukai, \emph{Classification of fano 3-folds with b2$\ge$2}, \href{https://doi.org/10.1007/BF01170131}{\emph{manuscripta mathematica} {\bfseries 36} (1981) 147}.

\bibitem{Mori_Mukai_2003}
S.~Mori and S.~Mukai, \emph{Classification of fano 3-folds with b2$\ge$2}, \href{https://doi.org/10.1007/s00229-002-0336-2}{\emph{manuscripta mathematica} {\bfseries 110} (2003) 407}.

\bibitem{Kriz:2004xj}
I.~Kriz and H.~Sati, \emph{{M theory, type IIA superstrings, and elliptic cohomology}}, \href{https://doi.org/10.4310/ATMP.2004.v8.n2.a3}{\emph{Adv. Theor. Math. Phys.} {\bfseries 8} (2004) 345} [\href{https://arxiv.org/abs/hep-th/0404013}{{\ttfamily hep-th/0404013}}].

\bibitem{Sati:2010ss}
H.~Sati, \emph{{Geometric and topological structures related to M-branes}}, {\emph{Proc. Symp. Pure Math.} {\bfseries 81} (2010) 181} [\href{https://arxiv.org/abs/1001.5020}{{\ttfamily 1001.5020}}].

\end{thebibliography}\endgroup


\end{document}